\documentclass[a4paper,12pt]{article}
\pdfoutput=1
\usepackage{epsfig,amsmath,amsfonts,amsthm}
\usepackage[figuresright]{rotating}
\usepackage[left=28mm,right=20mm,top=30mm,bottom=20mm]{geometry}
\usepackage{subfig}
\usepackage{graphicx}
\usepackage{rotating}
\usepackage{booktabs}
\usepackage{moreverb}

\DeclareGraphicsExtensions{.pdf,.png,.jpg}
\usepackage{natbib}
\numberwithin{equation}{section}

\def\e{\hbox{E}}

\def\var{\hbox{Var}}

\def\Bin{\hbox{Bin}}

\providecommand{\keywords}[1]{\textbf{{Keywords}} #1}
\usepackage[nokeyprefix]{refstyle}
\usepackage{varioref}
\usepackage{xr}
\usepackage{hyperref}

\begin{document}

\title{Simulations for the $Q$ statistic with constant and inverse variance weights for binary effect measures}

\author{Elena Kulinskaya and  David C. Hoaglin }

\date{\today}

\maketitle

\begin{abstract}
Cochran's $Q$ statistic is routinely used for testing heterogeneity in meta-analysis. Its expected value (under an incorrect null distribution) is part of several popular estimators of the between-study variance, $\tau^2$. Those applications generally do not account for the studies' use of estimated variances in the inverse-variance weights that define $Q$ (more explicitly, $Q_{IV}$). Importantly, those weights make approximating the distribution of $Q_{IV}$ rather complicated.

 As an alternative, we are investigating a $Q$ statistic,  $Q_F$, whose constant weights use only the studies' arm-level sample sizes. For  log-odds-ratio, log-relative-risk, and risk difference as the measure of effect, these simulations study approximations to the distributions of $Q_F$ and $Q_{IV}$, as the basis for tests of heterogeneity.

We present the results in  132 Figures, 153 pages in total.

\end{abstract}


\section{Introduction}

When the individual studies in a meta-analysis report binary outcomes in the treatment and control arms, the most common measure of effect is the odds ratio (OR) or its log (LOR).   The LOR is popular in medical research, but some substantive arguments favor the relative risk or risk ratio (RR). Popular measures of effect also include the risk difference (RD).

In studying estimation of the overall effect in random-effects meta-analyses of the mean difference (MD),  the standardized mean difference (SMD), and LOR, we found that SSW, a weighted mean whose constant weights involve only the studies' arm-level sample sizes, performed well, avoiding shortcomings associated with estimators that use inverse-variance weights based on estimated variances (\cite{BHK2018SMD, BHK2020LOR}).

We also previously studied $Q_F$, a version of Cochran's $Q$ statistic \citep{cochran1954combination} for assessment of heterogeneity  that uses those constant weights. That work produced favorable results for the mean difference (\cite{Qfixed}) and the standardized mean difference (\cite{QfixedSMD}). Here we investigate $Q_F$ for LOR, the log-relative-risk (LRR), and RD.

Simulation of the actual distribution of $Q$ for LOR, RR, and RD enables us to study the accuracy of approximations for its null distribution ($\tau^2 = 0$) and  the empirical level when $\tau^2 = 0$.

User-friendly R programs implementing all methods  are available in  \cite{Kulinskaya_Hoaglin_2022}.


\section{Study-level estimation of log-odds-ratio, log-relative-risk, and risk difference} \label{sec:EffectLOR}

Consider $K$ studies that used a particular individual-level binary outcome.
Study $i$ ($i = 1, \ldots, K$) reports $X_{iT}$ and $X_{iC}$, the numbers of events in the $n_{iT}$ subjects in the Treatment arm and the $n_{iC}$ subjects in the Control arm. It is customary to treat $X_{iT}$ and $X_{iC}$ as independent binomial variables:
\begin{equation} \label{eq:binomialXs}
X_{iT}\sim {\Bin}(n_{iT},p_{iT})\qquad \text{and}\qquad X_{iC}\sim {\Bin}(n_{iC},p_{iC}).
\end{equation}
The log-odds-ratio for Study $i$ is
\begin{equation} \label{eq:psi}
\theta_{i} = \log_{e} \left(\frac{p_{iT}(1 - p_{iC})} {p_{iC}(1 - p_{iT})}\right) \qquad \text{estimated by} \qquad
\check\theta_{i} = \log_{e} \left(\frac{\check p_{iT}(1 - \check p_{iC})} {\check p_{iC}(1 - \check p_{iT})}\right),
\end{equation}
where $\check{p}_{ij}$ is an estimate of $p_{ij}$.

As inputs, a two-stage meta-analysis uses estimates of the $\theta_i$ ($\hat{\theta}_i$) and estimates of their variances ($\hat{v}_i^2$). It is helpful to have an unbiased estimator of $\theta$.
The use of  $\hat p = (X + 1/2)/(n + 1)$ eliminates $O(1/n)$ bias and provides the least biased estimator of LOR (\cite{G-P-T-1985}).
We use $\hat p$ when estimating LOR in $Q_F$.

The (conditional, given the $p_{ij}$ and $n_{ij}$) asymptotic variance of $\hat{\theta}_i$, derived by the delta method, is
\begin{equation} \label{eq:sigma}
v_{i}^2 = {\var}(\hat{\theta}_{i}) = \frac{1} {n_{iT} {p}_{iT} (1 - {p}_{iT})} + \frac{1}{n_{iC} {p}_{iC} (1 - {p}_{iC})},
\end{equation}
estimated by substituting $\hat{p}_{ij}$ for $p_{ij}$ and $n_{ij} + 1$ for $n_{ij}$. This estimator of the variance is unbiased in large samples, but it overestimates the variance for small sample sizes (\cite{G-P-T-1985}).



The log-relative-risk for Study $i$ is
\begin{equation} \label{eq:rho}
\rho_{i} = \log_{e}(p_{iT}) - \log_{e}(p_{iC}) \quad \text{estimated by} \quad
\hat\rho_{i} = \log_{e}(\check{p}_{iT}) - \log_{e}(\check{p}_{iC}),
\end{equation}
where $\check{p} = (X + 1/2) / (n + 1/2)$ provides  an unbiased (to order $O(n^{-2})$) estimate of $\log(p)$  \citep{P-G-T-1986}.  An unbiased (to $O(n^{-3})$)  estimate of the variance of $\hat{\rho}$ \citep{P-G-T-1986} is
\begin{equation} \label{eq:Var_rho}
\widehat\var(\hat \rho) = \frac{1}{X_T + 1/2} - \frac{1}{n_T + 1/2} + \frac{1}{X_C + 1/2} - \frac{1}{n_C + 1/2},
\end{equation}
and \cite{P-G-T-1986} also give approximate conditional higher moments for $\log(\hat p)$.

The risk difference for Study $i$ is
\begin{equation} \label{eq:Delta}
\Delta_{i} = p_{iT} - p_{iC} \quad \text{estimated by} \quad
\hat\Delta_{i} = \tilde p_{iT} - \tilde p_{iC}.
\end{equation}
Its variance is $\var(\hat\Delta_i) = p_{iT} (1 - p_{iT}) / n_{iT} + p_{iC} (1 - p_{iC}) / n_{iC}$, estimated by substituting the maximum-likelihood (ML) estimator  $\tilde p=X/n$ for $p$.

Our simulations yield an exact calculation of conditional central moments    for all three effect measures, similar to the implementation of \cite{kulinskaya2015accurate} for LOR.

\section{Random-effects model and the $Q$ statistic} \label{sec:REMandQ}

We consider a generic random-effects model (REM): For Study $i$ ($i = 1,\ldots,K$)  the estimate of the effect is $\hat\theta_i \sim G(\theta_i, v_i^2)$, where the effect-measure-specific distribution $G$ has mean $\theta_i$ and variance $v_i^2$, and $\theta_i \sim N(\theta, \tau^2)$.
Thus, the $\hat{\theta}_i$ are unbiased estimates of the true conditional effects $\theta_i$, and the $v_i^2 = \var(\hat{\theta}_i | \theta_i)$ are the true conditional variances.

Cochran's $Q$ statistic is a weighted sum of the squared deviations of the estimated effects $\hat\theta_i$ from their weighted mean $\bar\theta_w = \sum w_i\hat\theta_i / \sum w_i$:
\begin{equation} \label{Q}
Q=\sum w_i (\hat{\theta}_i - \bar{\theta}_w)^2.
\end{equation}
In \cite{cochran1954combination} $w_i$ is the reciprocal of the \textit{estimated} variance of $\hat{\theta}_i$, hence the notation $Q_{IV}$. In meta-analysis those $w_i$ come from the fixed-effect model.
In what follows, we examine $Q_F$, discussed by \cite{dersimonian2007random} and further studied by \cite{Qfixed}, in which the $w_i$ are arbitrary positive constants. A particular choice of the fixed weights, which we usually specify in $Q_F$,  is $w_i = \tilde{n}_i = n_{iC} n_{iT} / n_i$, the effective sample size in Study $i$ ($n_i = n_{iC} + n_{iT}$).

Define $W = \sum w_i$,  $q_i = w_i / W$, and $\Theta_i = \hat\theta_i - \theta$.  In this notation, and expanding $\bar\theta_w$, Equation (\ref{Q}) can be written  as
\begin{equation} \label{Q1}
Q = W \left[ \sum q_i (1 - q_i) \Theta_i^2 - \sum_{i \not = j} q_i q_j \Theta_i \Theta_j \right].
\end{equation}
We distinguish between the conditional distribution of $Q$ (given the $\theta_i$) and the unconditional distribution, and the corresponding moments of $\Theta_i$. For instance, the conditional second moment of $\Theta_i$ is $M_{2i}^c = v_i^2$, and the unconditional second moment  is  $M_{2i} = \e(\Theta_i^2) = \var(\hat{\theta}_i) = \e(v_i^2) + \tau^2$.

Under the above REM, it is straightforward to obtain the first moment of $Q_F$ as
\begin{equation} \label{M1Q}
\e(Q_F) = W \left[ \sum q_i (1 - q_i) \var(\Theta_i) \right] = W \left[ \sum q_i (1 - q_i) (\e(v_i^2) + \tau^2) \right].
\end{equation}
This expression is similar to Equation (4) in \cite{dersimonian2007random};  they use the conditional variance  $v_i^2$ instead of its unconditional mean $\e(v_i^2)$.

\cite{Qfixed} also provide expressions for the second and third moments of $Q_F$, but these moments require higher moments of $\Theta$, up to the fourth and the sixth moments, respectively.

In fixed-intercept models (i.e., when the $p_{iC}$ are fixed), assuming also homogeneity of effects ($\tau^2 = 0$), the unconditional and conditional moments of each binary effect measure coincide. Therefore, the conditional moments of $Q$  are sufficient to obtain a moment-based approximation to the distribution of $Q_F$ under homogeneity.


\section{Approximations to the distributions of $Q_F$ and $Q_{IV}$} \label{sec:Approx}

For meta-analysis of mean differences, \cite{Qfixed} considered the distribution of $Q_F$, a quadratic form in normal variables, which has the form $Q = \Theta^{T}A\Theta$ for a symmetric matrix $A$ of rank $K - 1$. Because, for MD,  the vector $\Theta$ has a multivariate normal distribution, $N(\mu,\Sigma)$, the distribution of $Q_F$ can be evaluated by the algorithm of \cite{Farebrother1984} (after determining the eigenvalues of $A \Sigma$ and some other inputs). 
 In practice (as in our simulations), it is necessary to plug in estimated variances. The resulting approximation is quite accurate for MD.  \cite{Qfixed} also considered a two-moment approximation and a three-moment approximation. 

For the binary effect measures, $Q_F$ is a quadratic form in asymptotically normal variables.  The Farebrother algorithm may provide a satisfactory approximation for larger sample sizes, though it may not behave well for small $n$. To apply it, we again plug in  estimated variances.   We investigate the quality of that approximation, which we denote by F SSW, and the two-moment  approximation (2M SSW), which is based on the gamma distribution.

For each of these approximations, we investigate two approaches to estimation of the $p_{iT}$ to plug into the calculation of the second and fourth central moments of the effect measure. The \lq\lq na\"{i}ve"  approach estimates $p_{iT}$ from $X_{iT}$ and $n_{iT}$. For the \lq\lq model-based" approach, we observe that each of LOR, LRR, and RD has the form $\eta = h(p_T) - h(p_C)$, which facilitates calculation of conditional moments of $\hat{\eta}$ from the moments of $h(p)$. We obtain estimated moments from the relation $\widehat{h(p_{iT})} = \widehat{h(p_{iC})} + \bar{\eta}$ for a fixed-weights mean effect $\bar{\eta}$. Thus, we study four new approximations to the null distribution of $Q_F$: F SSW na\"{i}ve, F SSW model, 2M SSW na\"{i}ve, and  2M SSW model.

For RD and LOR, \cite{kulinskaya2011moments} and \cite{kulinskaya2015accurate}, respectively,  provided an improved approximation to the null distribution of $Q_{IV}$ based on a two-moment gamma approximation; we also study these approximations, denoted by KDB and KD. \cite{biggerstaff2008exact} used the Farebrother approximation to the distribution of a quadratic form in normal variables as the ``exact'' distribution of $Q_{IV}$.  \cite{Jacksen2014metaregr} extended this approach to a $Q$ with  arbitrary weights in a meta-regression setting. 
When $\tau^2 = 0$, the \cite{biggerstaff2008exact} approximation to the distribution of $Q_{IV}$ is the $\chi^2_{K - 1}$  distribution; we denote this approximation by ChiSq.

\section{Simulation design and the outcome measures } \label{simul}

\subsection{Simulation design}

Our simulation design follows that described in \cite{BHK2020LOR}. Briefly, we varied five parameters: the overall true effect ($\theta$, $\rho$, or $\Delta$), the between-studies variance ($\tau^2$), the number of studies ($K$), the studies' total sample size ($n$ or $\bar{n}$, the average of the $n_i$), and the probability in the control arm ($p_{iC}$). Table~\ref{tab:design} gives the details.

For LOR the values of $\theta$ (0, 0.1, 0.5, 1, 1.5, and 2) aim to represent the range containing most values encountered in practice. LOR is a symmetric effect measure, so positive values suffice.
However, for LRR  we considered both negative and positive values of $\rho$, from $-1.5$ to $1.5$ in steps of $0.5$.  For RD, for comparative purposes, we used the same pairs $(p_{iC}, p_{iT})$ as for LRR.  

The values of $\tau^2$ (0(0.1)1) systematically cover a reasonable range. For RD and LRR, we generated only the null distribution of $Q$ ($\tau^2 = 0$).

The numbers of studies ($K$ = 5, 10, and 30) reflect the sizes of many meta-analyses and have yielded valuable insights in previous work.

In practice, many studies' total sample sizes fall in the ranges covered by our choices ($n$ = 20, 40, 100, and 250 when all studies have the same $n$, and $\bar{n}$ = 30, 60, 100, and 160 when sample sizes vary among studies). The choices of sample sizes corresponding to the values of $\bar{n}$ follow a suggestion of \cite{sanches-2000}, who constructed the studies' sample sizes to have skewness 1.464, which they regarded as typical in behavioral and health sciences. For $K = 5$, Table~\ref{tab:design} lists the sets of five sample sizes. The simulations for $K = 10$ and $K = 30$ used each set of unequal sample sizes twice and six times, respectively.

The values of $p_{iC}$ were .1, .2, and .5.

We kept the proportion of observations in the control arm ($f$) at $1/2$. Many studies allocate subjects equally to the two groups ($f = 1/2$), and rough equality holds more widely.

We generated the true effect sizes for LOR, LRR and RD  from a normal distribution:  $\theta_{i} \sim N(\theta, \tau^2)$.
The  values of $p_{iC}$ and true effect ($\theta_i$,  $\rho_i$  or $\Delta_i$) defined the probabilities $p_{iT}$, and the counts $X_{iC}$ and $X_{iT}$ were generated from the respective binomial distributions.

We used a total of $10,000$ repetitions for each combination of parameters. 
We discarded \lq\lq double-zero" or \lq\lq double-$n$" studies and reduced the observed  value of $K$ accordingly. Then we discarded repetitions with $K<3$ and used the resulting repetitions for analysis.

R statistical software \citep{rrr} was used for simulations. 



\begin{table}[ht]
	\caption{ \label{tab:design} \emph{Values of parameters in the simulations}}
	\begin{footnotesize}
		\begin{center}
			\begin{tabular}
				{|l|l|l|}
				\hline
				Parameter & Equal study sizes & Unequal study sizes \\
                                	& & \\
				\hline
				$K$ (number of studies) & 5, 10, 30 & 5, 10, 30 \\
				\hline
				$n$ or $\bar{n}$ (average (individual) study size ---  & 20, 40, 100, 250 & 30 (12,16,18,20,84), \\
				total of the two arms) &  & 60 (24,32,36,40,168), \\
				For  $K = 10$ and $K = 30$, the same set of unequal & & 100 (64,72,76,80,208), \\
				 study sizes is used twice or six times, respectively. & &160 (124,132,136,140,268) \\
                			\hline
				$f$ (proportion of observations in the control arm) & 1/2 &1/2  \\
				\hline
                			$p_{iC}$ (probability in the control arm) & .1, .2, .5 & .1, .2, .5 \\
                			\hline
				$\theta$ (true value of LOR) & 0, 0.1, 0.5, 1, 1.5, 2 &  0, 0.1, 0.5, 1, 1.5, 2 \\
                			$\tau^{2}$ (variance of random effects for LOR) & 0(0.1)1& 0(0.1)1 \\
                			\hline
                			$\rho$ (fixed value of LRR) & & \\
                			For $p_{iC} = .1$ or $.2$  &  $-0.5$, 0, 0.5, 1, 1.5 &  $-0.5$, 0, 0.5, 1, 1.5  \\
                			For $p_{iC} = .5$   & $-1.5$, $-1$, $-0.5$, 0, 0.5 &  $-1.5$, $-1$, $-0.5$, 0, 0.5  \\
                			\hline
                			$p_T$ (fixed probability in the treatment arm) && \\
                			for RD (and for RR),	& & \\
                			when $p_{iC} = .1$ & $.06, .10, .16, .27, .44$ & $.06, .10, .16, .27, .44$\\
                			when $p_{iC} = .2$ & $.12, .20, .33, .54, .90$ & $.12, .20, .33, .54, .90$\\
                			when $p_{iC} = .5$ & $.12, .18, .30, .50, .82$ & $.12, .18, .30, .50, .82$\\
                			\hline
			\end{tabular}
		\end{center}
	\end{footnotesize}
\end{table}

\subsection{Evaluating the goodness of fit of competing approximations to the null distribution of $Q_F$ and $Q_{IV}$}

Under the null hypothesis the p-values of a parametric test, obtained from the (continuous) distribution function of the test statistic, are uniformly distributed on $[0,\;1]$. Our simulations produce information on the accuracy of an approximation, $\hat{F}$, for the distribution function of $Q$. From the value of $Q$ in each of $M$ iterations, we calculate the p-value, $\tilde{p} = 1 - \hat{F}(Q)$. For selected values of the upper tail area $p$ ($p$ = .001, .0025, .005, .01, .025, .05, .1, .25, .5 and the complementary values .75, \ldots, .999), the results of the $M$ iterations yield $\hat{p}(\hat{F},p) = \#(\tilde{p} < p)/M$, which estimates $P(1 - \hat{F}(Q) < p)$, the actual level of the approximate test based on $Q$, at nominal level $p$. Conveniently, this approach does not require the true distribution function of $Q$, which is generally not available in closed form.

We can examine these results by plotting $\hat{p}(\hat{F},p)$ versus $p$, a type of probability--probability (P--P) plot \citep{Wilk_1968_BMTA_1}. To focus on the difference, we flatten the P--P plot by plotting the error, $\hat{p}(\hat{F},p) - p$ versus $p$. The importance of a given error varies with $p$ (e.g., the error cannot be more negative than $-p$), so a further step (not taken here) would plot the relative error, $(\hat{p}(\hat{F},p) - p) / p$ versus $p$.  Because of the use of the values of $\tilde{p}$ in assessing heterogeneity, we judge the performance of the approximations by their errors when $p$ is in the usual range, say from .01 to .1.

The flattened P--P plots in Appendices A, D, and F show the departures of the approximate distributions from the empirical distribution of $Q$ over the whole range $[0,\;1]$. To show the performance of various tests of heterogeneity, Appendices B, E, and G also plot achieved empirical levels of the corresponding approximations at the nominal .05 level versus $\theta$, $\rho$, and $\Delta$, respectively.
Empirical power of various tests of heterogeneity of LOR at the nominal .05 level is plotted vs $\tau^2$ in Appendix C (equal sample sizes only).

\section{Summary of simulation results}

In the summaries that follow, we aim to provide a general indication of the performance of the various approximations and tests. We give greater weight to behavior at values of $p$ in the usual range and, sometimes, little weight to behavior in the lower tail (i.e., large values of $p$). Thus, necessarily, our summaries do not capture complexity present in some areas.

Although the summaries do not separate them, the 2M SSW and F SSW approximations pertain to the distribution of $Q_F$, whereas ChiSq, MD, and MDB pertain to the distribution of $Q_{IV}$.

\subsection{Approximations to the distributions of $Q_F$ and $Q_{IV}$ for LOR (Appendices A, B  and C)}

The plots in Appendix A  show that none of the six approximations has smaller error than any of the others. ChiSq, however, often has the largest error (in magnitude), at values of $p$ that matter in practice. Also, situations with $K = 30$ and small $n$ (or $\bar{n}$) are especially challenging for all six approximations.

When $p_{iC} = .1$ and $\theta\leq 0.5$, 2M SSW na\"{i}ve has error closest to 0 when $n = 20$, KD is closest when $n = 40$, and 2M SSW model and F SSW model (whose traces nearly coincide) are the best approximation when $n \geq 100$. However, when  $\theta \geq 1$, ChiSq is always the worst, and the performance of the other approximations varies with $n$ and $\theta$. For example, F SSW na\"{i}ve is no better than ChiSq when $n \leq 40$, is slightly better when $n = 100$, and has very little error when $n = 250$; KD has substantial negative error when $n = 20$, is inferior to 2M SSW na\"{i}ve when $n = 40$, and has small positive error (increasing with $K$) when $n = 100$. As a single choice when $\theta \geq 1$, 2M SSW na\"{i}ve seems satisfactory.

When $p_{iC} = .2$, 2M SSW na\"{i}ve works well for $n \leq 40$, but F SSW na\"{i}ve is best when $n \geq 100$.

When $p_{iC} = .5$, F SSW na\"{i}ve is the best approximation for $n \leq 100$ when $\theta \leq 1$. For  $\theta \geq 1.5$,  FM SSW model works better than F SSW na\"{i}ve for $n<100$.  Results are similar for equal and unequal sample sizes.

In the above summary, the dependence on $p_{iC}$ and $\theta$ shows the challenge of choosing an approximation for the null distribution of $Q$ for LOR. We can readily exclude ChiSq, which never fits the distribution of $Q_{IV}$ well. Otherwise, the best approach is unclear.

The plots in Appendix B show that, for small sample sizes, the estimated level of 2M SSW na\"{i}ve usually comes closest to the nominal .05 (the level can be lower or higher by about .02 when $\theta$ is small).
For large sample sizes, when LOR is approximately normal, F SSW na\"{i}ve works well. Levels close to .05 are achieved when $n \geq 100$ for $p_{iC} = .2$ and only at $n = 250$ for $p_{iC} = .1$ and $p_{iC} = .5$. KD is also a good choice when $n \geq 100$ (except for $p_{iC} = .1$ and $K = 30$).

The plots in Appendix C show that the empirical power is rather low for $n\leq 40$. It increases in $n$,  $K$, $\theta$ and $p_{iC}$. The ordering of the actual levels for all tests  mostly defines the ordering of their  power at all $\tau^2$ values. The only exception is the sometimes crossing power curves of ChiSq and F SSW na\"{i}ve.

ChiSq has the lowest empirical power, and F SSW na\"{i}ve is the second worst; this is probably due to their too low actual levels. Both have extremely low power when $n \leq 40$. KD also has very low power for $n=20$, but its power improves, starting at $n = 40$, and it often has the highest power when $n \geq 100$; however, its levels are also too high then.

Overall, when $p_{iC}=.1$, power is reasonable when $n \geq 100$ and $K = 30$ or when $n = 250$ for smaller $K$. For $p_{iC}\geq .2$, the power improves from $n\geq 40$, and all methods have similar power when $n \geq 100$.

\subsection{ Approximations to the distributions of $Q_F$ and $Q_{IV}$ for LRR (Appendices D and E)}

The plots in Appendix D show that, when $p_{iC} = 0.1$, four of the five approximations are completely unsatisfactory for very small sample sizes. 2M SSW na\"{i}ve is better than the others for $p$ between .01 and .1, but far from usable. For $n \geq 100$, F SSW model, 2M SSW model, and 2M SSW na\"{i}ve usually provide a reasonable fit when $\rho \leq 1$, and 2M SSW na\"{i}ve is better than the other two for larger $\rho$. Overall, the quality of all approximations deteriorates as $K$ increases.

When $p_{iC} = 0.2$, 2M SSW na\"{i}ve is best for $n = 20$, and both 2M approximations work well for $n = 40$. For $n \geq 100$,  F SSW na\"{i}ve is the best when $\rho \leq 1$, but ChiSq is the best when $\rho = 1.5$.

When $p_{iC} = .5$, 2M SSW na\"{i}ve  is a good choice for small $n$ when $\rho \leq 0$, but F SSW model appears to fit better when $\rho > 0$ or $n \geq 100$.  Uncharacteristically, ChiSq is also  not a bad choice when $p_{iC} = .5$ and $\rho > 0$.

For  heterogeneity testing (Appendix E) we recommend 2M SSW na\"{i}ve for $n < 100$ and F SSW na\"{i}ve for $n \geq 100$. The choice of approximation for $n=100$ depends on the value of $p_{iC}$: 2M SSW na\"{i}ve when $p_{iC} = .1$, F SSW na\"{i}ve when $p_{iC} = .2$, and F SSW model when $p_{iC} = .5$.

\subsection{ Approximations to the distributions of $Q_F$ and $Q_{IV}$ for RD (Appendices F and G)}

Only KDB had convergence issues for very small sample sizes combined with small probabilities. The worst convergence, only 35.3\%, occurred for $p_{iC} = .1$, $\Delta = -0.04$ and $K=5$. For $n=40$, the same configuration resulted in 83.8\% convergence. The only other problematic configuration was $p_{iC} = .2$, $\Delta = -0.08$, $n=20$  and $K=5$, with a convergence rate of 86.1\%.

The plots in Appendix F show that the 2M SSW model and F SSW model approximations fit well when $p_{iC} = .1$, starting from $n=20$. When $p_{iC} = .2$, 2M SSW model and F SSW model work well for small sample sizes, and so does KDB unless $\Delta \geq 0.7$, making $p_{iT} \leq .1$. In that case, both SSW  na\"{i}ve  approximations work well. When $p_{iC} = .5$, 2M SSW model and F SSW model work well for $n \geq 20$. All approximations fit reasonably well for $n \geq 100$.

For heterogeneity testing (Appendix G) we recommend the F SSW model approximation to the distribution of $Q_F$; it provides very good results for very small sample sizes, $n = 20$ or $\bar{n} = 30$.

\clearpage
\bibliography{Qfixed_LOR.bib}

\section*{Appendices}
\begin{itemize}
\item Appendix A: Plots of error in the level of the test for heterogeneity of LOR for six approximations for the null distribution of $Q$
\item Appendix B: Empirical level at $\alpha = .05$, vs $\theta$, of the test for heterogeneity of LOR ($\tau^2 = 0$ versus $\tau^2 > 0$) based on approximations for the null distribution of $Q$
\item Appendix C: Empirical power at nominal level  $\alpha = .05$, vs $\tau^2$, of the test for heterogeneity of LOR ($\tau^2 = 0$ versus $\tau^2 > 0$) based on approximations for the null distribution of $Q$
\item Appendix D:  Plots of error in the level of the test for heterogeneity of LRR for five approximations for the null distribution of $Q$
\item Appendix E: Empirical level at $\alpha = .05$, vs $\rho$, of the test for heterogeneity of LRR ($\tau^2 = 0$ versus $\tau^2 > 0$) based on approximations for the null distribution of $Q$
\item Appendix F: Plots of error in the level of the test for heterogeneity of RD for six approximations for the null distribution of $Q$
\item Appendix G: Empirical level at $\alpha = .05$, vs $\Delta$, of the test for heterogeneity of RD ($\tau^2 = 0$ versus $\tau^2 > 0$) based on approximations for the null distribution of $Q$
\end{itemize}

\clearpage

\setcounter{figure}{0}
\setcounter{section}{0}
\renewcommand{\thefigure}{A.\arabic{figure}}

\section*{Appendix A: Plots of error in the level of the test for heterogeneity of LOR for six approximations for the null distribution of $Q$}

Each figure corresponds to a value of the probability of an event in the Control arm $p_{iC}$  (= .1, .2, .5), a value of the overall log-odds-ratio $\theta$ (= 0, 0.1, 0.5, 1, 1.5, 2), and a choice of equal or unequal sample sizes ($n$ or $bar{n}$). \\
The fraction of each study's sample size in the Control arm  $f$ is held constant at 0.5.

For each combination of a value of $n$ (= 20, 40, 100, 250) or $\bar{n}$ (= 30, 60, 100, 160) and a value of $K$ (= 5, 10, 30), a panel plots, versus the nominal upper tail area ( = .001, .0025, .005, .01, .025, .05, .1, .25, .5 and the complementary values .75, \ldots, .999), the difference between the achieved level and the nominal level for six approximations to the null distribution of Q: \\
\begin{itemize}
\item ChiSq (Chi-square approximation with $K-1$ df, inverse-variance weights)
\item KD (Kulinskaya-Dollinger (2015) approximation, inverse-variance weights)
\item 2M SSW na\"{i}ve (Two-moment gamma approximation, na\"{i}ve estimation of $p_{iT}$ from $X_{iT}$ and $n_{iT}$, effective-sample-size weights)
\item 2M SSW model (Two-moment gamma approximation, model-based estimation of $p_{iT}$, effective-sample-size weights)
\item F SSW  na\"{i}ve (Farebrother approximation, na\"{i}ve estimation of $p_{iT}$ from $X_{iT}$ and $n_{iT}$, effective-sample-size weights)
\item F SSW model (Farebrother approximation, model-based estimation of $p_{iT}$, effective-sample-size weights)
\end{itemize}

\clearpage

\begin{figure}[ht]
	\centering
	\includegraphics[scale=0.33]{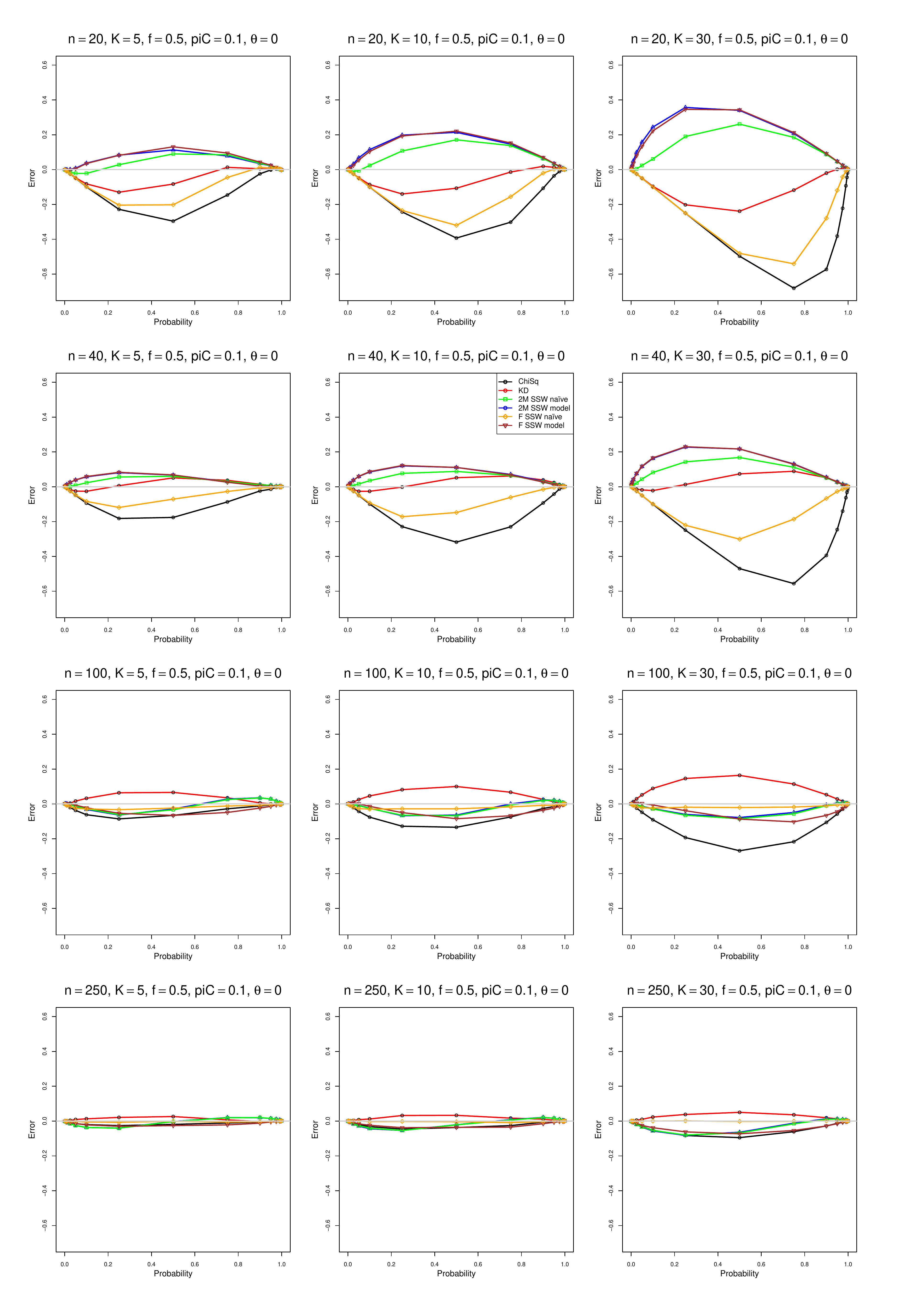}
	\caption{Plots of error in the level of the test for heterogeneity of LOR for six approximations for the null distribution of $Q$, $p_{iC} = .1$, $f = .5$, and $\theta = 0$, equal sample sizes}
	\label{PPplot_piC_01theta=0_LOR_equal_sample_sizes}
\end{figure}

\begin{figure}[ht]
	\centering
	\includegraphics[scale=0.33]{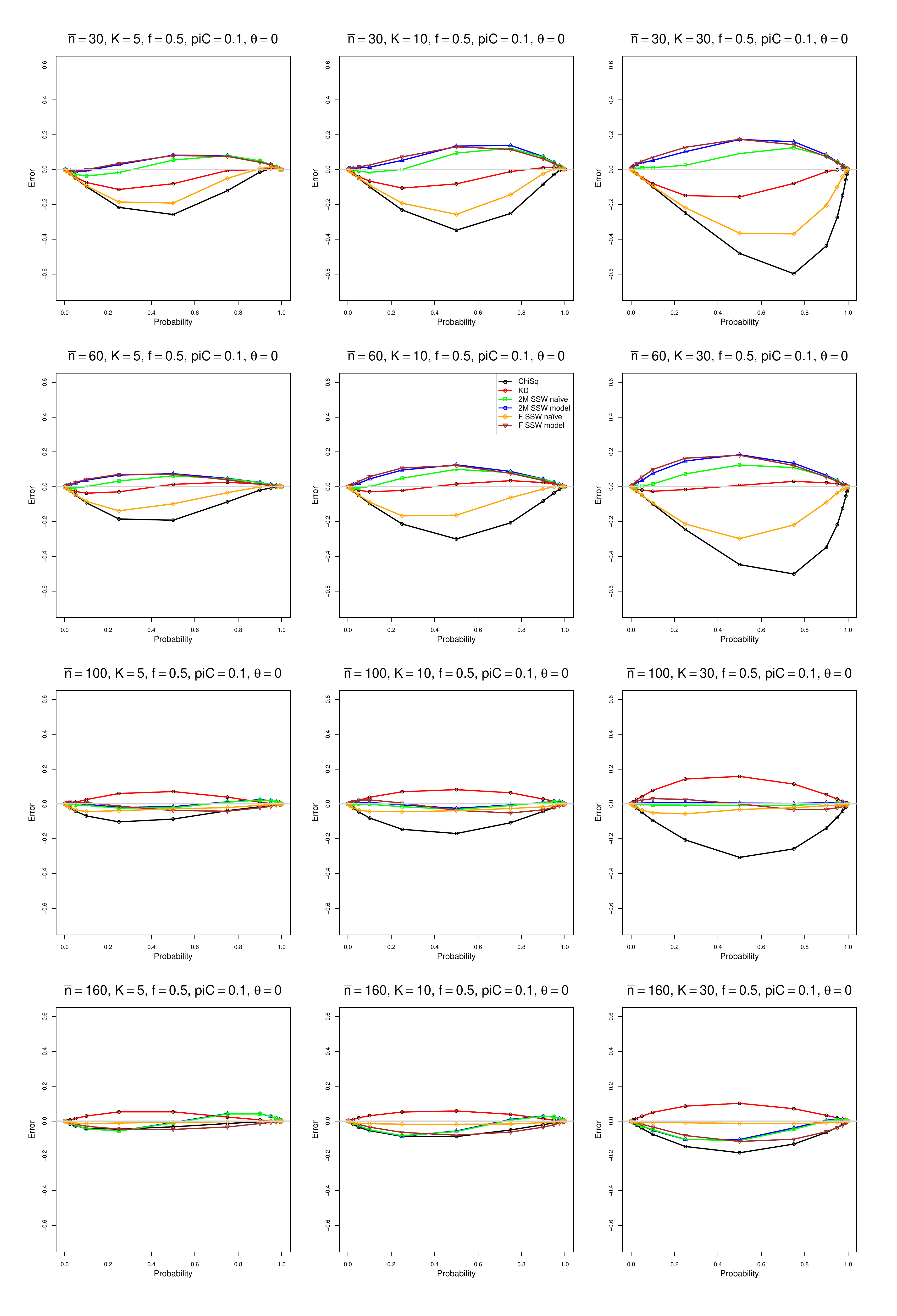}
	\caption{Plots of error in the level of the test for heterogeneity of LOR for six approximations for the null distribution of $Q$, $p_{iC} = .1$, $f = .5$, and $\theta = 0$, unequal sample sizes}
	\label{PPplot_piC_01theta=0_LOR_unequal_sample_sizes}
\end{figure}

\begin{figure}[ht]
	\centering
	\includegraphics[scale=0.33]{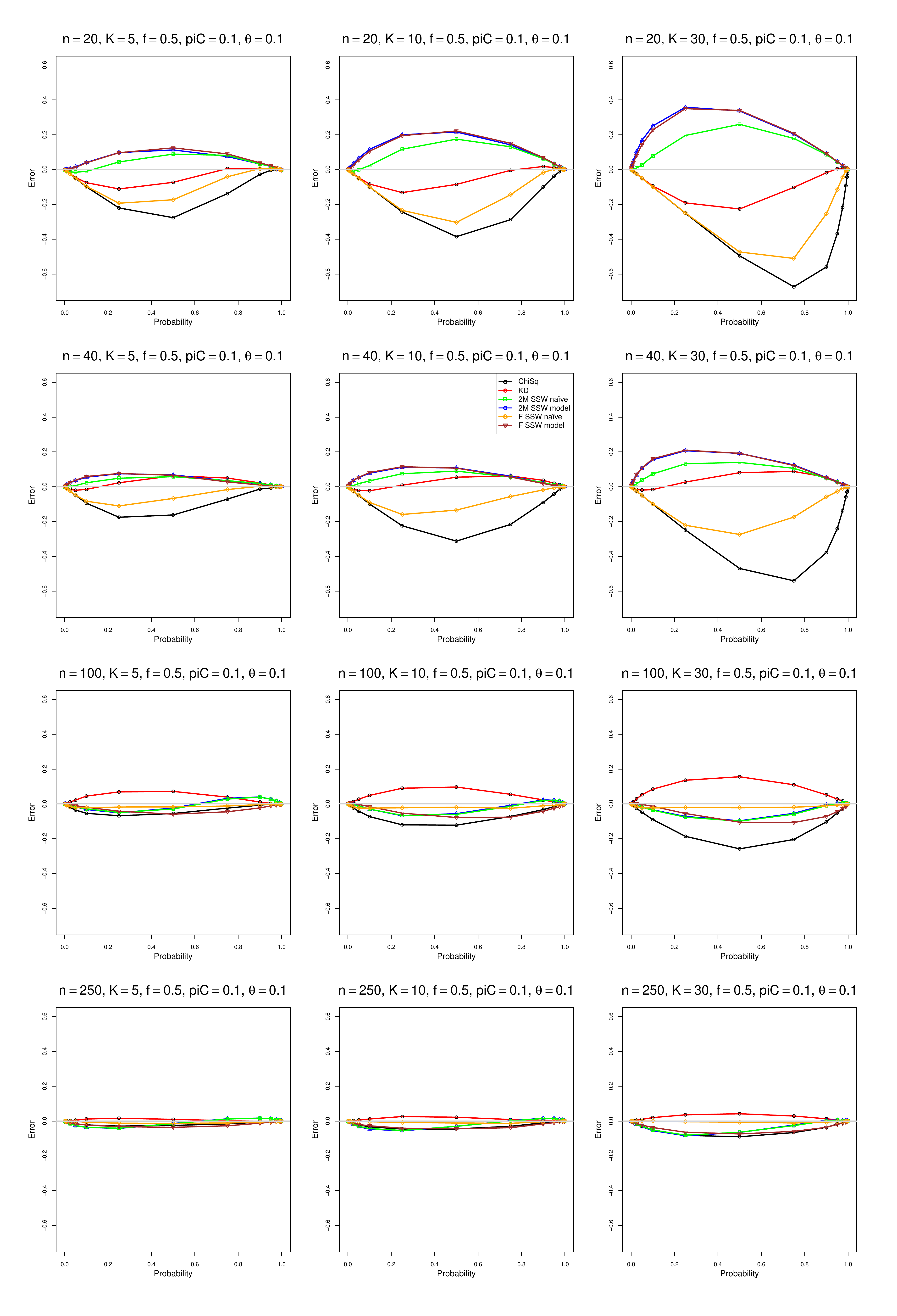}
	\caption{Plots of error in the level of the test for heterogeneity of LOR for six approximations for the null distribution of $Q$, $p_{iC} = .1$, $f = .5$, and $\theta = 0.1$, equal sample sizes}
	\label{PPplot_piC_01theta=0.1_LOR_equal_sample_sizes}
\end{figure}

\begin{figure}[ht]
	\centering
	\includegraphics[scale=0.33]{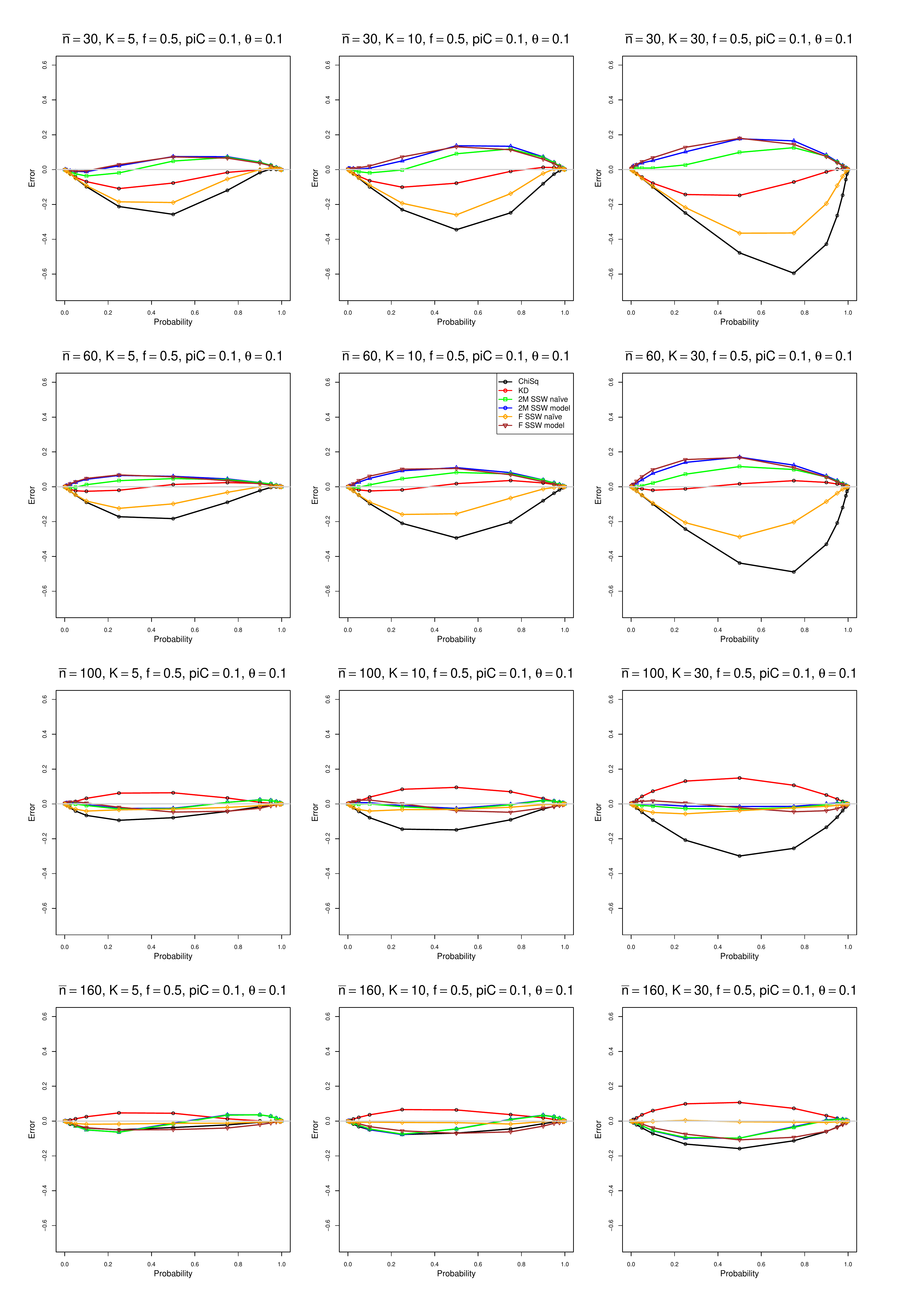}
	\caption{Plots of error in the level of the test for heterogeneity of LOR for six approximations for the null distribution of $Q$, $p_{iC} = .1$, $f = .5$, and $\theta = 0.1$, unequal sample sizes}
	\label{PPplot_piC_01theta=0.1_LOR_unequal_sample_sizes}
\end{figure}

\begin{figure}[ht]
	\centering
	\includegraphics[scale=0.33]{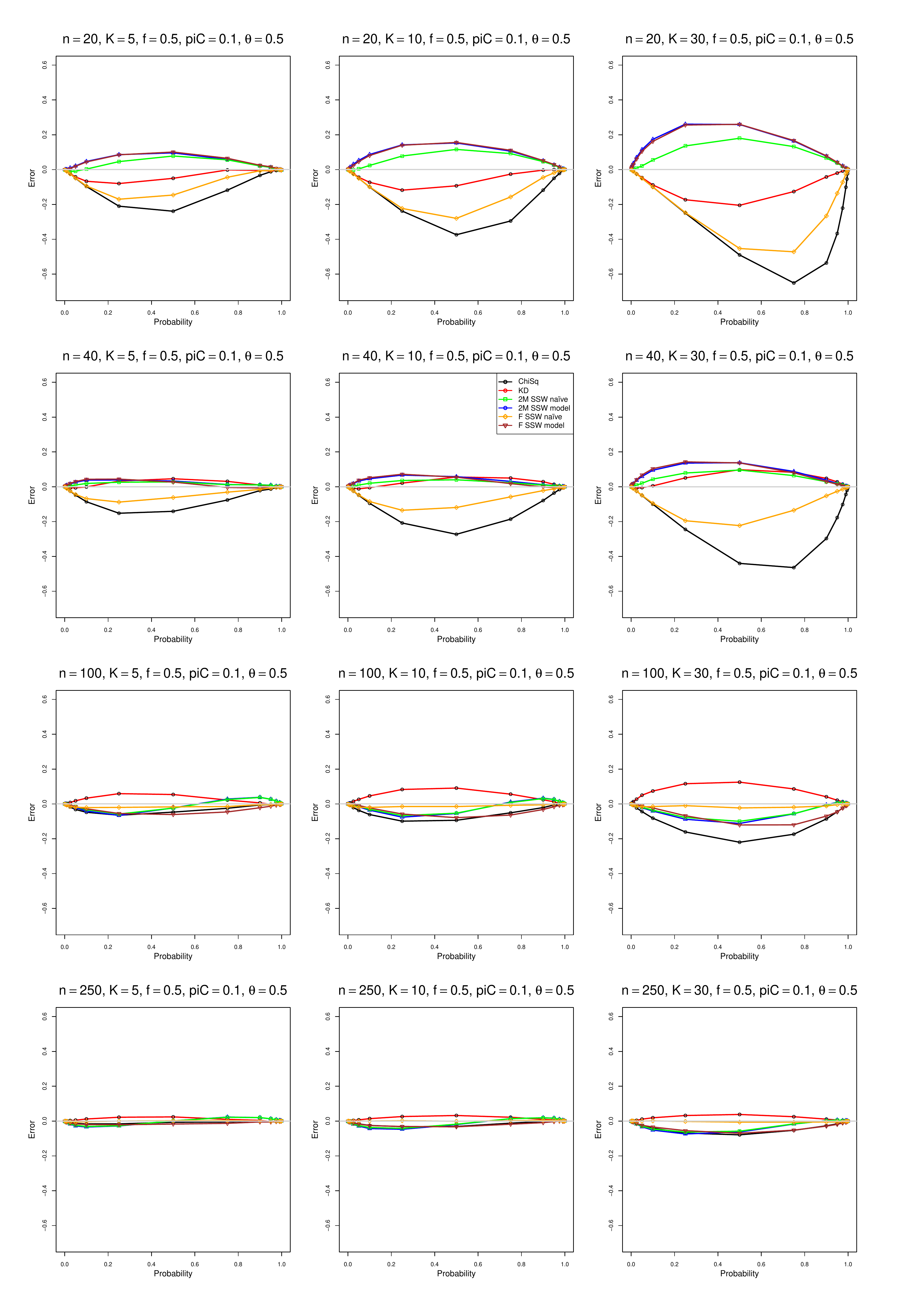}
	\caption{Plots of error in the level of the test for heterogeneity of LOR for six approximations for the null distribution of $Q$, $p_{iC} = .1$, $f = .5$, and $\theta = 0.5$, equal sample sizes}
	\label{PPplot_piC_01theta=0.5_LOR_equal_sample_sizes}
\end{figure}

\begin{figure}[ht]
	\centering
	\includegraphics[scale=0.33]{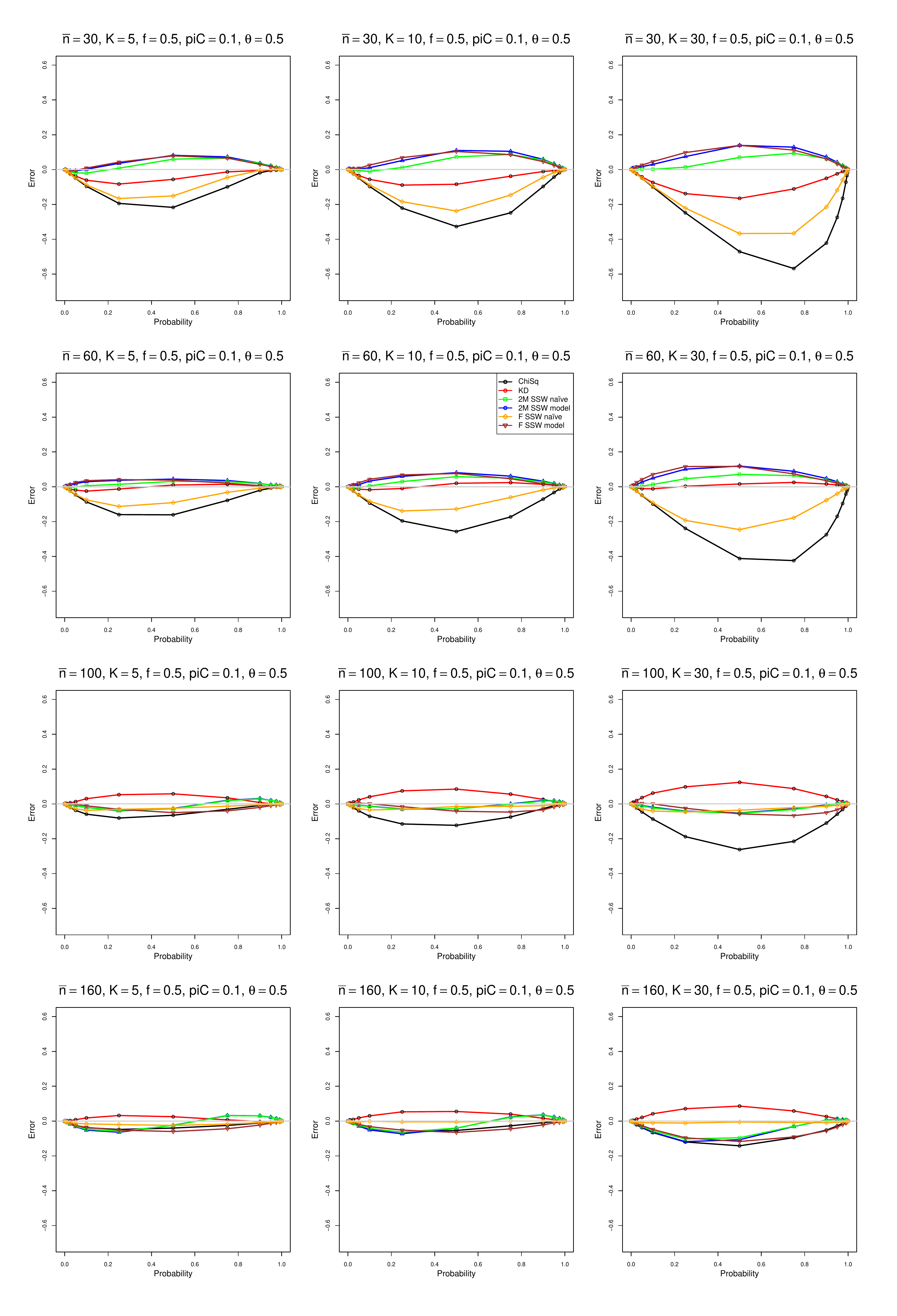}
	\caption{Plots of error in the level of the test for heterogeneity of LOR for six approximations for the null distribution of $Q$, $p_{iC} = .1$, $f = .5$, and $\theta = 0.5$, unequal sample sizes }
	\label{PPplot_piC_01theta=0.5_LOR_unequal_sample_sizes}
\end{figure}

\begin{figure}[ht]
	\centering
	\includegraphics[scale=0.33]{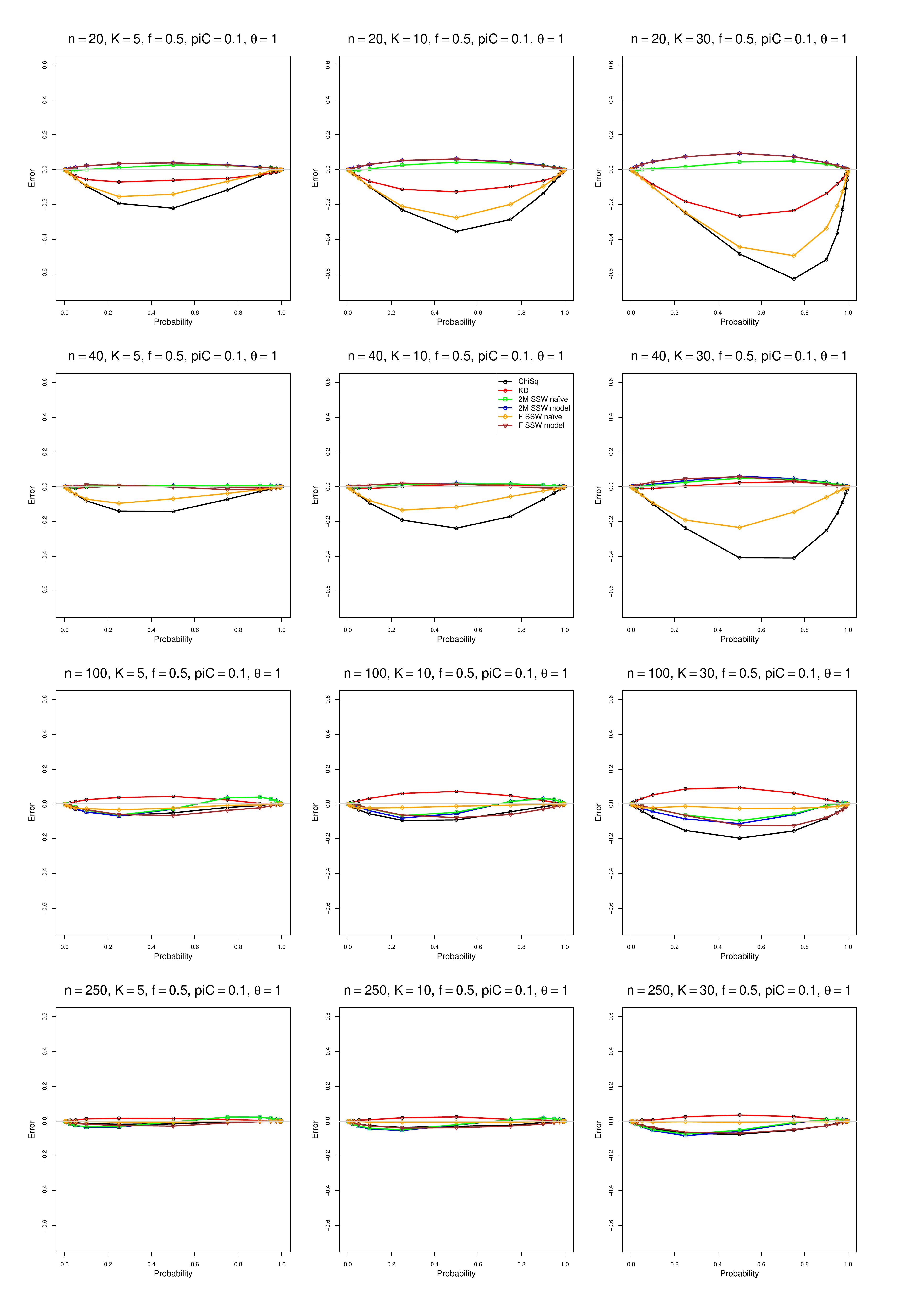}
	\caption{Plots of error in the level of the test for heterogeneity of LOR for six approximations for the null distribution of $Q$, $p_{iC} = .1$, $f = .5$, and $\theta = 1$, equal sample sizes }
	\label{PPplot_piC_01theta=1_LOR_equal_sample_sizes}
\end{figure}

\begin{figure}[ht]
	\centering
	\includegraphics[scale=0.33]{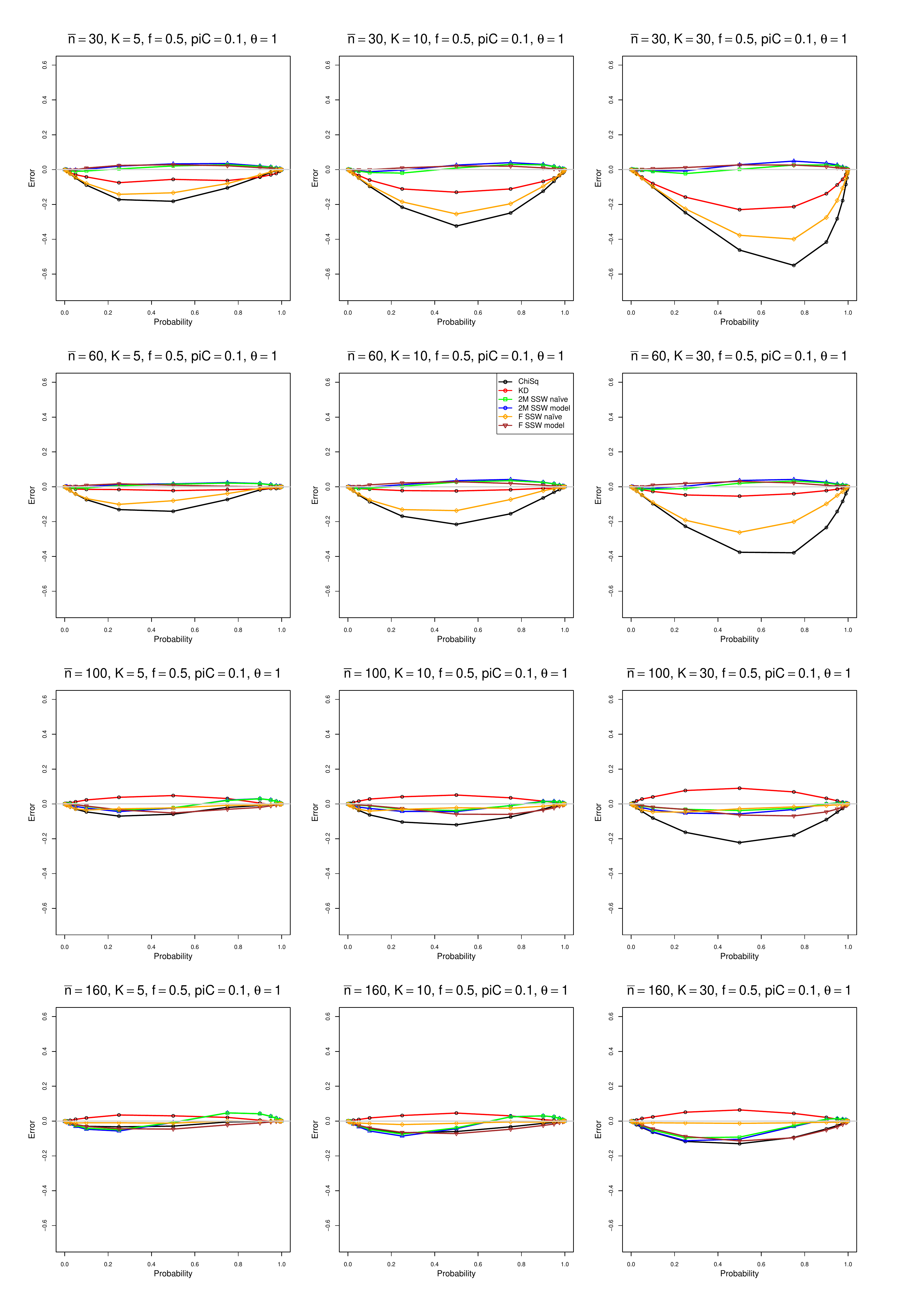}
	\caption{Plots of error in the level of the test for heterogeneity of LOR for six approximations for the null distribution of $Q$, $p_{iC} = .1$, $f = .5$, and $\theta = 1$, unequal sample sizes }
	\label{PPplot_piC_01theta=1_LOR_unequal_sample_sizes}
\end{figure}

\begin{figure}[ht]
	\centering
	\includegraphics[scale=0.33]{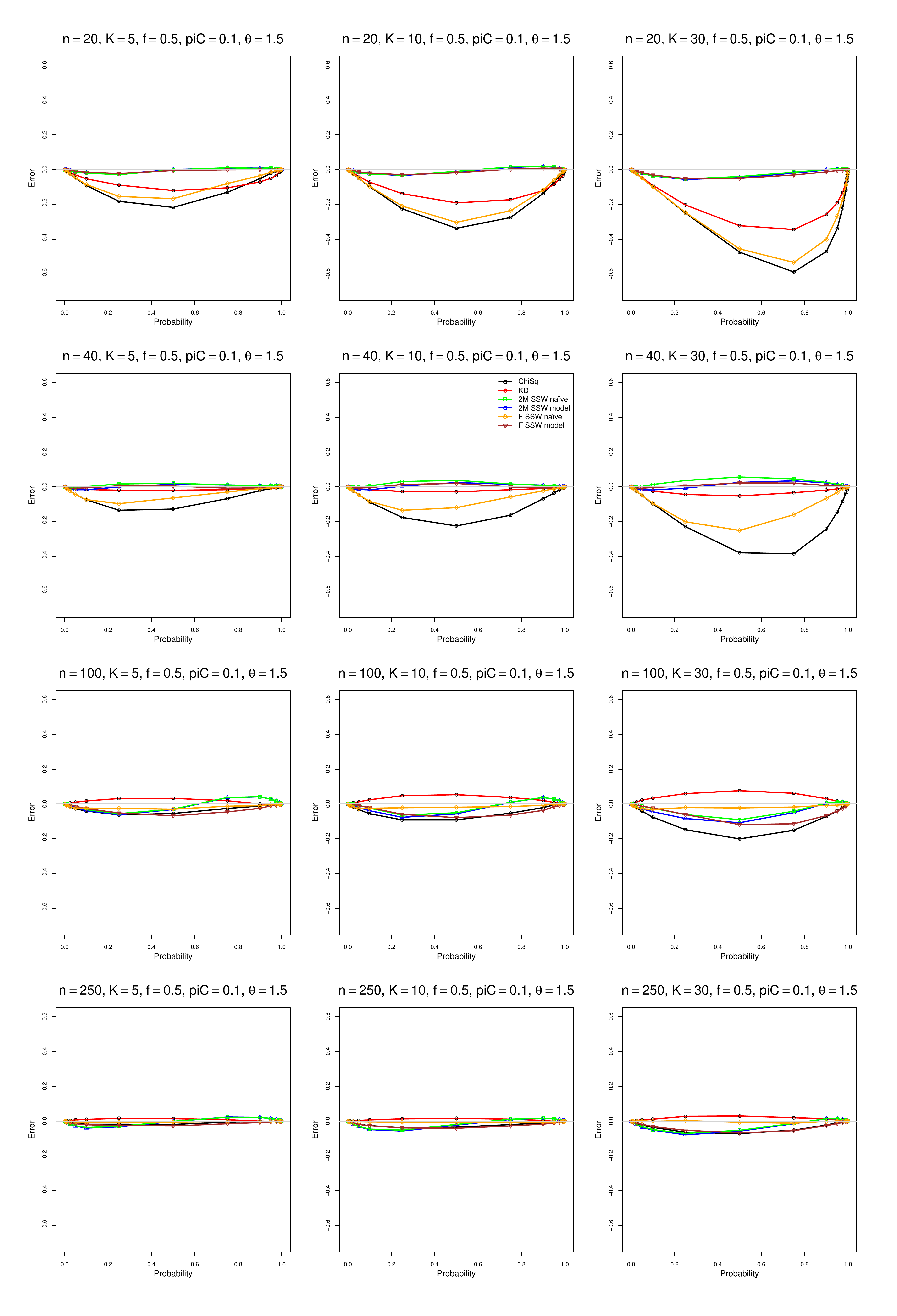}
	\caption{Plots of error in the level of the test for heterogeneity of LOR for six approximations for the null distribution of $Q$, $p_{iC} = .1$, $f = .5$, and $\theta = 1.5$, equal sample sizes}
	\label{PPplot_piC_01theta=1.5_LOR_equal_sample_sizes}
\end{figure}

\begin{figure}[ht]
	\centering
	\includegraphics[scale=0.33]{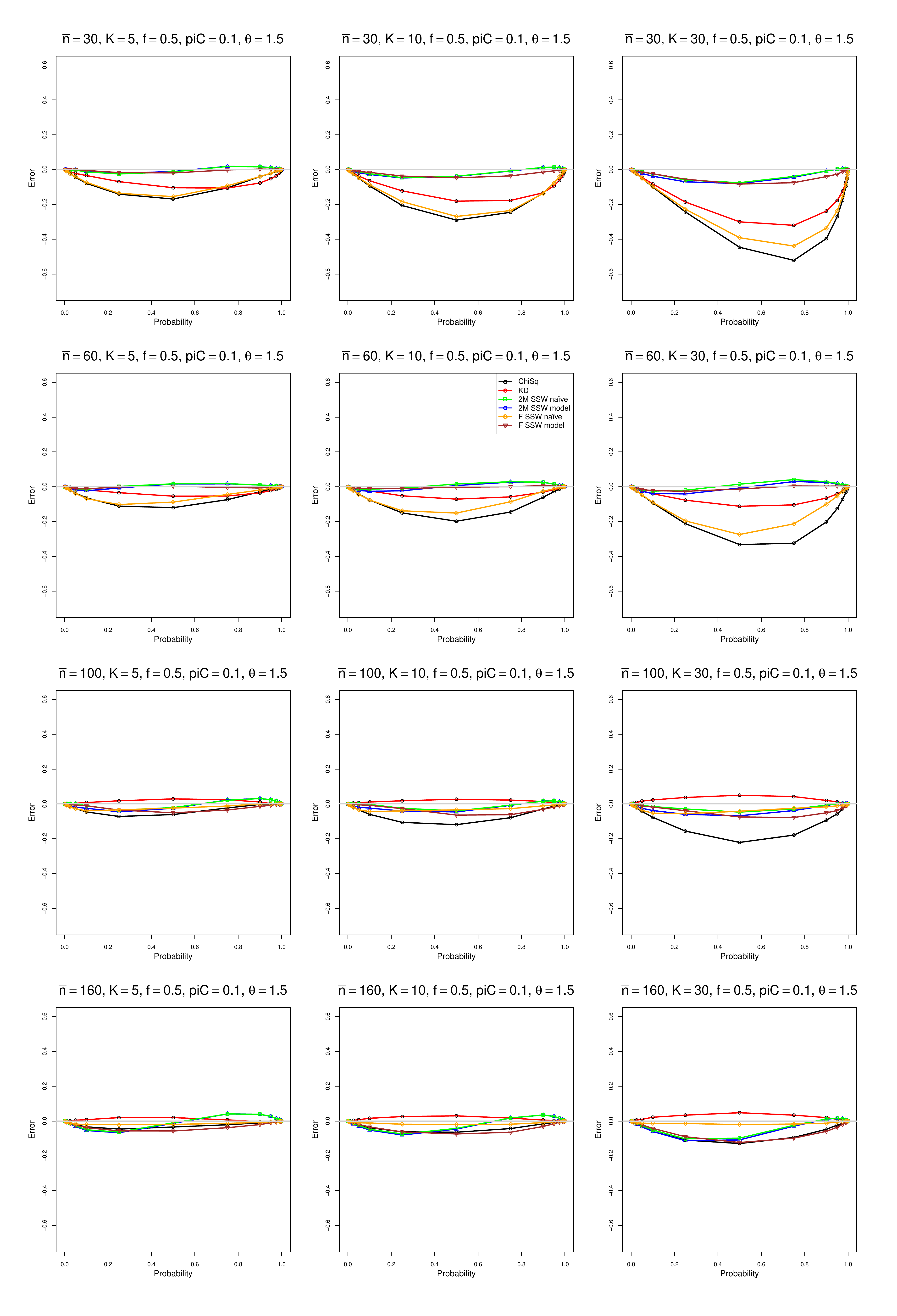}
	\caption{Plots of error in the level of the test for heterogeneity of LOR for six approximations for the null distribution of $Q$, $p_{iC} = .1$, $f = .5$, and $\theta = 1.5$, unequal sample sizes }
	\label{PPplot_piC_01theta=1.5_LOR_unequal_sample_sizes}
\end{figure}

\begin{figure}[ht]
	\centering
	\includegraphics[scale=0.33]{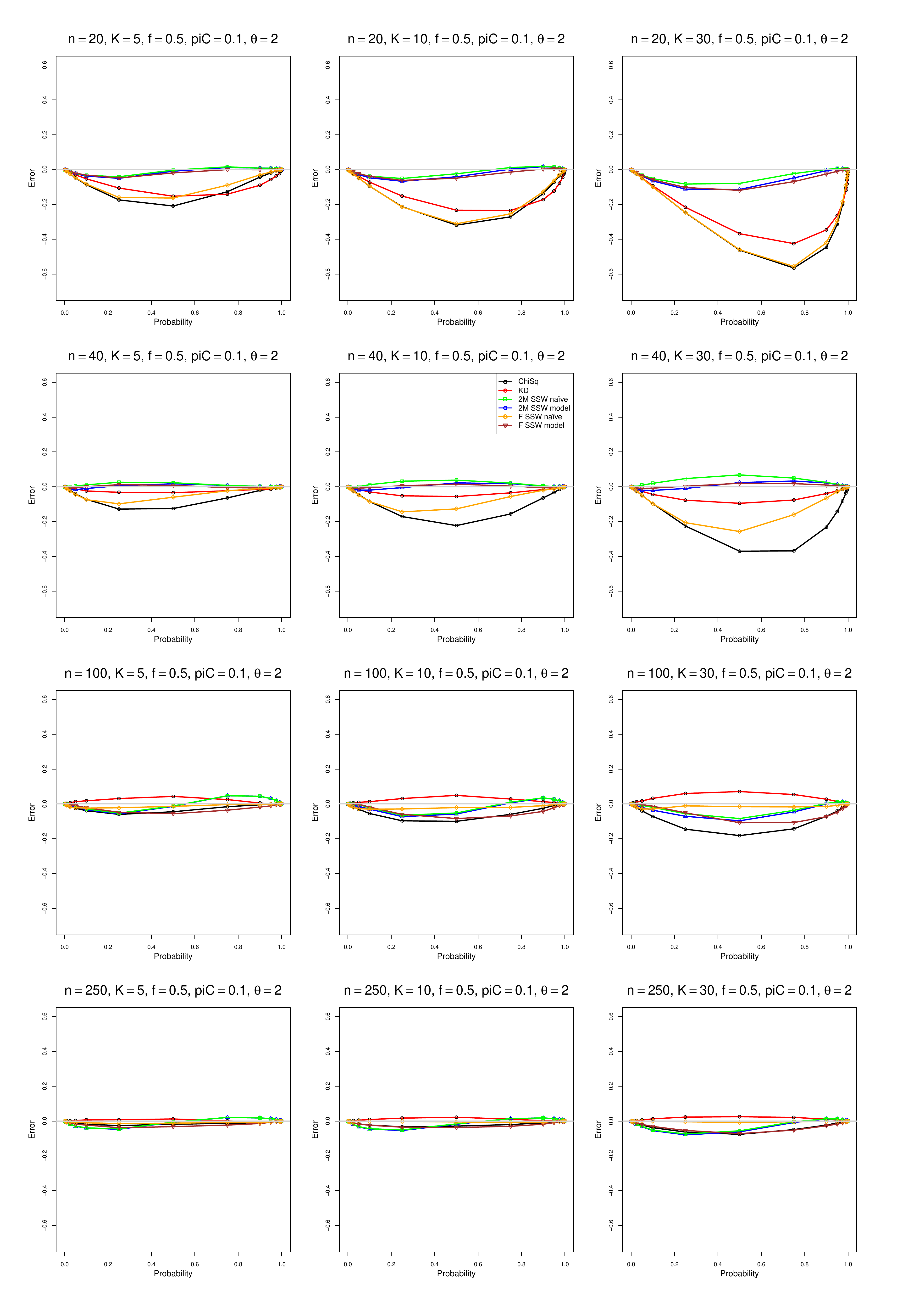}
	\caption{Plots of error in the level of the test for heterogeneity of LOR for six approximations for the null distribution of $Q$, $p_{iC} = .1$, $f = .5$, and $\theta = 2$, equal sample sizes }
	\label{PPplot_piC_01theta=2_LOR_equal_sample_sizes}
\end{figure}

\begin{figure}[ht]
	\centering
	\includegraphics[scale=0.33]{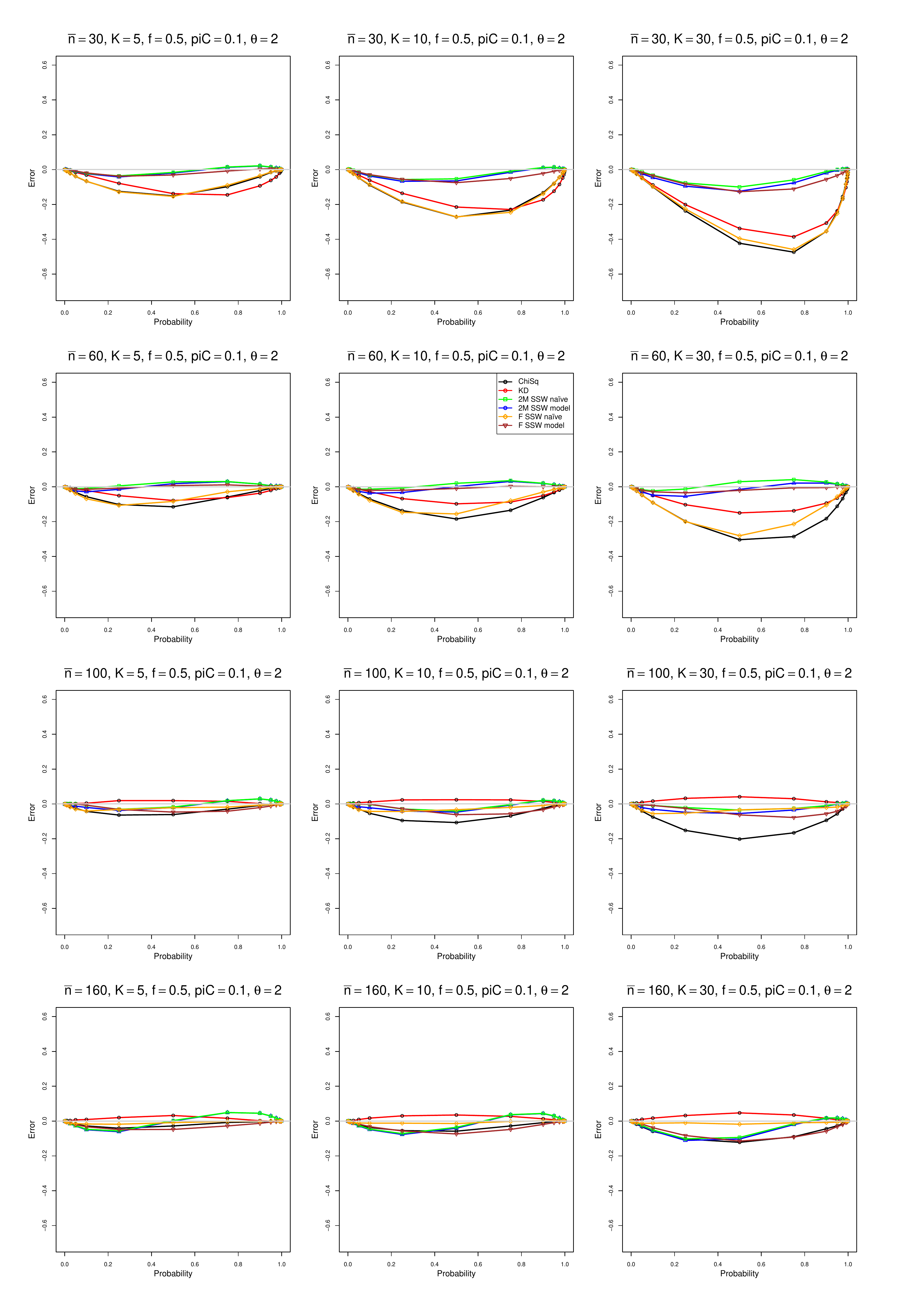}
	\caption{Plots of error in the level of the test for heterogeneity of LOR for six approximations for the null distribution of $Q$, $p_{iC} = .1$, $f = .5$, and $\theta = 2$, unequal sample sizes}
	\label{PPplot_piC_01theta=2_LOR_unequal_sample_sizes}
\end{figure}
\begin{figure}[ht]
	\centering
	\includegraphics[scale=0.33]{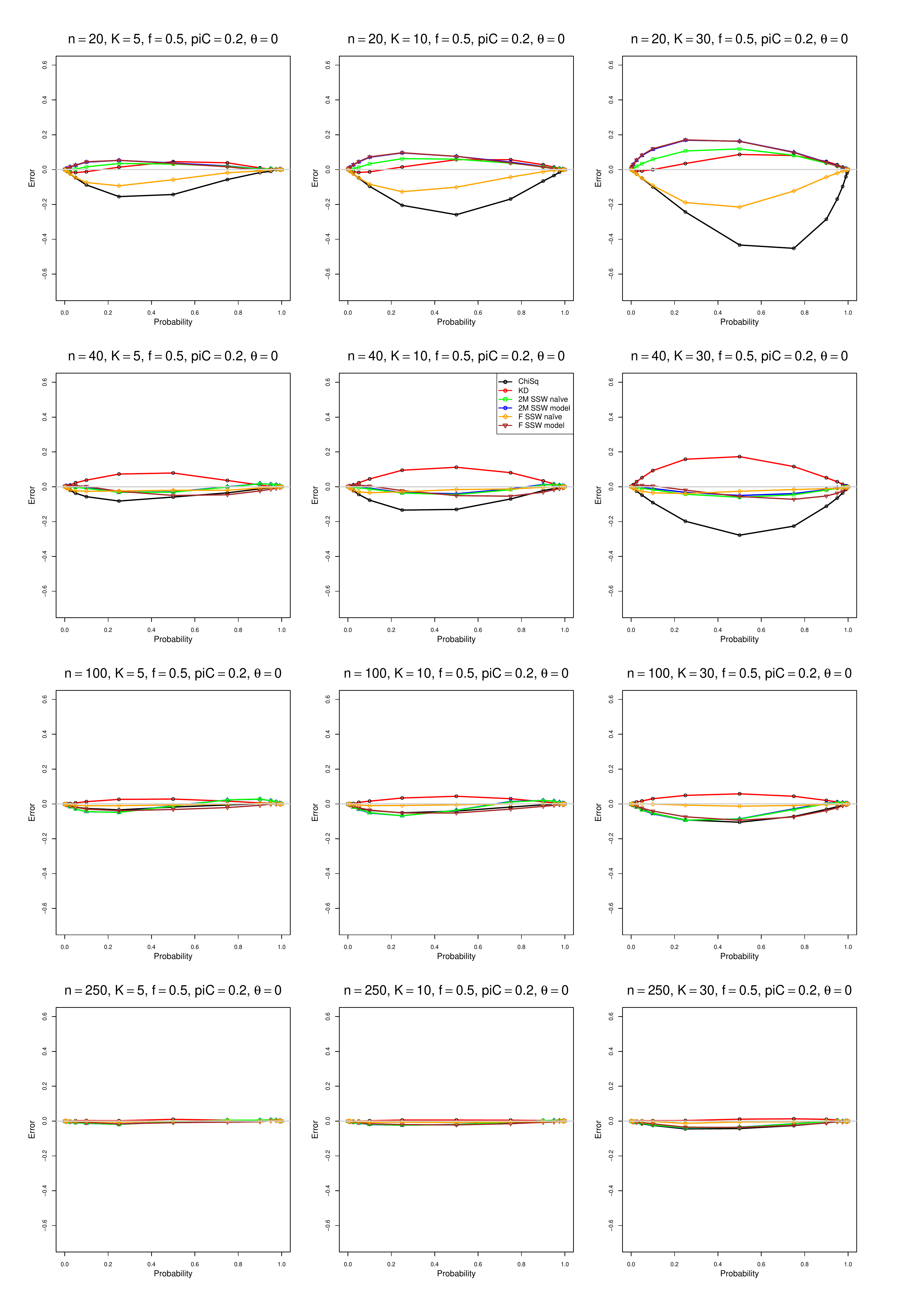}
	\caption{Plots of error in the level of the test for heterogeneity of LOR for six approximations for the null distribution of $Q$, $p_{iC} = .2$, $f = .5$, and $\theta = 0$, equal sample sizes}
	\label{PPplot_piC_02theta=0_LOR_equal_sample_sizes}
\end{figure}

\begin{figure}[ht]
	\centering
	\includegraphics[scale=0.33]{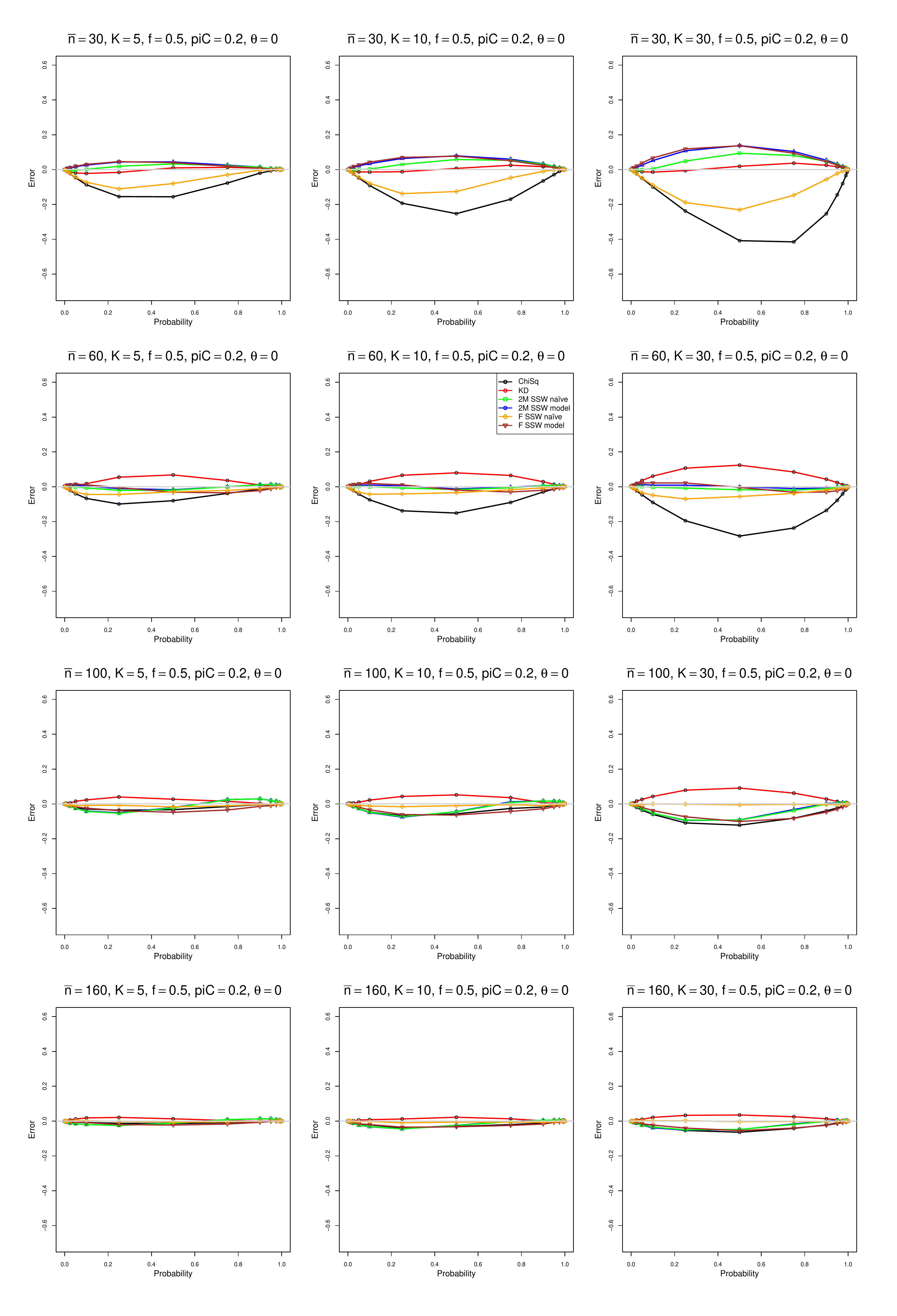}
	\caption{Plots of error in the level of the test for heterogeneity of LOR for six approximations for the null distribution of $Q$, $p_{iC} = .2$, $f = .5$, and $\theta = 0$, unequal sample sizes}
	\label{PPplot_piC_02theta=0_LOR_unequal_sample_sizes}
\end{figure}

\begin{figure}[ht]
	\centering
	\includegraphics[scale=0.33]{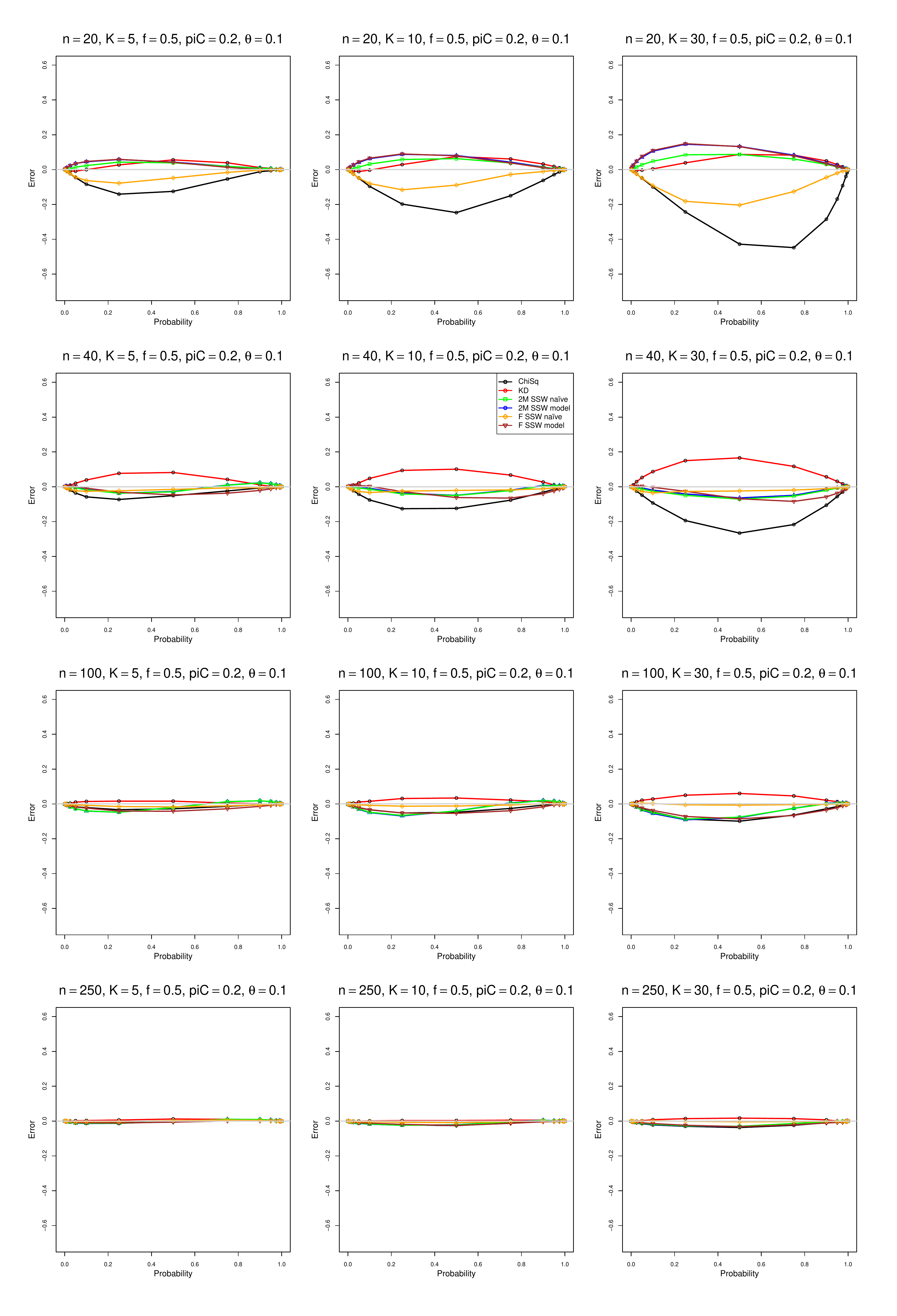}
	\caption{Plots of error in the level of the test for heterogeneity of LOR for six approximations for the null distribution of $Q$, $p_{iC} = .2$, $f = .5$, and $\theta = 0.1$, equal sample sizes}
	\label{PPplot_piC_02theta=0.1_LOR_equal_sample_sizes}
\end{figure}

\begin{figure}[ht]
	\centering
	\includegraphics[scale=0.33]{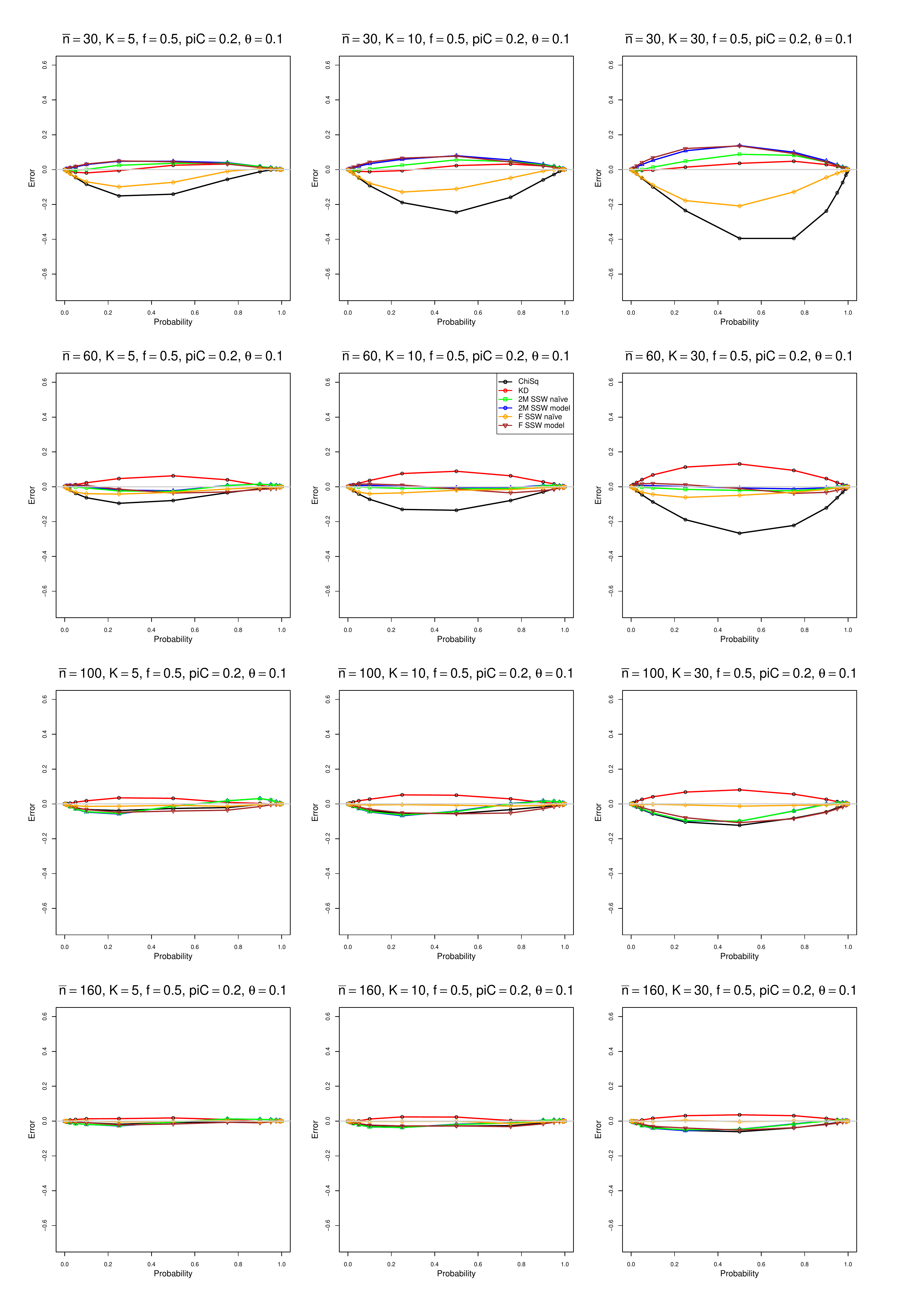}
	\caption{Plots of error in the level of the test for heterogeneity of LOR for six approximations for the null distribution of $Q$, $p_{iC} = .2$, $f = .5$, and $\theta = 0.1$, unequal sample sizes}
	\label{PPplot_piC_02theta=0.1_LOR_unequal_sample_sizes}
\end{figure}

\begin{figure}[ht]
	\centering
	\includegraphics[scale=0.33]{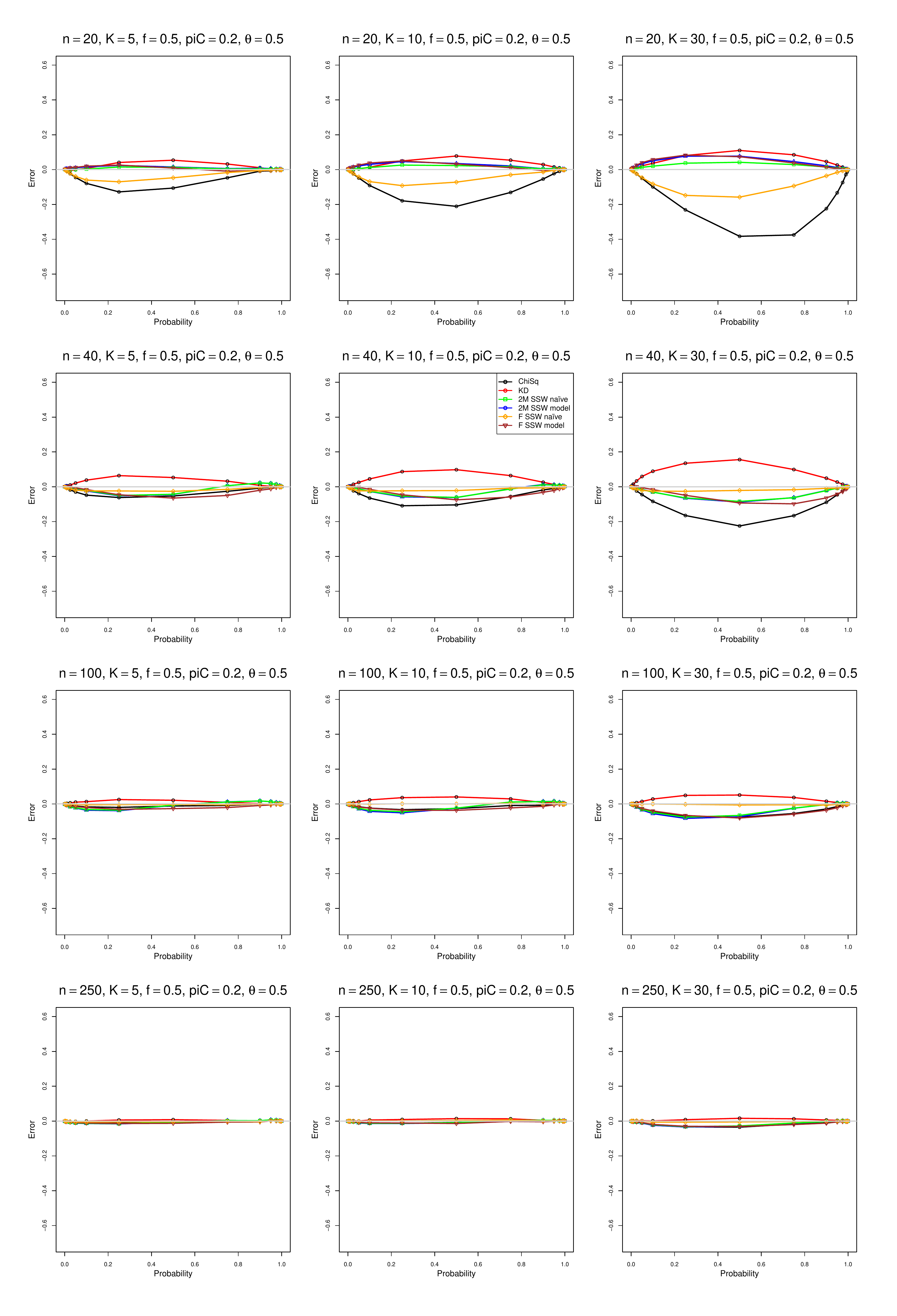}
	\caption{Plots of error in the level of the test for heterogeneity of LOR for six approximations for the null distribution of $Q$, $p_{iC} = .2$, $f = .5$, and $\theta = 0.5$, equal sample sizes}
	\label{PPplot_piC_02theta=0.5_LOR_equal_sample_sizes}
\end{figure}

\begin{figure}[ht]
	\centering
	\includegraphics[scale=0.33]{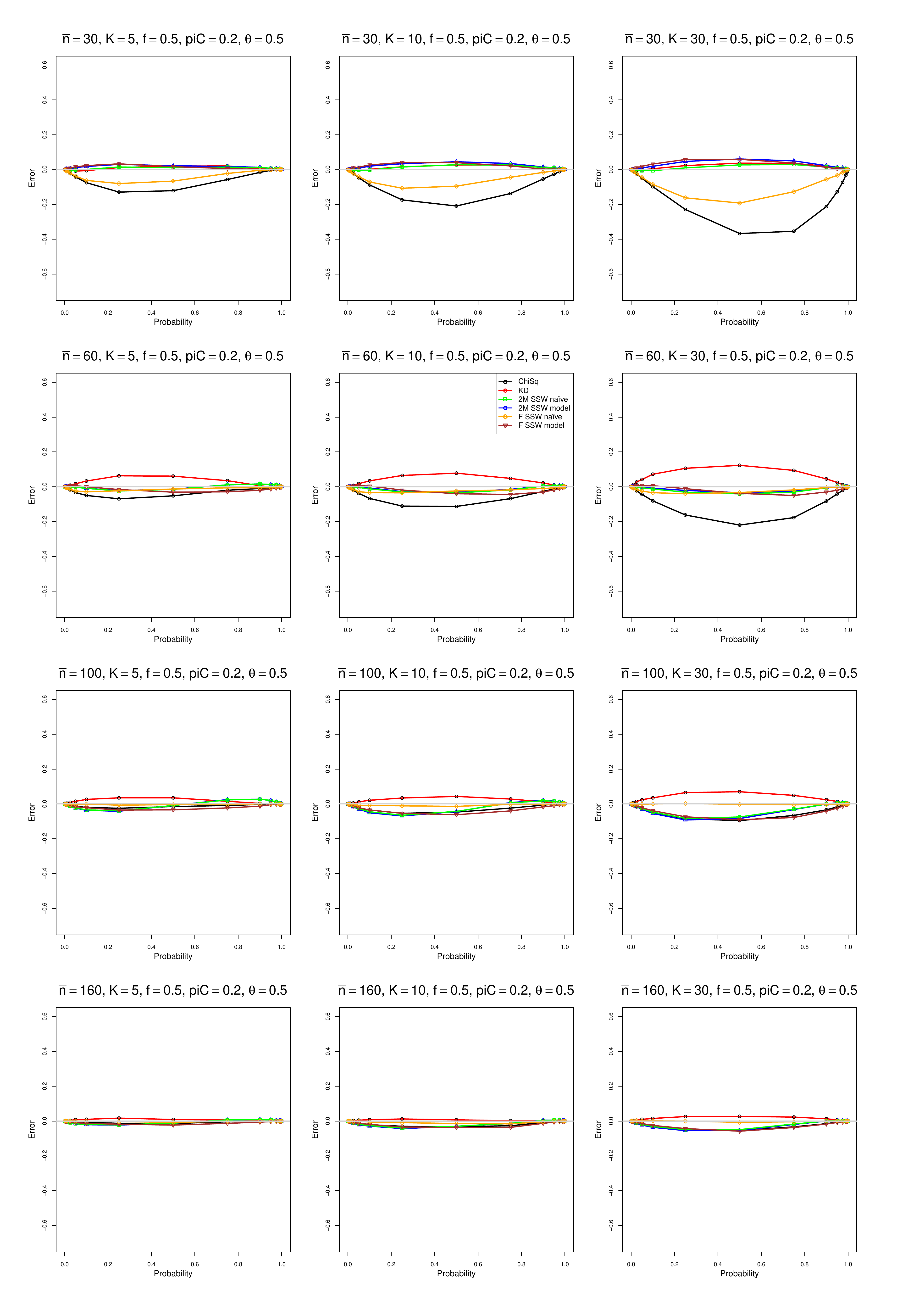}
	\caption{Plots of error in the level of the test for heterogeneity of LOR for six approximations for the null distribution of $Q$, $p_{iC} = .2$, $f = .5$, and $\theta = 0.5$, unequal sample sizes}
	\label{PPplot_piC_02theta=0.5_LOR_unequal_sample_sizes}
\end{figure}

\begin{figure}[ht]
	\centering
	\includegraphics[scale=0.33]{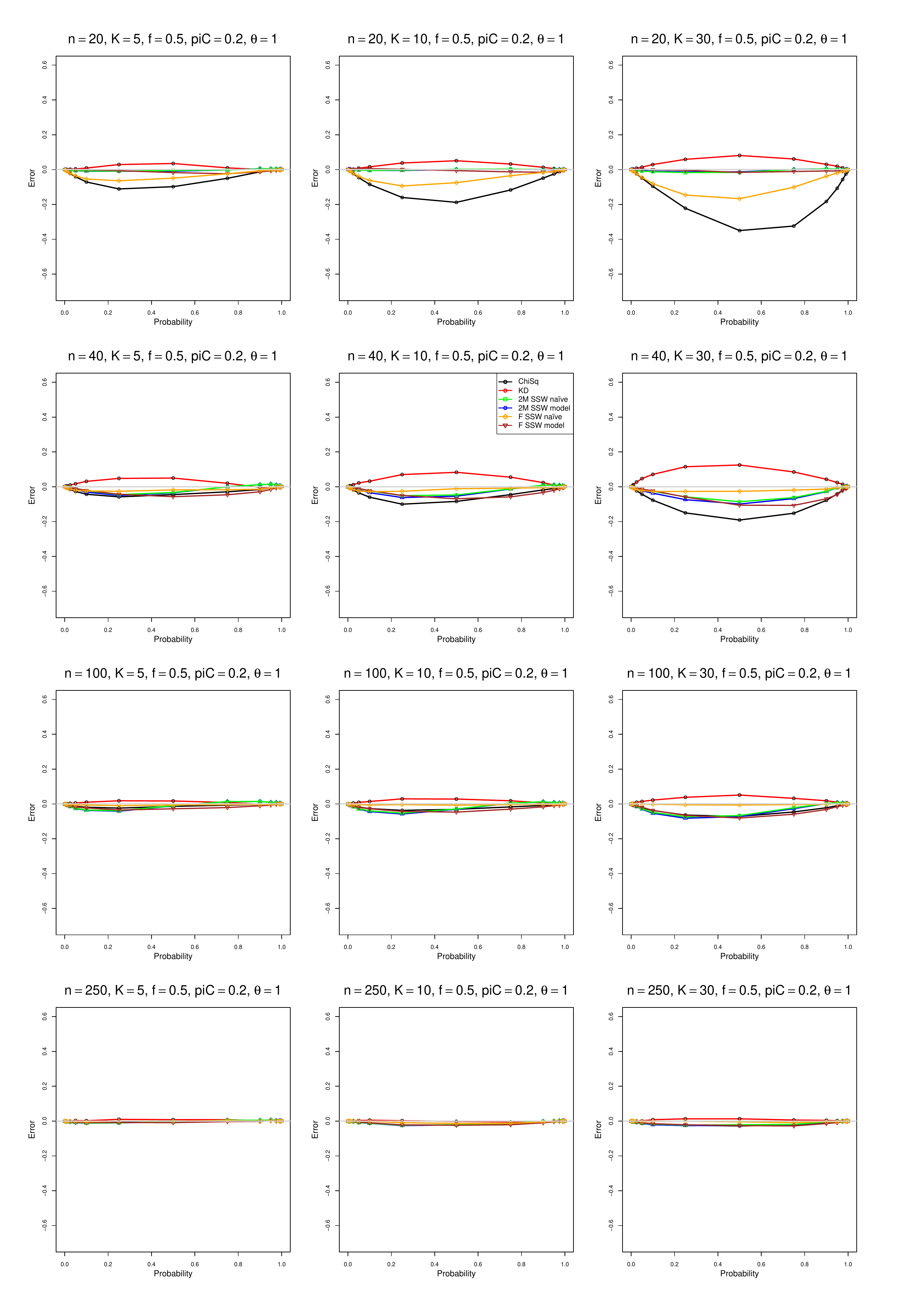}
	\caption{Plots of error in the level of the test for heterogeneity of LOR for six approximations for the null distribution of $Q$, $p_{iC} = .2$, $f = .5$, and $\theta = 1$, equal sample sizes}
	\label{PPplot_piC_02theta=1_LOR_equal_sample_sizes}
\end{figure}

\begin{figure}[ht]
	\centering
	\includegraphics[scale=0.33]{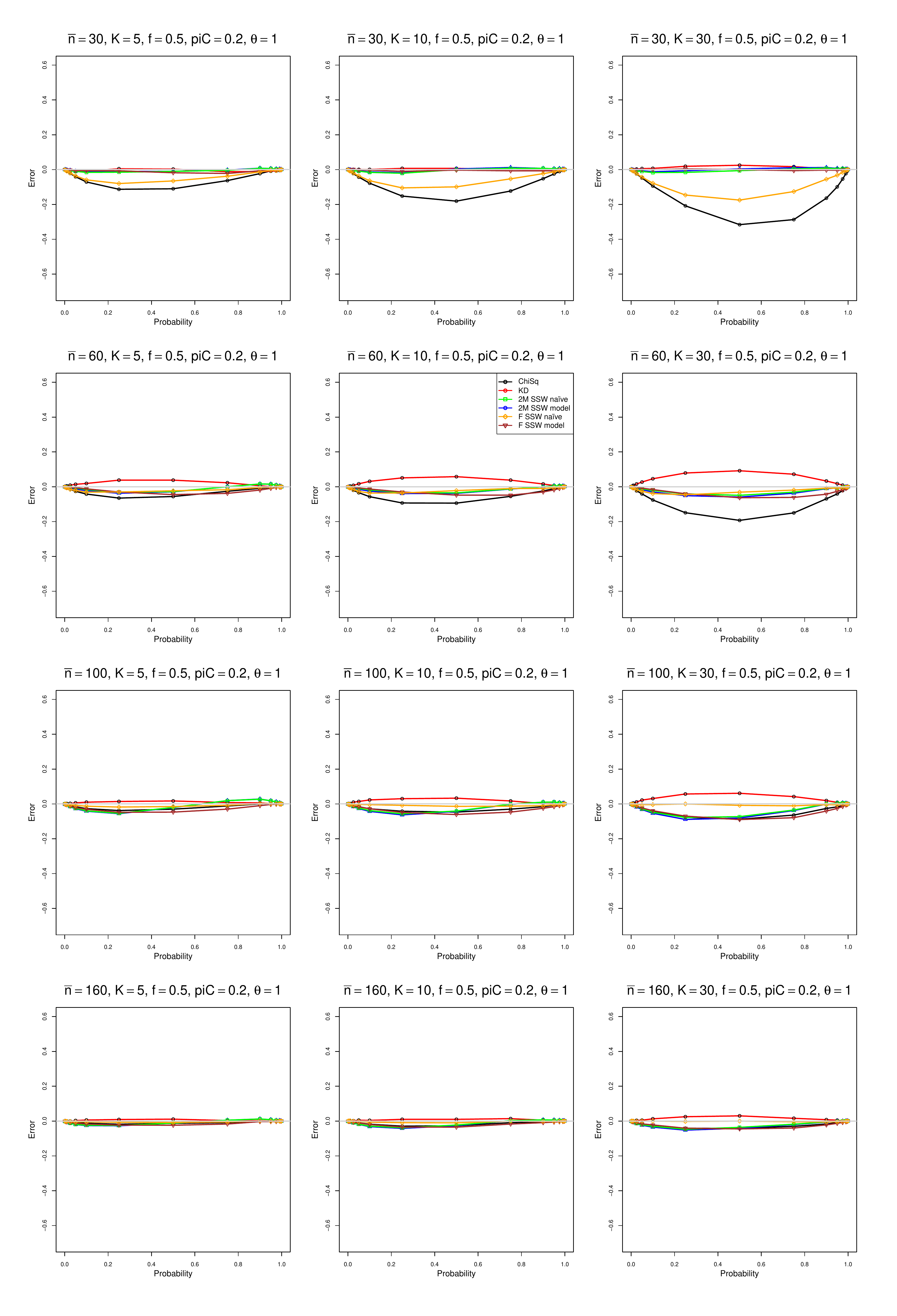}
	\caption{Plots of error in the level of the test for heterogeneity of LOR for six approximations for the null distribution of $Q$, $p_{iC} = .2$, $f = .5$, and $\theta = 1$, unequal sample sizes}
	\label{PPplot_piC_02theta=1_LOR_unequal_sample_sizes}
\end{figure}

\begin{figure}[ht]
	\centering
	\includegraphics[scale=0.33]{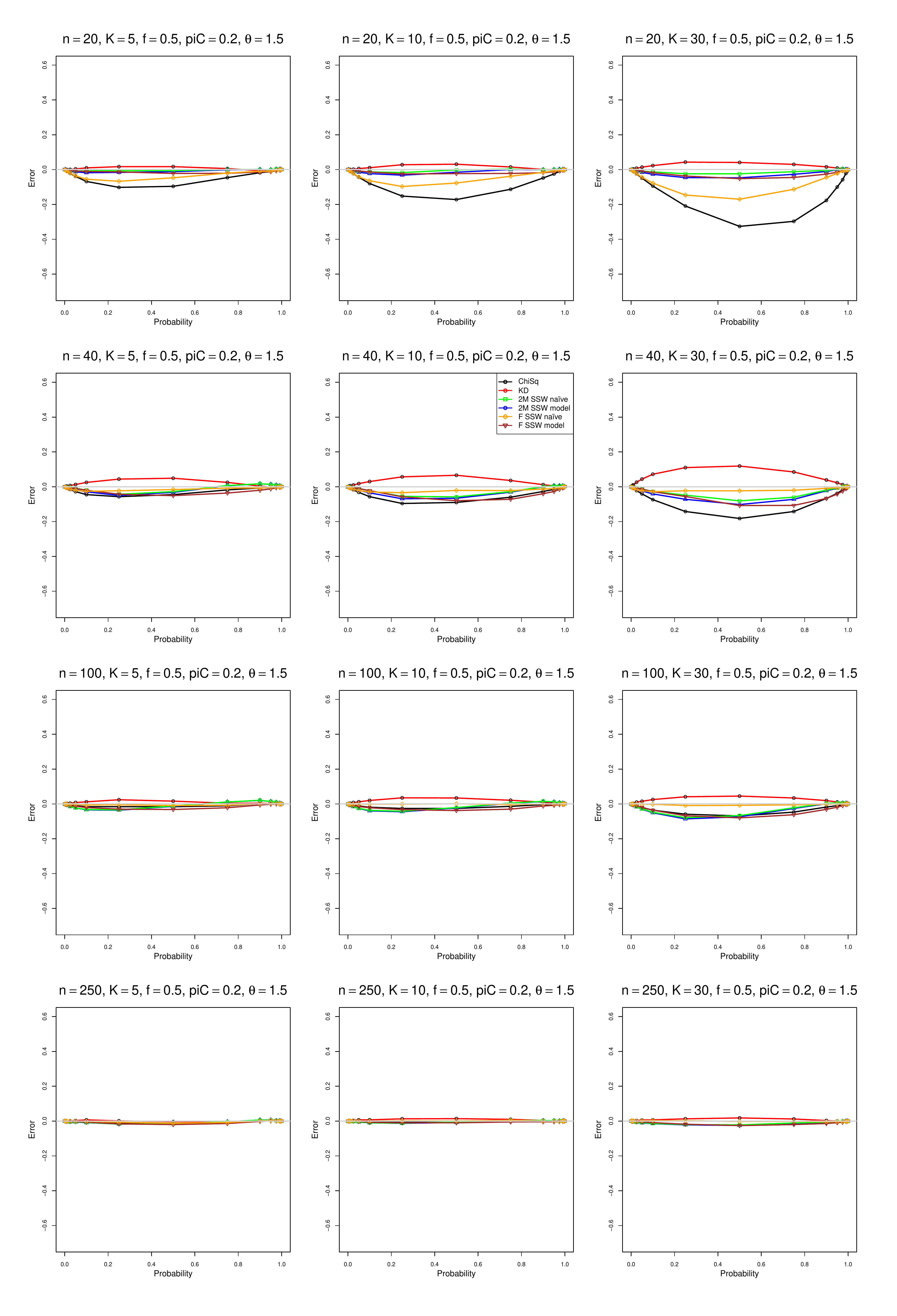}
	\caption{Plots of error in the level of the test for heterogeneity of LOR for six approximations for the null distribution of $Q$, $p_{iC} = .2$, $f = .5$, and $\theta = 1.5$, equal sample sizes}
	\label{PPplot_piC_02theta=1.5_LOR_equal_sample_sizes}
\end{figure}

\begin{figure}[ht]
	\centering
	\includegraphics[scale=0.33]{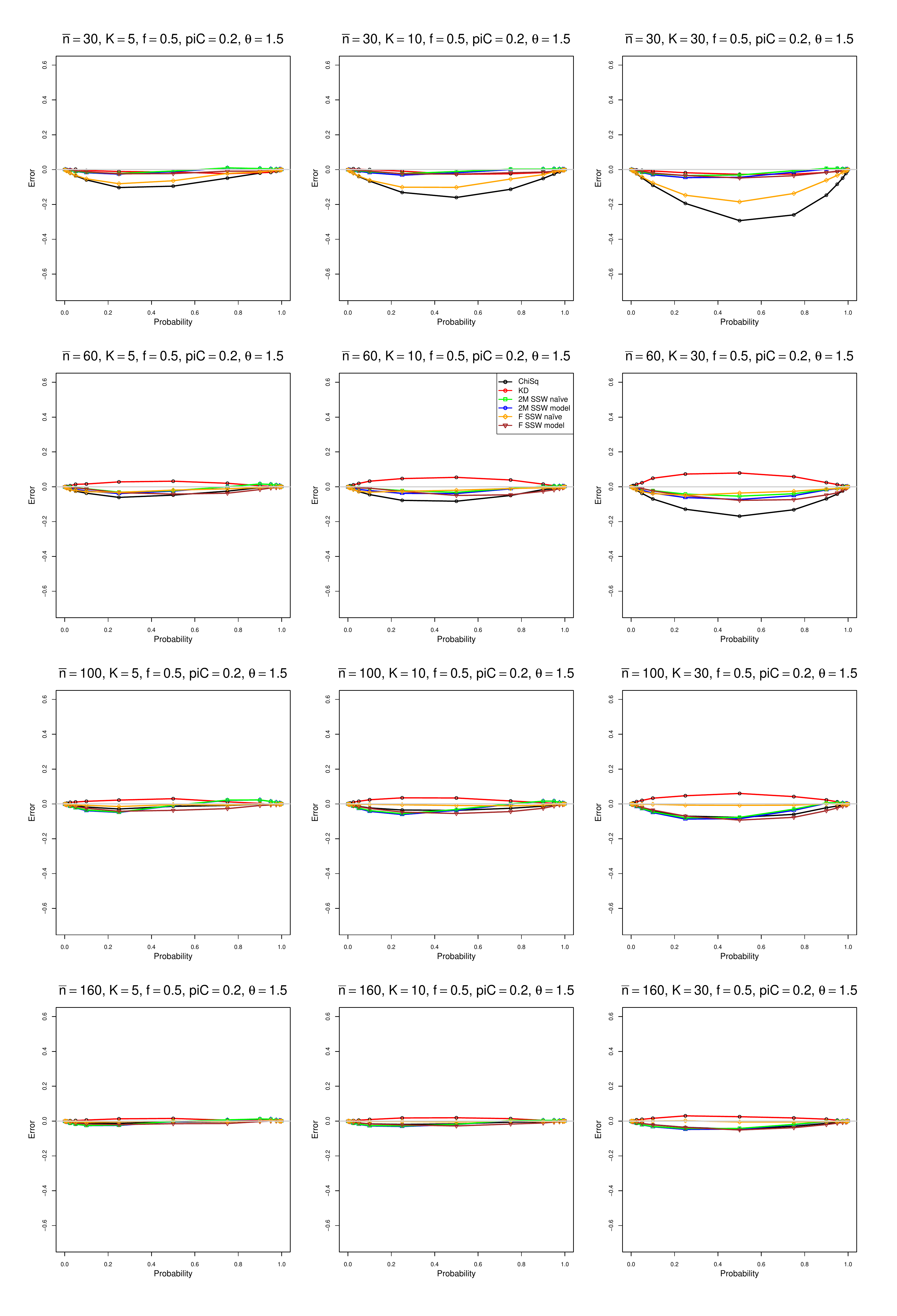}
	\caption{Plots of error in the level of the test for heterogeneity of LOR for six approximations for the null distribution of $Q$, $p_{iC} = .2$, $f = .5$, and $\theta = 1.5$, unequal sample sizes}
	\label{PPplot_piC_02theta=1.5_LOR_unequal_sample_sizes}
\end{figure}

\begin{figure}[ht]
	\centering
	\includegraphics[scale=0.33]{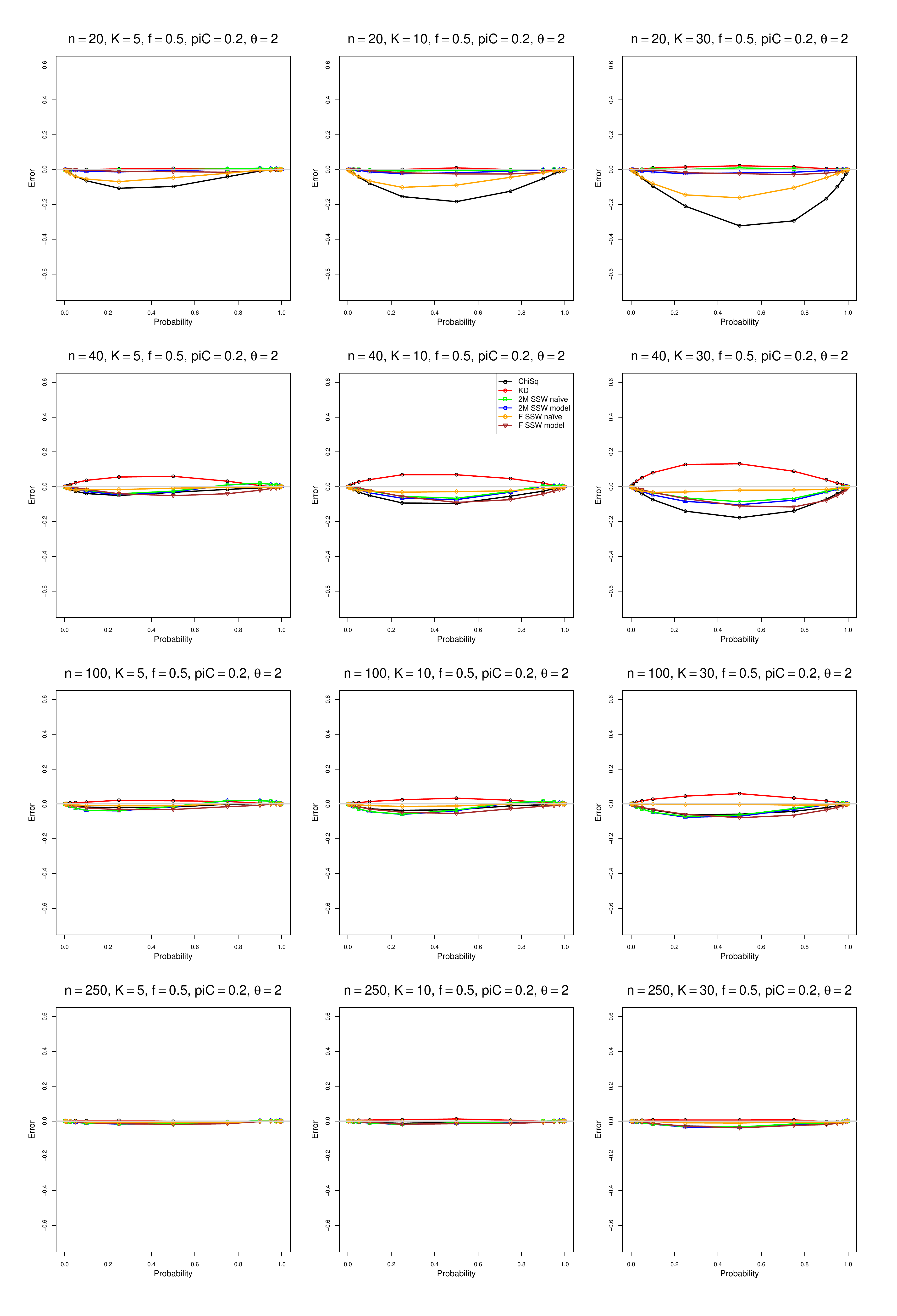}
	\caption{Plots of error in the level of the test for heterogeneity of LOR for six approximations for the null distribution of $Q$, $p_{iC} = .2$, $f = .5$, and $\theta = 2$, equal sample sizes}
	\label{PPplot_piC_02theta=2_LOR_equal_sample_sizes}
\end{figure}

\begin{figure}[ht]
	\centering
	\includegraphics[scale=0.33]{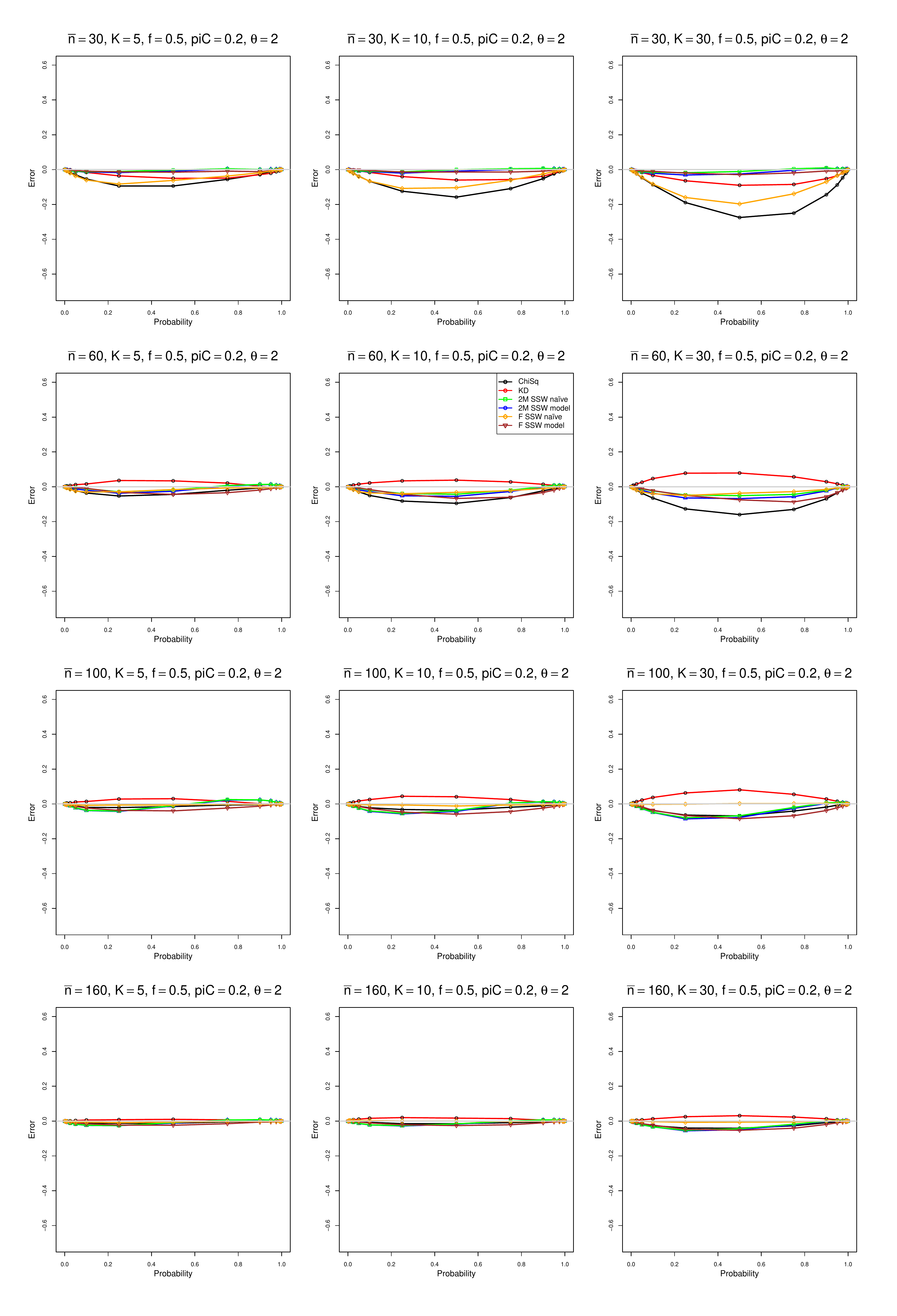}
	\caption{Plots of error in the level of the test for heterogeneity of LOR for six approximations for the null distribution of $Q$, $p_{iC} = .2$, $f = .5$, and $\theta = 2$, unequal sample sizes}
	\label{PPplot_piC_02theta=2_LOR_unequal_sample_sizes}
\end{figure}

\begin{figure}[ht]
	\centering
	\includegraphics[scale=0.33]{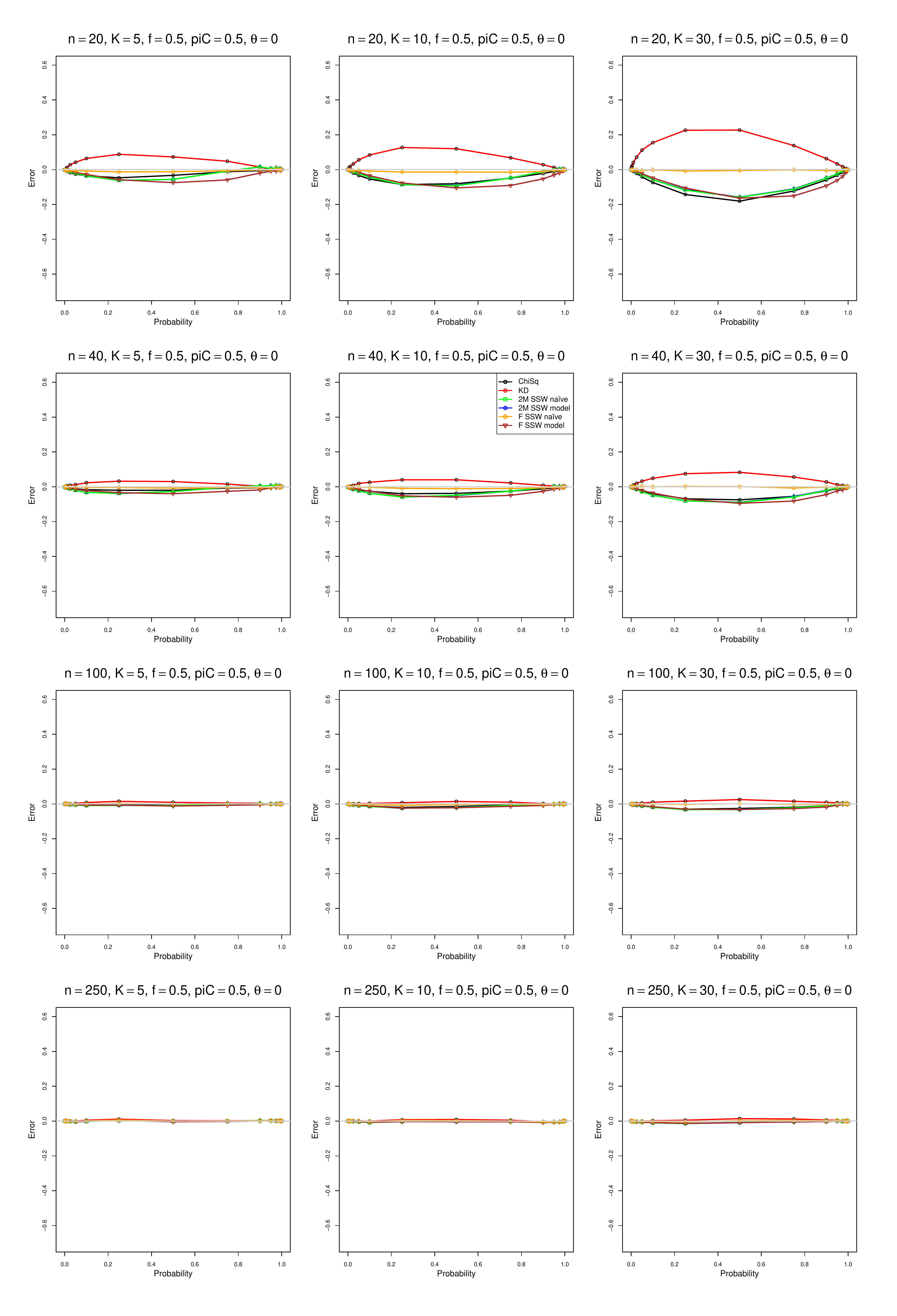}
	\caption{Plots of error in the level of the test for heterogeneity of LOR for six approximations for the null distribution of $Q$, $p_{iC} = .5$, $f = .5$, and $\theta = 0$, equal sample sizes}
	\label{PPplot_piC_05theta=0_LOR_equal_sample_sizes}
\end{figure}

\begin{figure}[ht]
	\centering
	\includegraphics[scale=0.33]{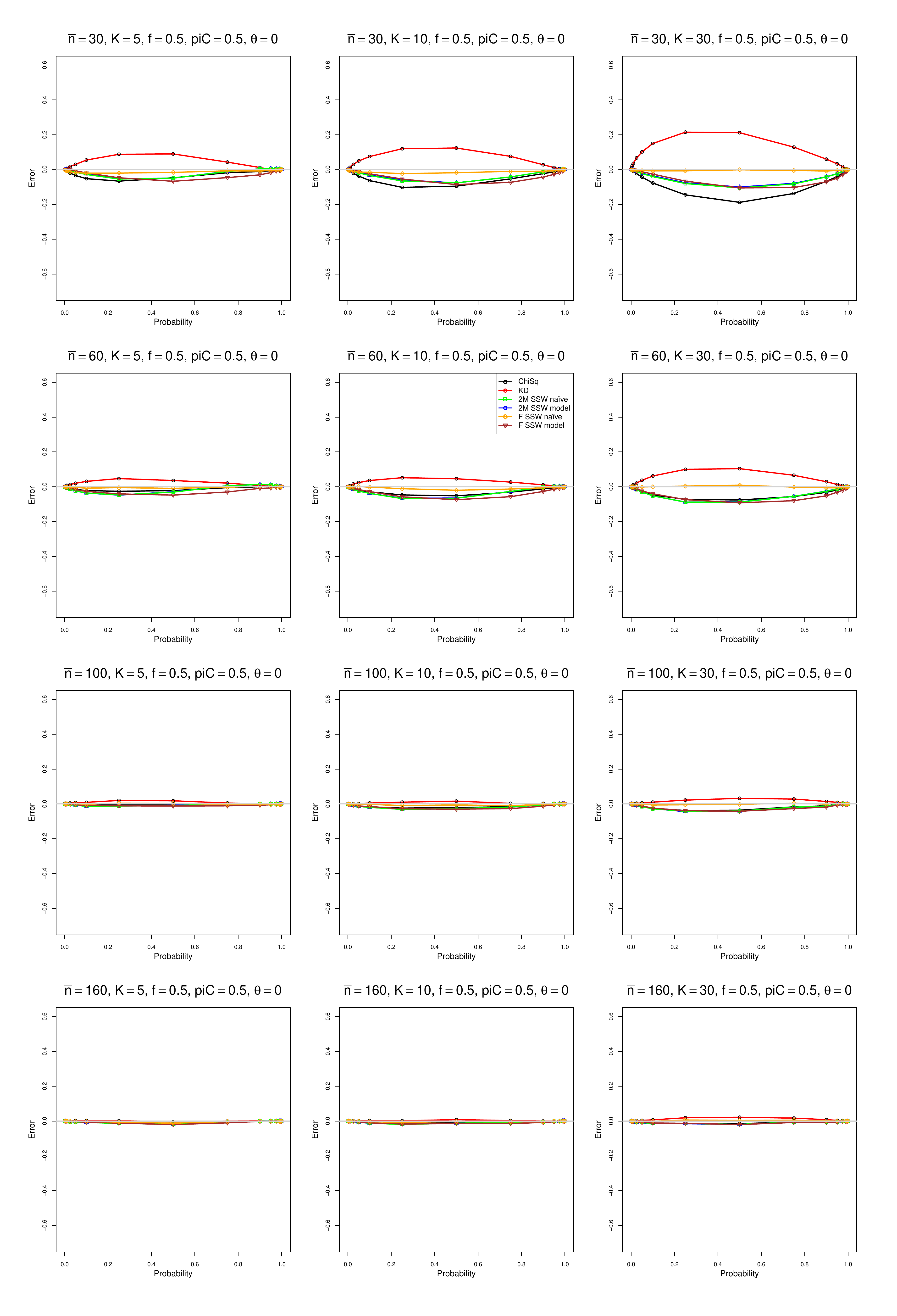}
	\caption{Plots of error in the level of the test for heterogeneity of LOR for six approximations for the null distribution of $Q$, $p_{iC} = .5$, $f = .5$, and $\theta = 0$, unequal sample sizes}
	\label{PPplot_piC_05theta=0_LOR_unequal_sample_sizes}
\end{figure}

\begin{figure}[ht]
	\centering
	\includegraphics[scale=0.33]{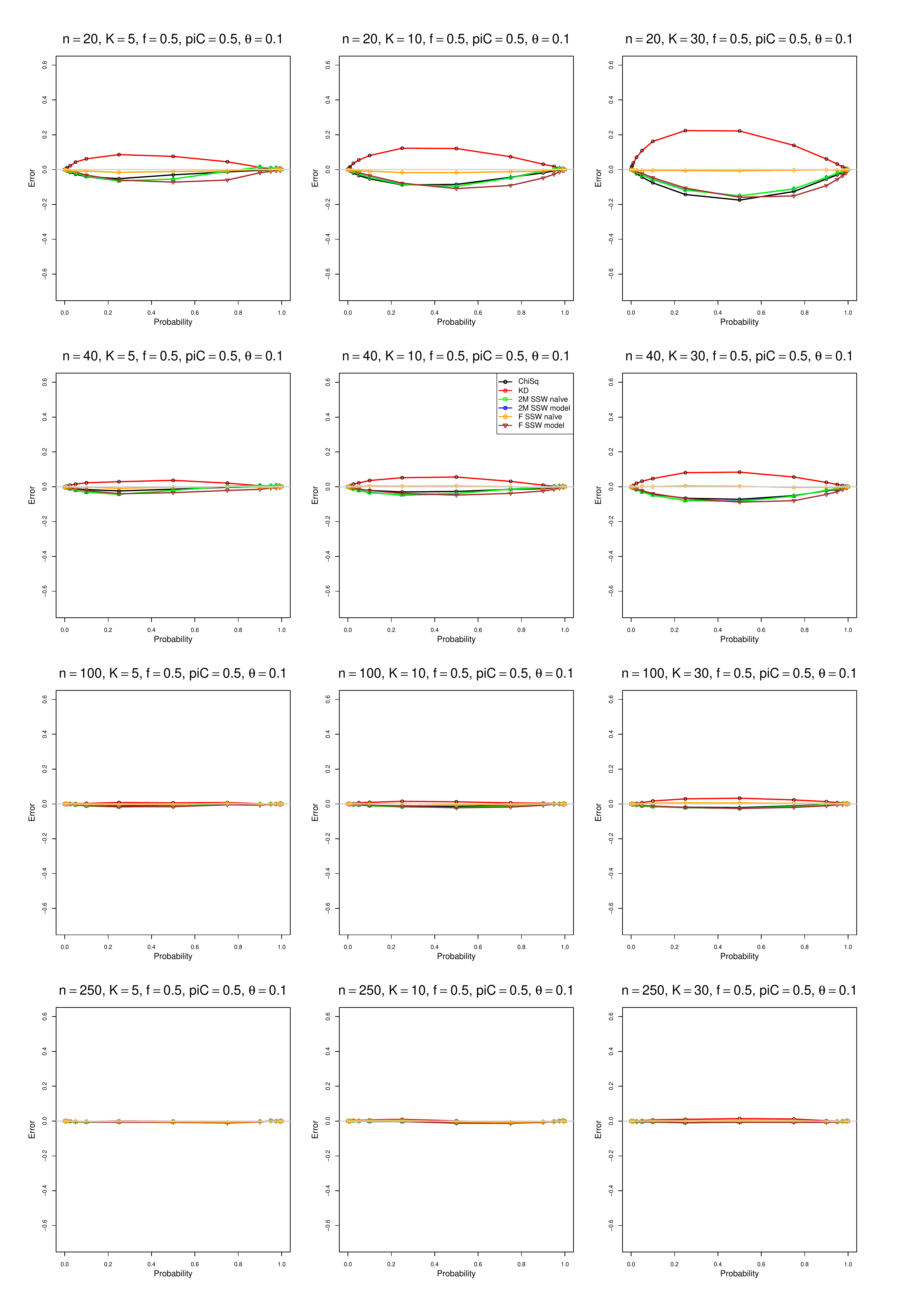}
	\caption{Plots of error in the level of the test for heterogeneity of LOR for six approximations for the null distribution of $Q$, $p_{iC} = .5$, $f = .5$, and $\theta = 0.1$, equal sample sizes}
	\label{PPplot_piC_05theta=0.1_LOR_equal_sample_sizes}
\end{figure}

\begin{figure}[ht]
	\centering
	\includegraphics[scale=0.33]{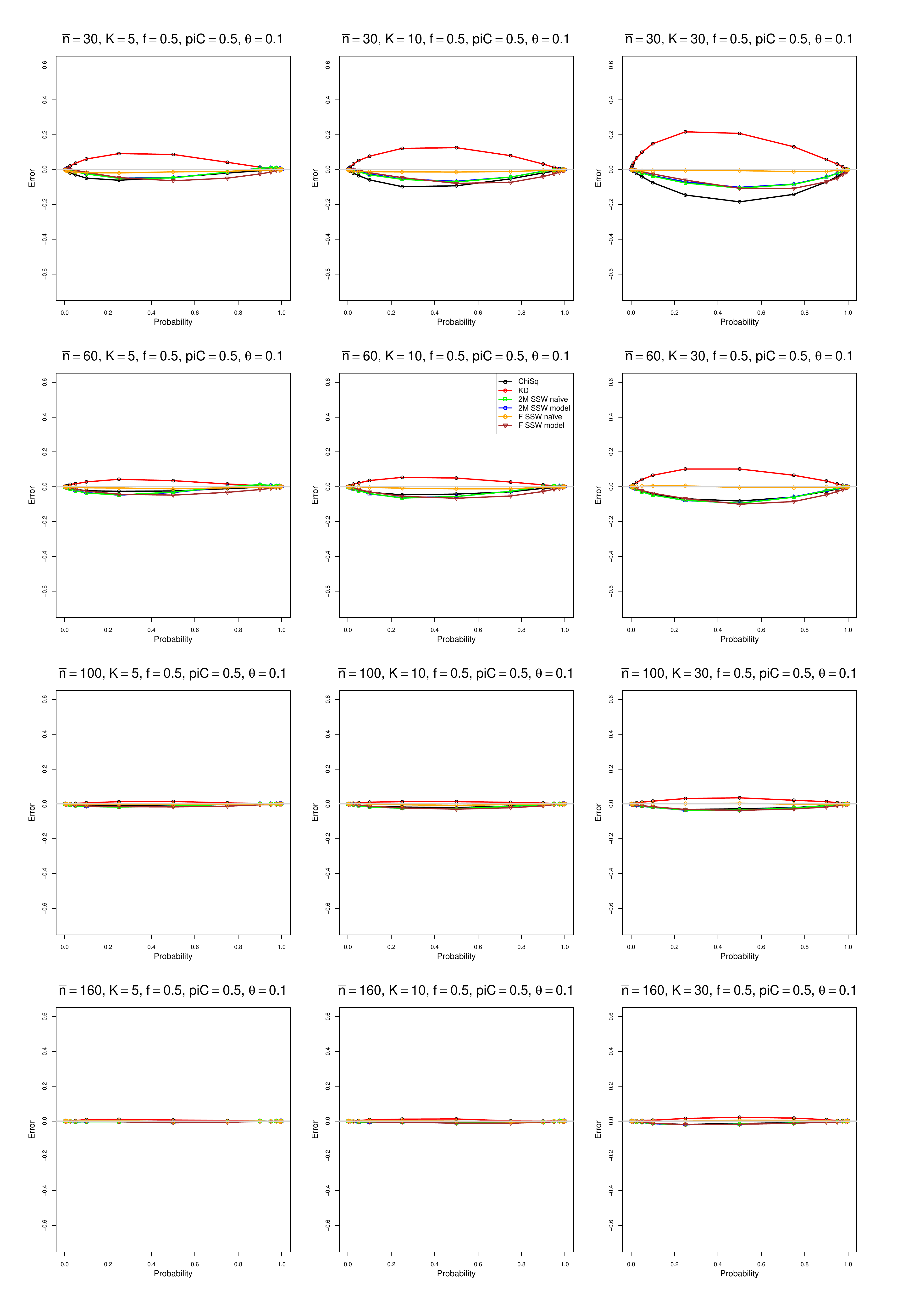}
	\caption{Plots of error in the level of the test for heterogeneity of LOR for six approximations for the null distribution of $Q$, $p_{iC} = .5$, $f = .5$, and $\theta = 0.1$, unequal sample sizes}
	\label{PPplot_piC_05theta=0.1_LOR_unequal_sample_sizes}
\end{figure}

\begin{figure}[ht]
	\centering
	\includegraphics[scale=0.33]{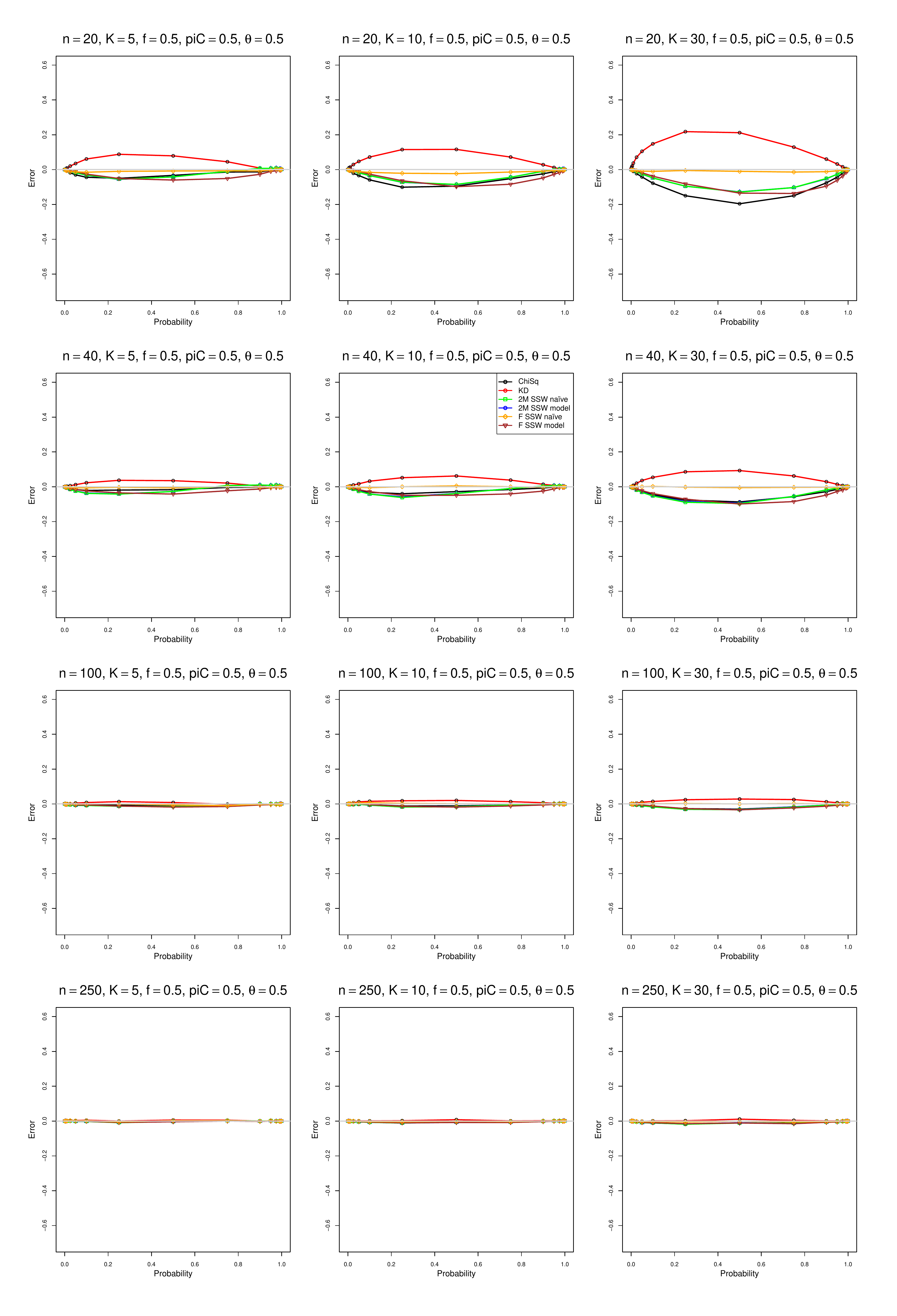}
	\caption{Plots of error in the level of the test for heterogeneity of LOR for six approximations for the null distribution of $Q$, $p_{iC} = .5$, $f = .5$, and $\theta = 0.5$, equal sample sizes}
	\label{PPplot_piC_05theta=0.5_LOR_equal_sample_sizes}
\end{figure}

\begin{figure}[ht]
	\centering
	\includegraphics[scale=0.33]{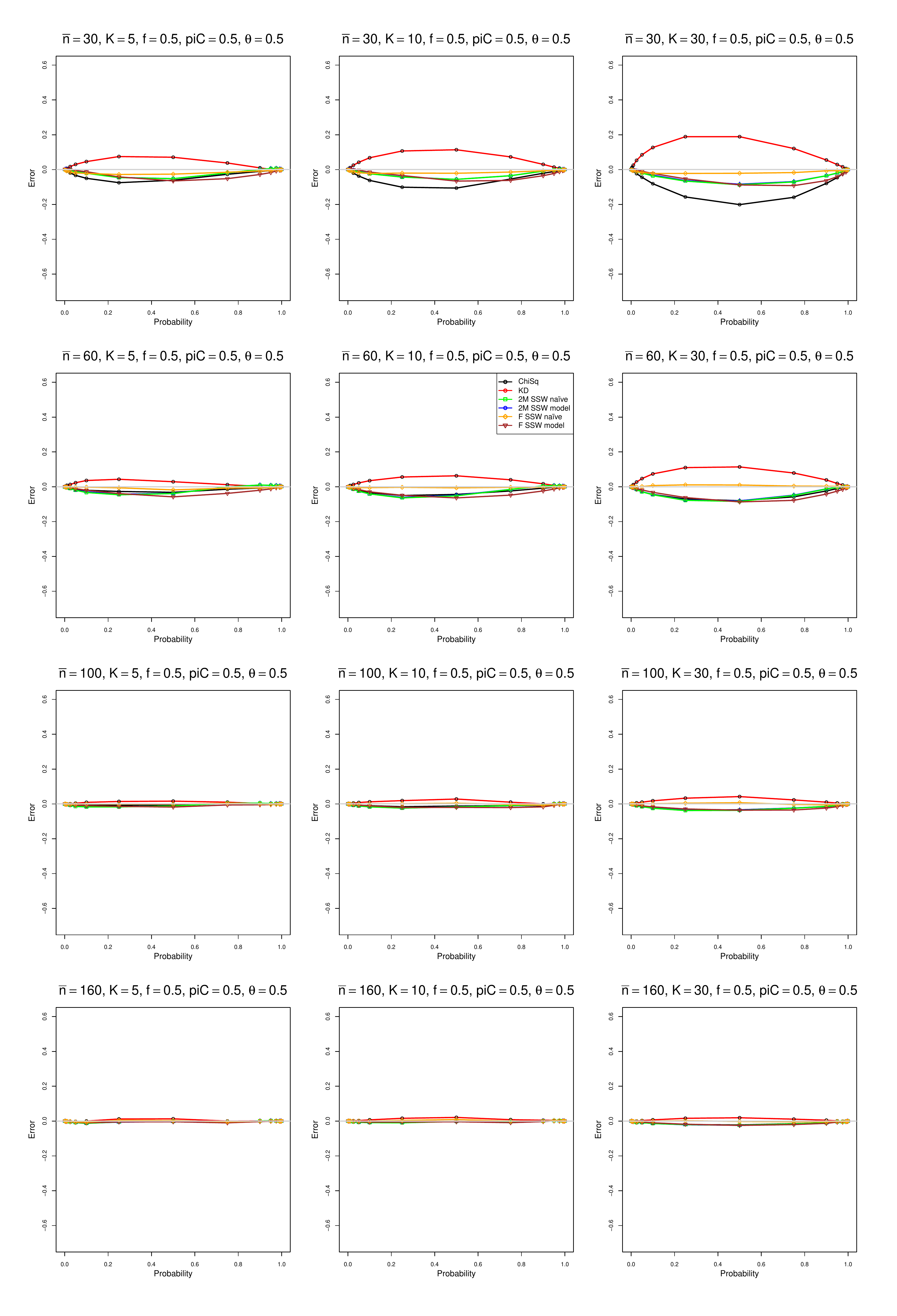}
	\caption{Plots of error in the level of the test for heterogeneity of LOR for six approximations for the null distribution of $Q$, $p_{iC} = .5$, $f = .5$, and $\theta = 0.5$, unequal sample sizes}
	\label{PPplot_piC_05theta=0.5_LOR_unequal_sample_sizes}
\end{figure}

\begin{figure}[ht]
	\centering
	\includegraphics[scale=0.33]{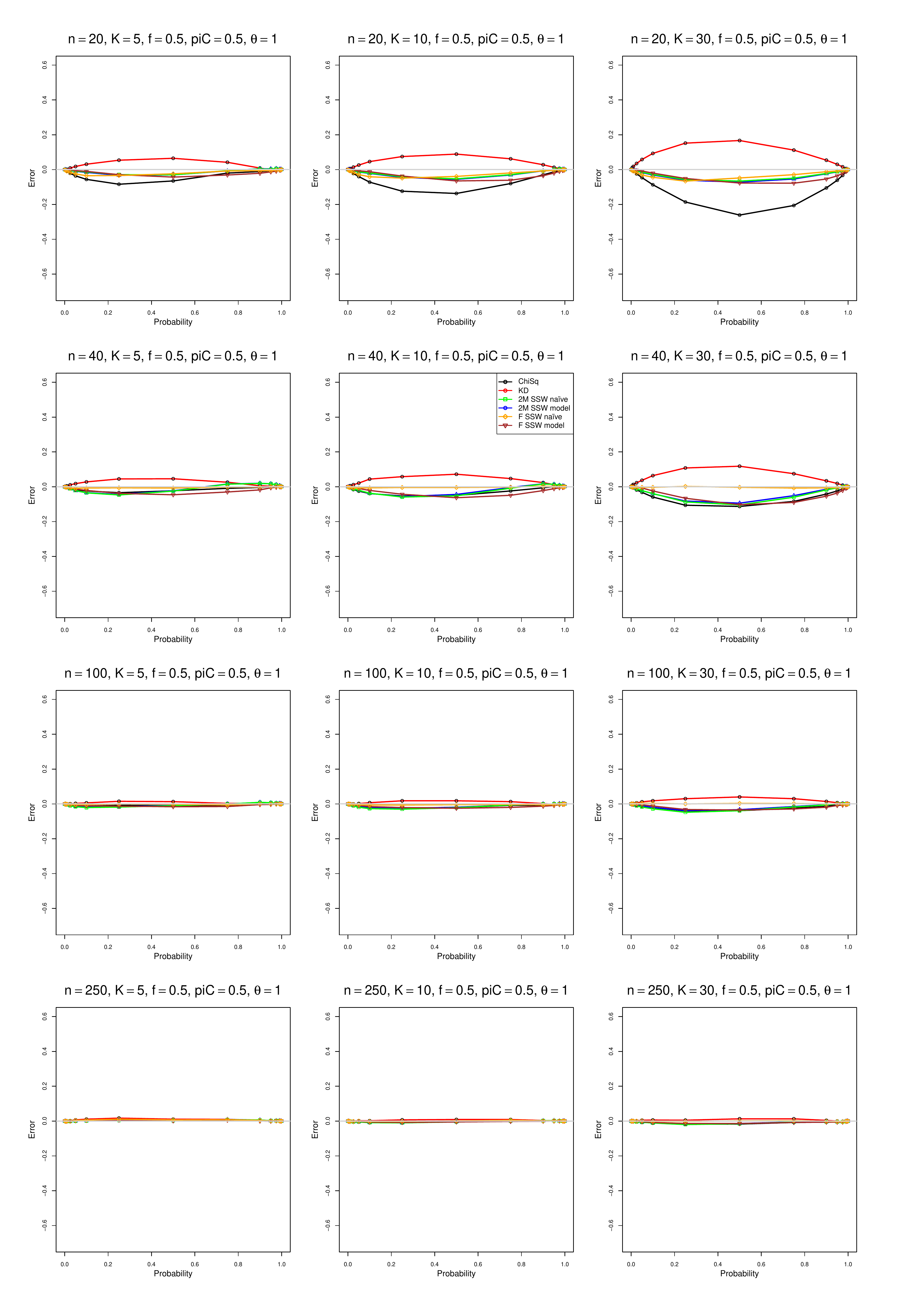}
	\caption{Plots of error in the level of the test for heterogeneity of LOR for six approximations for the null distribution of $Q$, $p_{iC} = .5$, $f = .5$, and $\theta = 1$, equal sample sizes}
	\label{PPplot_piC_05theta=1_LOR_equal_sample_sizes}
\end{figure}

\begin{figure}[ht]
	\centering
	\includegraphics[scale=0.33]{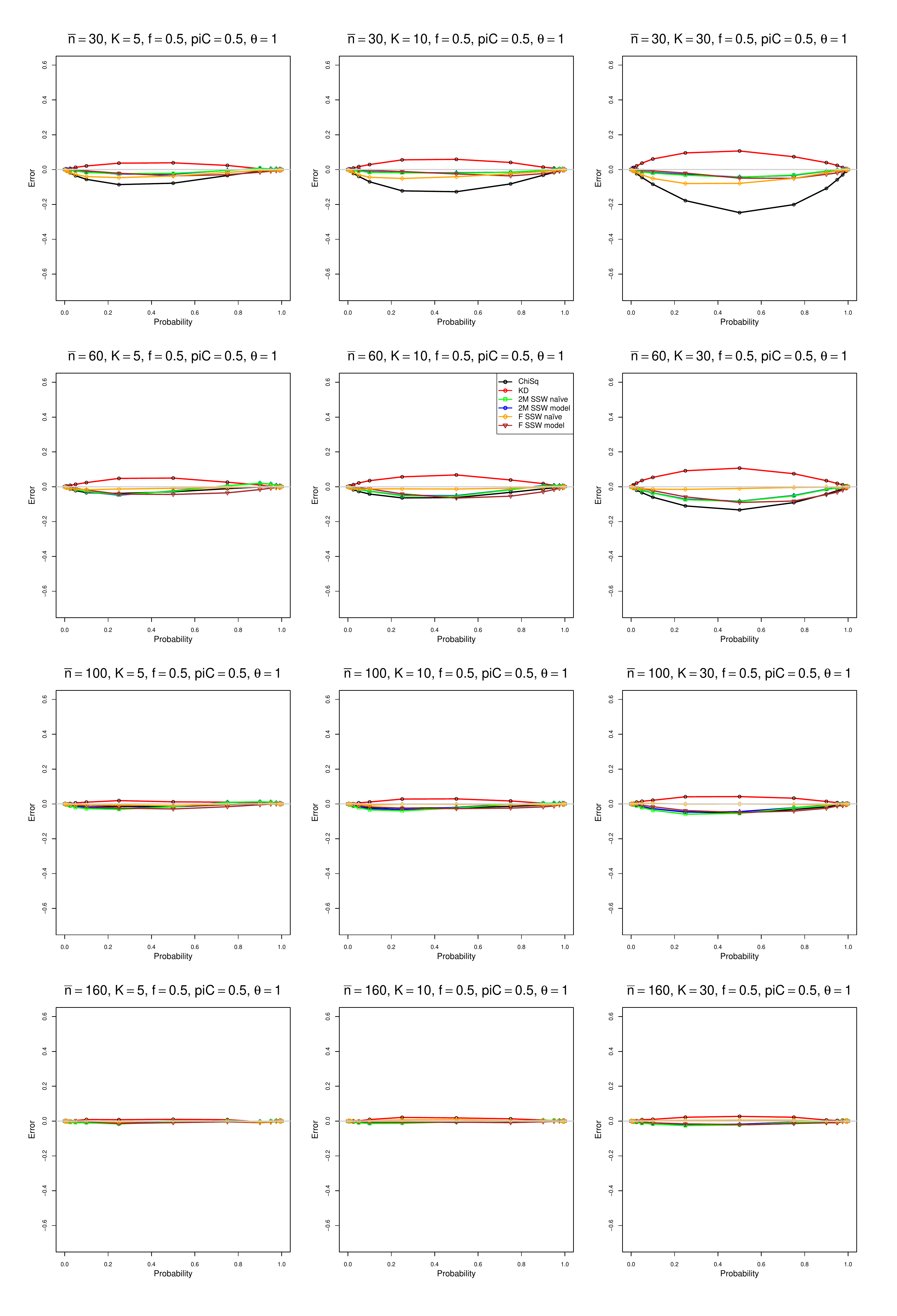}
	\caption{Plots of error in the level of the test for heterogeneity of LOR for six approximations for the null distribution of $Q$, $p_{iC} = .5$, $f = .5$, and $\theta = 1$, unequal sample sizes}
	\label{PPplot_piC_05theta=1_LOR_unequal_sample_sizes}
\end{figure}

\begin{figure}[ht]
	\centering
	\includegraphics[scale=0.33]{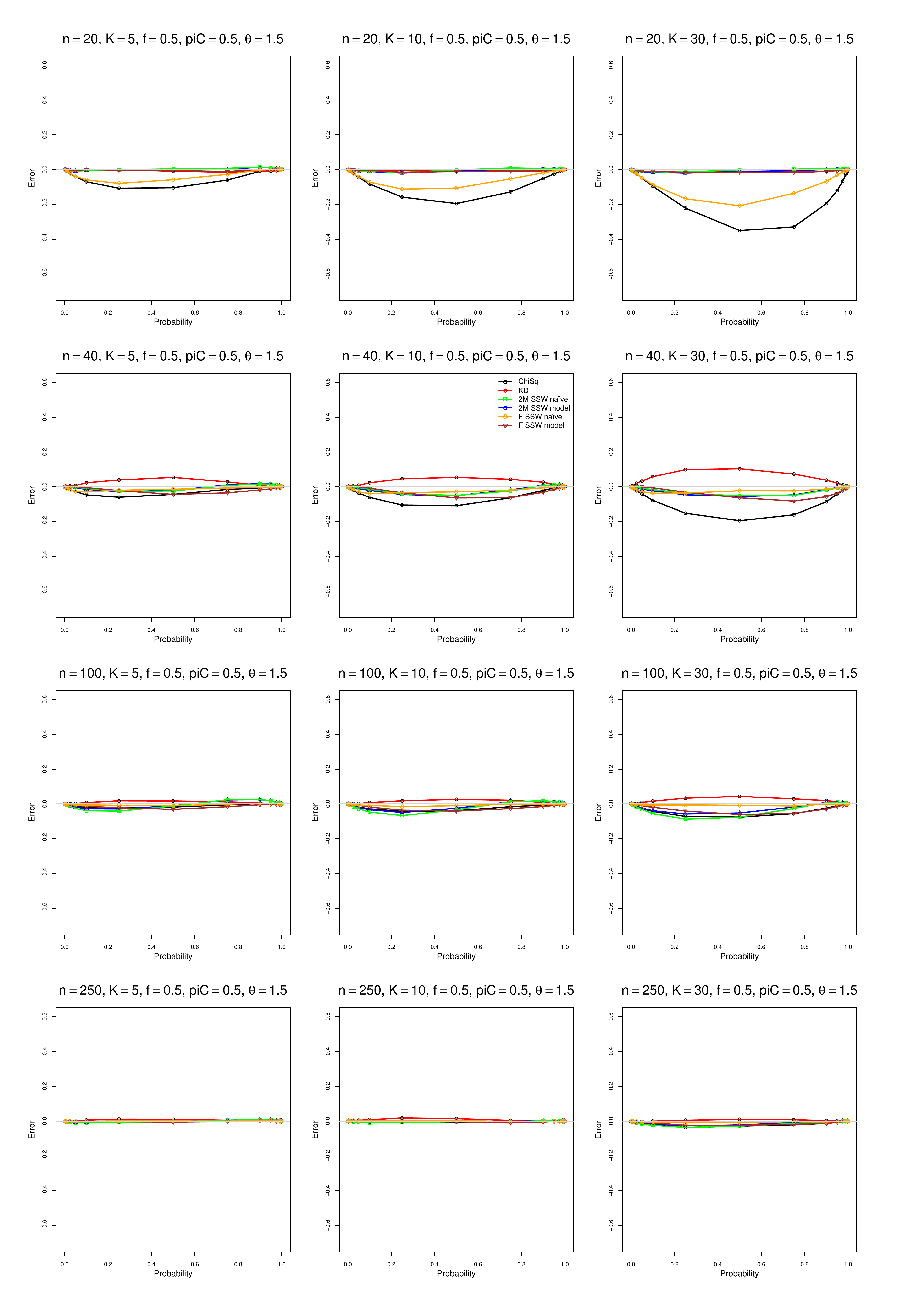}
	\caption{Plots of error in the level of the test for heterogeneity of LOR for six approximations for the null distribution of $Q$, $p_{iC} = .5$, $f = .5$, and $\theta = 1.5$, equal sample sizes}
	\label{PPplot_piC_05theta=1.5_LOR_equal_sample_sizes}
\end{figure}

\begin{figure}[ht]
	\centering
	\includegraphics[scale=0.33]{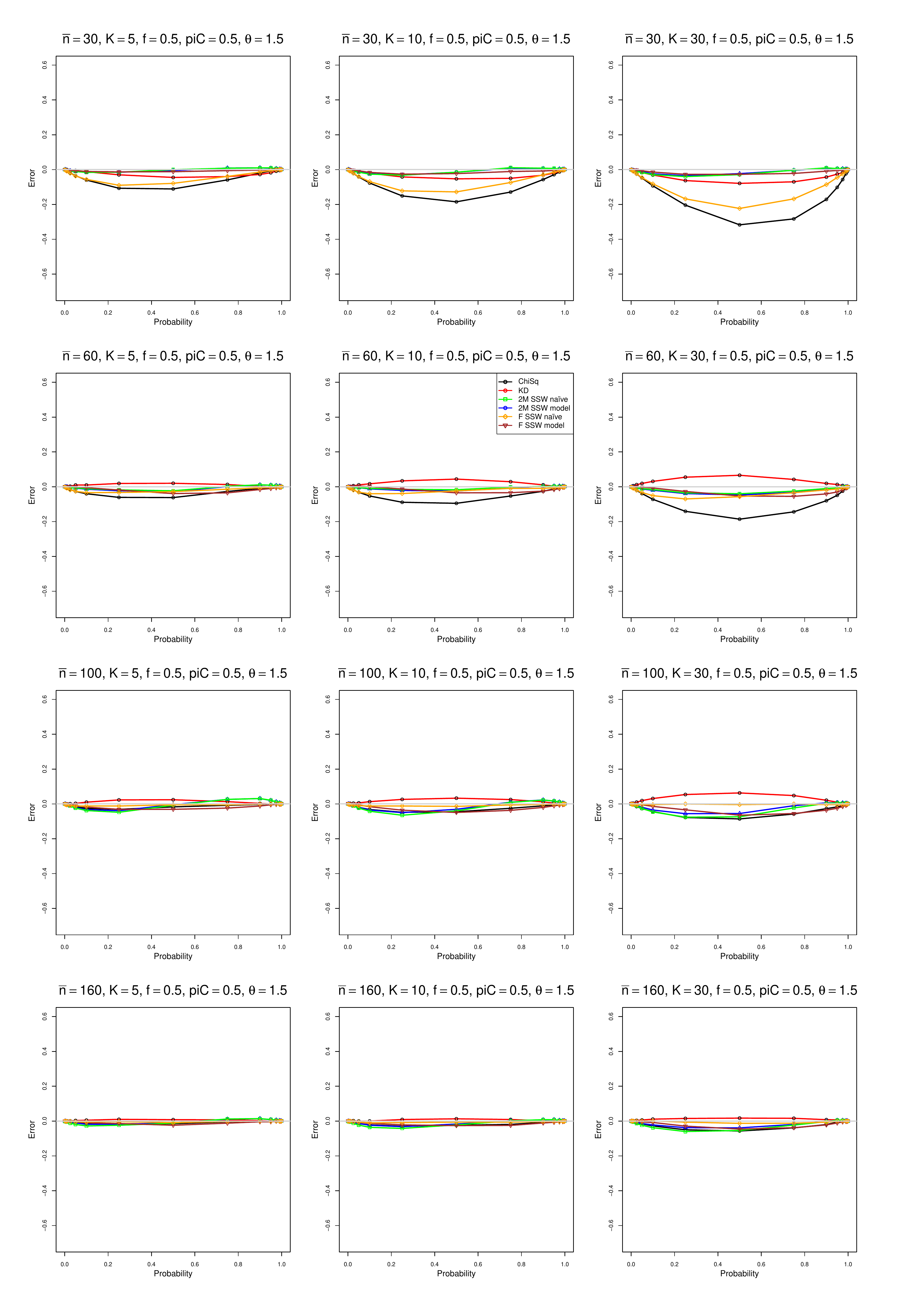}
	\caption{Plots of error in the level of the test for heterogeneity of LOR for six approximations for the null distribution of $Q$, $p_{iC} = .5$, $f = .5$, and $\theta = 1.5$, unequal sample sizes}
	\label{PPplot_piC_05theta=1.5_LOR_unequal_sample_sizes}
\end{figure}

\begin{figure}[ht]
	\centering
	\includegraphics[scale=0.33]{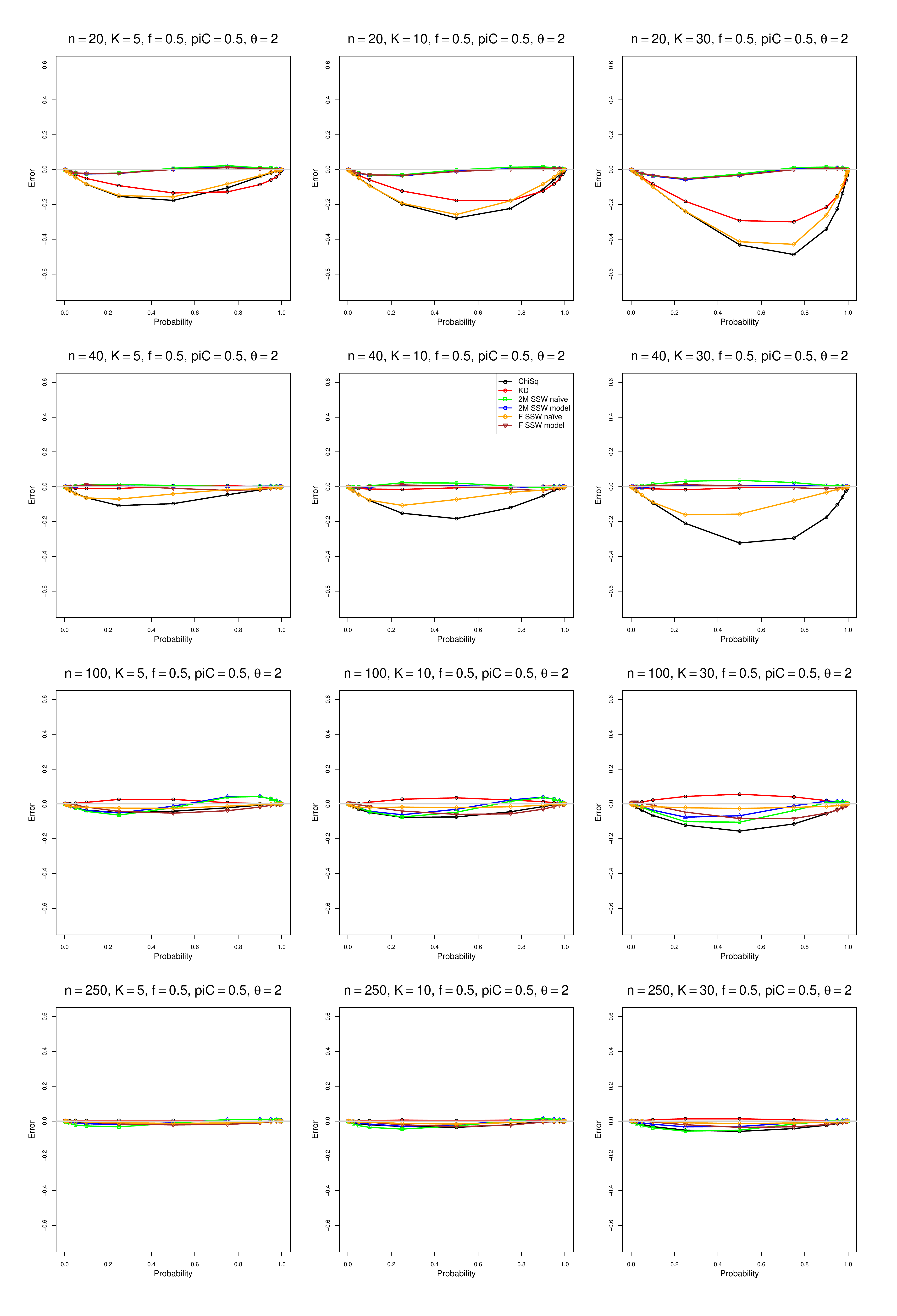}
	\caption{Plots of error in the level of the test for heterogeneity of LOR for six approximations for the null distribution of $Q$, $p_{iC} = .5$, $f = .5$, and $\theta = 2$, equal sample sizes}
	\label{PPplot_piC_05theta=2_LOR_equal_sample_sizes}
\end{figure}

\begin{figure}[ht]
	\centering
	\includegraphics[scale=0.33]{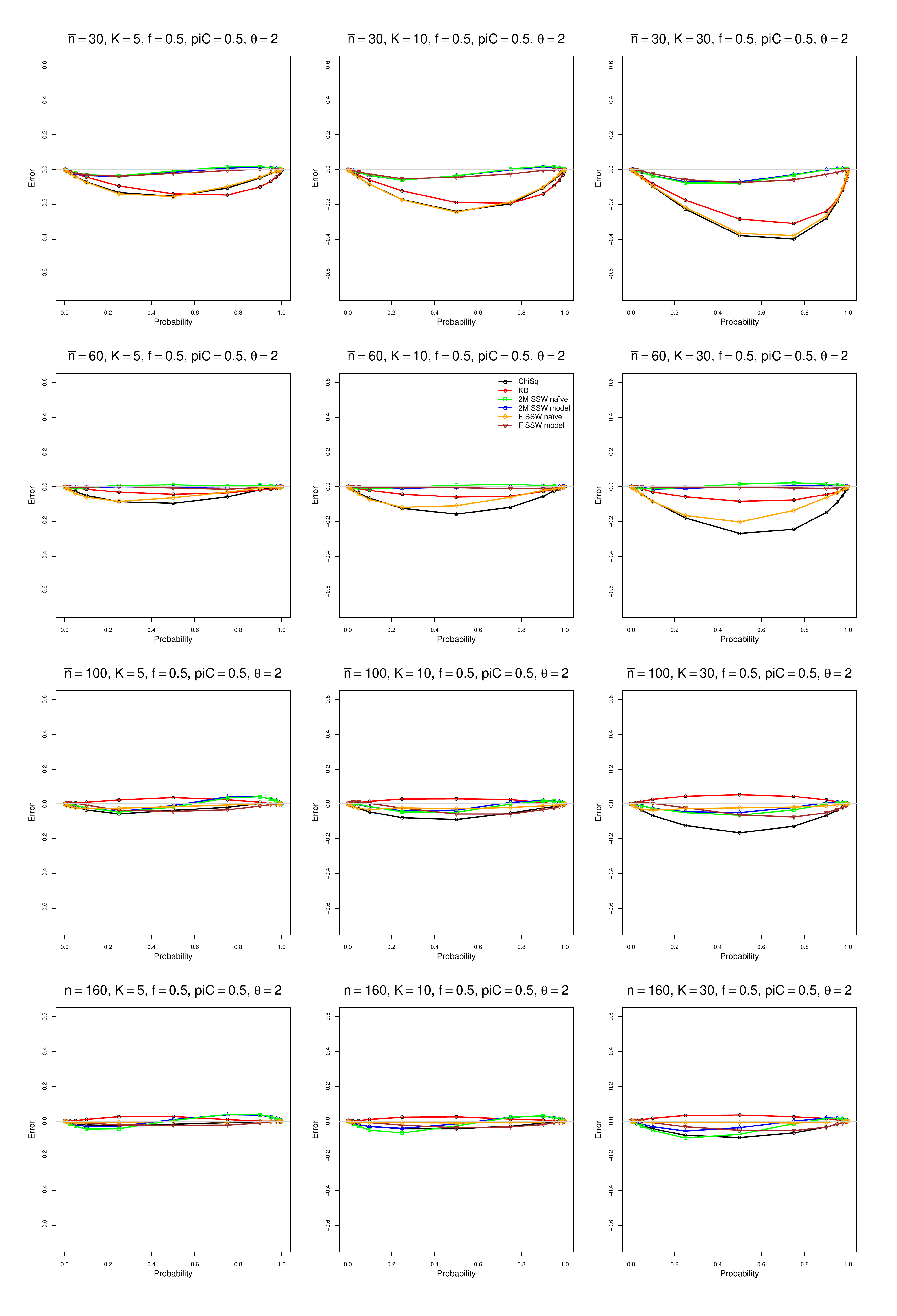}
	\caption{Plots of error in the level of the test for heterogeneity of LOR for six approximations for the null distribution of $Q$, $p_{iC} = .5$, $f = .5$, and $\theta = 2$, unequal sample sizes}
	\label{PPplot_piC_05theta=2_LOR_unequal_sample_sizes}
\end{figure}

\renewcommand{\thefigure}{B.\arabic{figure}}

\setcounter{figure}{0}
\setcounter{section}{0}
\clearpage

\section*{Appendix B: Empirical level at $\alpha = .05$, vs $\theta$, of the test for heterogeneity of LOR ($\tau^2 = 0$ versus $\tau^2 > 0$) based on approximations for the null distribution of $Q$}

Each figure corresponds to a value of the probability of an event in the Control arm $p_{iC}$  (= .1, .2, .5) and a choice of equal or unequal sample sizes ($n$ or $bar{n}$). \\
The fraction of each study's sample size in the Control arm  $f$ is held constant at 0.5.

For each combination of a value of $n$ (= 20, 40, 100, 250) or $\bar{n}$ (= 30, 60, 100, 160) and a value of $K$ (= 5, 10, 30), a panel plots the empirical level versus $\theta$ (= 0.0, 0.1, 0.5, 1, 1.5, 2).\\
The approximations for the null distribution of $Q$ are
\begin{itemize}
\item ChiSq (Chi-square approximation with $K-1$ df, inverse-variance weights)
\item KD (Kulinskaya-Dollinger (2015) approximation, inverse-variance weights)
\item 2M SSW na\"{i}ve (Two-moment gamma approximation, na\"{i}ve estimation of $p_{iT}$ from $X_{iT}$ and $n_{iT}$, effective-sample-size weights)
\item 2M SSW model (Two-moment gamma approximation, model-based estimation of $p_{iT}$, effective-sample-size weights)
\item F SSW  na\"{i}ve (Farebrother approximation, na\"{i}ve estimation of $p_{iT}$ from $X_{iT}$ and $n_{iT}$, effective-sample-size weights)
\item F SSW model (Farebrother approximation, model-based estimation of $p_{iT}$, effective-sample-size weights)
\end{itemize}

\clearpage
\renewcommand{\thefigure}{B.\arabic{figure}}

\begin{figure}[ht]
	\centering
	\includegraphics[scale=0.33]{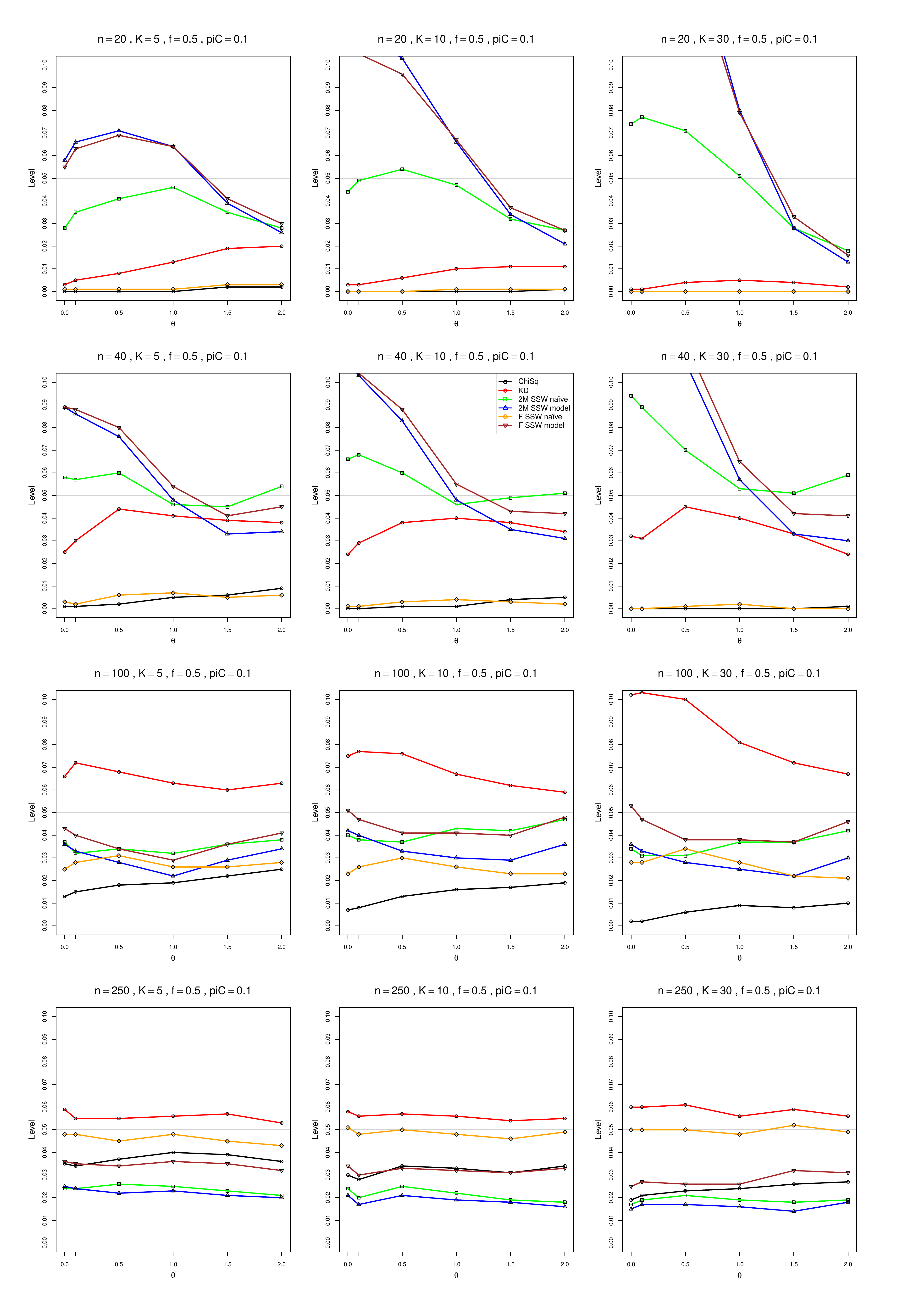}
	\caption{Q for LOR: actual level at $\alpha = .05$ for $p_{iC} = .1$ and $f = .5$, equal sample sizes
		\label{plotsPvalue_piC01andq05LOR_equal_sample_sizes}}
\end{figure}

\begin{figure}[ht]
	\centering
	\includegraphics[scale=0.33]{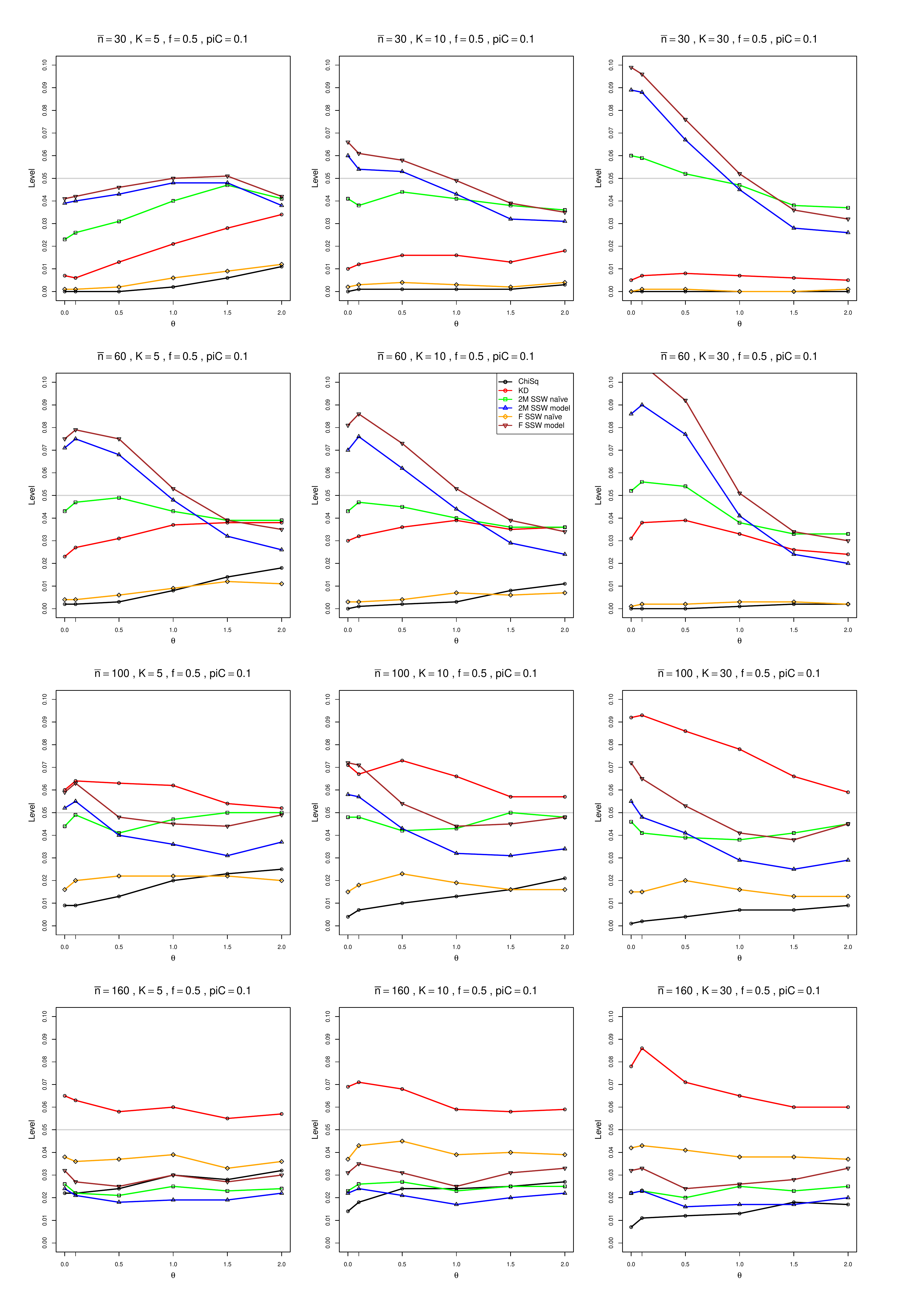}
	\caption{Q for LOR: actual level at $\alpha = .05$ for $p_{iC} = .1$ and $f = .5$, unequal sample sizes
		\label{plotsPvalue_piC01andq05LOR_unequal_sample_sizes}}
\end{figure}

\begin{figure}[t]
	\centering
	\includegraphics[scale=0.33]{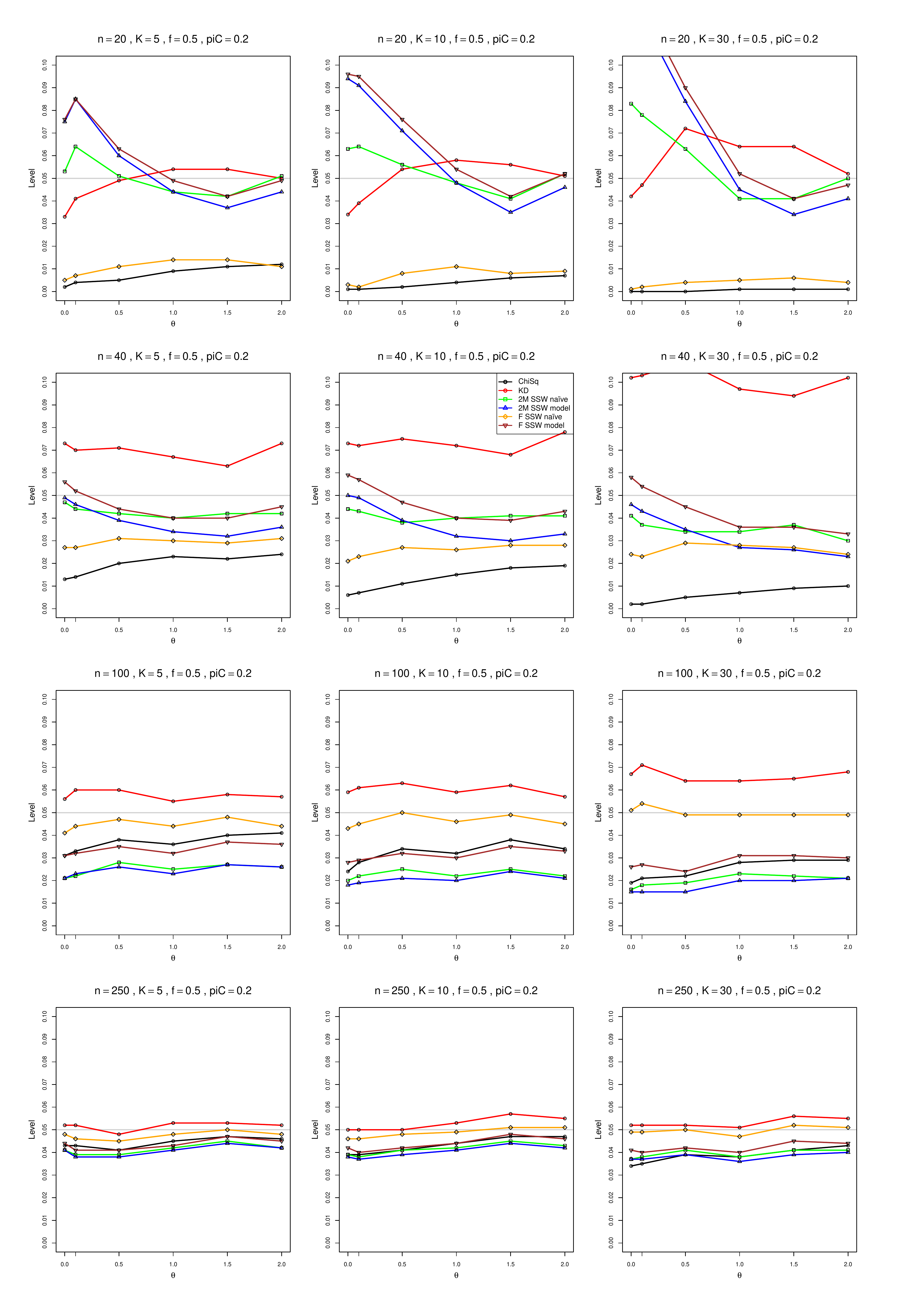}
	\caption{Q for LOR: actual level at $\alpha = .05$ for $p_{iC} = .2$ and $f = .5$, equal sample sizes
		\label{plotsPvalue_piC02andq05LOR_equal_sample_sizes}}
\end{figure}

\begin{figure}[ht]
	\centering
	\includegraphics[scale=0.33]{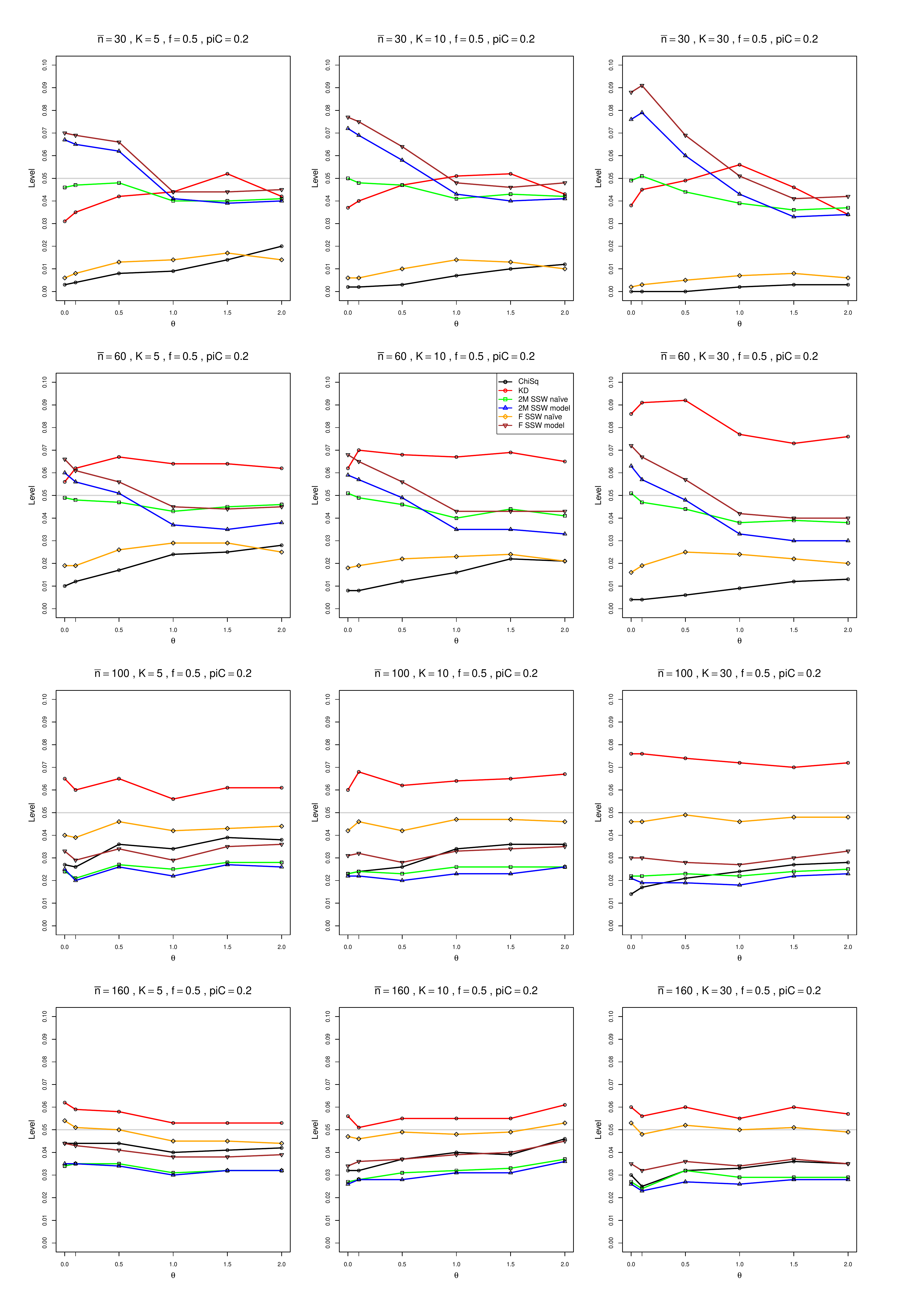}
	\caption{Q for LOR: actual level at $\alpha = .05$ for $p_{iC} = .2$ and $f = .5$, unequal sample sizes
		\label{plotsPvalue_piC02andq05LOR_unequal_sample_sizes}}
\end{figure}

\begin{figure}[ht]
	\centering
	\includegraphics[scale=0.33]{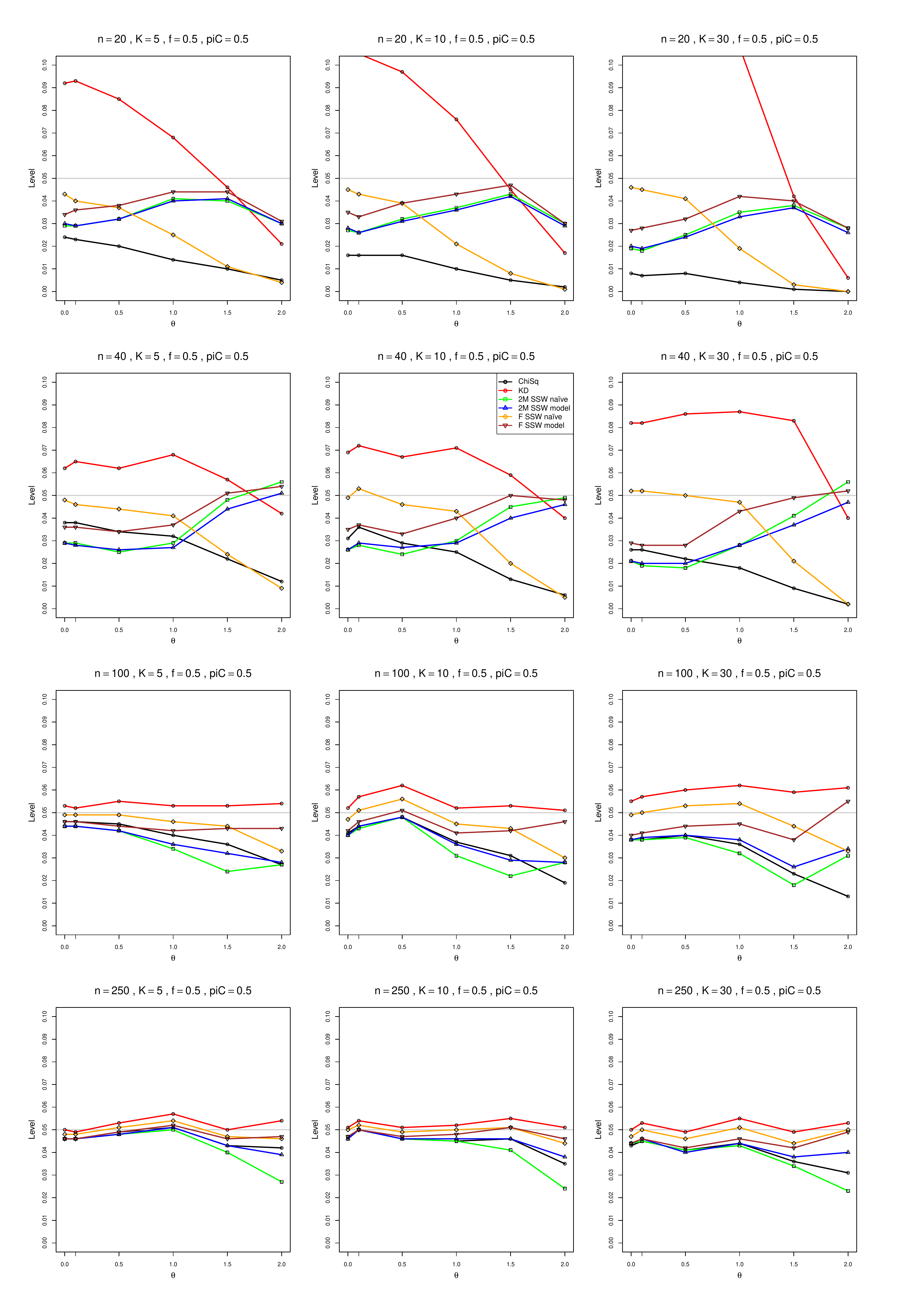}
	\caption{Q for LOR: actual level at $\alpha = .05$ for $p_{iC} = .5$ and $f = .5$, equal sample sizes
		\label{plotsPvalue_piC05andq05LOR_equal_sample_sizes}}
\end{figure}

\begin{figure}[ht]
	\centering
	\includegraphics[scale=0.33]{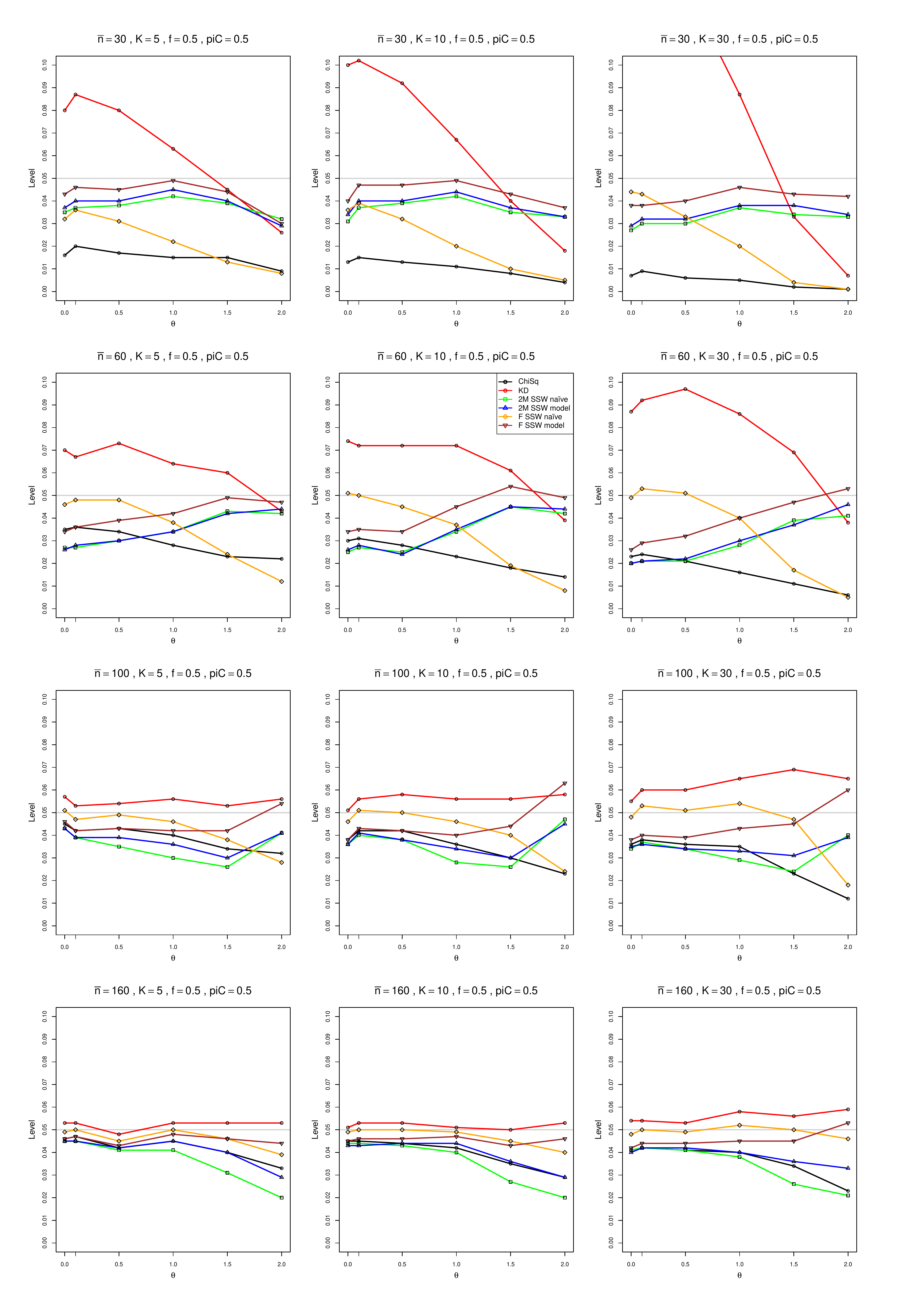}
	\caption{Q for LOR: actual level at $\alpha = .05$ for $p_{iC} = .5$ and $f = .5$, unequal sample sizes
		\label{plotsPvalue_piC05andq05LOR_unequal_sample_sizes}}
\end{figure}

\renewcommand{\thefigure}{C.\arabic{figure}}

\setcounter{figure}{0}
\setcounter{section}{0}
\clearpage

\section*{Appendix C: Empirical power at nominal level  $\alpha = .05$, vs $\tau^2$, of the test for heterogeneity of LOR ($\tau^2 = 0$ versus $\tau^2 > 0$) based on approximations for the null distribution of $Q$}

Each figure corresponds to a value of the probability of an event in the Control arm $p_{iC}$  (= .1, .2, .5) and a value of the overall LOR $\theta$ (= 0.0, 0.1, 0.5, 1, 1.5, 2). \\
The fraction of each study's sample size in the Control arm  $f$ is held constant at 0.5.

For each combination of a value of $n$ (= 20, 40, 100, 250)  and a value of $K$ (= 5, 10, 30), a panel plots the actual level versus $\tau^2$ (= 0.0, 0.1, 0.2, 0.3, 0.4, 0.5, 0.6, 0.7, 0.8, 0.9,1.0).\\
The approximations for the null distribution of $Q$ are
\begin{itemize}
\item ChiSq (Chi-square approximation with $K-1$ df, inverse-variance weights)
\item KD (Kulinskaya-Dollinger (2015) approximation, inverse-variance weights)
\item 2M SSW na\"{i}ve (Two-moment gamma approximation, na\"{i}ve estimation of $p_{iT}$ from $X_{iT}$ and $n_{iT}$, effective-sample-size weights)
\item 2M SSW model (Two-moment gamma approximation, model-based estimation of $p_{iT}$, effective-sample-size weights)
\item F SSW  na\"{i}ve (Farebrother approximation, na\"{i}ve estimation of $p_{iT}$ from $X_{iT}$ and $n_{iT}$, effective-sample-size weights)
\item F SSW model (Farebrother approximation, model-based estimation of $p_{iT}$, effective-sample-size weights)
\end{itemize}

\clearpage
\renewcommand{\thefigure}{C.\arabic{figure}}

\begin{figure}[ht]
	\centering
	\includegraphics[scale=0.33]{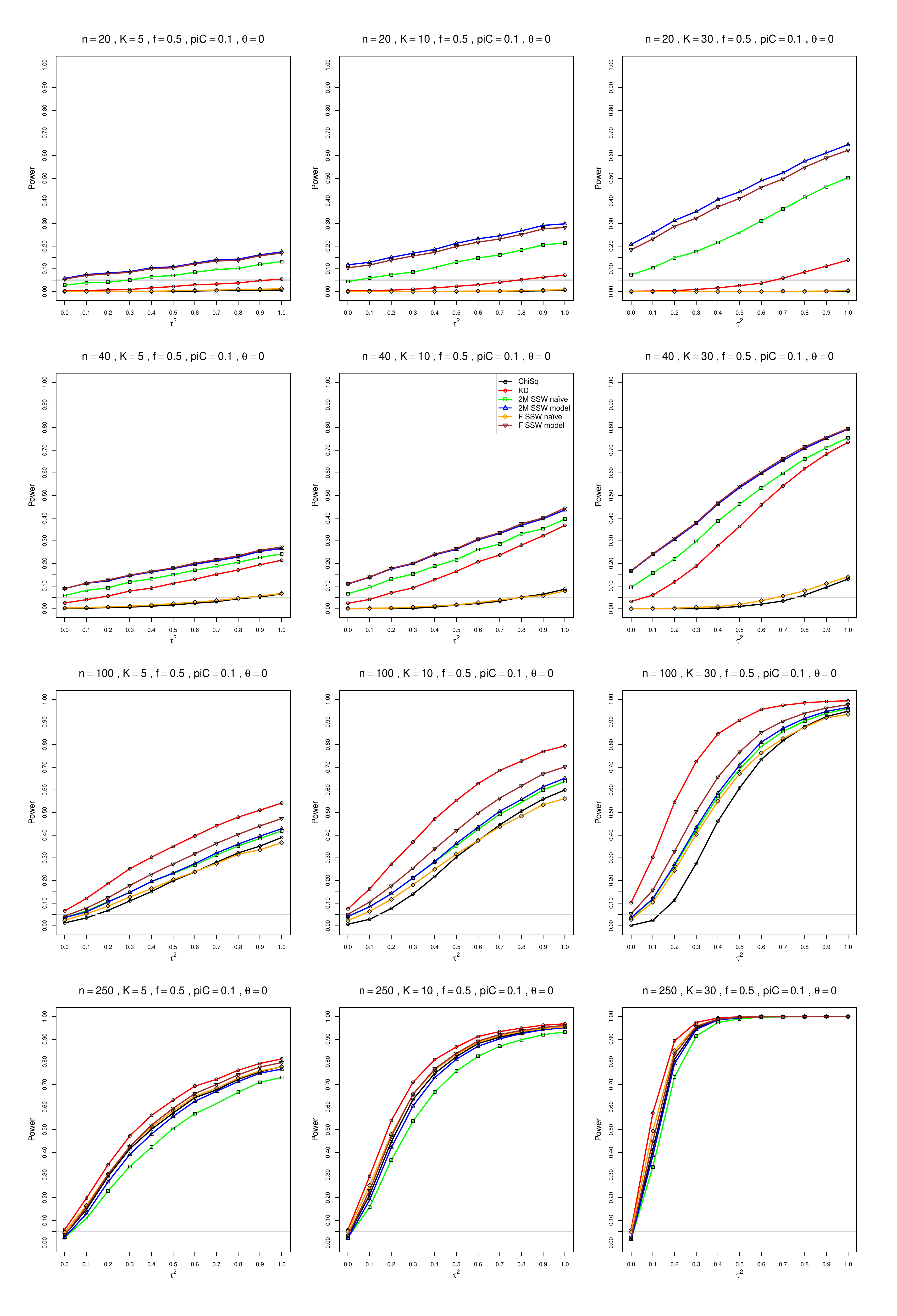}
	\caption{Q for LOR: empirical power at nominal level $\alpha = .05$ for $p_{iC} = .1$, $\theta=0$ and $f = .5$, equal sample sizes
		\label{PowerPlotAtNominal005_piC01andTheta0LOR_equal_sample_sizes}}
\end{figure}

\begin{figure}[t]
	\centering
	\includegraphics[scale=0.33]{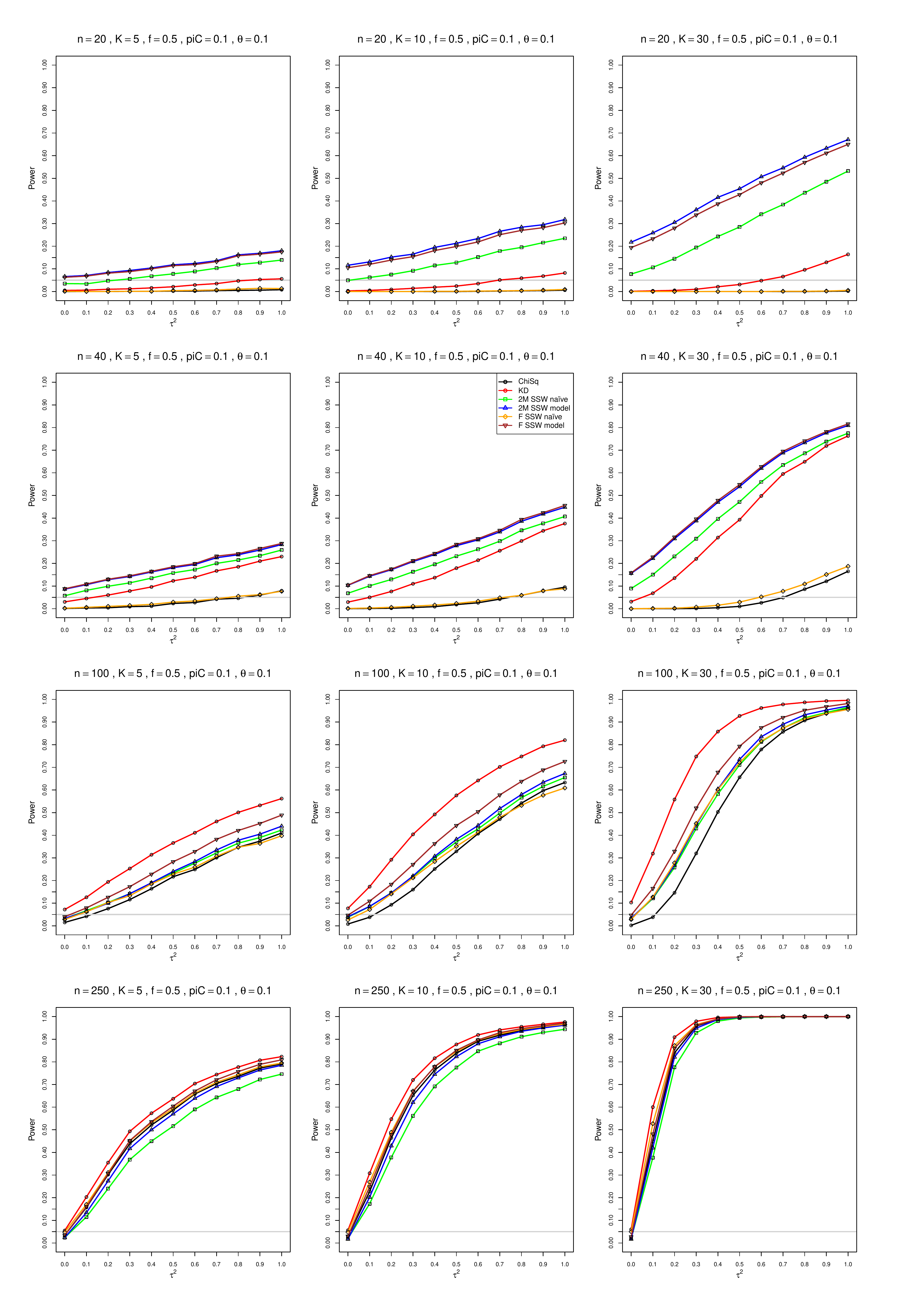}
	\caption{Q for LOR: empirical power at nominal level $\alpha = .05$ for $p_{iC} = .1$, $\theta=0.1$ and $f = .5$, equal sample sizes
		\label{PowerPlotAtNominal005_piC01andTheta01LOR_equal_sample_sizes}}
\end{figure}

\begin{figure}[ht]
	\centering
	\includegraphics[scale=0.33]{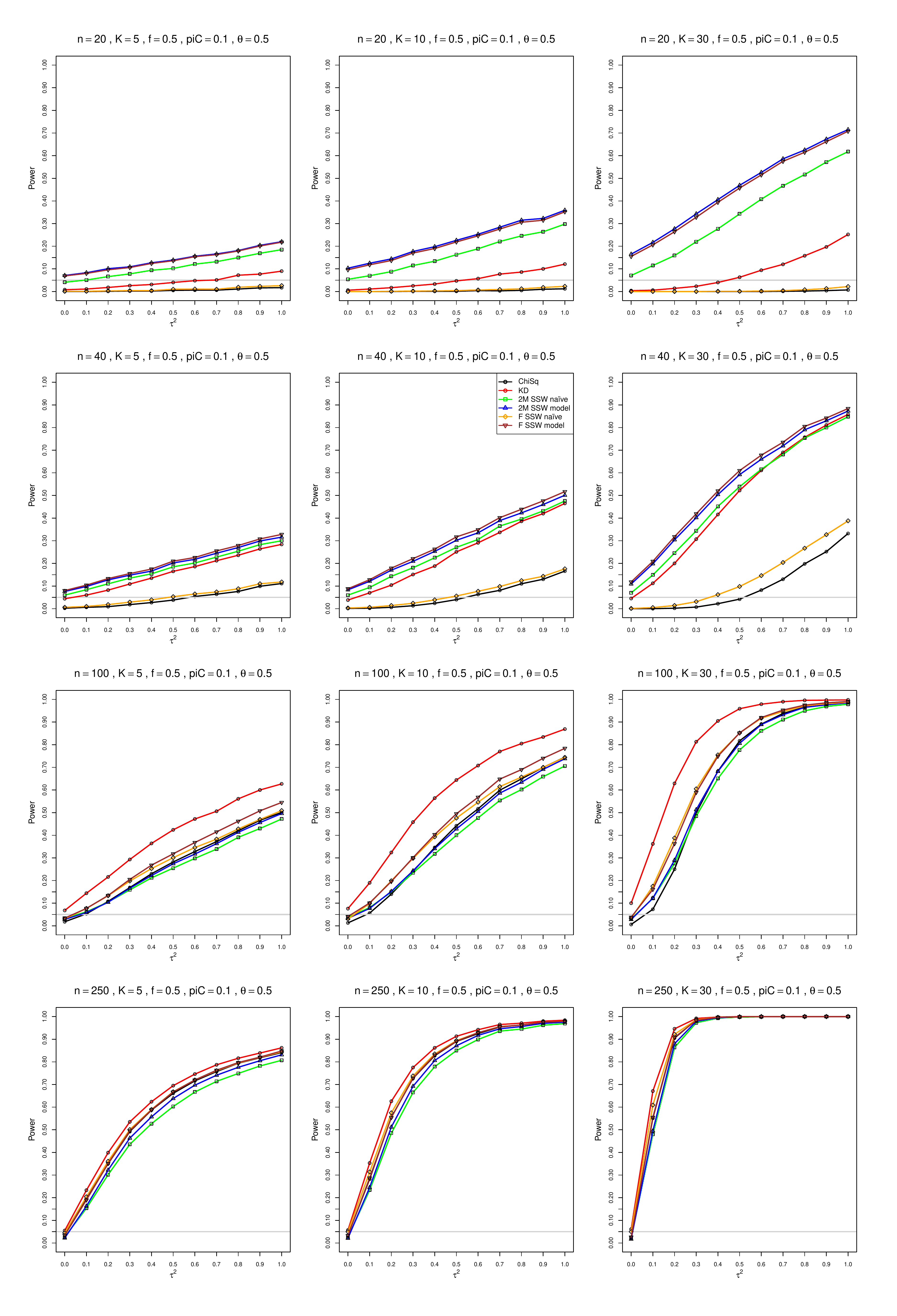}
	\caption{Q for LOR: empirical power at nominal level $\alpha = .05$ for $p_{iC} = .1$, $\theta=0.5$ and $f = .5$, equal sample sizes
		\label{PowerPlotAtNominal005_piC01andTheta05LOR_equal_sample_sizes}}
\end{figure}

\begin{figure}[ht]
	\centering
	\includegraphics[scale=0.33]{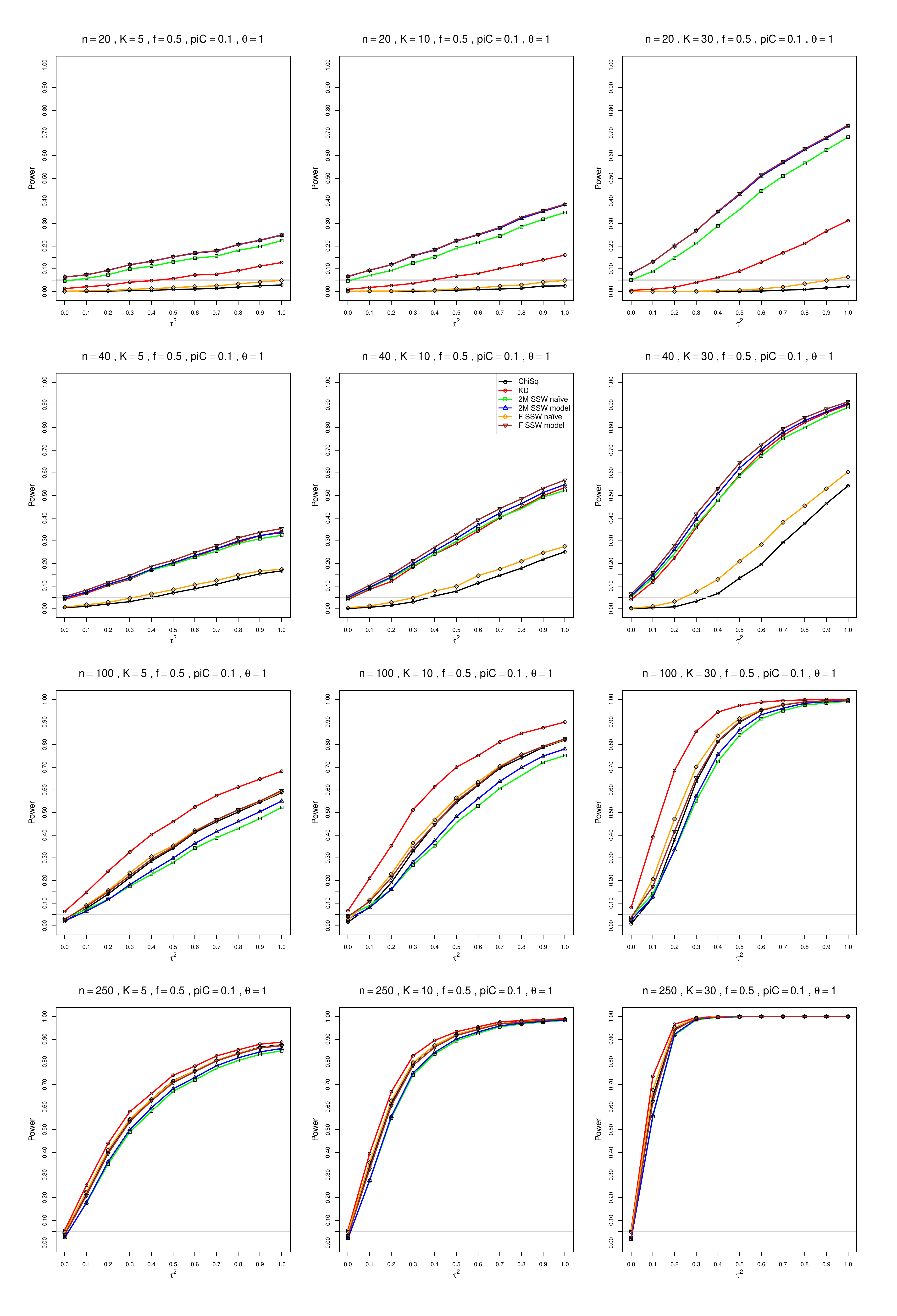}
	\caption{Q for LOR: empirical power at nominal level $\alpha = .05$ for $p_{iC} = .1$, $\theta=1.0$ and $f = .5$, equal sample sizes
		\label{PowerPlotAtNominal005_piC01andTheta1LOR_equal_sample_sizes}}
\end{figure}

\begin{figure}[ht]
	\centering
	\includegraphics[scale=0.33]{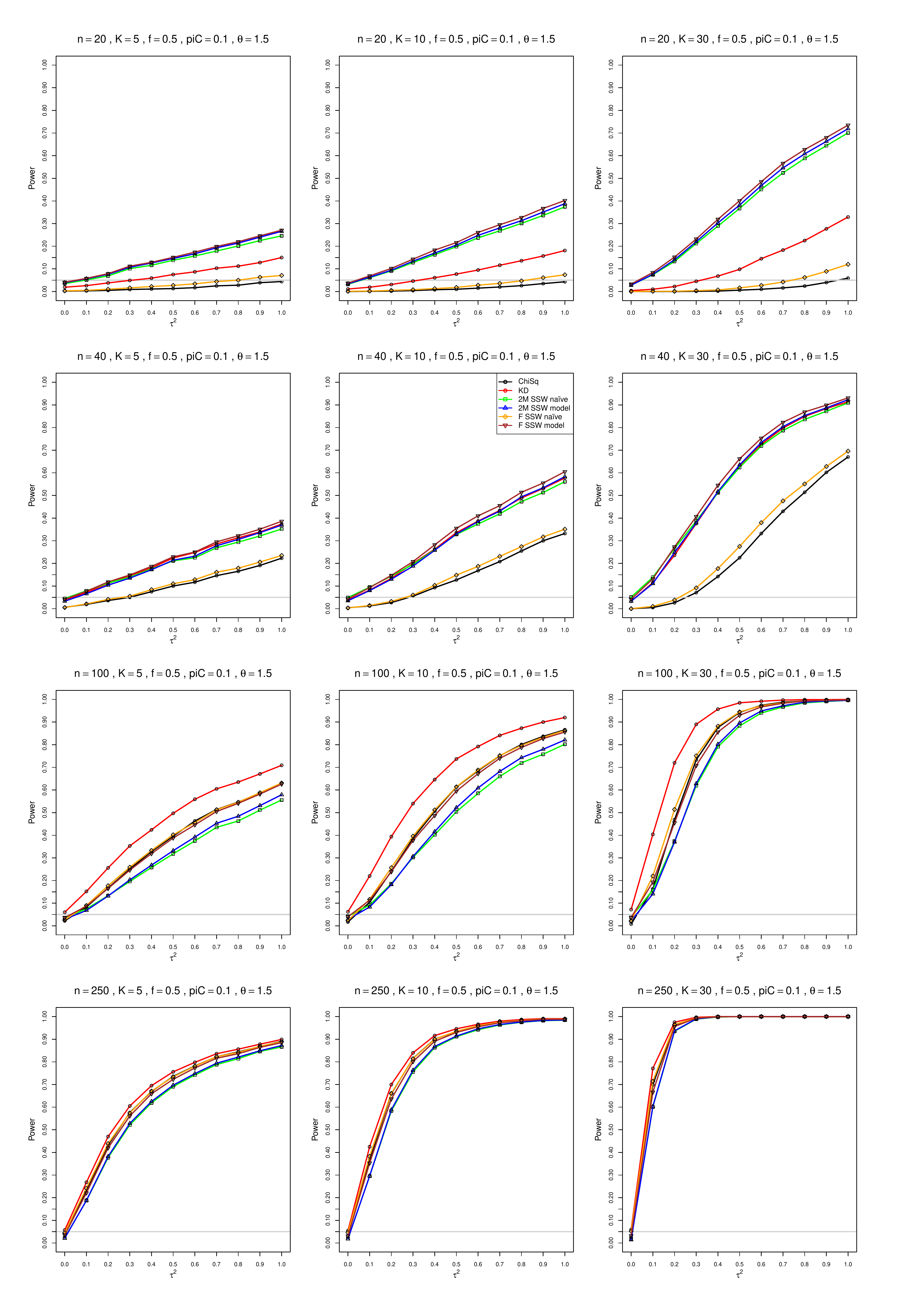}
	\caption{Q for LOR: empirical power at nominal level $\alpha = .05$ for $p_{iC} = .1$, $\theta=1.5$ and $f = .5$, equal sample sizes
		\label{PowerPlotAtNominal005_piC01andTheta1.5LOR_equal_sample_sizes}}
\end{figure}

\begin{figure}[ht]
	\centering
	\includegraphics[scale=0.33]{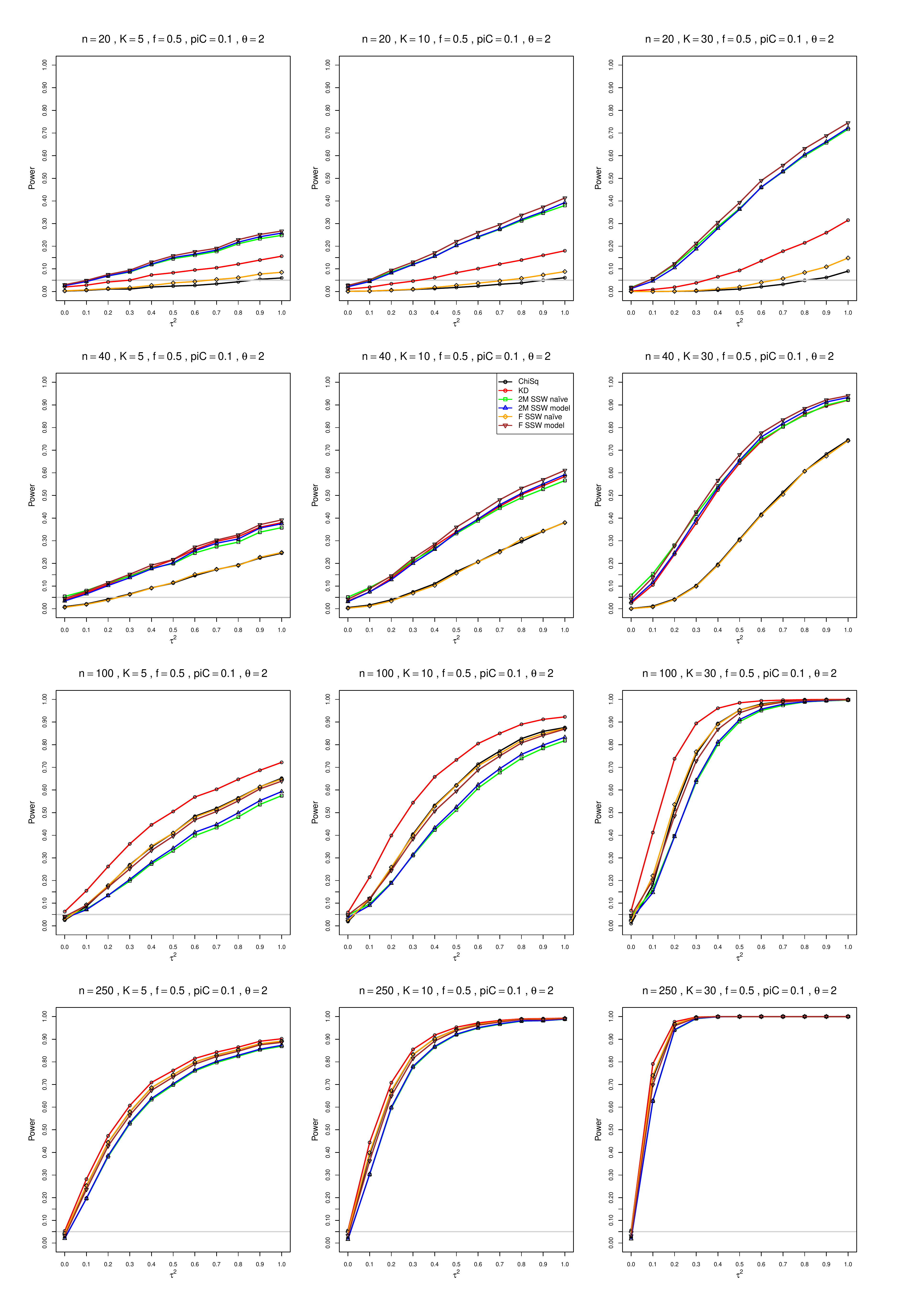}
	\caption{Q for LOR: empirical power at nominal level $\alpha = .05$ for $p_{iC} = .1$, $\theta=2.0$ and $f = .5$, equal sample sizes
		\label{PowerPlotAtNominal005_piC01andTheta2LOR_equal_sample_sizes}}
\end{figure}
\begin{figure}[ht]
	\centering
	\includegraphics[scale=0.33]{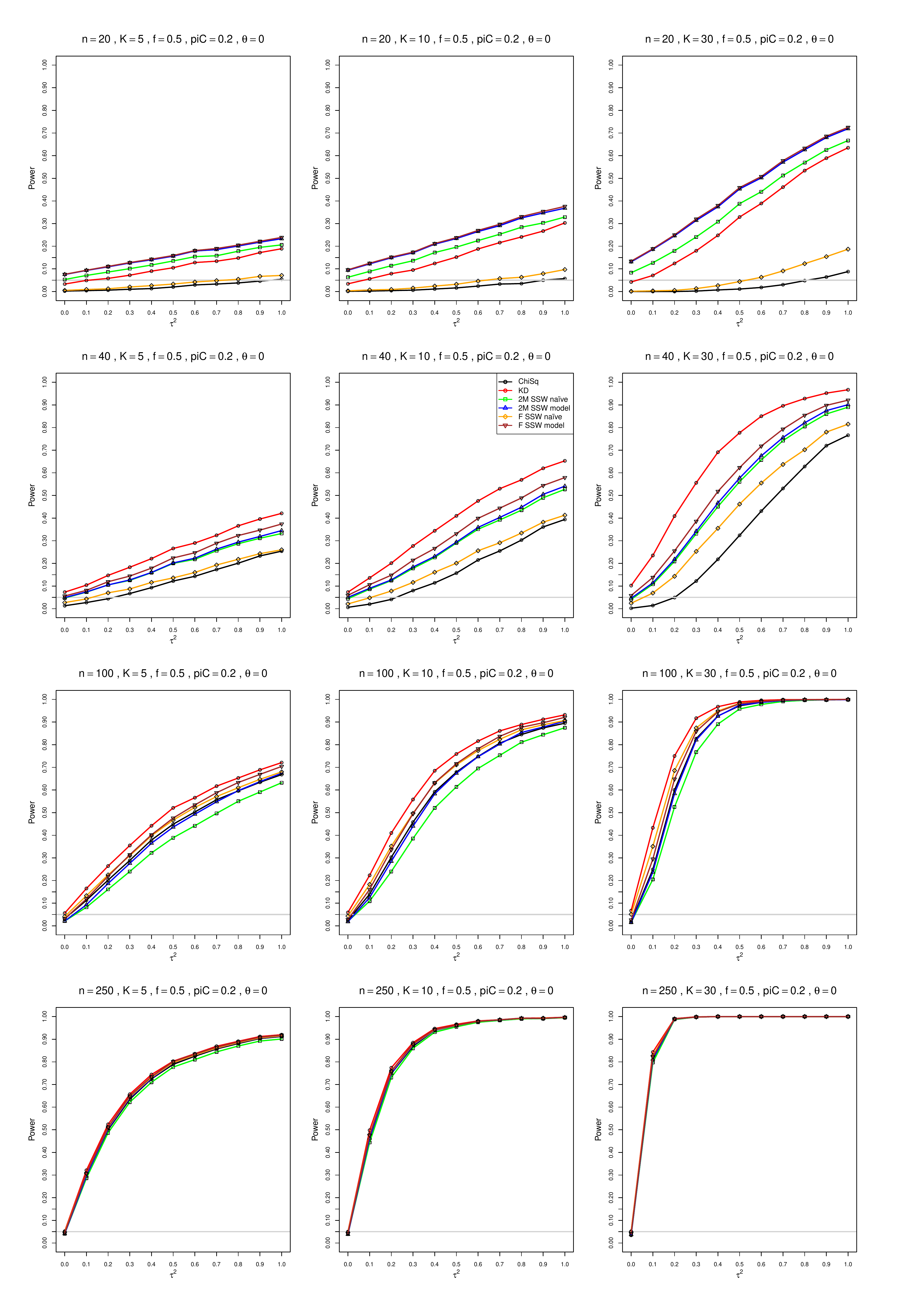}
	\caption{Q for LOR: empirical power at nominal level $\alpha = .05$ for $p_{iC} = .2$, $\theta=0$ and $f = .5$, equal sample sizes
		\label{PowerPlotAtNominal005_piC02andTheta0LOR_equal_sample_sizes}}
\end{figure}

\begin{figure}[t]
	\centering
	\includegraphics[scale=0.33]{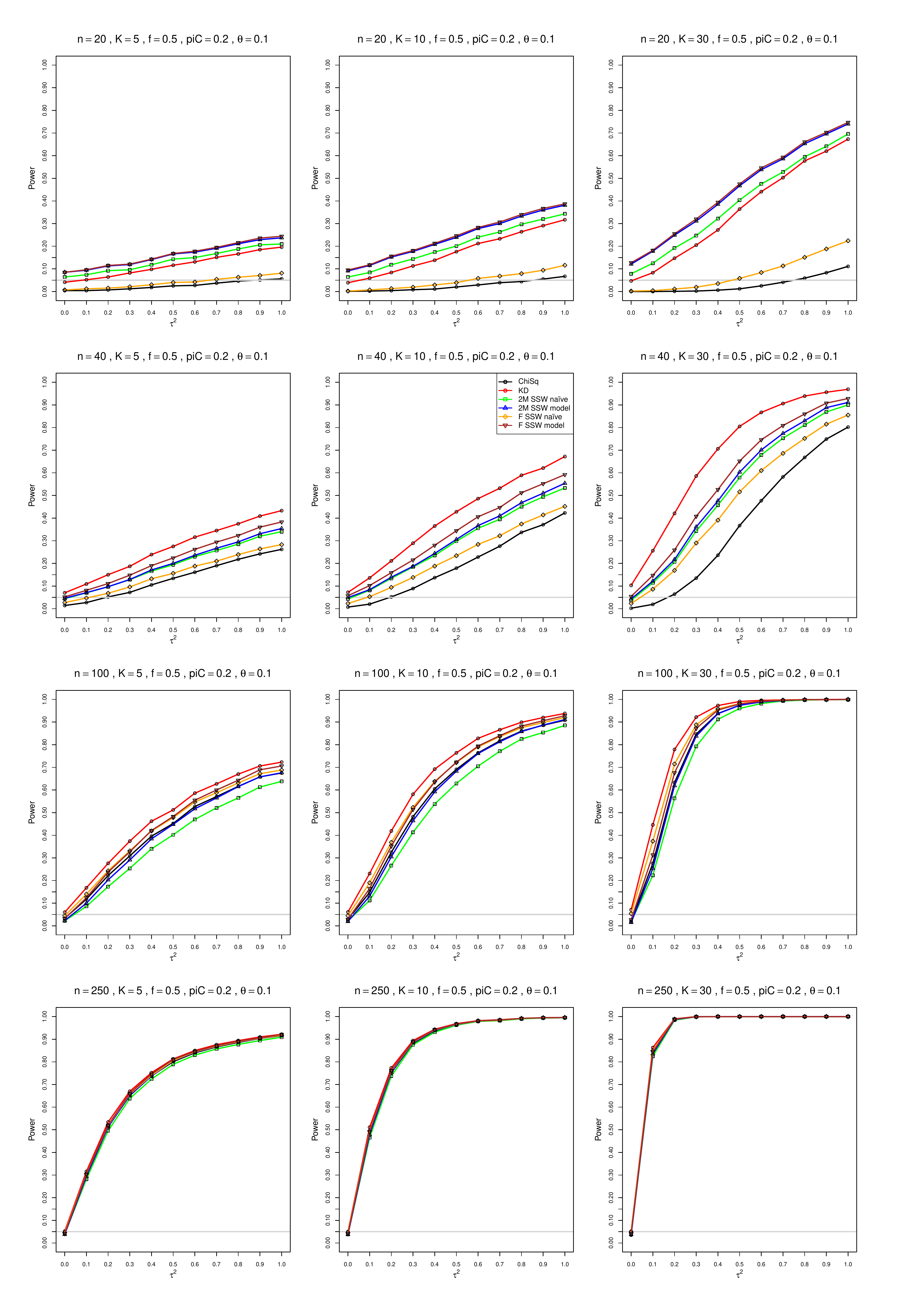}
	\caption{Q for LOR: empirical power at nominal level $\alpha = .05$ for $p_{iC} = .2$, $\theta=0.1$ and $f = .5$, equal sample sizes
		\label{PowerPlotAtNominal005_piC02andTheta01LOR_equal_sample_sizes}}
\end{figure}

\begin{figure}[ht]
	\centering
	\includegraphics[scale=0.33]{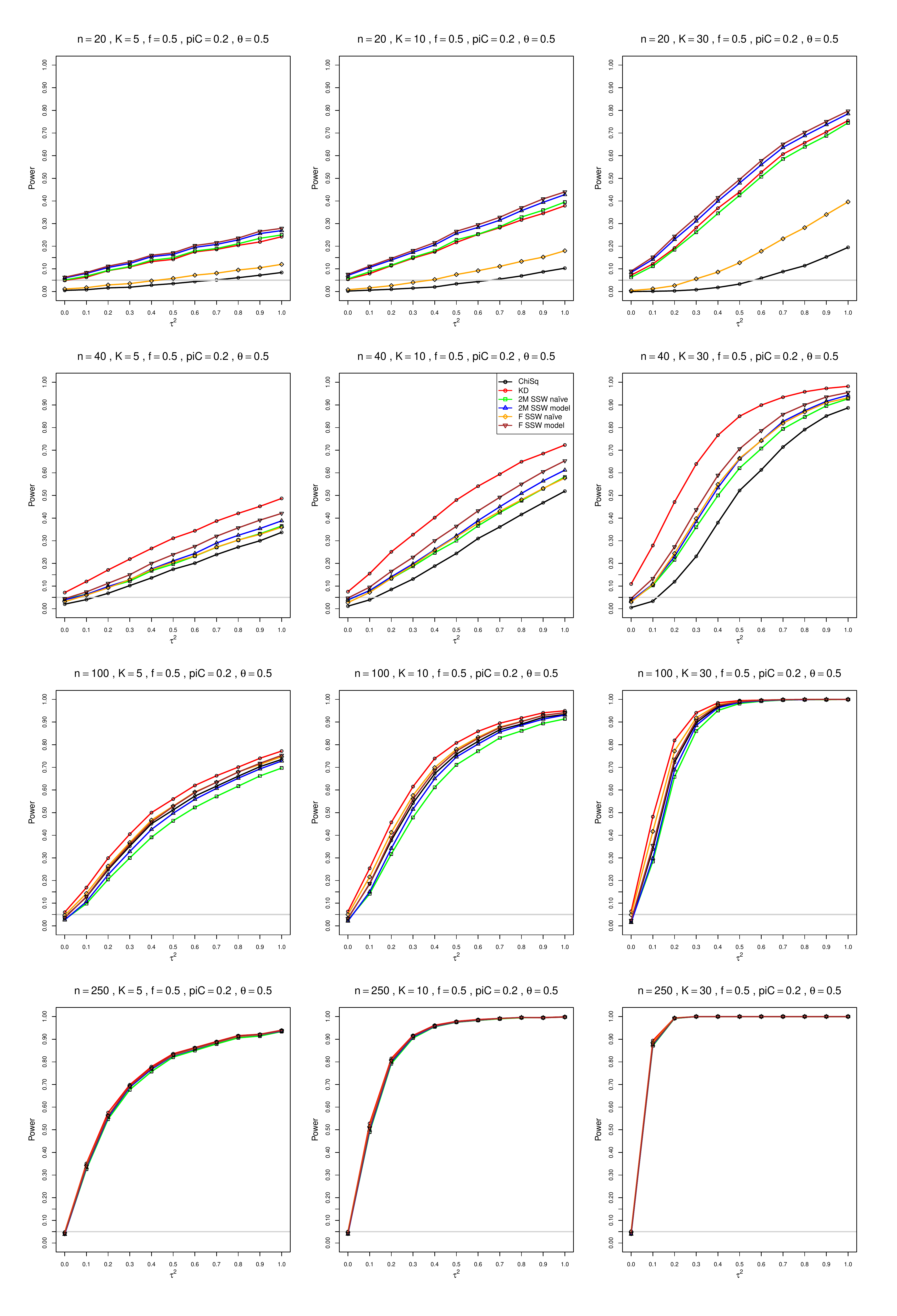}
	\caption{Q for LOR: empirical power at nominal level $\alpha = .05$ for $p_{iC} = .2$, $\theta=0.5$ and $f = .5$, equal sample sizes
		\label{PowerPlotAtNominal005_piC02andTheta05LOR_equal_sample_sizes}}
\end{figure}

\begin{figure}[ht]
	\centering
	\includegraphics[scale=0.33]{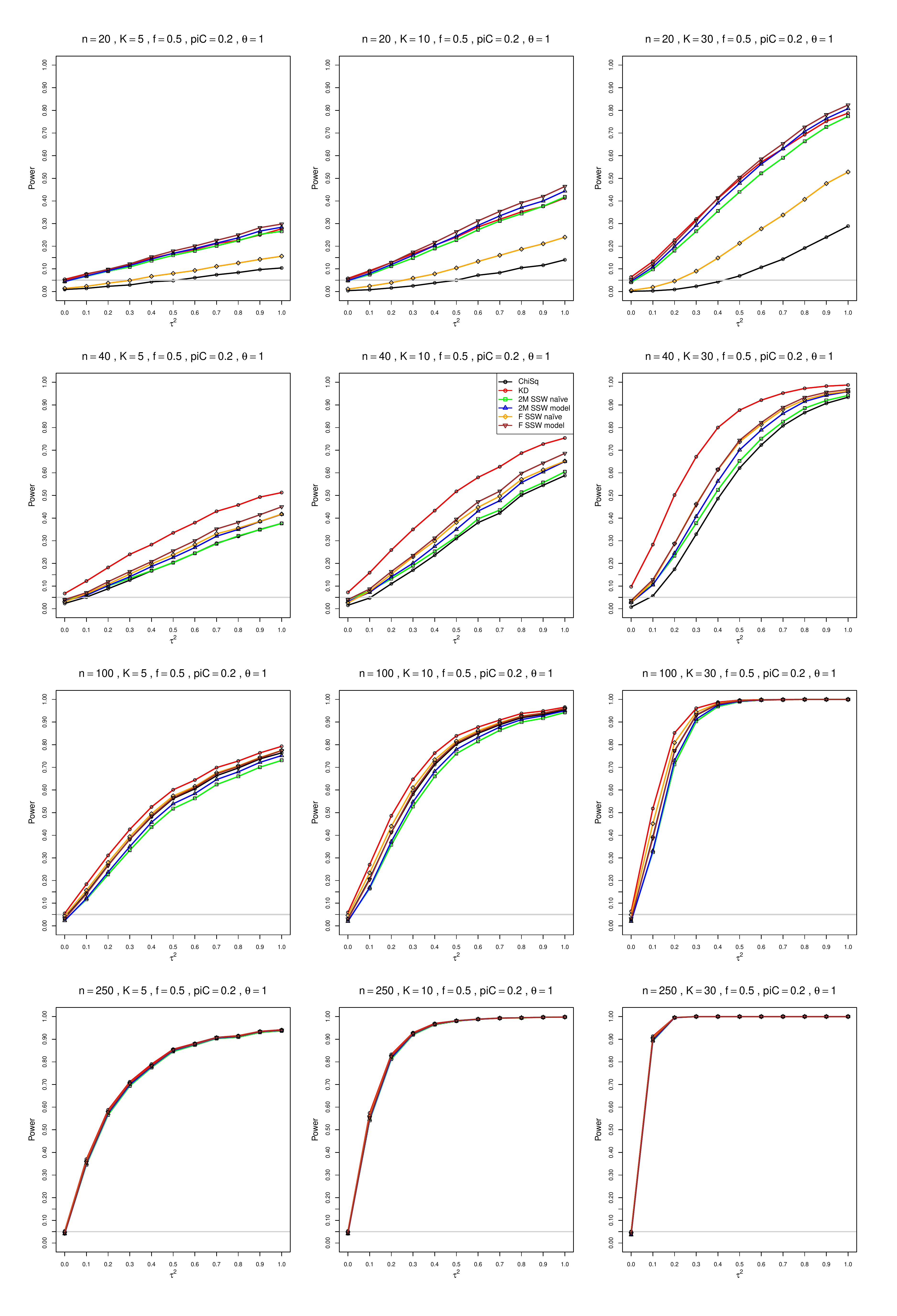}
	\caption{Q for LOR: empirical power at nominal level $\alpha = .05$ for $p_{iC} = .2$, $\theta=1.0$ and $f = .5$, equal sample sizes
		\label{PowerPlotAtNominal005_piC02andTheta1LOR_equal_sample_sizes}}
\end{figure}

\begin{figure}[ht]
	\centering
	\includegraphics[scale=0.33]{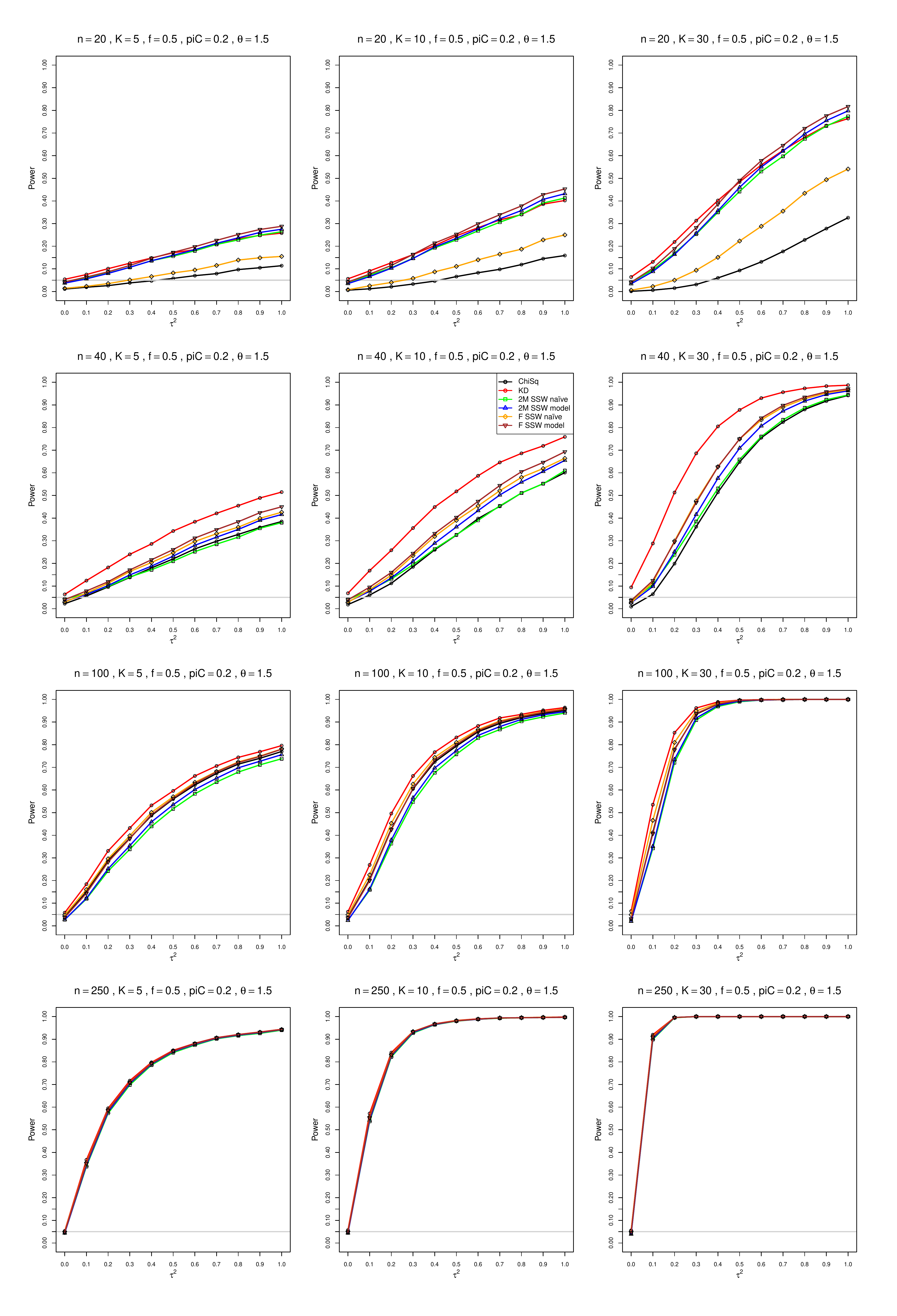}
	\caption{Q for LOR: empirical power at nominal level $\alpha = .05$ for $p_{iC} = .2$, $\theta=1.5$ and $f = .5$, equal sample sizes
		\label{PowerPlotAtNominal005_piC02andTheta1.5LOR_equal_sample_sizes}}
\end{figure}

\begin{figure}[ht]
	\centering
	\includegraphics[scale=0.33]{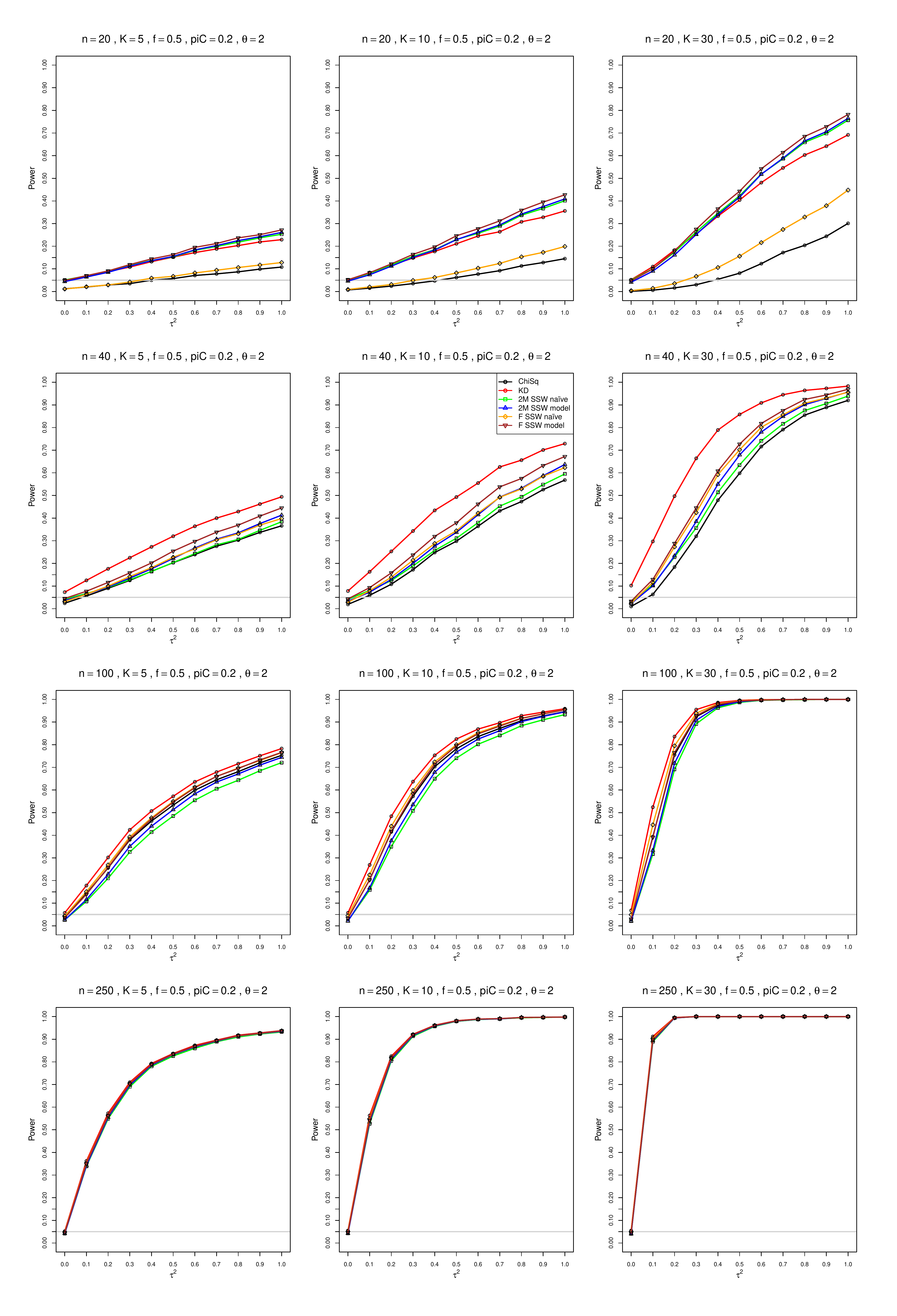}
	\caption{Q for LOR: empirical power at nominal level $\alpha = .05$ for $p_{iC} = .2$, $\theta=2.0$ and $f = .5$, equal sample sizes
		\label{PowerPlotAtNominal005_piC02andTheta2LOR_equal_sample_sizes}}
\end{figure}

\begin{figure}[ht]
	\centering
	\includegraphics[scale=0.33]{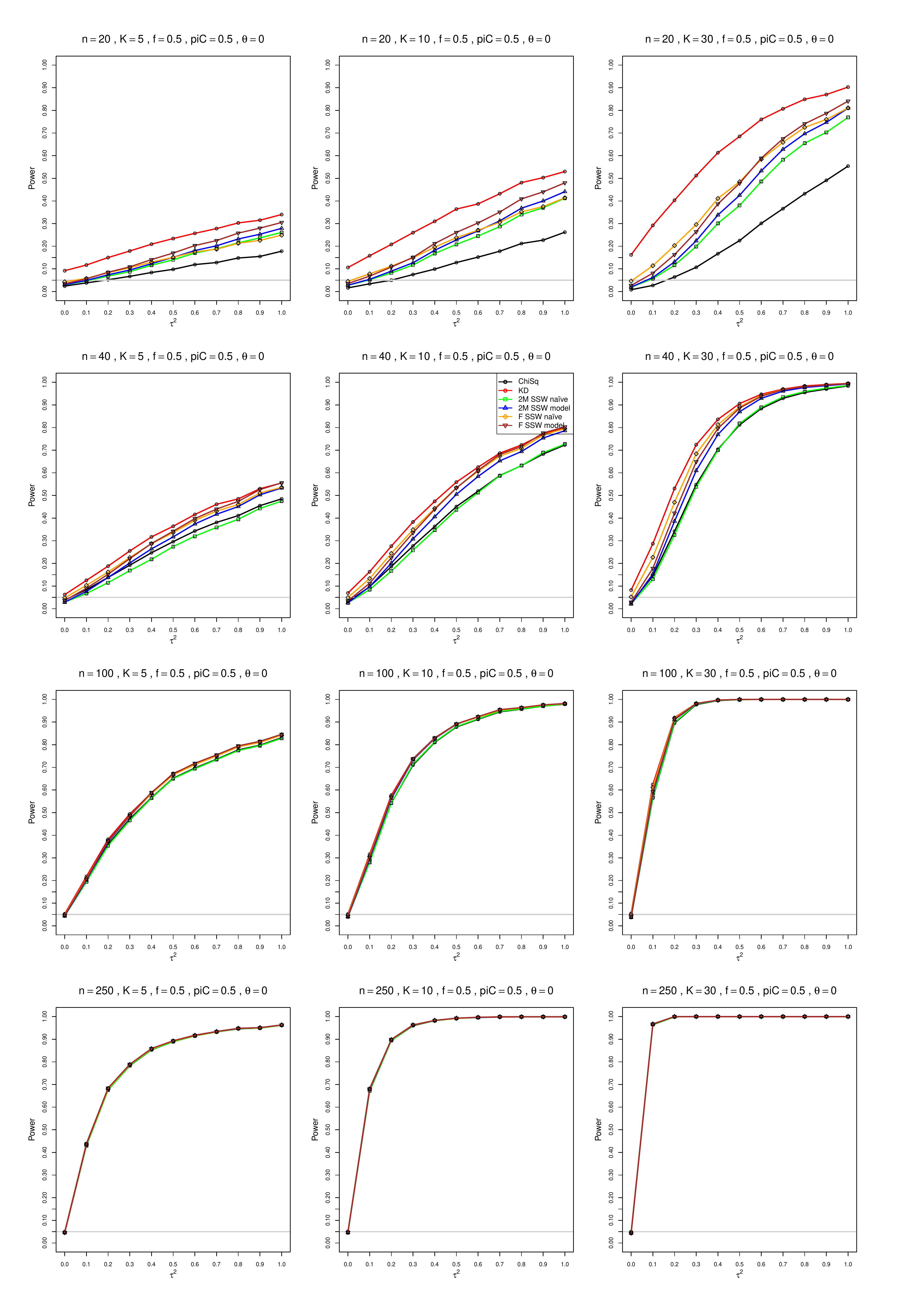}
	\caption{Q for LOR: empirical power at nominal level $\alpha = .05$ for $p_{iC} = .5$, $\theta=0$ and $f = .5$, equal sample sizes
		\label{PowerPlotAtNominal005_piC05andTheta0LOR_equal_sample_sizes}}
\end{figure}

\begin{figure}[t]
	\centering
	\includegraphics[scale=0.33]{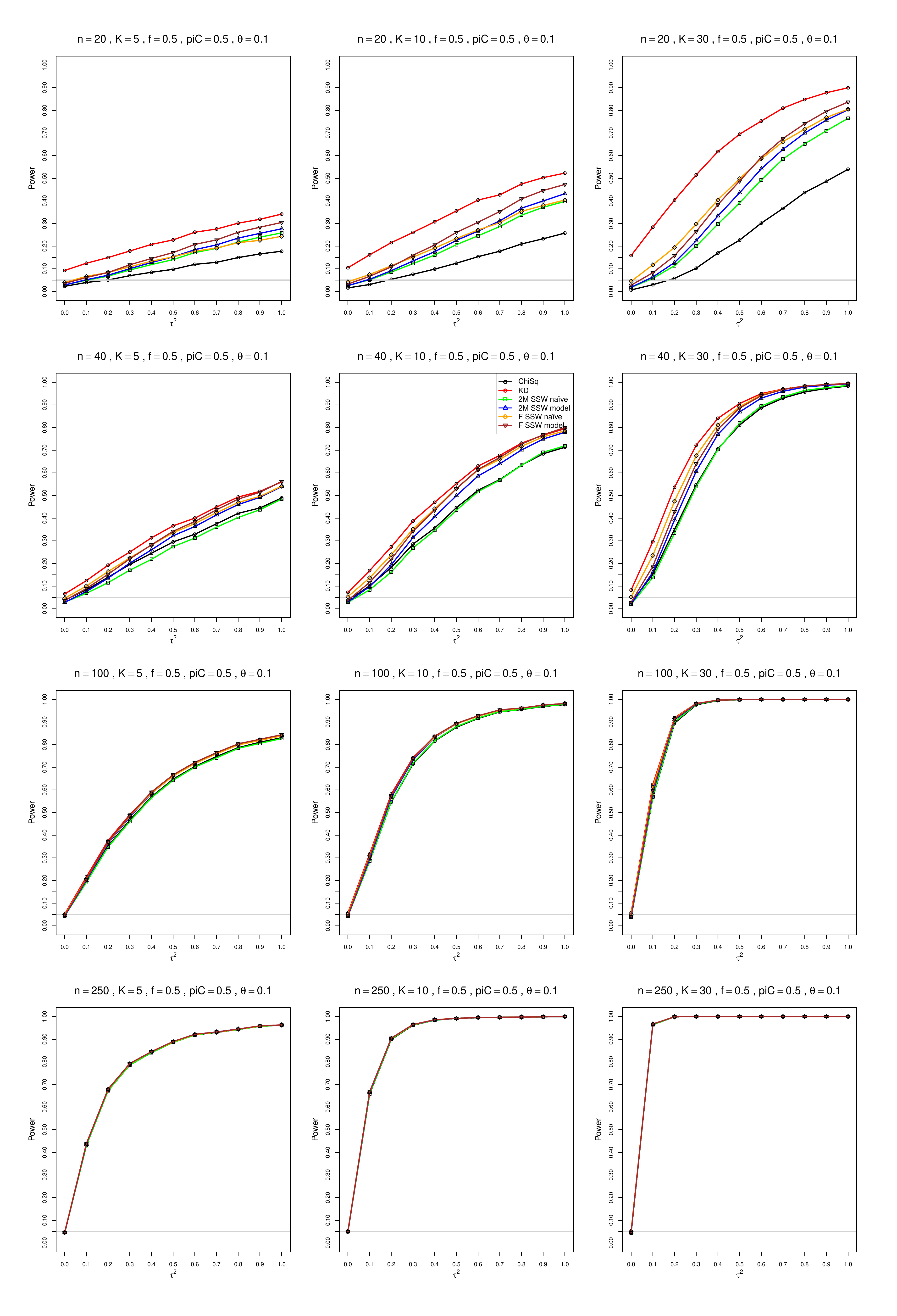}
	\caption{Q for LOR: empirical power at nominal level $\alpha = .05$ for $p_{iC} = .5$, $\theta=0.1$ and $f = .5$, equal sample sizes
		\label{PowerPlotAtNominal005_piC05andTheta01LOR_equal_sample_sizes}}
\end{figure}

\begin{figure}[ht]
	\centering
	\includegraphics[scale=0.33]{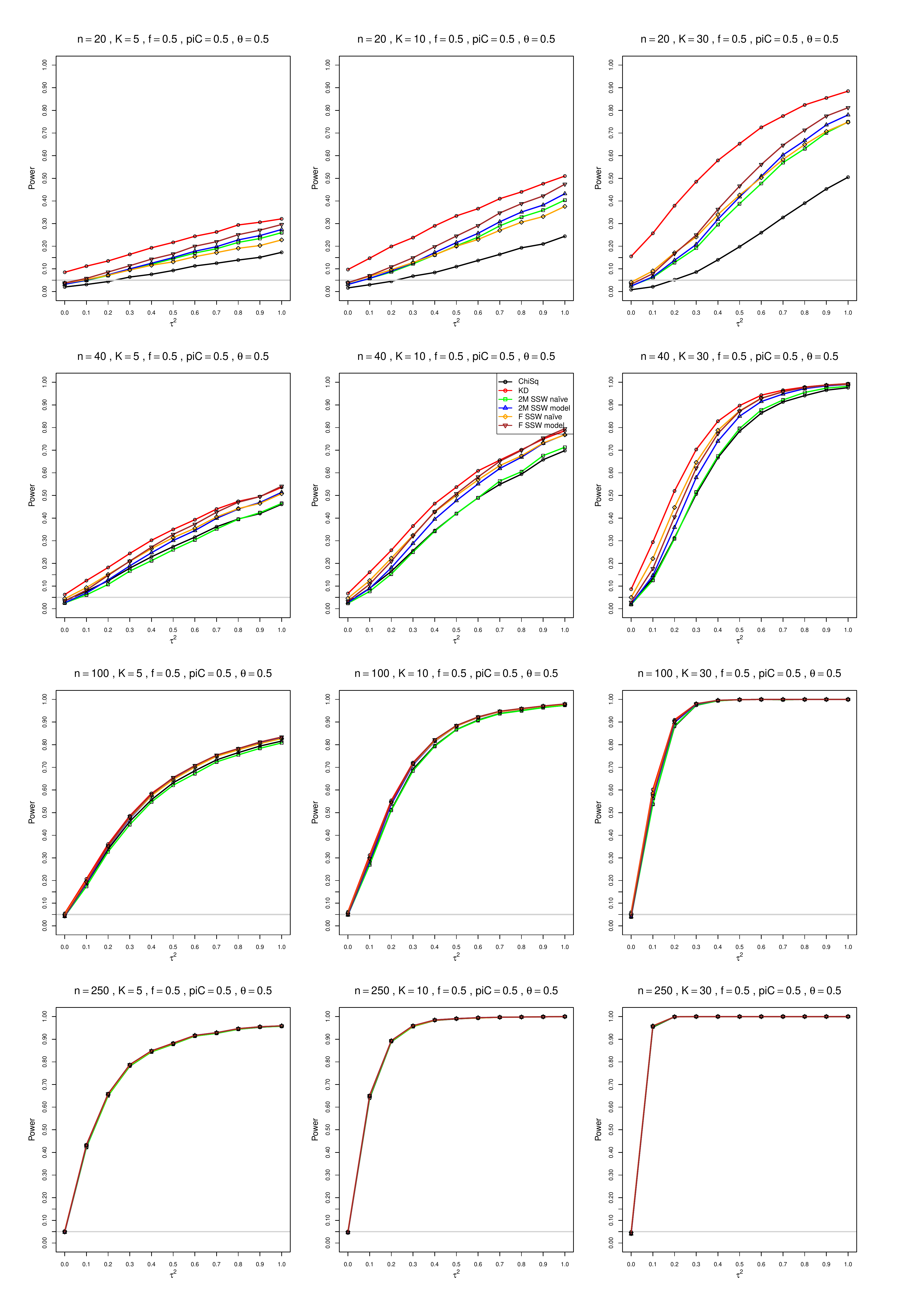}
	\caption{Q for LOR: empirical power at nominal level $\alpha = .05$ for $p_{iC} = .5$, $\theta=0.5$ and $f = .5$, equal sample sizes
		\label{PowerPlotAtNominal005_piC05andTheta05LOR_equal_sample_sizes}}
\end{figure}

\begin{figure}[ht]
	\centering
	\includegraphics[scale=0.33]{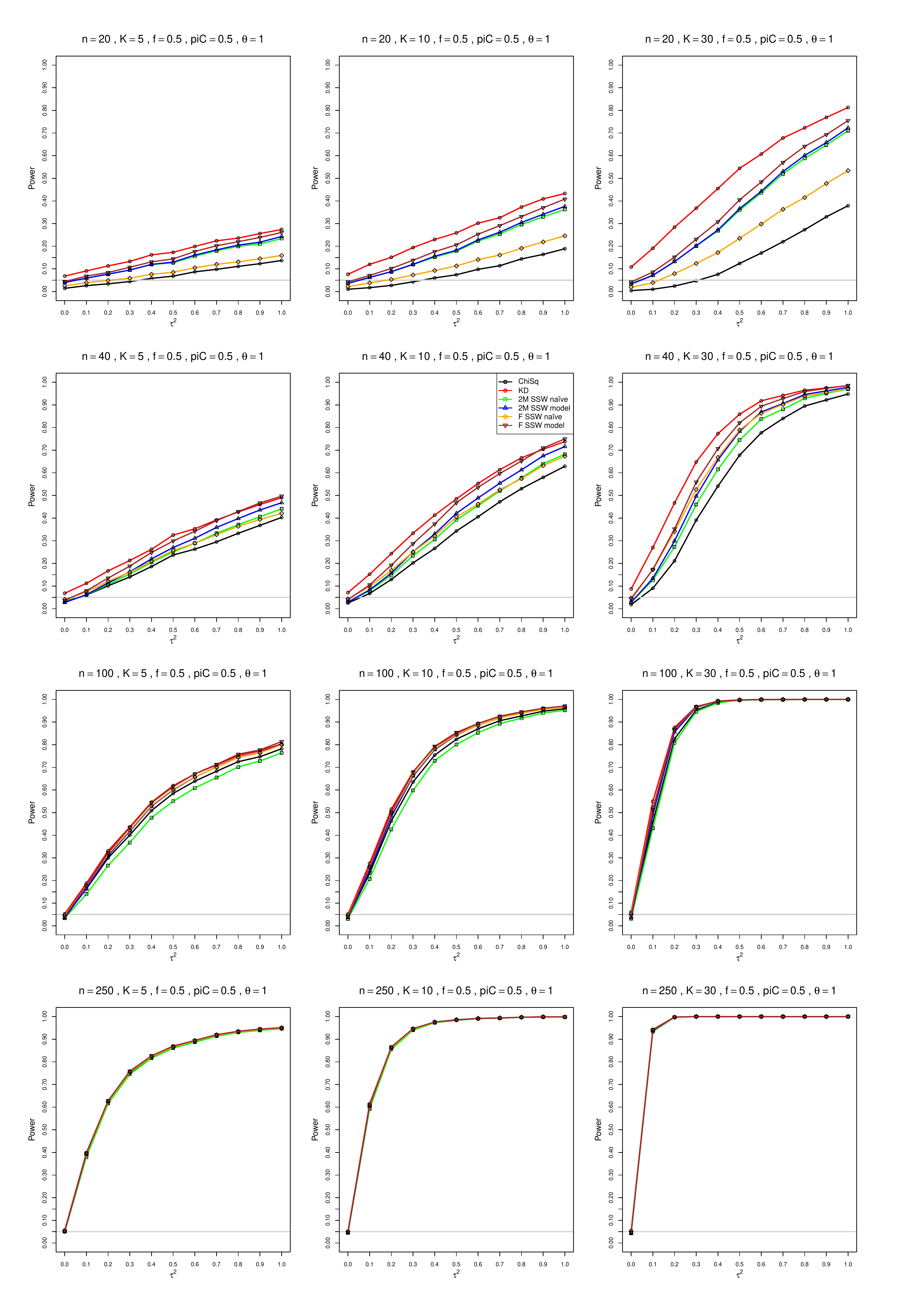}
	\caption{Q for LOR: empirical power at nominal level $\alpha = .05$ for $p_{iC} = .1$, $\theta=1.0$ and $f = .5$, equal sample sizes
		\label{PowerPlotAtNominal005_piC01andTheta1LOR_equal_sample_sizes}}
\end{figure}

\begin{figure}[ht]
	\centering
	\includegraphics[scale=0.33]{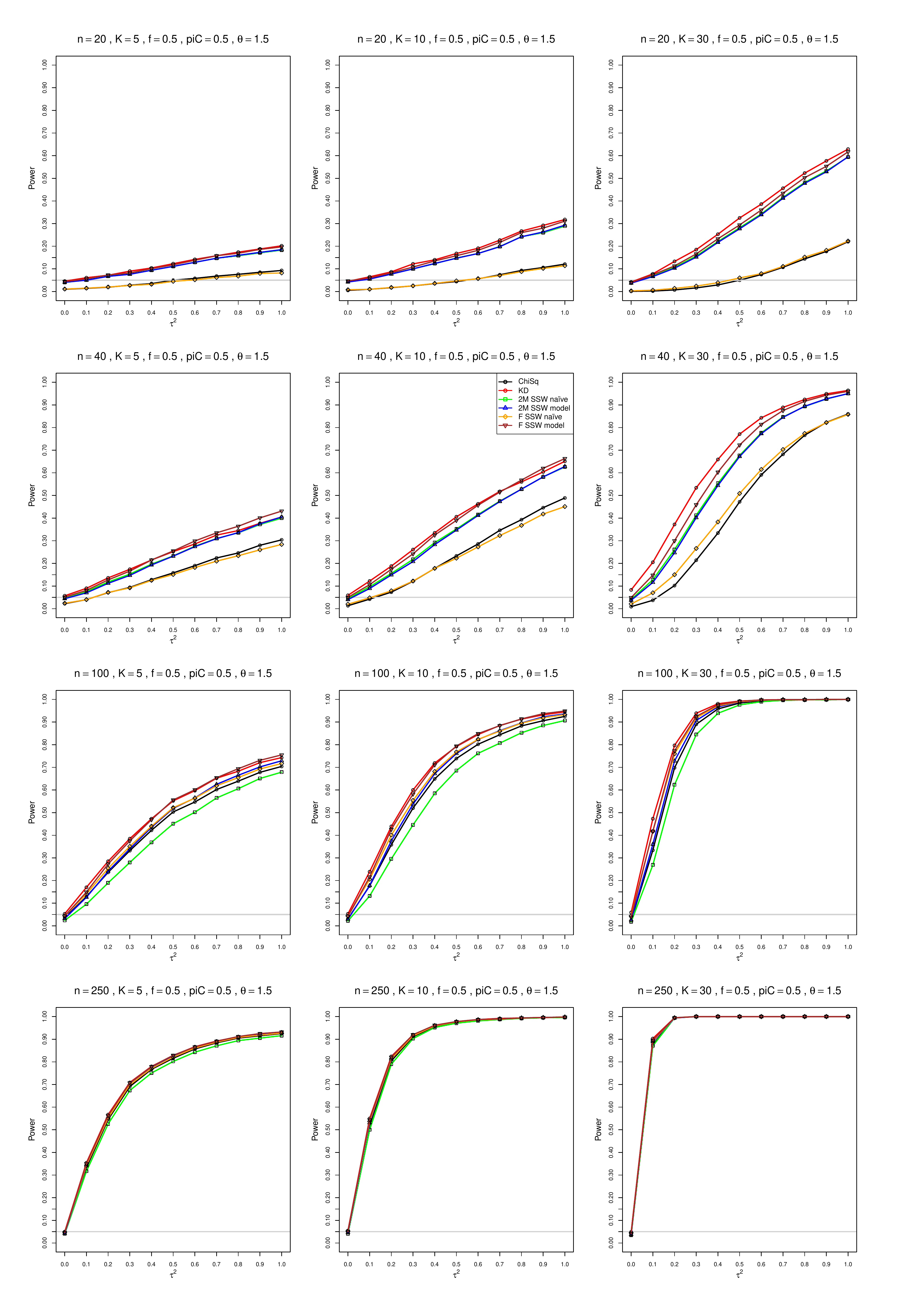}
	\caption{Q for LOR: empirical power at nominal level $\alpha = .05$ for $p_{iC} = .1$, $\theta=1.5$ and $f = .5$, equal sample sizes
		\label{PowerPlotAtNominal005_piC05andTheta1.5LOR_equal_sample_sizes}}
\end{figure}

\begin{figure}[ht]
	\centering
	\includegraphics[scale=0.33]{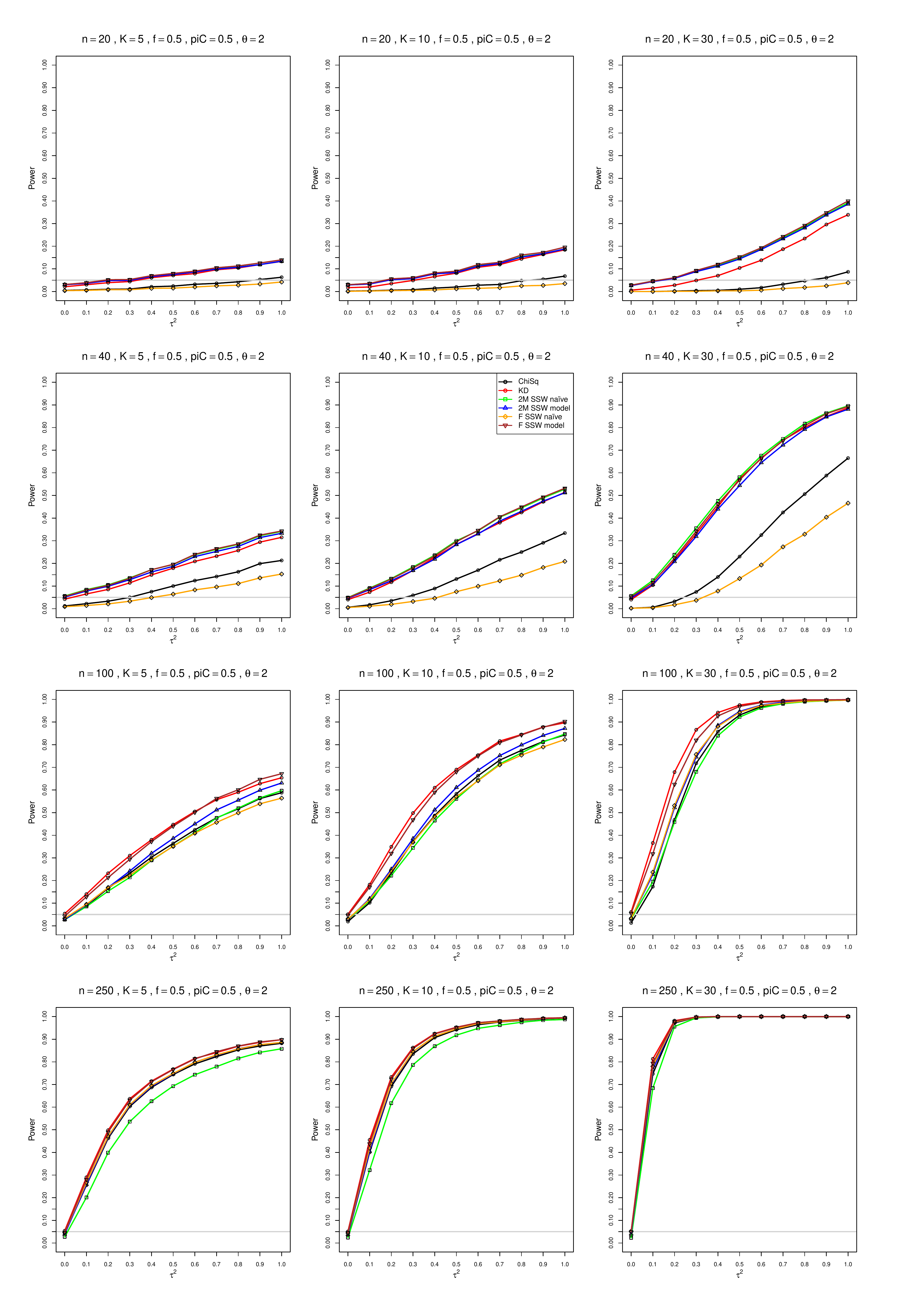}
	\caption{Q for LOR: empirical power at nominal level $\alpha = .05$ for $p_{iC} = .5$, $\theta=2.0$ and $f = .5$, equal sample sizes
		\label{PowerPlotAtNominal005_piC05andTheta2LOR_equal_sample_sizes}}
\end{figure}


\clearpage

\setcounter{figure}{0}
\setcounter{section}{0}
\renewcommand{\thefigure}{D.\arabic{figure}}

\section*{Appendix D: Plots of error in the level of the test for heterogeneity of LRR for five approximations for the null distribution of $Q$}

Each figure corresponds to a value of the probability of an event in the Control arm $p_{iC}$  (= .1, .2, .5), a value of the overall LRR $\rho$, and a choice of equal or unequal sample sizes ($n$ or $bar{n}$). For $p_{iC} = .1$ or $.2$, $\rho =$
$-0.5, 0, 0.5, 1, 1.5$.
For $p_{iC} = .5$, $\rho = \;-1.5, -1, -0.5, 0, 0.5.$ \\
The fraction of each study's sample size in the Control arm $f$ is held constant at 0.5.

For each combination of a value of $n$ (= 20, 40, 100, 250) or  $\bar{n}$ (= 30, 60, 100, 160) and a value of $K$ (= 5, 10, 30), a panel plots, versus the nominal upper tail areas  (= .001, .0025, .005, .01, .025, .05, .1, .25, .5 and the complementary values .75, \ldots, .999), the difference between the achieved level and the nominal level for five approximations to the null distribution of Q: \\
\begin{itemize}
\item ChiSq (Chi-square approximation with $K-1$ df, inverse-variance weights)
\item 2M SSW na\"{i}ve (Two-moment gamma approximation, na\"{i}ve estimation of $p_{iT}$ from $X_{iT}$ and $n_{iT}$, effective-sample-size weights)
\item 2M SSW model (Two-moment gamma approximation, model-based estimation of $p_{iT}$, effective-sample-size weights)
\item F SSW na\"{i}ve (Farebrother approximation, na\"{i}ve estimation of $p_{iT}$ from $X_{iT}$ and $n_{iT}$, effective-sample-size weights)
\item F SSW model (Farebrother approximation, model-based estimation of $p_{iT}$, effective-sample-size weights)
\end{itemize}

\clearpage

\begin{figure}[ht]
	\centering
	\includegraphics[scale=0.33]{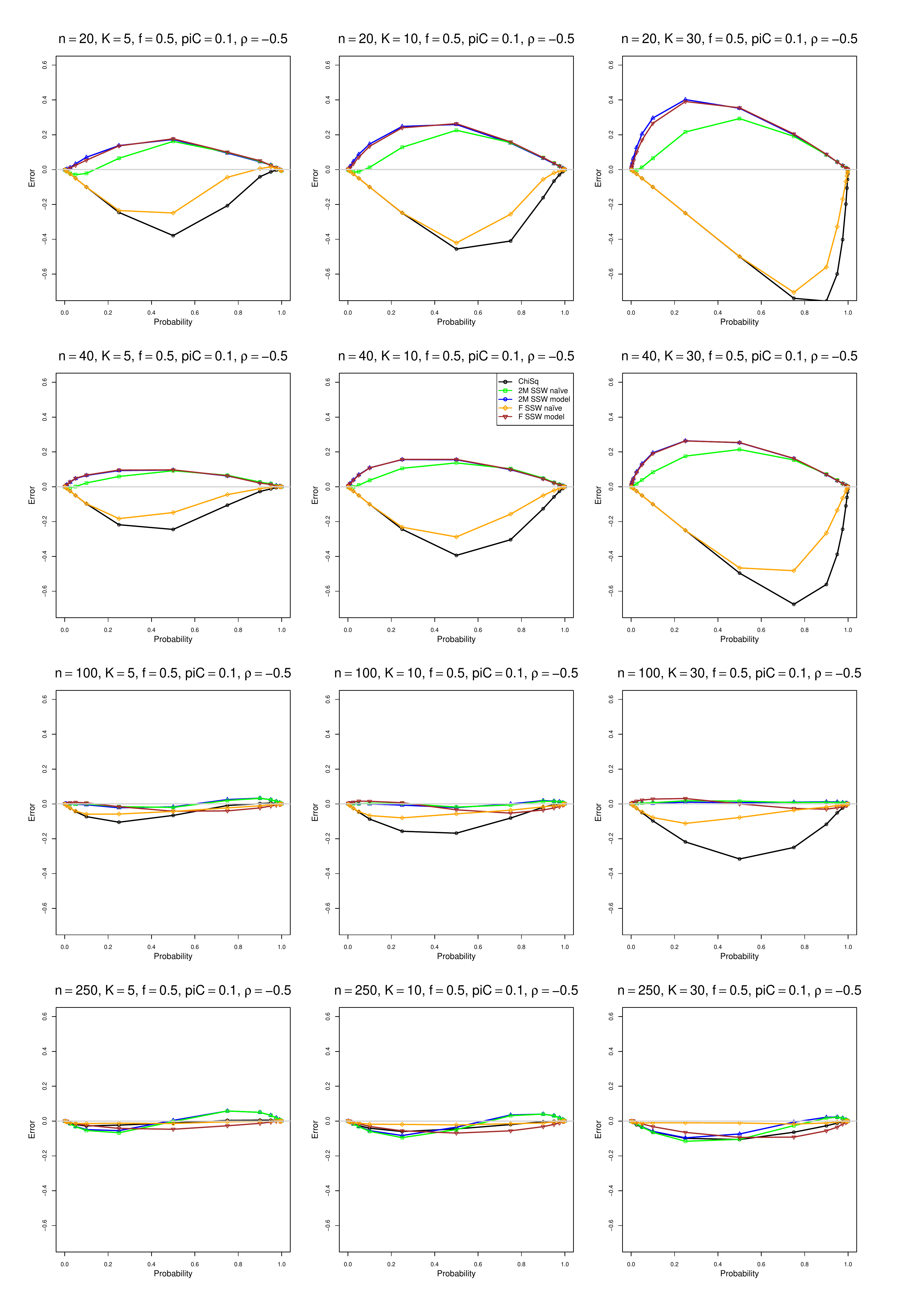}
	\caption{Plots of error in the level of the test for heterogeneity of LRR for five approximations for the null distribution of $Q$, $p_{iC} = .1$, $f = .5$, and $\rho = -0.5$, equal sample sizes}
	\label{PPplot_piC_01theta=-0.5_LRR_equal_sample_sizes}
\end{figure}
\begin{figure}[ht]
		\centering
	\includegraphics[scale=0.33]{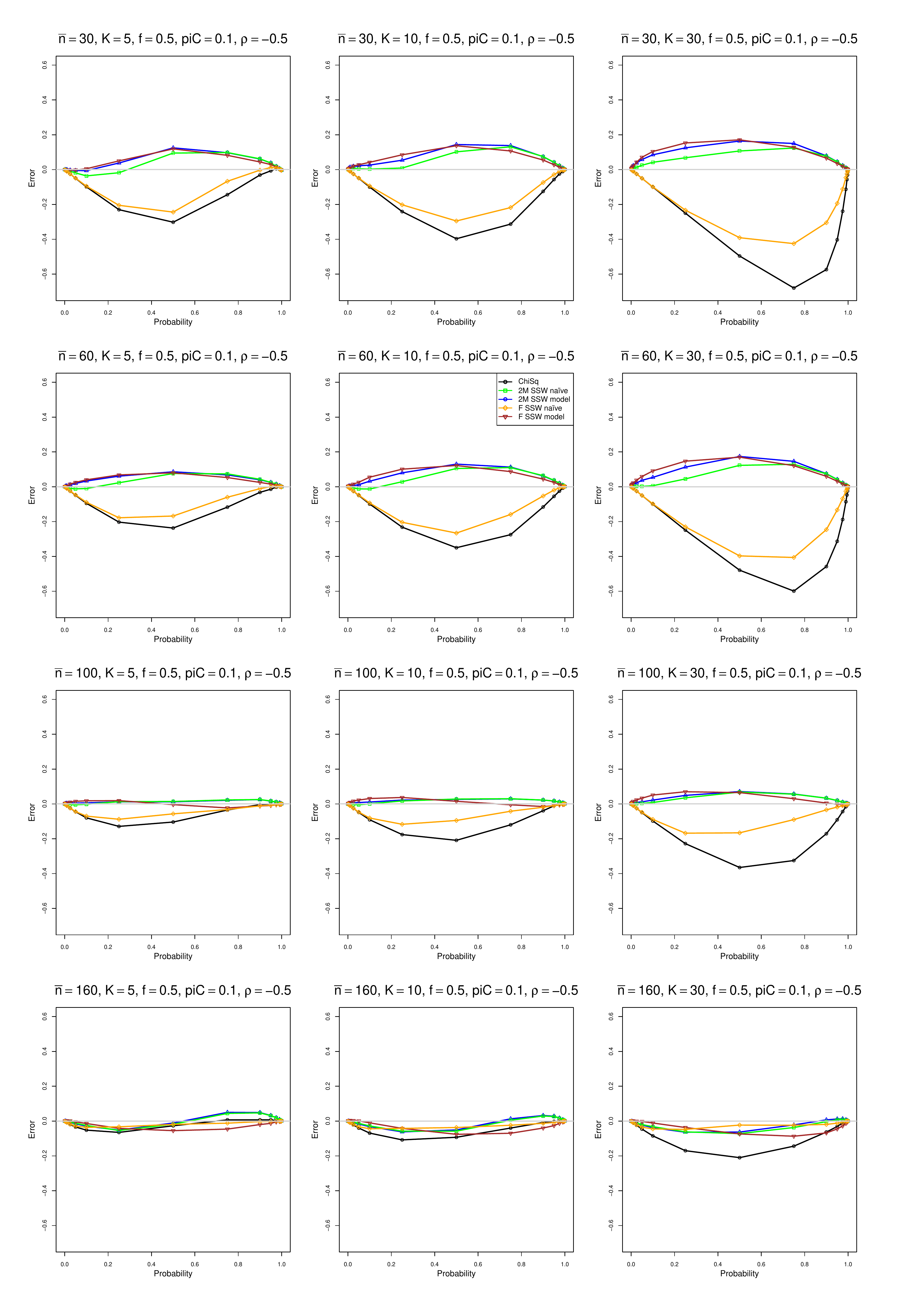}
	\caption{Plots of error in the level of the test for heterogeneity of LRR for five approximations for the null distribution of $Q$, $p_{iC} = .1$, $f = .5$, and $\rho = -0.5$, unequal sample sizes}
	\label{PPplot_piC_01theta=-0.5_LRR_unequal_sample_sizes}
\end{figure}

\begin{figure}[ht]
	\centering
	\includegraphics[scale=0.33]{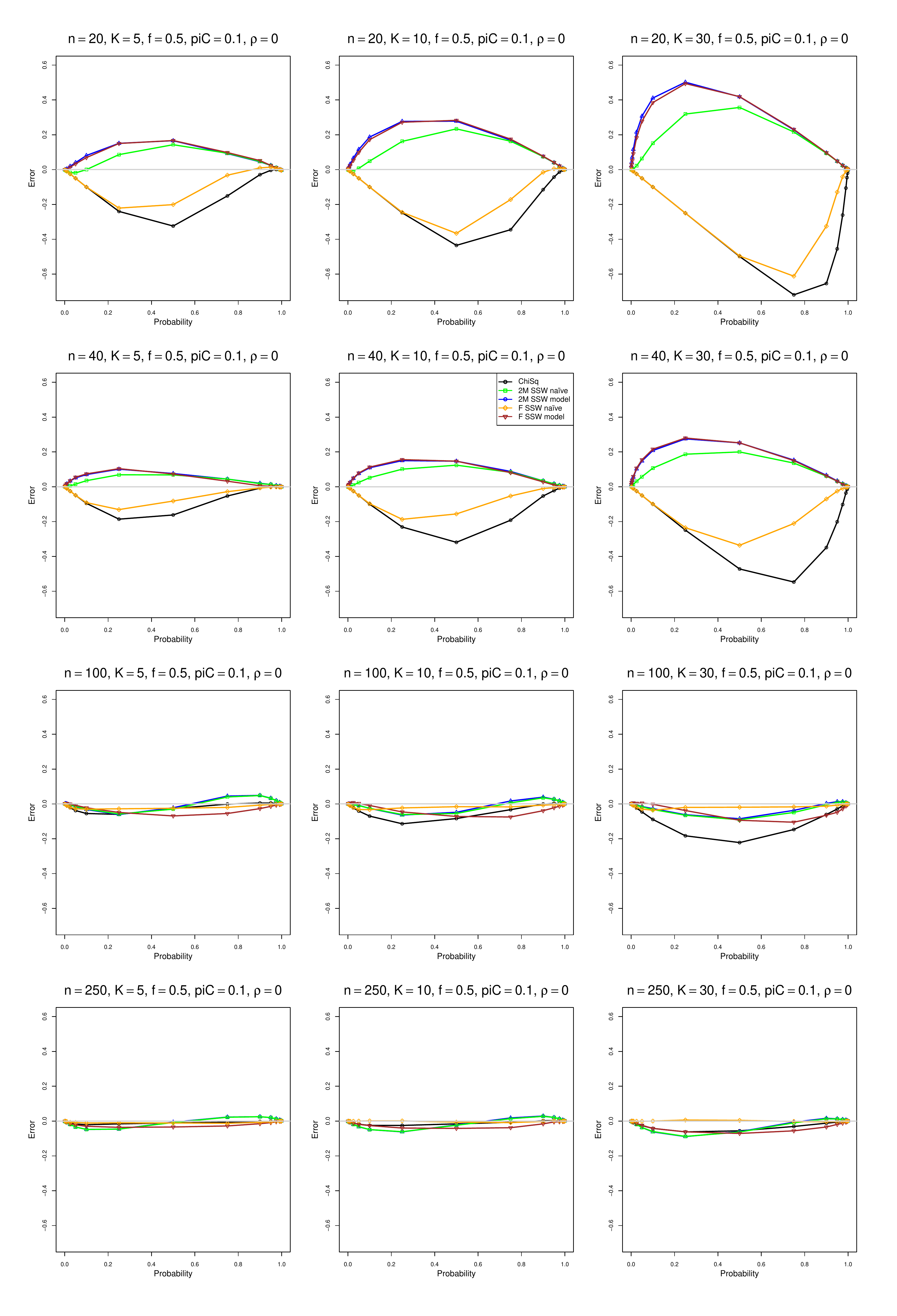}
	\caption{Plots of error in the level of the test for heterogeneity of LRR for five approximations for the null distribution of $Q$, $p_{iC} = .1$, $f = .5$, and $\rho = 0$, equal sample sizes}
	\label{PPplot_piC_01theta=0_LRR_equal_sample_sizes}
\end{figure}
\begin{figure}[ht]
		\centering
	\includegraphics[scale=0.33]{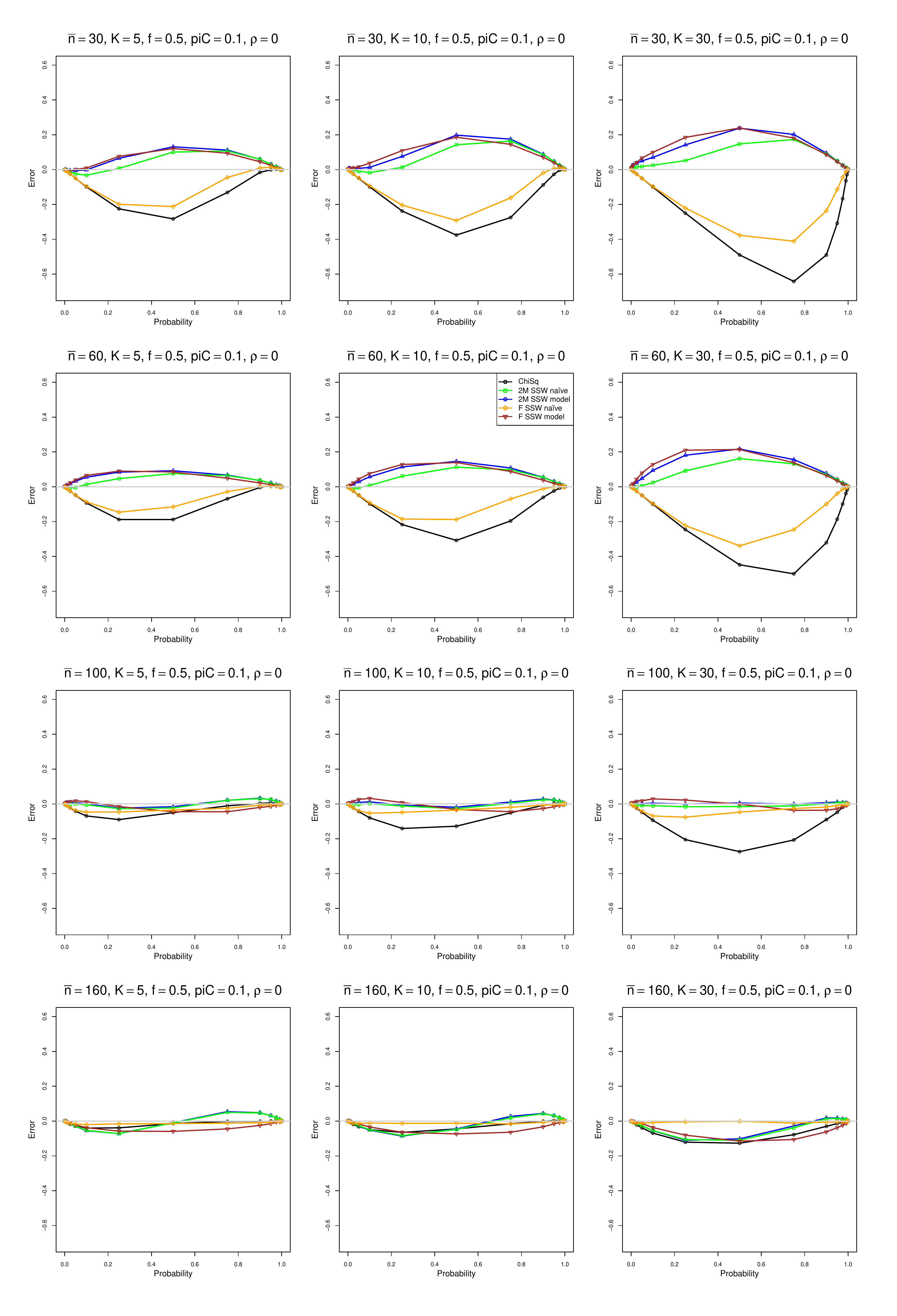}
	\caption{Plots of error in the level of the test for heterogeneity of LRR for five approximations for the null distribution of $Q$, $p_{iC} = .1$, $f = .5$, and $\rho = 0$, unequal sample sizes}
	\label{PPplot_piC_01theta=0_LRR_unequal_sample_sizes}
\end{figure}

\begin{figure}[ht]
	\centering
	\includegraphics[scale=0.33]{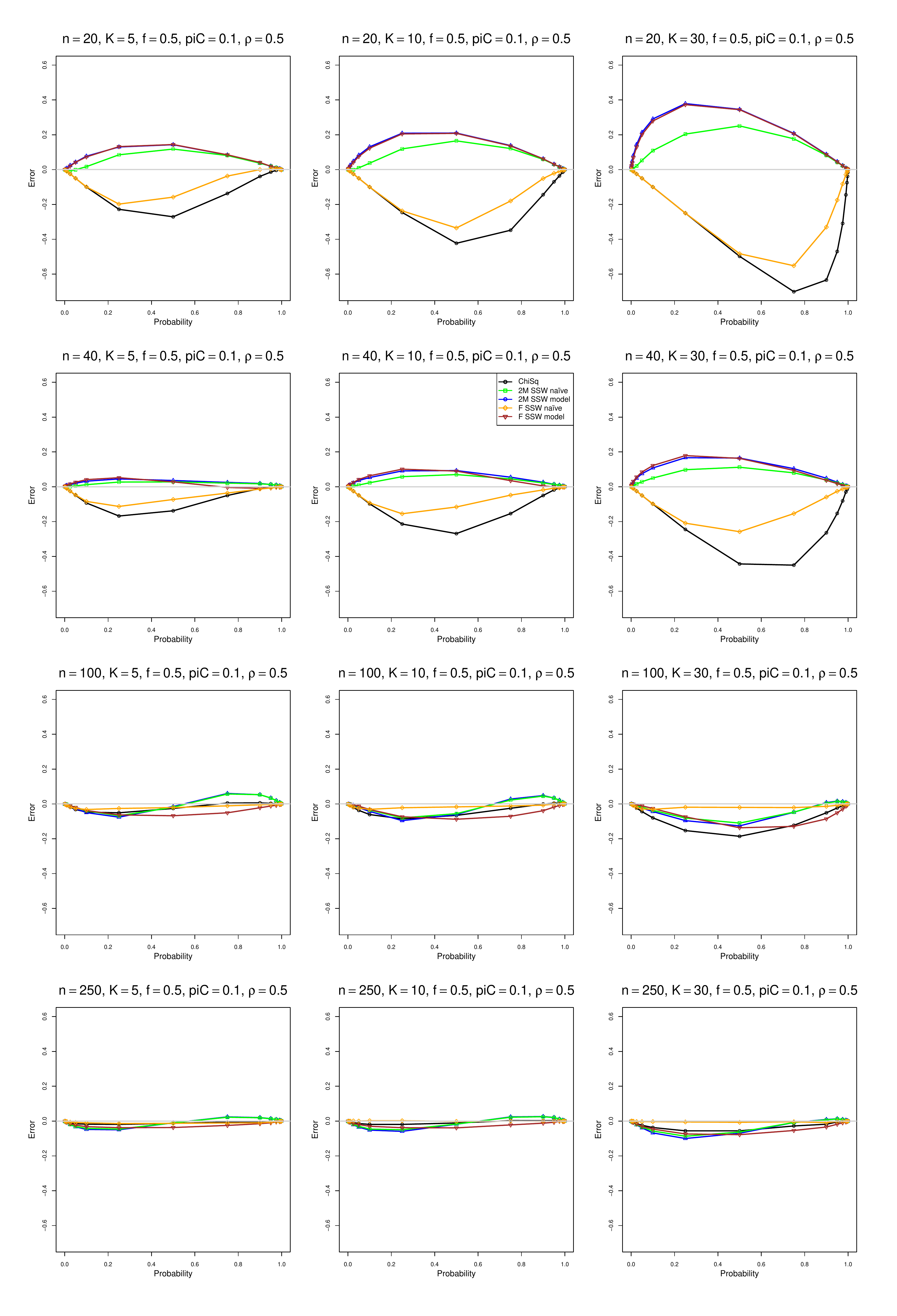}
	\caption{Plots of error in the level of the test for heterogeneity of LRR for five approximations for the null distribution of $Q$, $p_{iC} = .1$, $f = .5$, and $\rho = 0.5$, equal sample sizes}
	\label{PPplot_piC_01theta=0.5_LRR_equal_sample_sizes}
\end{figure}
\begin{figure}[ht]
		\centering
	\includegraphics[scale=0.33]{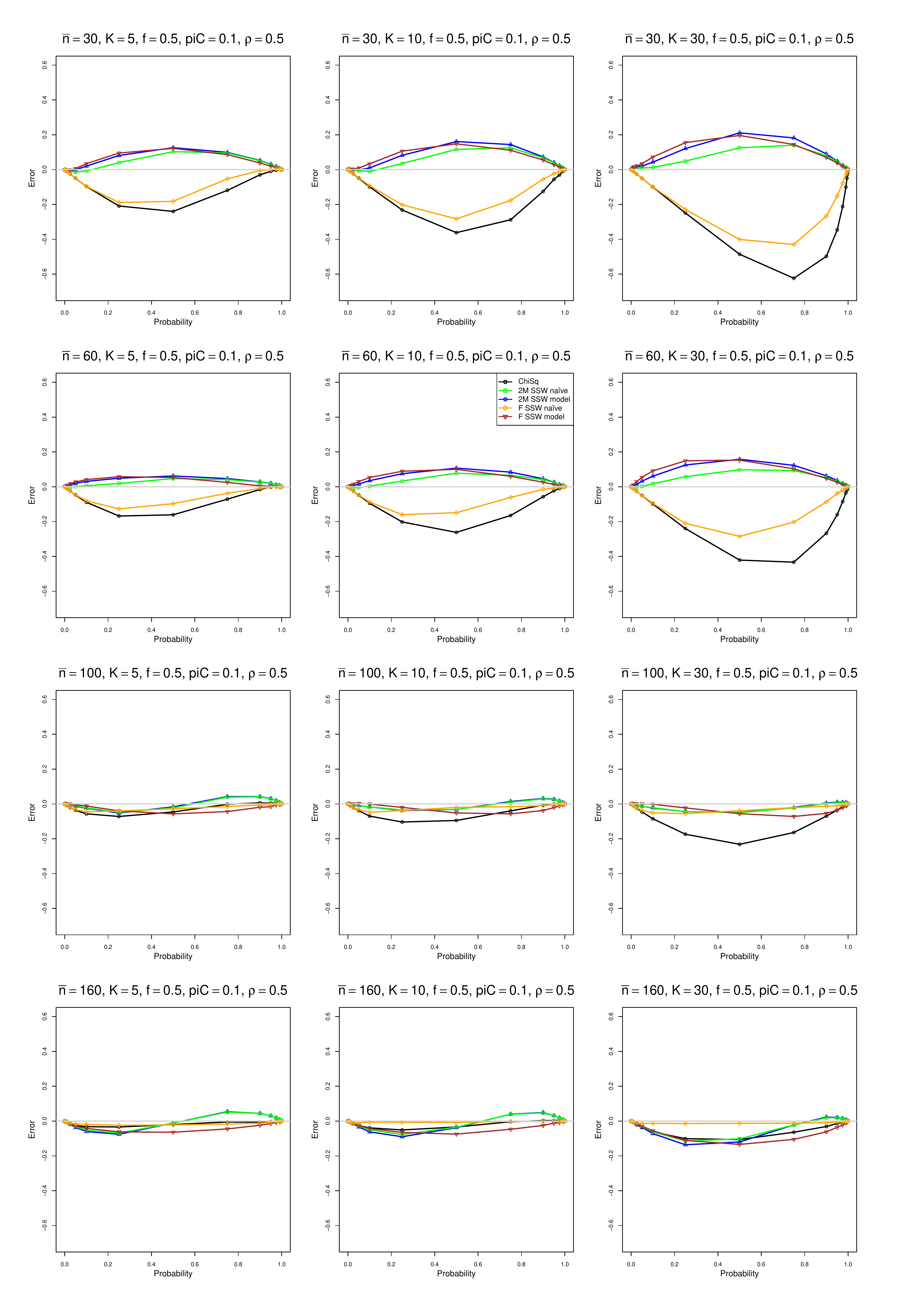}
	\caption{Plots of error in the level of the test for heterogeneity of LRR for five approximations for the null distribution of $Q$, $p_{iC} = .1$, $f = .5$, and $\rho = 0.5$, unequal sample sizes}
	\label{PPplot_piC_01theta=0.5_LRR_unequal_sample_sizes}
\end{figure}

\begin{figure}[ht]
	\centering
	\includegraphics[scale=0.33]{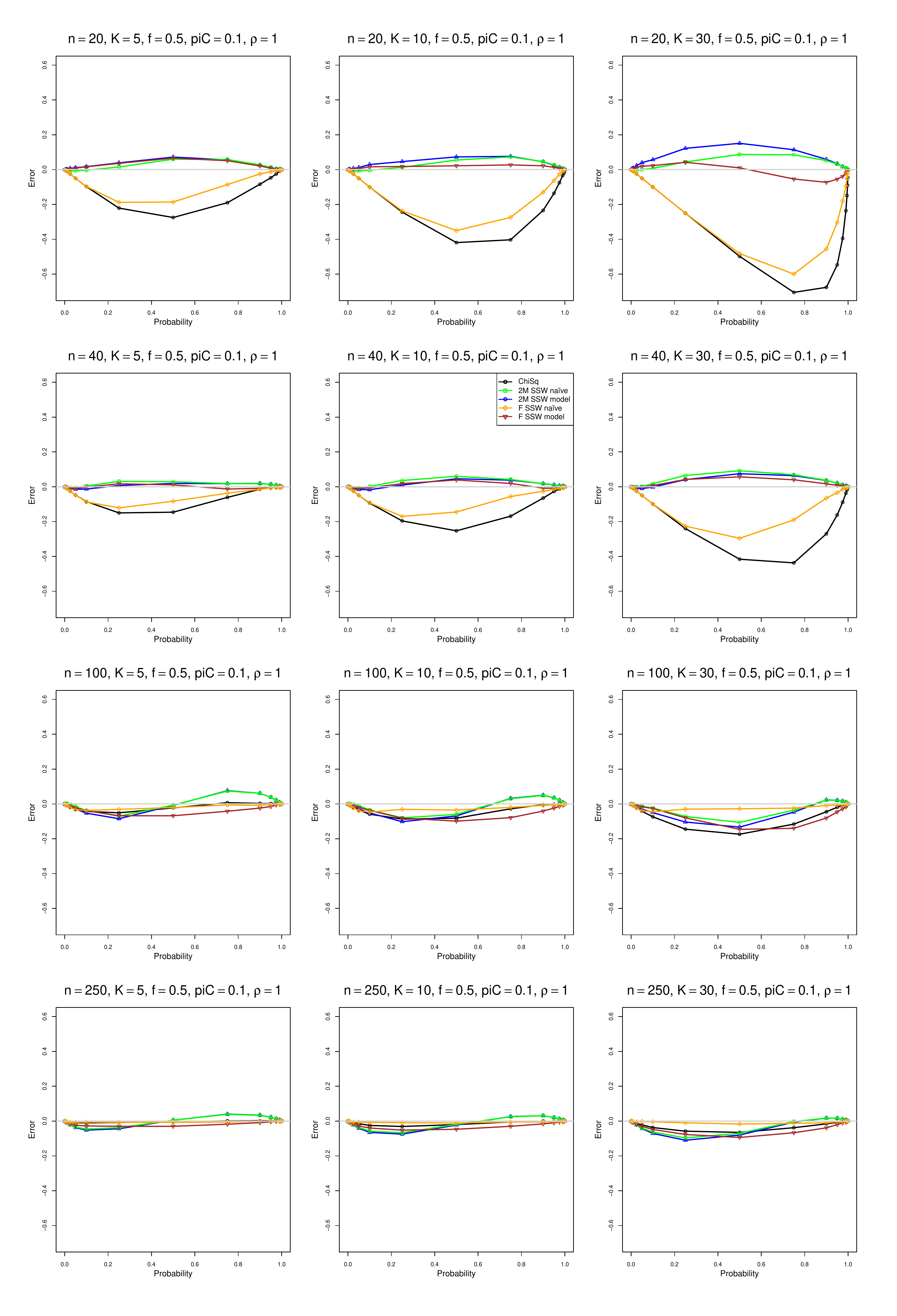}
	\caption{Plots of error in the level of the test for heterogeneity of LRR for five approximations for the null distribution of $Q$, $p_{iC} = .1$, $f = .5$, and $\rho = 1$, equal sample sizes}
	\label{PPplot_piC_01theta=1_LRR_equal_sample_sizes}
\end{figure}
\begin{figure}[ht]
		\centering
	\includegraphics[scale=0.33]{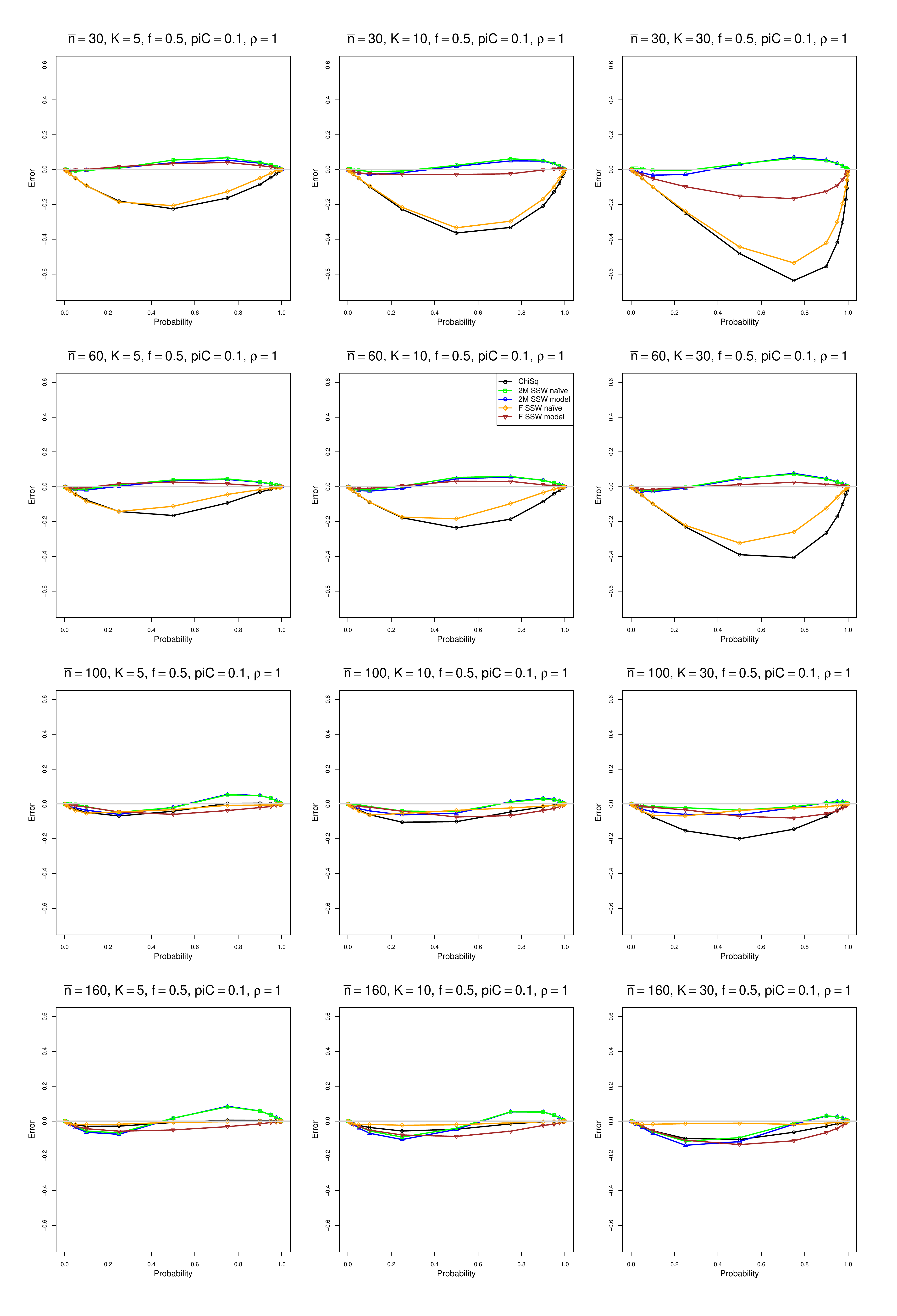}
	\caption{Plots of error in the level of the test for heterogeneity of LRR for five approximations for the null distribution of $Q$, $p_{iC} = .1$, $f = .5$, and $\rho = 1$, unequal sample sizes}
	\label{PPplot_piC_01theta=1_LRR_unequal_sample_sizes}
\end{figure}

\begin{figure}[ht]
	\centering
	\includegraphics[scale=0.33]{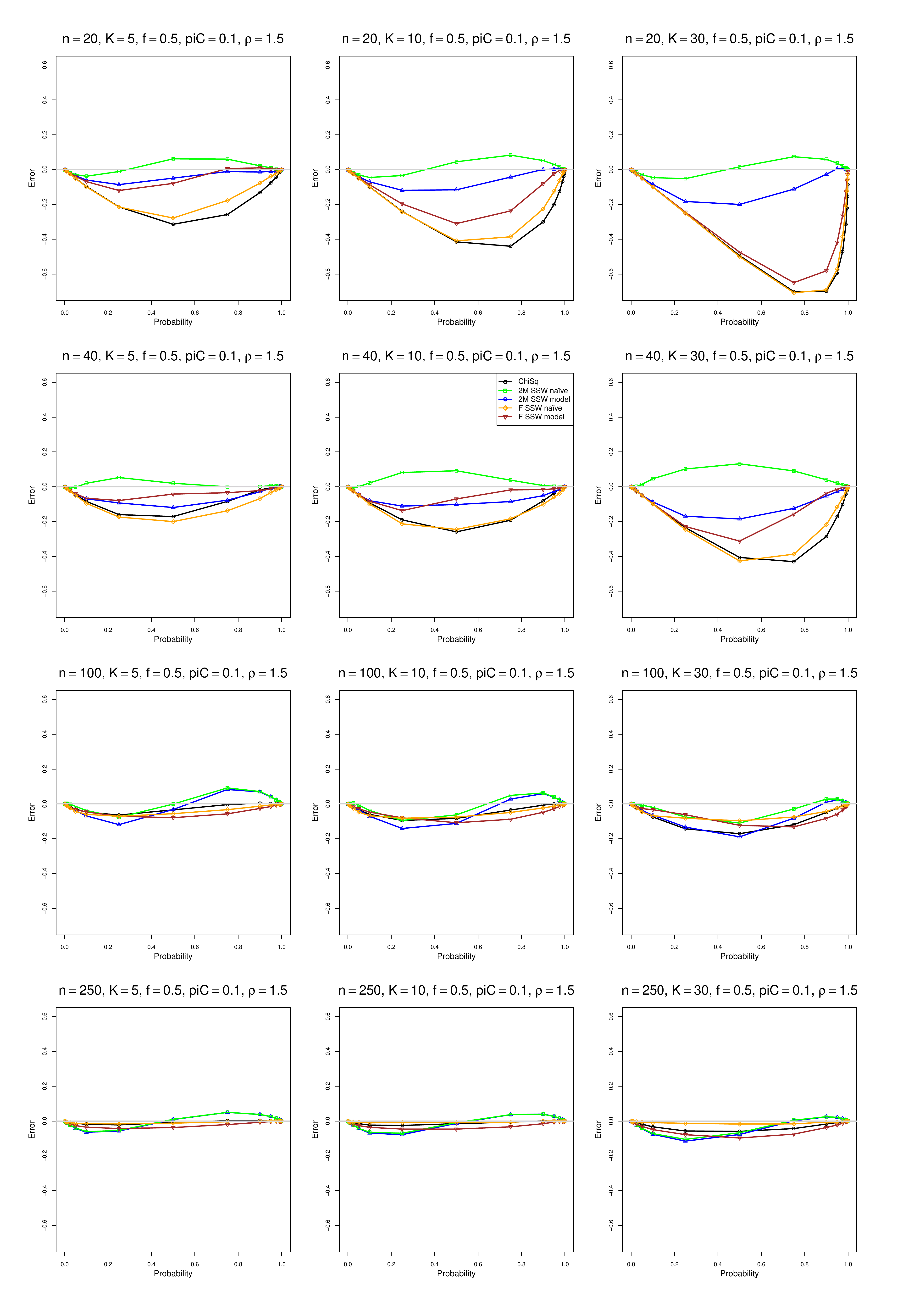}
	\caption{Plots of error in the level of the test for heterogeneity of LRR for five approximations for the null distribution of $Q$, $p_{iC} = .1$, $f = .5$, and $\rho = 1.5$, equal sample sizes}
	\label{PPplot_piC_01theta=1.5_LRR_equal_sample_sizes}
\end{figure}
\begin{figure}[ht]
		\centering
	\includegraphics[scale=0.33]{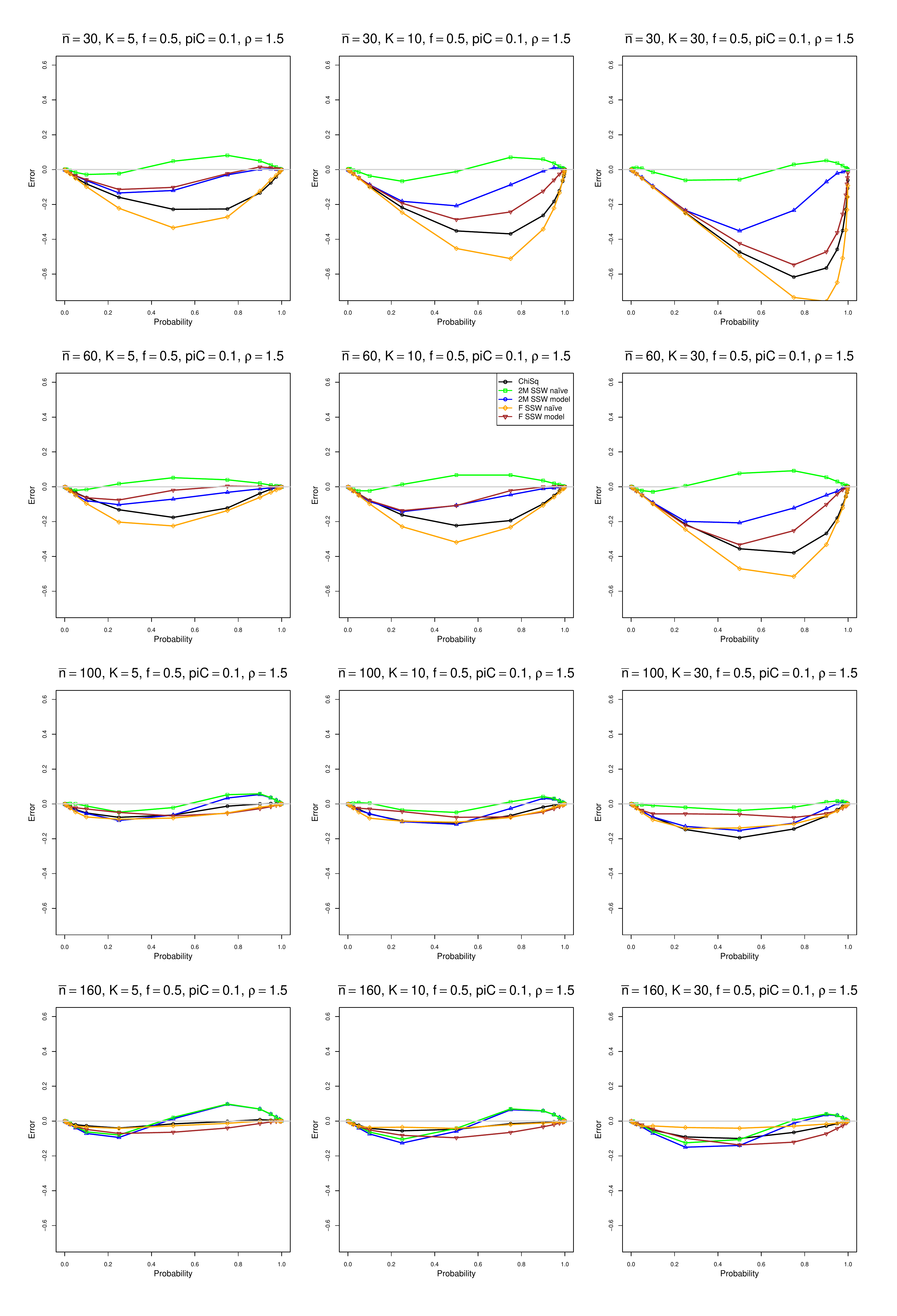}
	\caption{Plots of error in the level of the test for heterogeneity of LRR for five approximations for the null distribution of $Q$, $p_{iC} = .1$, $f = .5$, and $\rho = 1.5$, unequal sample sizes}
	\label{PPplot_piC_01theta=1.5_LRR_unequal_sample_sizes}
\end{figure}


\begin{figure}[ht]
	\centering
	\includegraphics[scale=0.33]{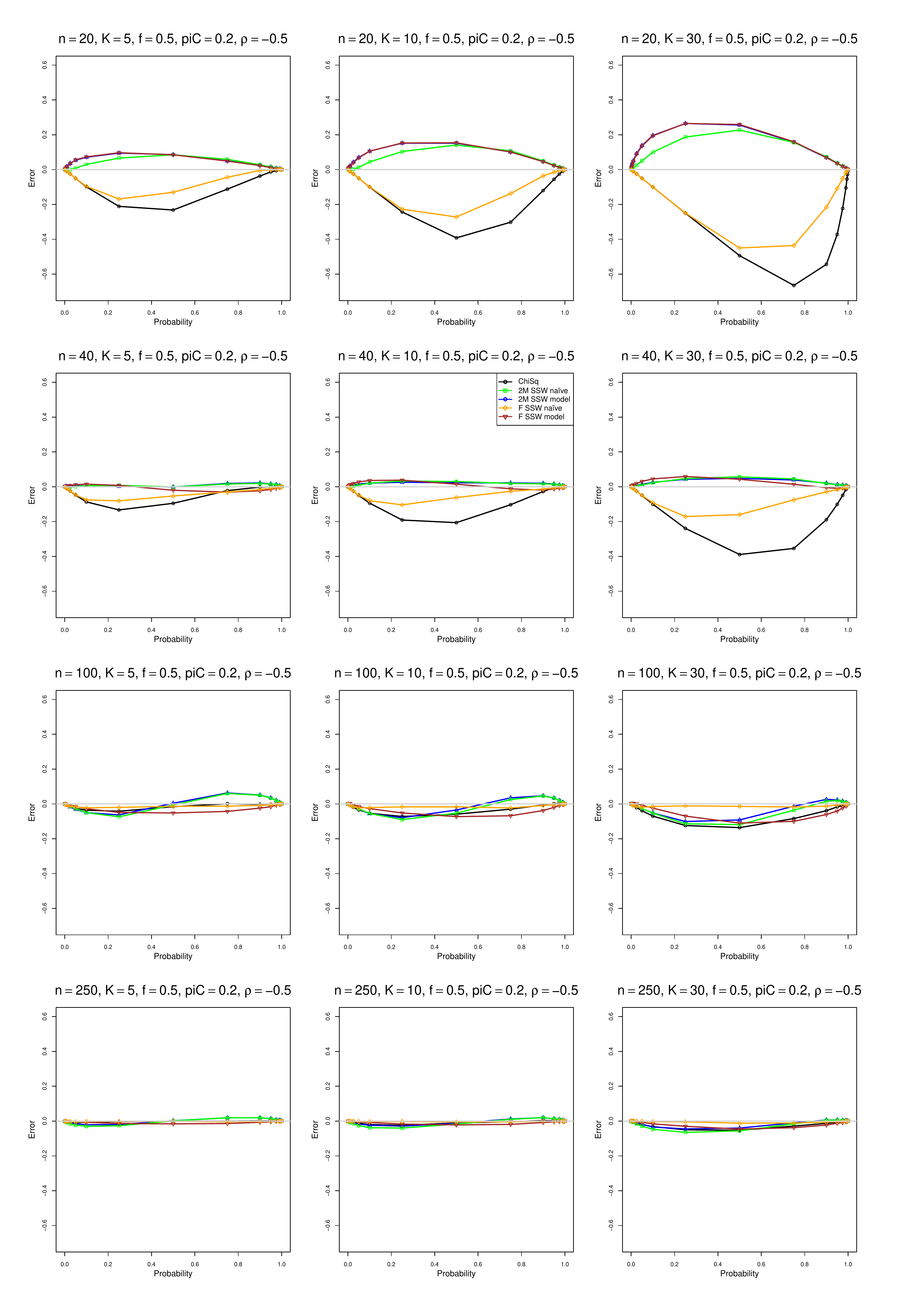}
	\caption{Plots of error in the level of the test for heterogeneity of LRR for five approximations for the null distribution of $Q$, $p_{iC} = .2$, $f = .5$, and $\rho = -0.5$, equal sample sizes}
	\label{PPplot_piC_02theta=-0.5_LRR_equal_sample_sizes}
\end{figure}
\begin{figure}[ht]
		\centering
	\includegraphics[scale=0.33]{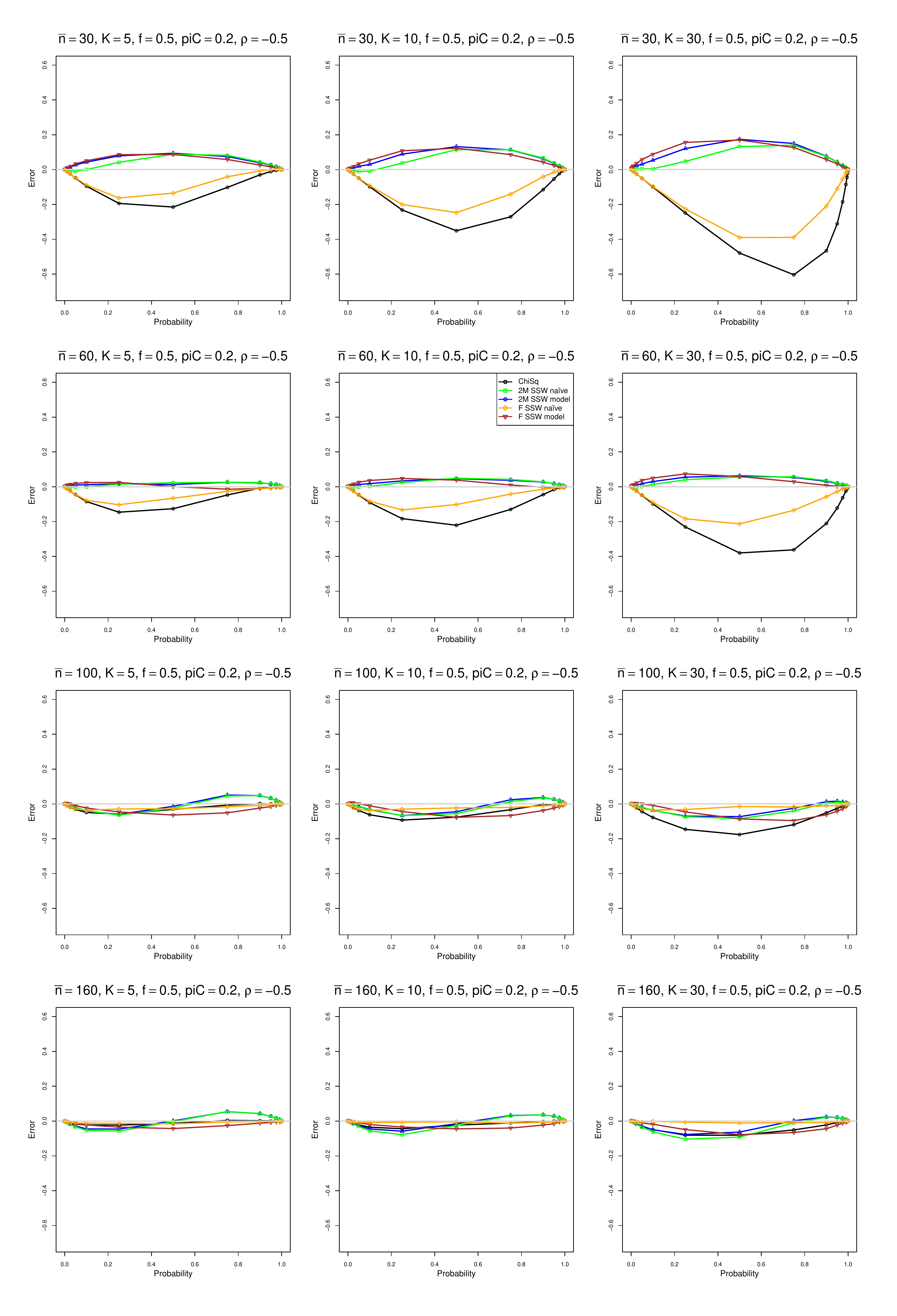}
	\caption{Plots of error in the level of the test for heterogeneity of LRR for five approximations for the null distribution of $Q$, $p_{iC} = .2$, $f = .5$, and $\rho = -0.5$, unequal sample sizes}
	\label{PPplot_piC_02theta=-0.5_LRR_unequal_sample_sizes}
\end{figure}

\begin{figure}[ht]
	\centering
	\includegraphics[scale=0.33]{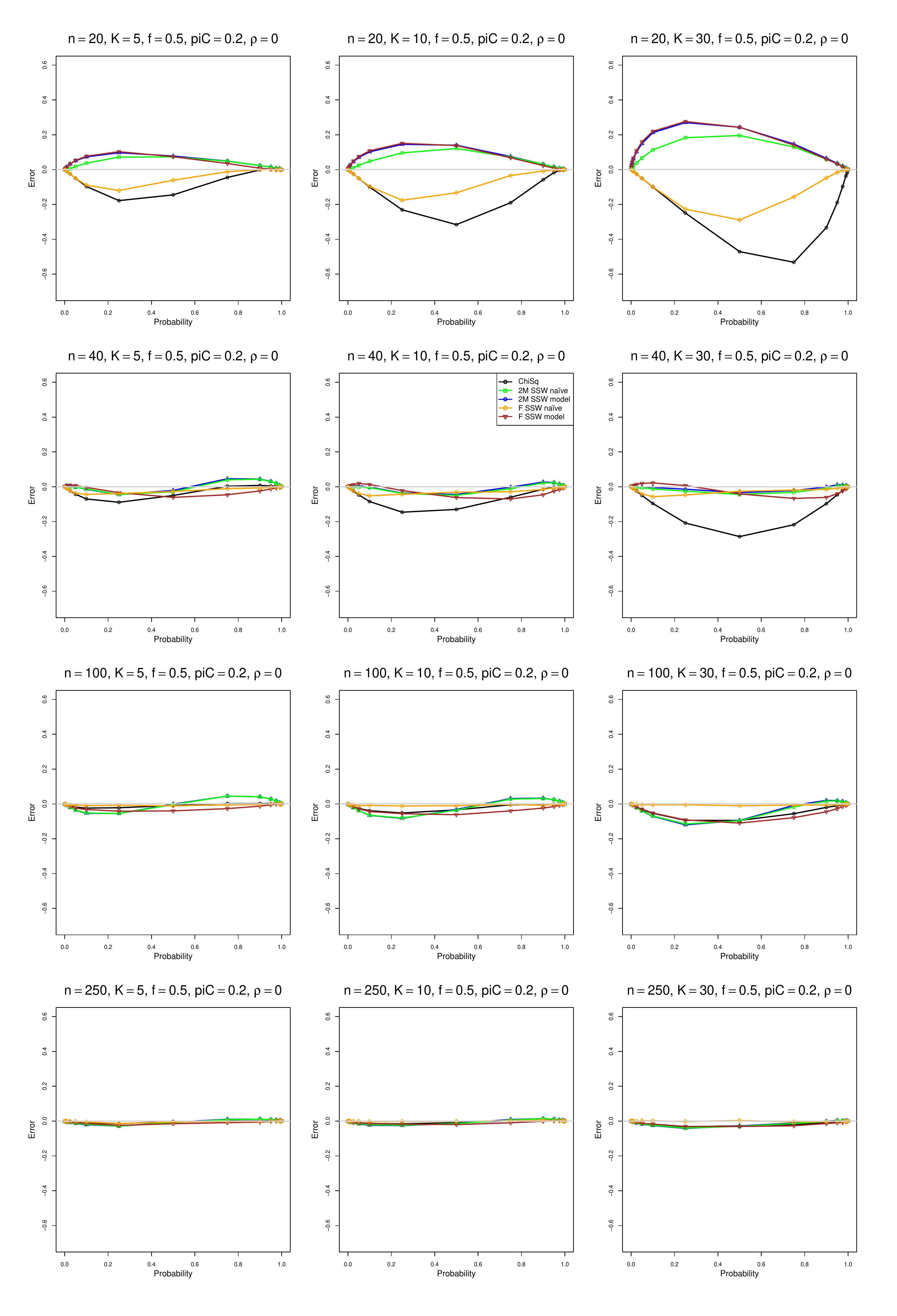}
	\caption{Plots of error in the level of the test for heterogeneity of LRR for five approximations for the null distribution of $Q$, $p_{iC} = .2$, $f = .5$, and $\rho = 0$, equal sample sizes}
	\label{PPplot_piC_02theta=0_LRR_equal_sample_sizes}
\end{figure}
\begin{figure}[ht]
		\centering
	\includegraphics[scale=0.33]{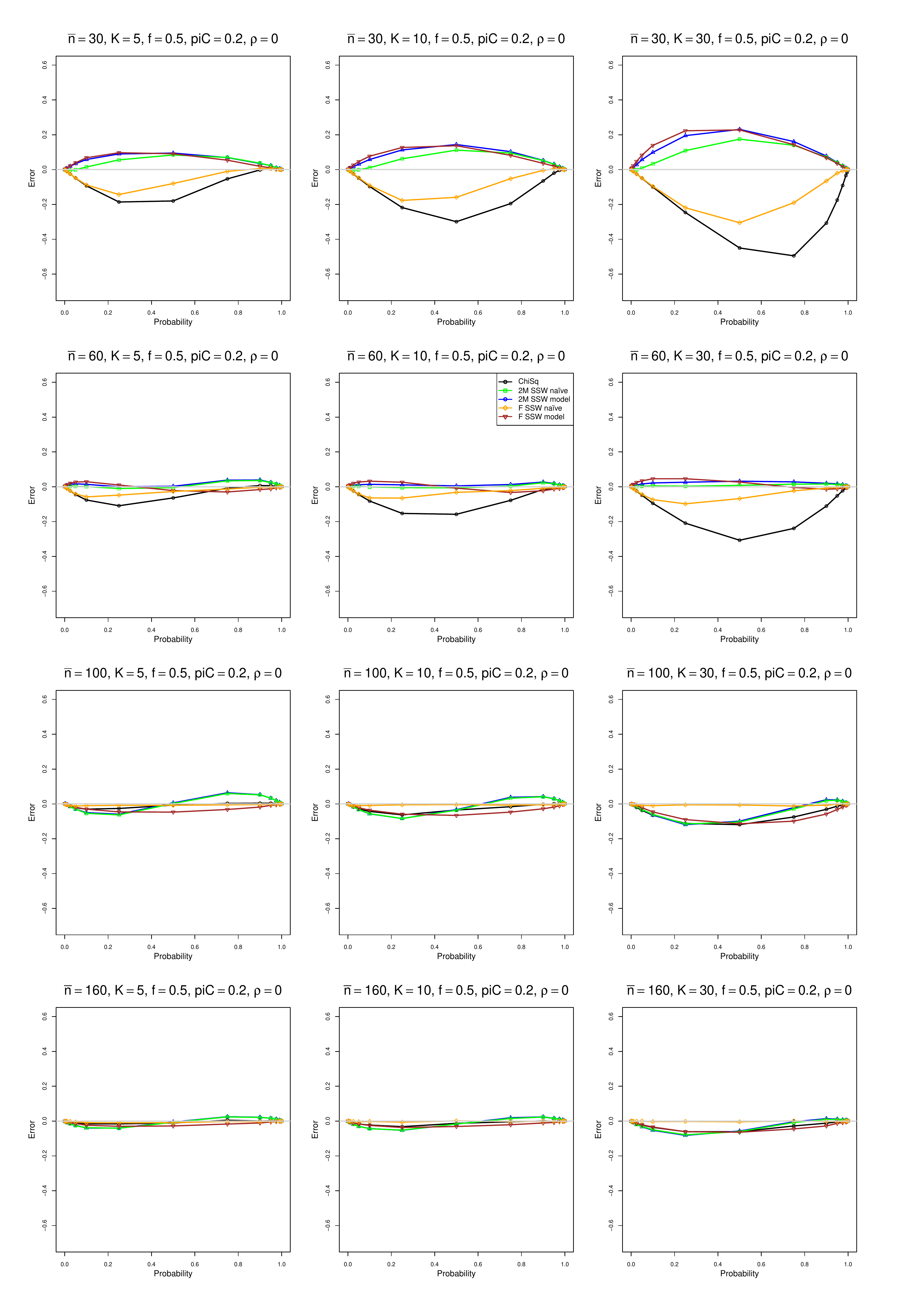}
	\caption{Plots of error in the level of the test for heterogeneity of LRR for five approximations for the null distribution of $Q$, $p_{iC} = .2$, $f = .5$, and $\rho = 0$, unequal sample sizes}
	\label{PPplot_piC_02theta=0_LRR_unequal_sample_sizes}
\end{figure}

\begin{figure}[ht]
	\centering
	\includegraphics[scale=0.33]{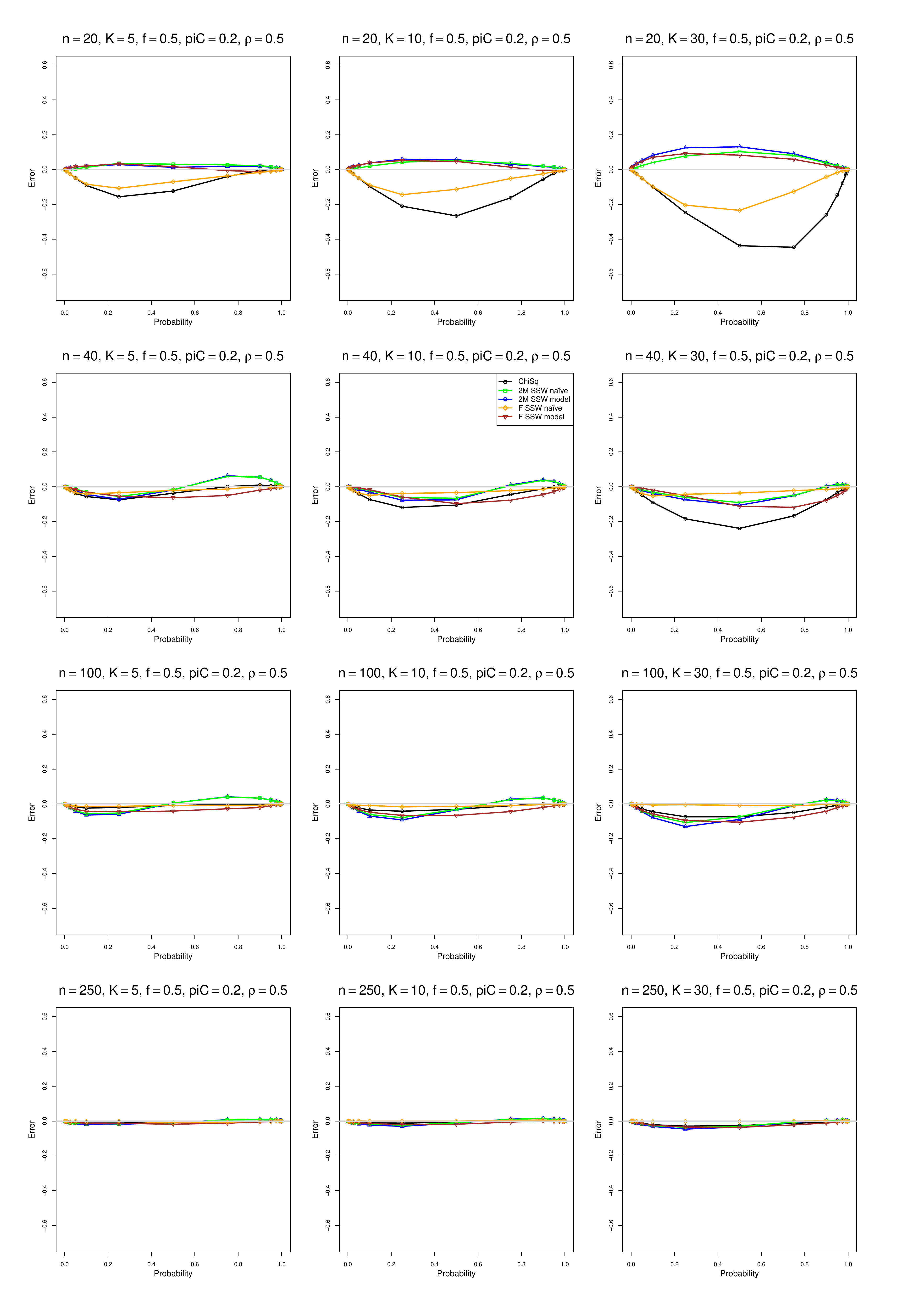}
	\caption{Plots of error in the level of the test for heterogeneity of LRR for five approximations for the null distribution of $Q$, $p_{iC} = .2$, $f = .5$, and $\rho = 0.5$, equal sample sizes}
	\label{PPplot_piC_02theta=0.5_LRR_equal_sample_sizes}
\end{figure}
\begin{figure}[ht]
		\centering
	\includegraphics[scale=0.33]{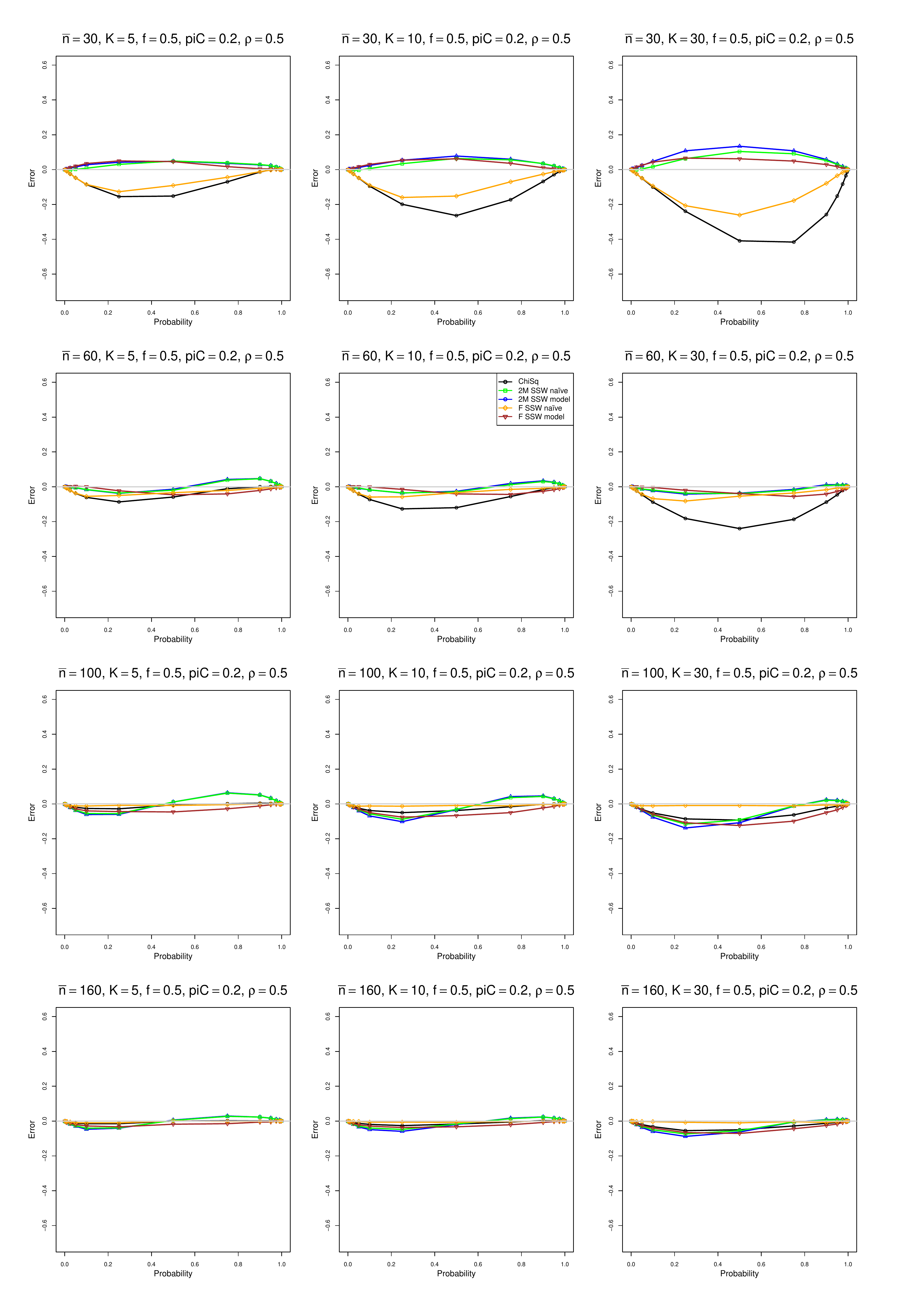}
	\caption{Plots of error in the level of the test for heterogeneity of LRR for five approximations for the null distribution of $Q$, $p_{iC} = .2$, $f = .5$, and $\rho = 0.5$, unequal sample sizes}
	\label{PPplot_piC_02theta=0.5_LRR_unequal_sample_sizes}
\end{figure}

\begin{figure}[ht]
	\centering
	\includegraphics[scale=0.33]{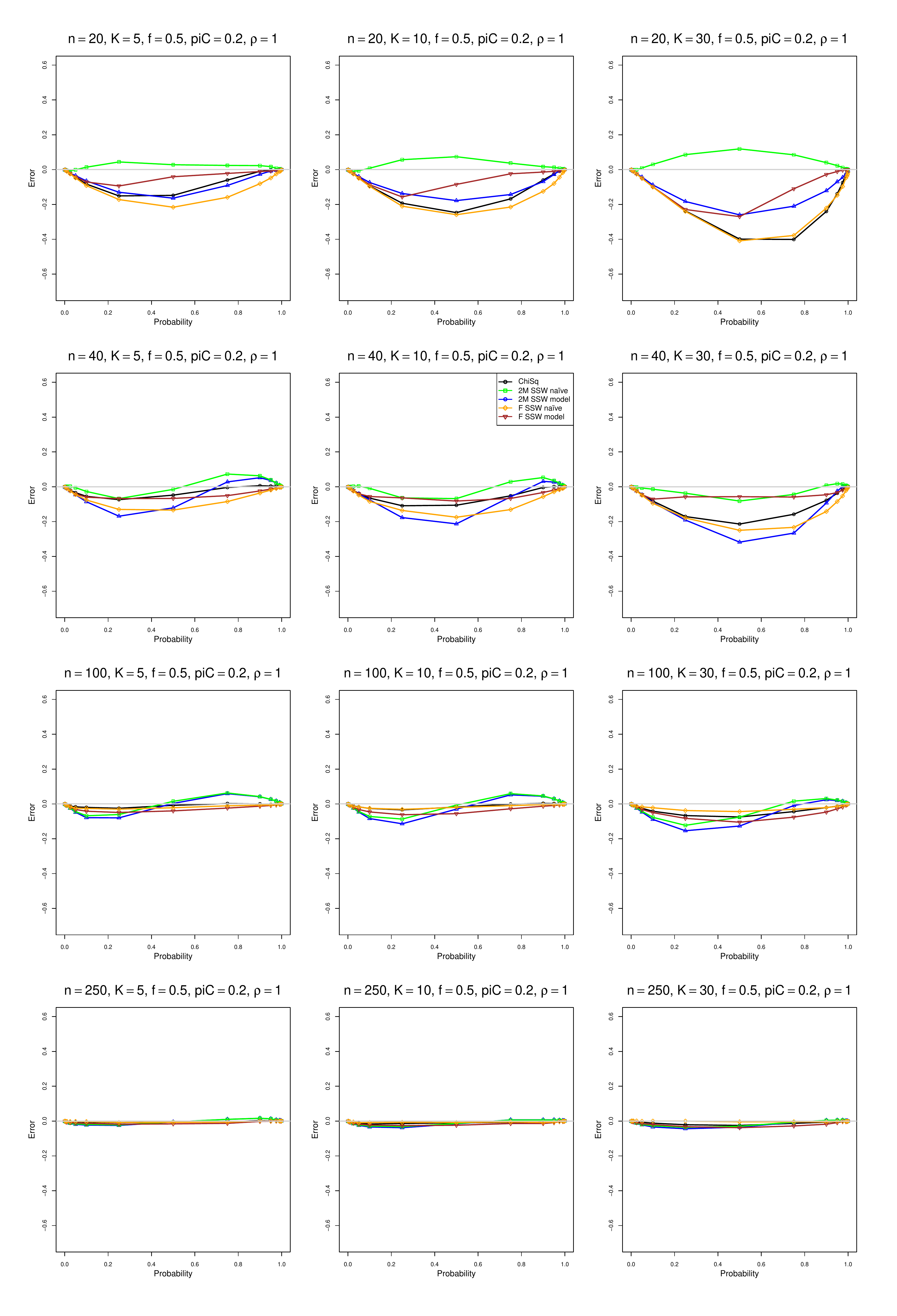}
	\caption{Plots of error in the level of the test for heterogeneity of LRR for five approximations for the null distribution of $Q$, $p_{iC} = .2$, $f = .5$, and $\rho = 1$, equal sample sizes}
	\label{PPplot_piC_02theta=1_LRR_equal_sample_sizes}
\end{figure}
\begin{figure}[ht]
		\centering
	\includegraphics[scale=0.33]{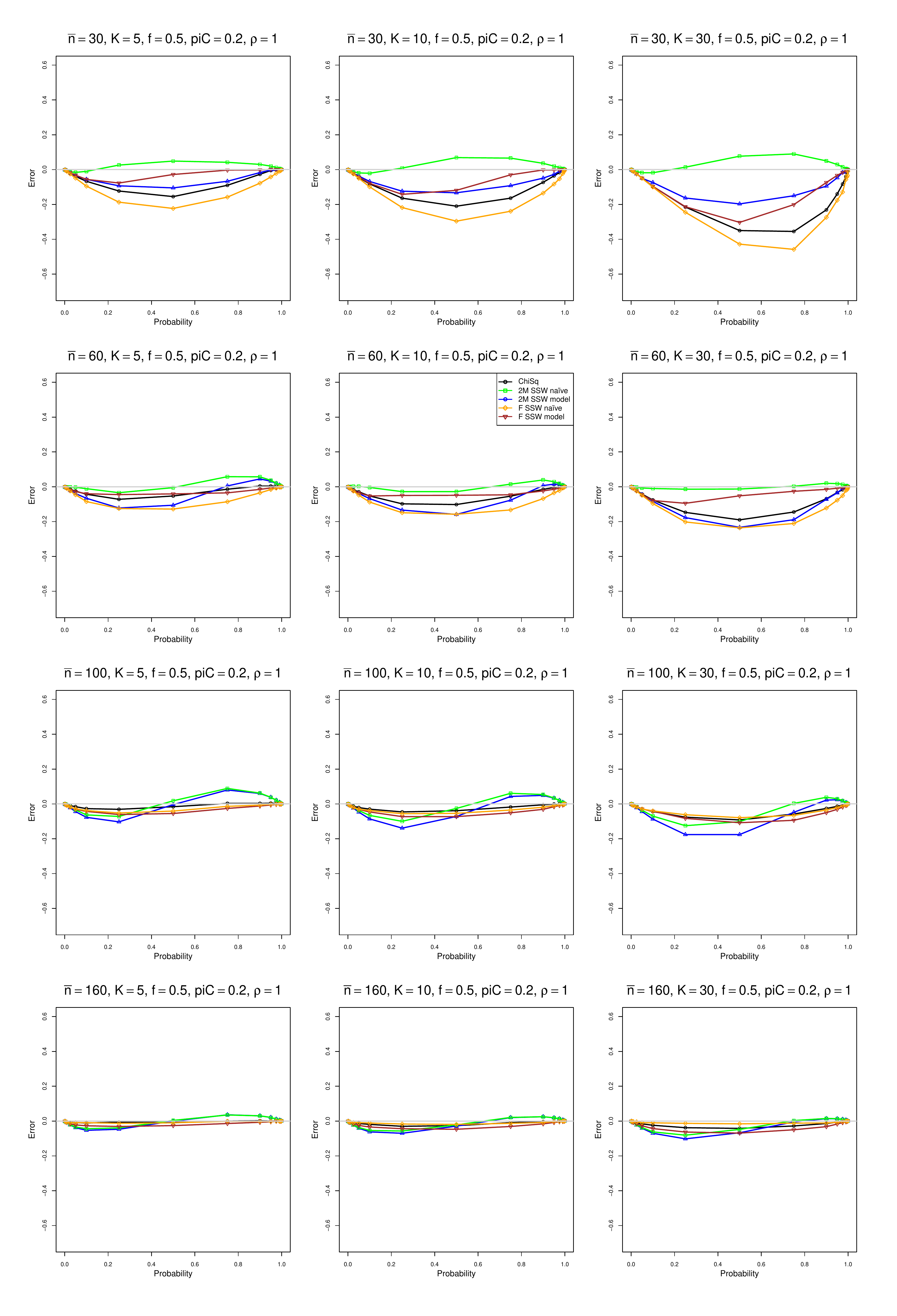}
	\caption{Plots of error in the level of the test for heterogeneity of LRR for five approximations for the null distribution of $Q$, $p_{iC} = .2$, $f = .5$, and $\rho = 1$, unequal sample sizes}
	\label{PPplot_piC_02theta=1_LRR_unequal_sample_sizes}
\end{figure}

\begin{figure}[ht]
	\centering
	\includegraphics[scale=0.33]{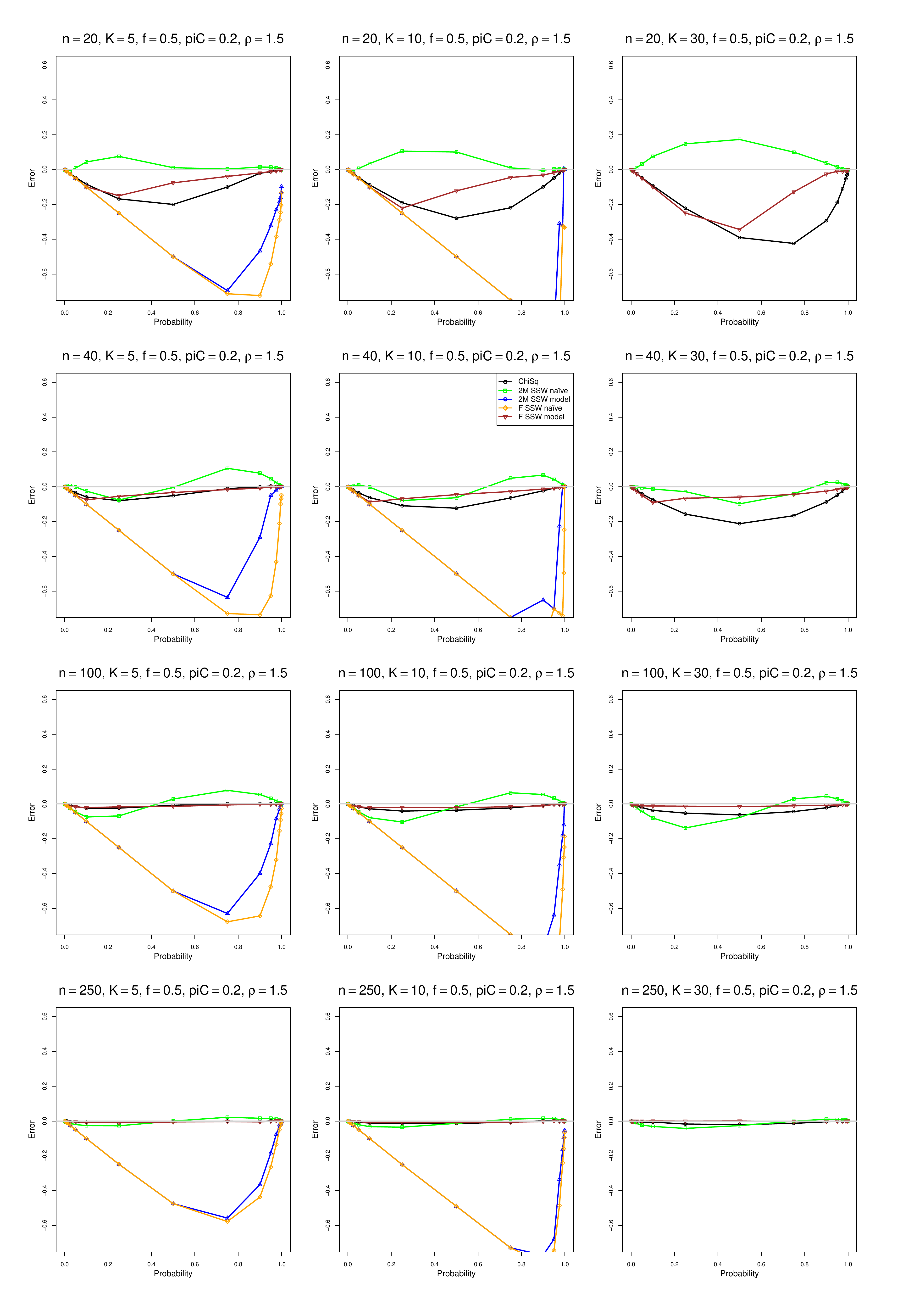}
	\caption{Plots of error in the level of the test for heterogeneity of LRR for five approximations for the null distribution of $Q$, $p_{iC} = .2$, $f = .5$, and $\rho = 1.5$, equal sample sizes}
	\label{PPplot_piC_02theta=1.5_LRR_equal_sample_sizes}
\end{figure}
\begin{figure}[ht]
		\centering
	\includegraphics[scale=0.33]{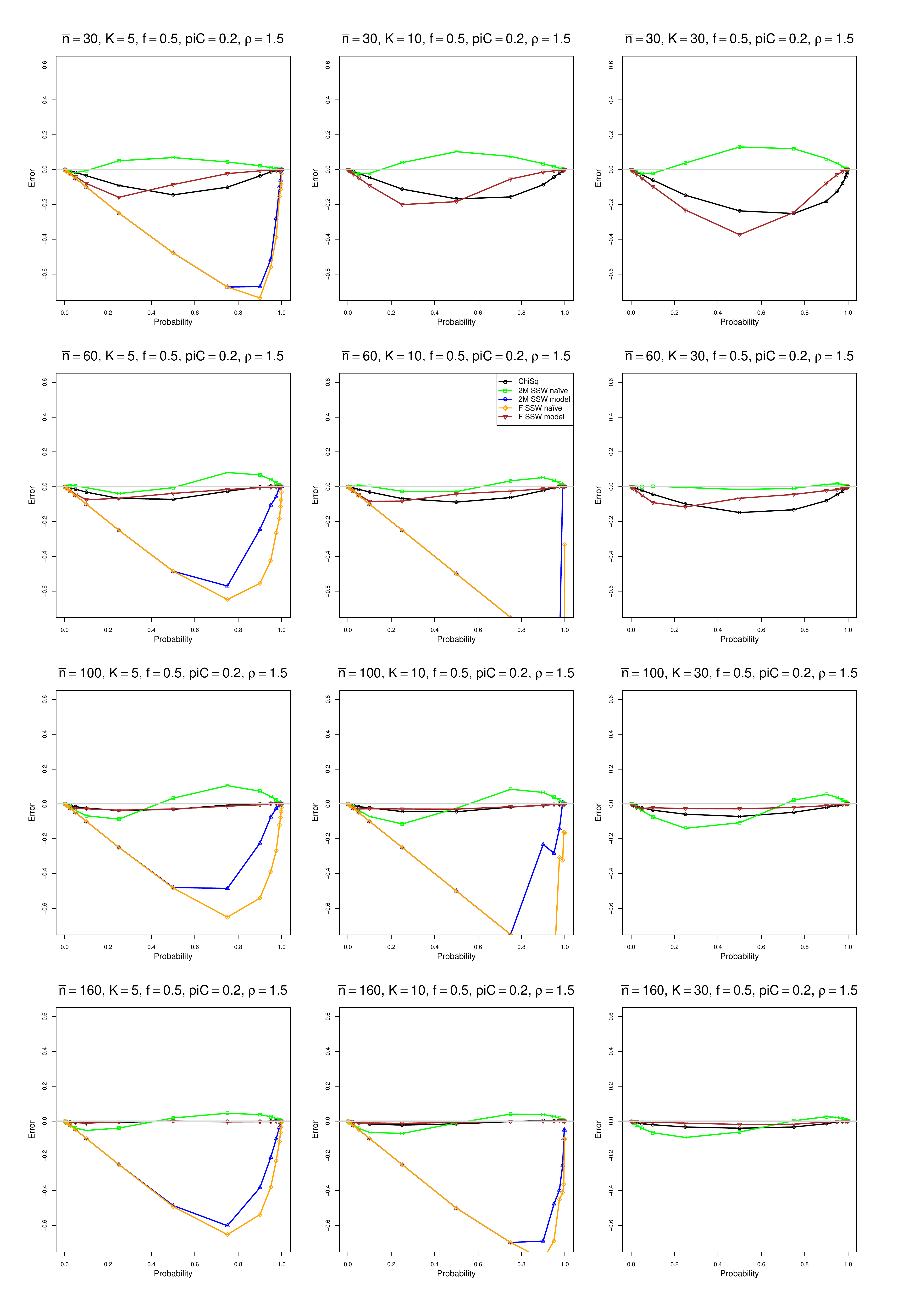}
	\caption{Plots of error in the level of the test for heterogeneity of LRR for five approximations for the null distribution of $Q$, $p_{iC} = .2$, $f = .5$, and $\rho = 1.5$, unequal sample sizes}
	\label{PPplot_piC_02theta=1.5_LRR_unequal_sample_sizes}
\end{figure}
\begin{figure}[ht]
	\centering
	\includegraphics[scale=0.33]{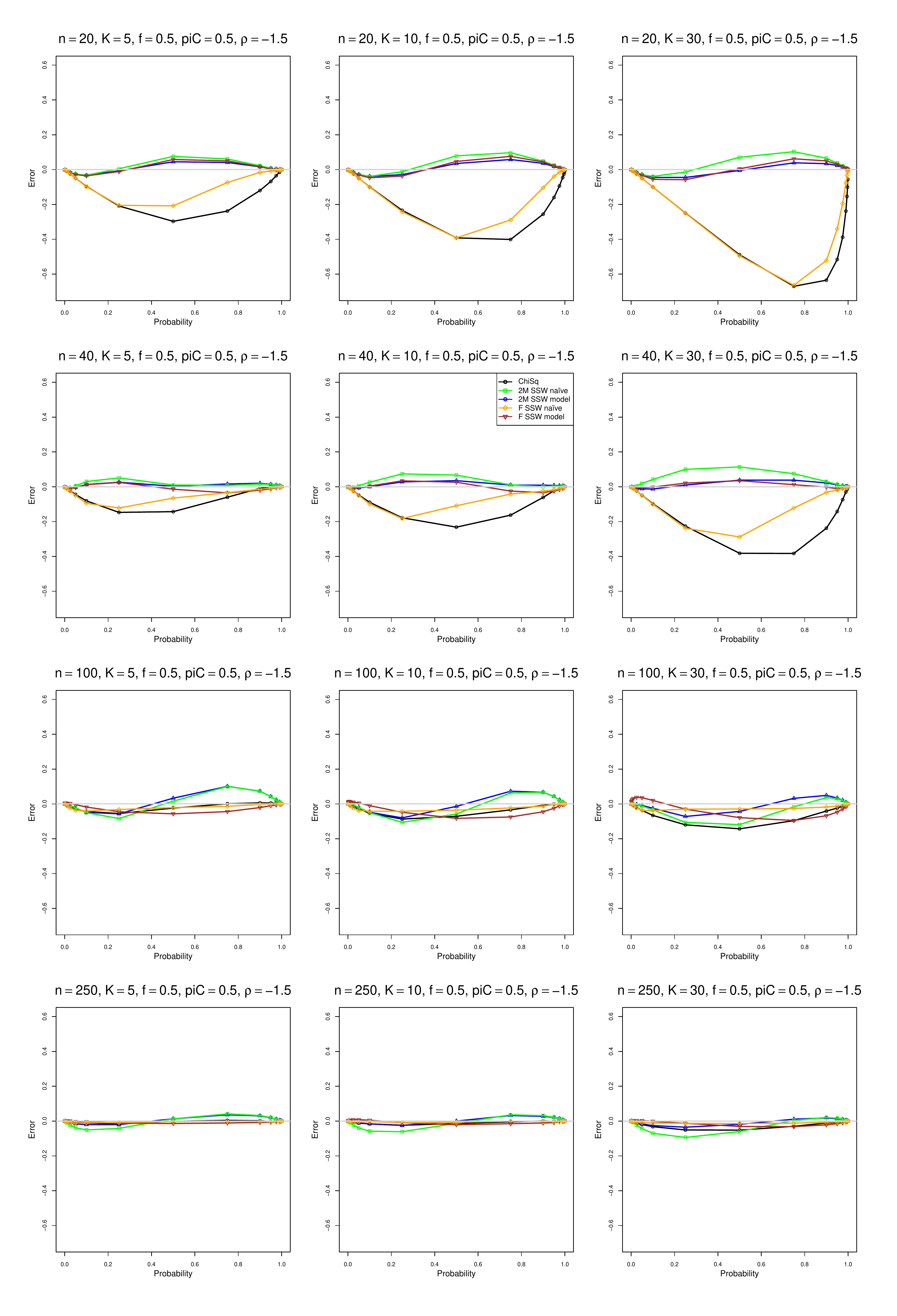}
	\caption{Plots of error in the level of the test for heterogeneity of LRR for five approximations for the null distribution of $Q$, $p_{iC} = .5$, $f = .5$, and $\rho = -1.5$, equal sample sizes}
	\label{PPplot_piC_05theta=-1.5_LRR_equal_sample_sizes}
\end{figure}
\begin{figure}[ht]
		\centering
	\includegraphics[scale=0.33]{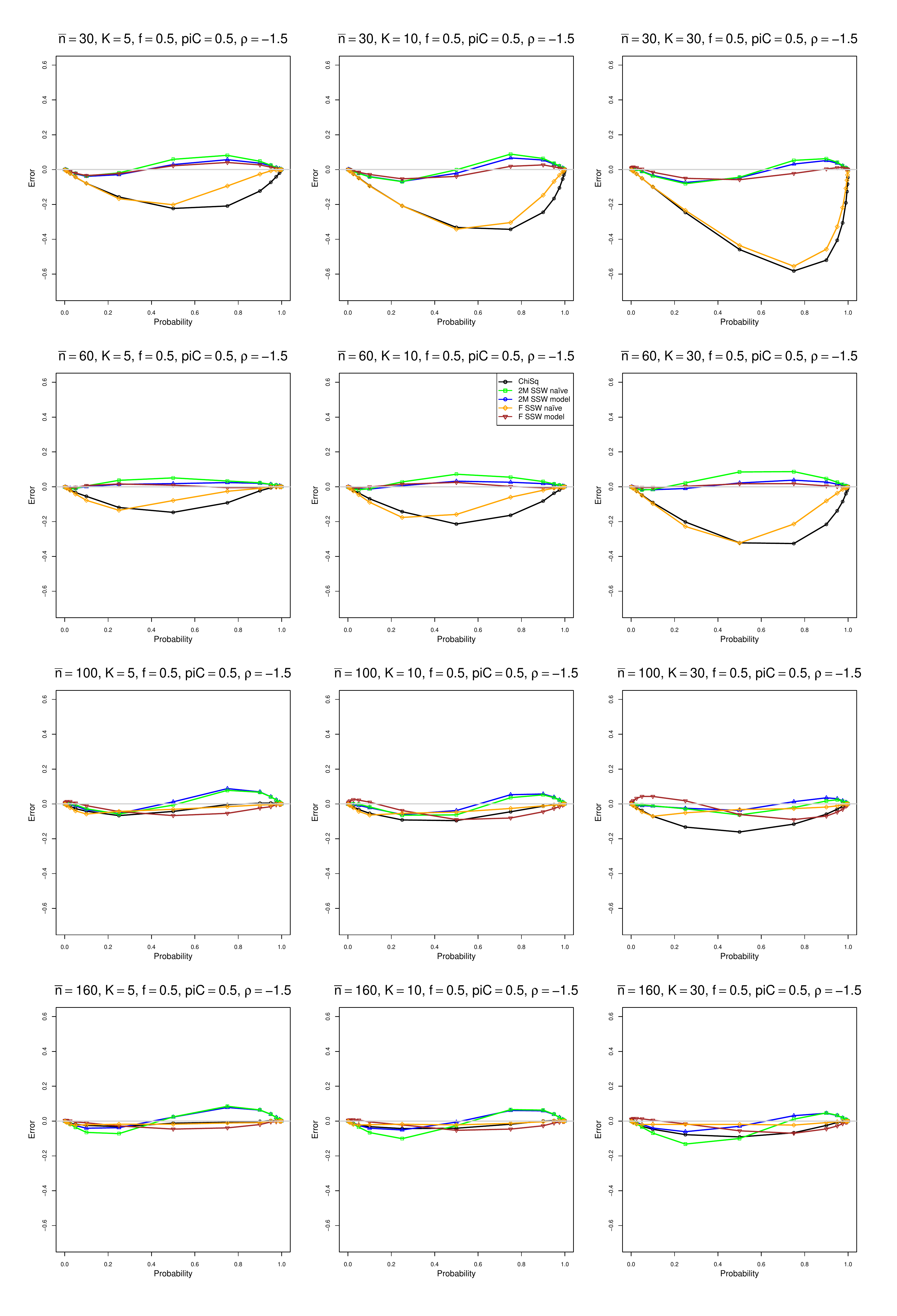}
	\caption{Plots of error in the level of the test for heterogeneity of LRR for five approximations for the null distribution of $Q$, $p_{iC} = .5$, $f = .5$, and $\rho = -1.5$, unequal sample sizes}
	\label{PPplot_piC_05theta=-1.5_LRR_unequal_sample_sizes}
\end{figure}

\begin{figure}[ht]
	\centering
	\includegraphics[scale=0.33]{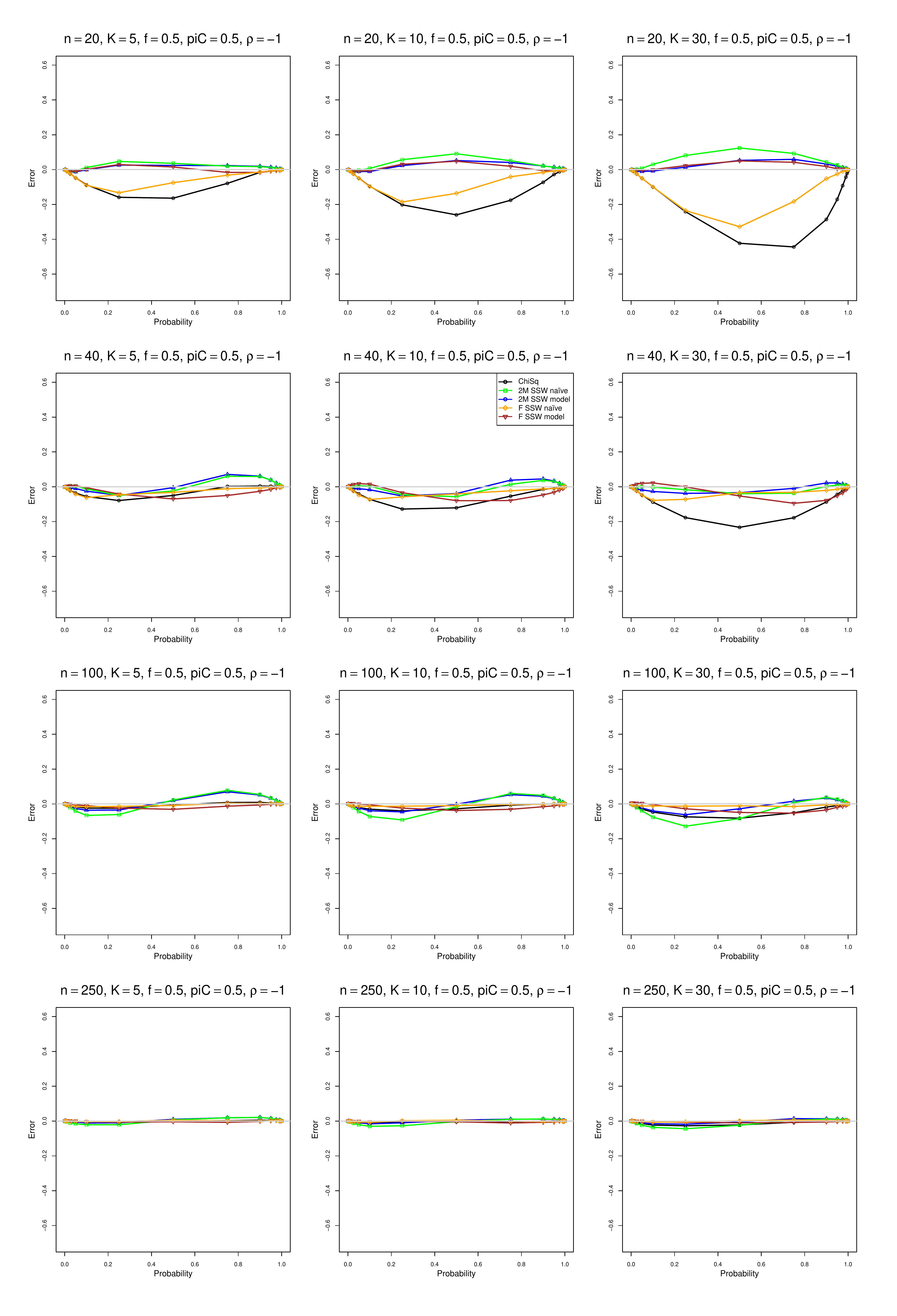}
	\caption{Plots of error in the level of the test for heterogeneity of LRR for five approximations for the null distribution of $Q$, $p_{iC} = .5$, $f = .5$, and $\rho = -1$, equal sample sizes}
	\label{PPplot_piC_05theta=-1_LRR_equal_sample_sizes}
\end{figure}
\begin{figure}[ht]
		\centering
	\includegraphics[scale=0.33]{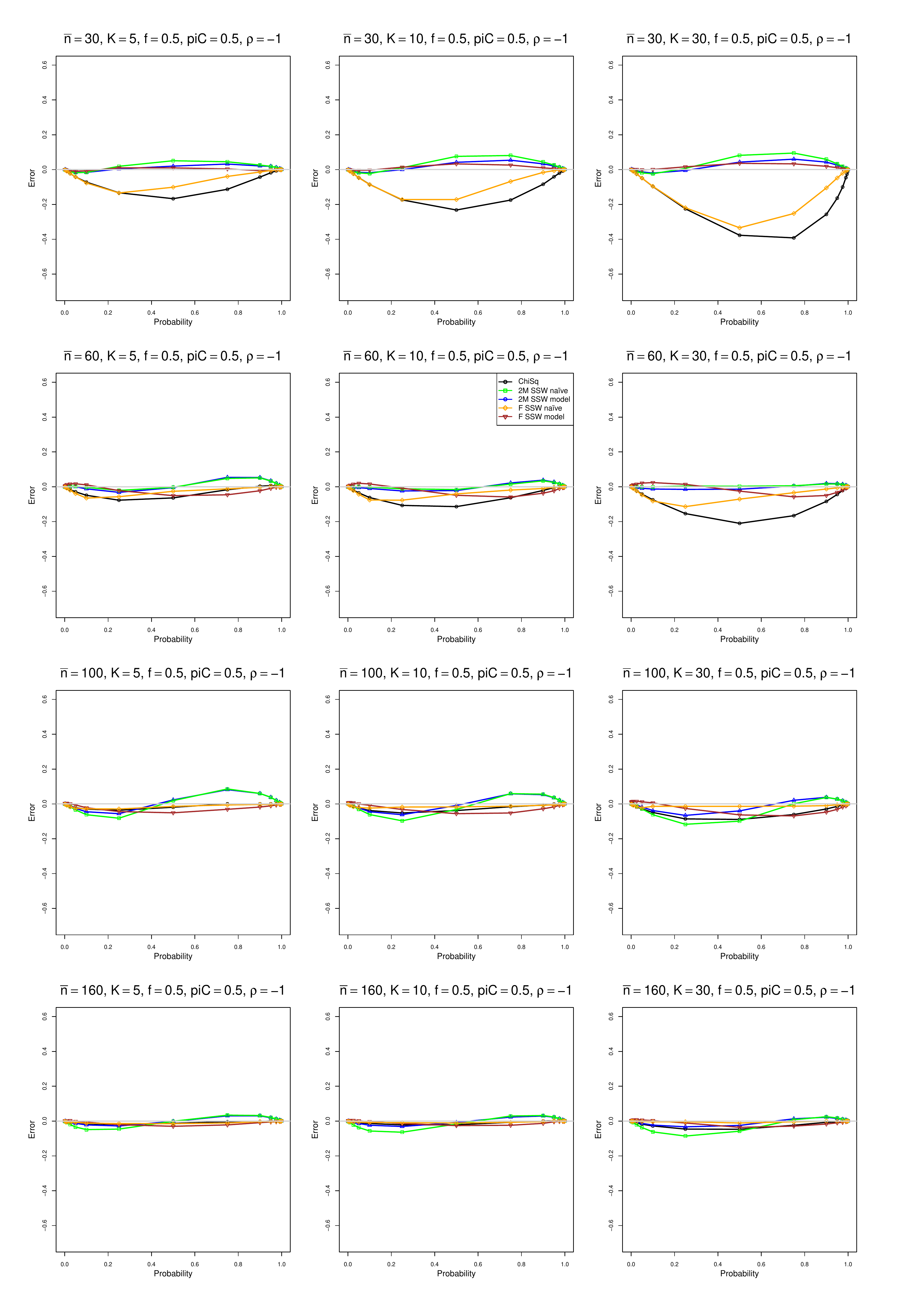}
	\caption{Plots of error in the level of the test for heterogeneity of LRR for five approximations for the null distribution of $Q$, $p_{iC} = .5$, $f = .5$, and $\rho = -1$, unequal sample sizes}
	\label{PPplot_piC_05theta=-1_LRR_unequal_sample_sizes}
\end{figure}

\begin{figure}[ht]
	\centering
	\includegraphics[scale=0.33]{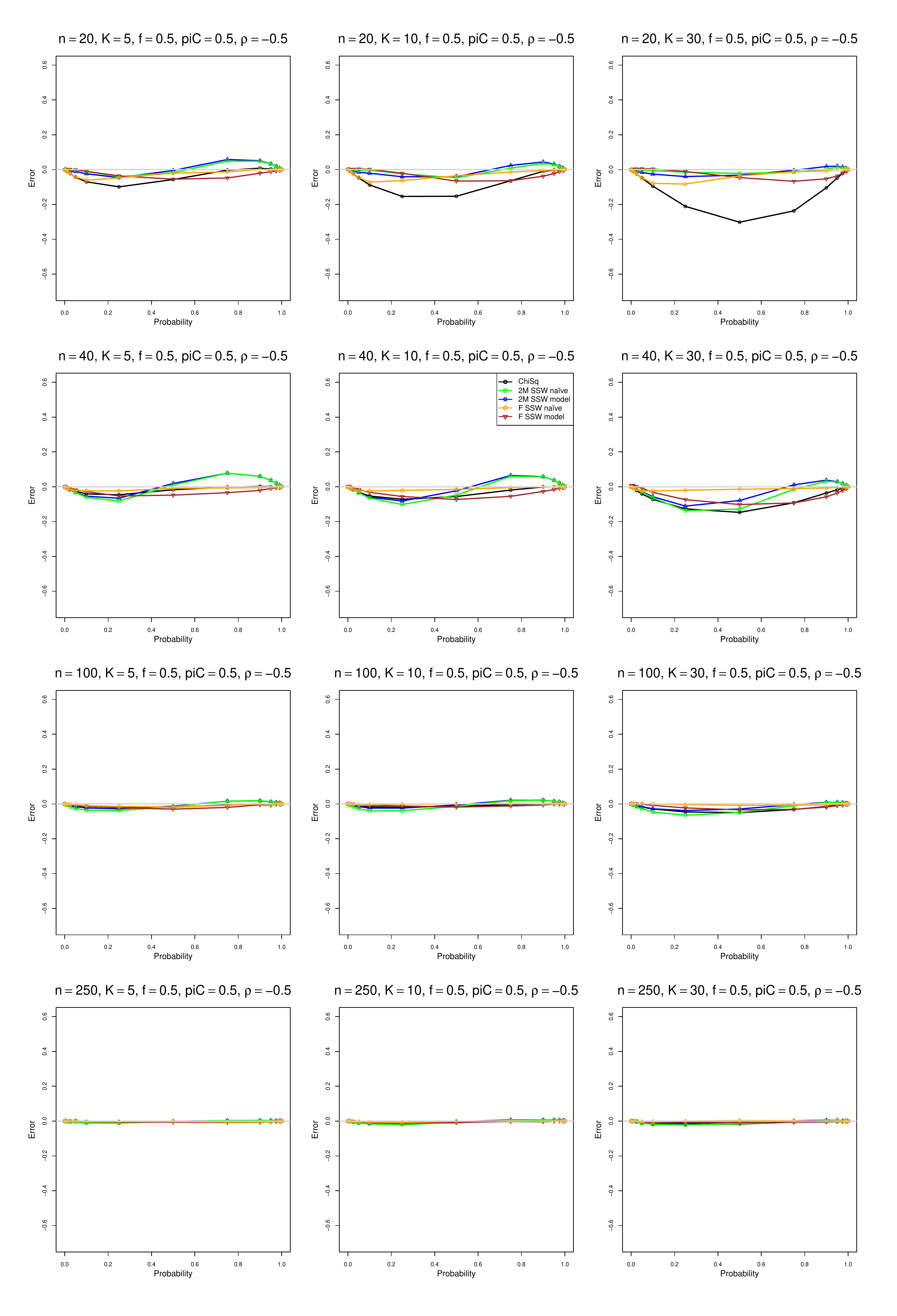}
	\caption{Plots of error in the level of the test for heterogeneity of LRR for five approximations for the null distribution of $Q$, $p_{iC} = .5$, $f = .5$, and $\rho = -0.5$, equal sample sizes}
	\label{PPplot_piC_05theta=-0.5_LRR_equal_sample_sizes}
\end{figure}
\begin{figure}[ht]
		\centering
	\includegraphics[scale=0.33]{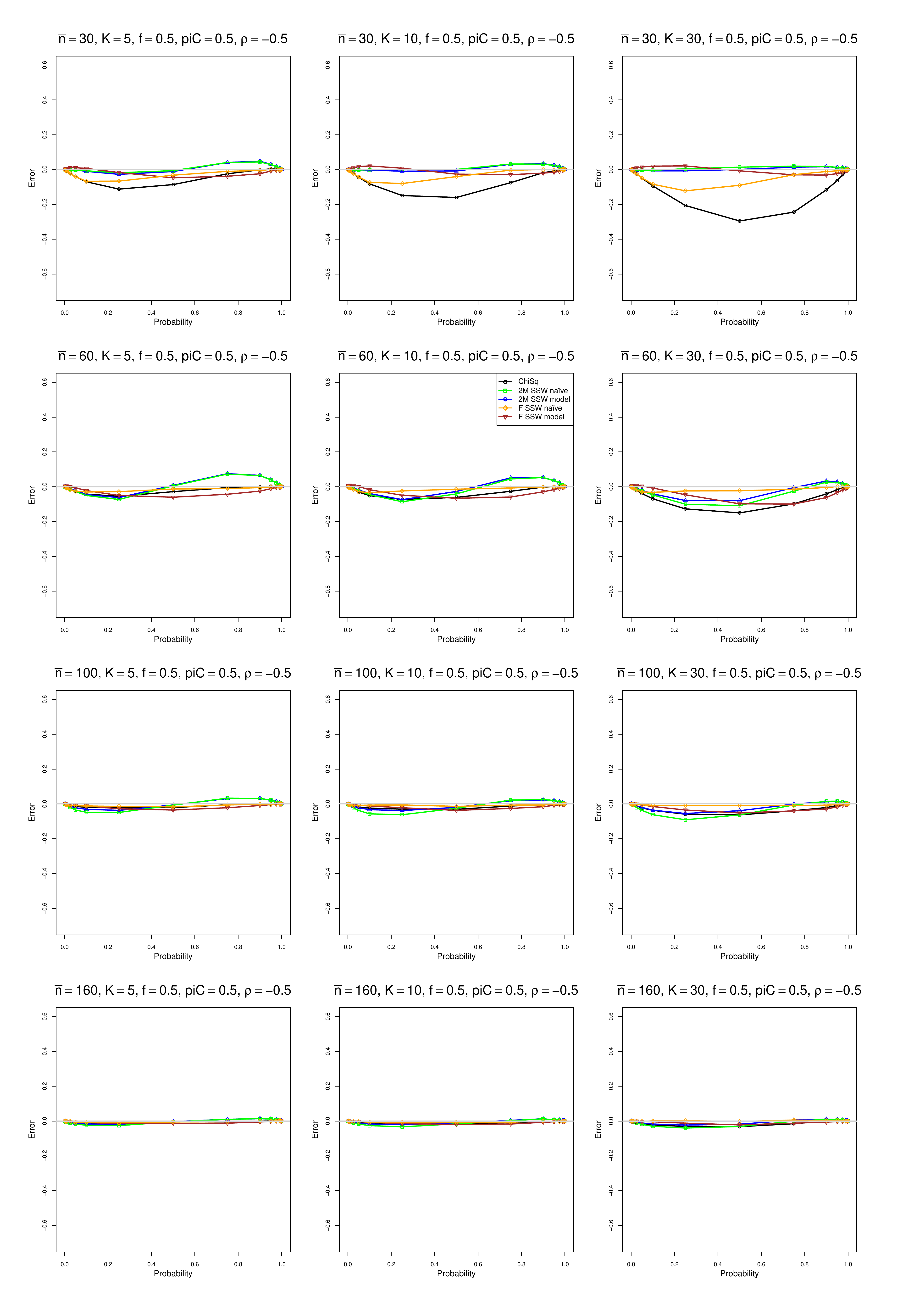}
	\caption{Plots of error in the level of the test for heterogeneity of LRR for five approximations for the null distribution of $Q$, $p_{iC} = .5$, $f = .5$, and $\rho = -0.5$, unequal sample sizes}
	\label{PPplot_piC_05theta=-0.5_LRR_unequal_sample_sizes}
\end{figure}

\begin{figure}[ht]
	\centering
	\includegraphics[scale=0.33]{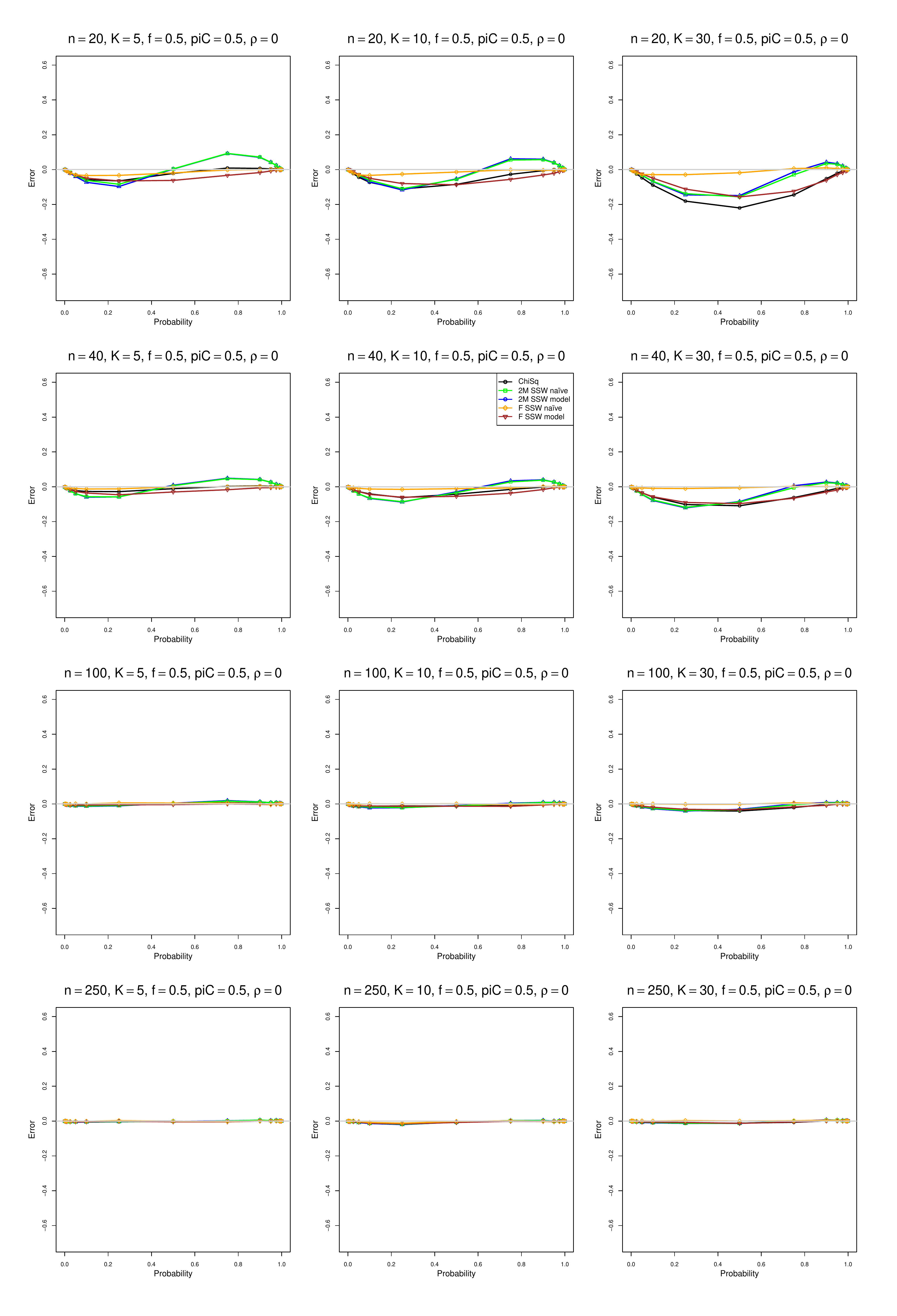}
	\caption{Plots of error in the level of the test for heterogeneity of LRR for five approximations for the null distribution of $Q$, $p_{iC} = .5$, $f = .5$, and $\rho = 0$, equal sample sizes}
	\label{PPplot_piC_05theta=0_LRR_equal_sample_sizes}
\end{figure}
\begin{figure}[ht]
		\centering
	\includegraphics[scale=0.33]{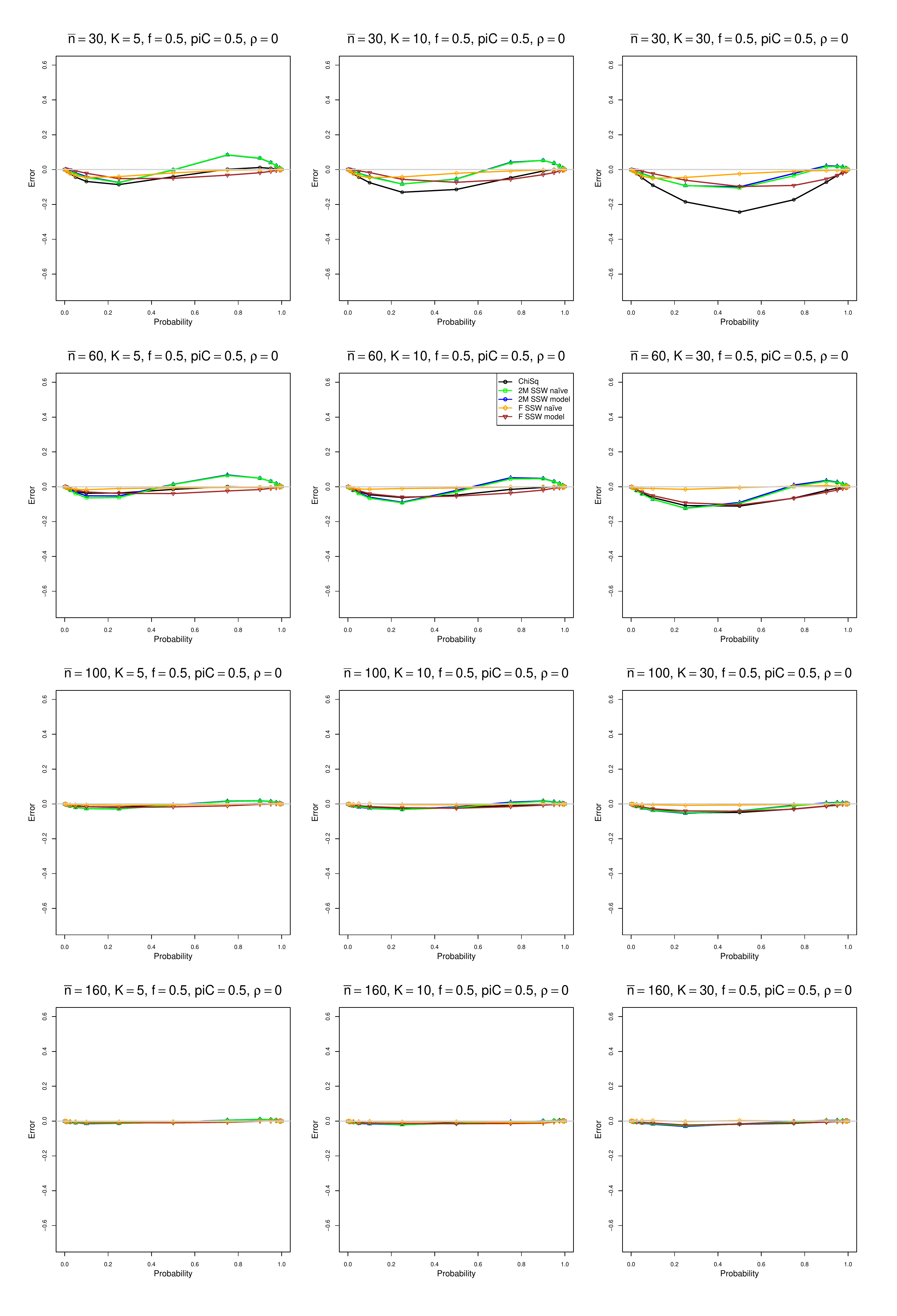}
	\caption{Plots of error in the level of the test for heterogeneity of LRR for five approximations for the null distribution of $Q$, $p_{iC} = .5$, $f = .5$, and $\rho = 0$, unequal sample sizes}
	\label{PPplot_piC_05theta=0_LRR_unequal_sample_sizes}
\end{figure}

\begin{figure}[ht]
	\centering
	\includegraphics[scale=0.33]{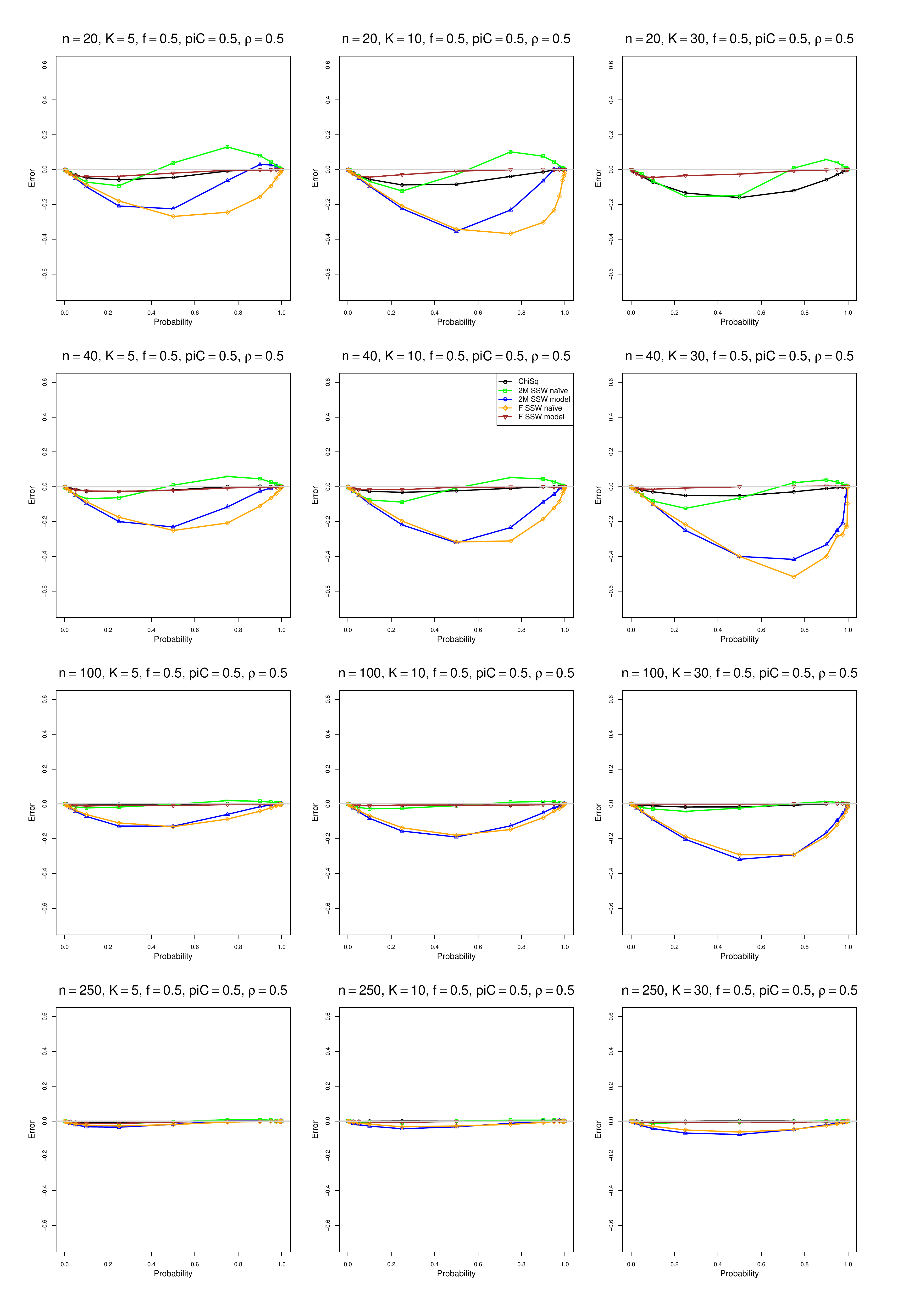}
	\caption{Plots of error in the level of the test for heterogeneity of LRR for five approximations for the null distribution of $Q$, $p_{iC} = .5$, $f = .5$, and $\rho = 0.5$, equal sample sizes}
	\label{PPplot_piC_05theta=0.5_LRR_equal_sample_sizes}
\end{figure}
\begin{figure}[ht]
		\centering
	\includegraphics[scale=0.33]{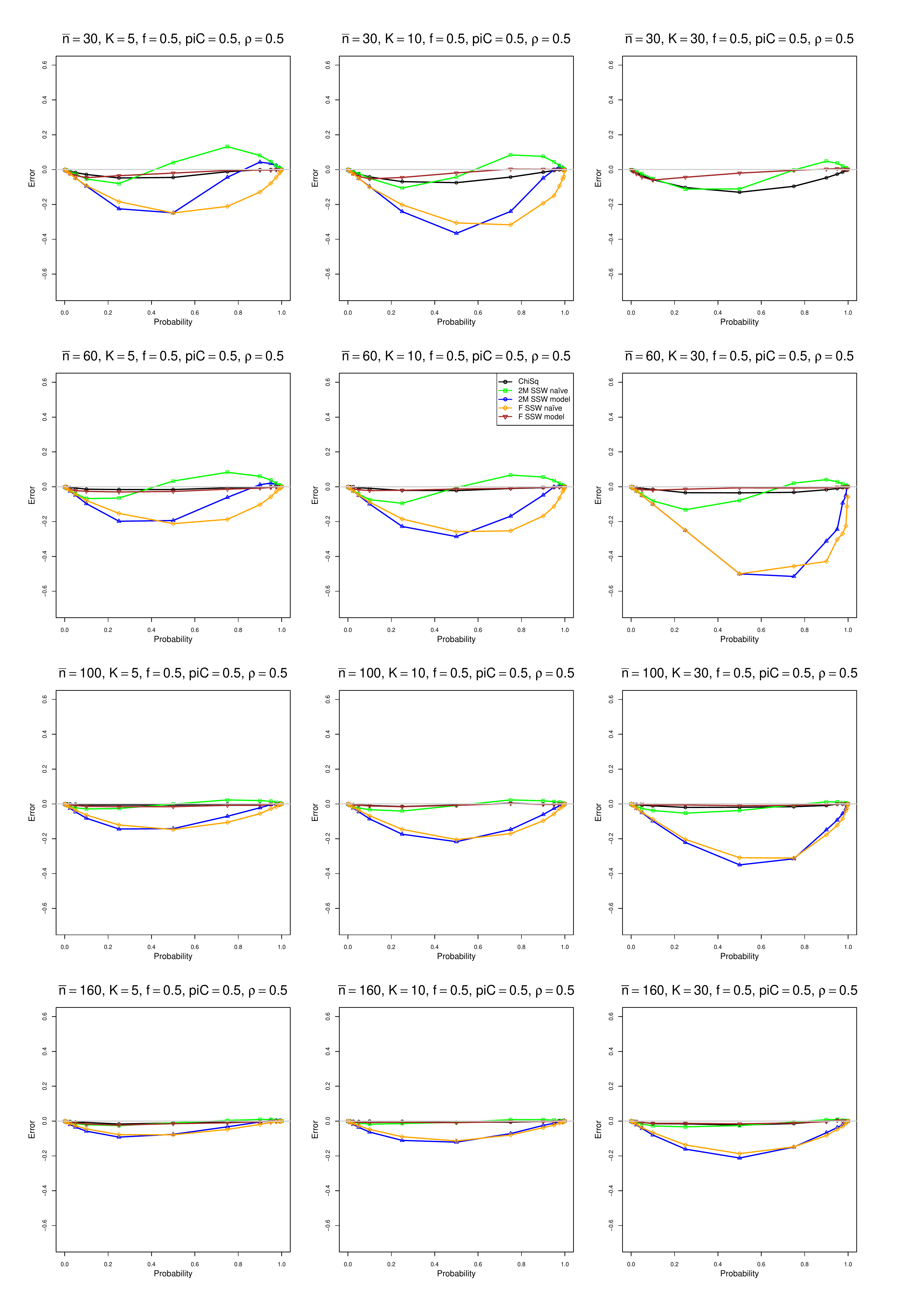}
	\caption{Plots of error in the level of the test for heterogeneity of LRR for five approximations for the null distribution of $Q$, $p_{iC} = .5$, $f = .5$, and $\rho = 0.5$, unequal sample sizes}
	\label{PPplot_piC_05theta=0.5_LRR_unequal_sample_sizes}
\end{figure}

\clearpage

\section*{Appendix E: Empirical level at $\alpha = .05$, vs $\rho$, of the test for heterogeneity of LRR ($\tau^2 = 0$ versus $\tau^2 > 0$) based on approximations for the null distribution of $Q$}

Each figure corresponds to a value of the probability of an event in the Control arm $p_{iC}$  (= .1, .2, .5) and a choice of equal or unequal sample sizes ($n$ or $bar{n}$). \\
The fraction of each study's sample size in the Control arm  $f$ is held constant at 0.5.

For each combination of a value of $n$ (= 20, 40, 100, 250) or $\bar{n}$ (= 30, 60, 100, 160) and a value of $K$ (= 5, 10, 30), a panel plots the empirical level versus $\rho$ ($\rho$ = $-0.5$, 0, 0.5, 1, 1.5 when $p_{iC} = .1$ or $.2$, and $\rho$ = $-1.5$, $-1$, $-0.5$, 0, 0.5 when $p_{iC} = .5$).\\
The approximations for the null distribution of $Q$ are
\begin{itemize}
\item ChiSq (Chi-square approximation with $K-1$ df, inverse-variance weights)
\item 2M SSW na\"{i}ve (Two-moment gamma approximation, na\"{i}ve estimation of $p_{iT}$ from $X_{iT}$ and $n_{iT}$, effective-sample-size weights)
\item 2M SSW model (Two-moment gamma approximation, model-based estimation of $p_{iT}$, effective-sample-size weights)
\item F SSW na\"{i}ve (Farebrother approximation, na\"{i}ve estimation of $p_{iT}$ from $X_{iT}$ and $n_{iT}$, effective-sample-size weights)
\item F SSW model (Farebrother approximation, model-based estimation of $p_{iT}$, effective-sample-size weights)
\end{itemize}

\clearpage
\setcounter{figure}{0}
\renewcommand{\thefigure}{E.\arabic{figure}}
\begin{figure}[t]
	\centering
	\includegraphics[scale=0.33]{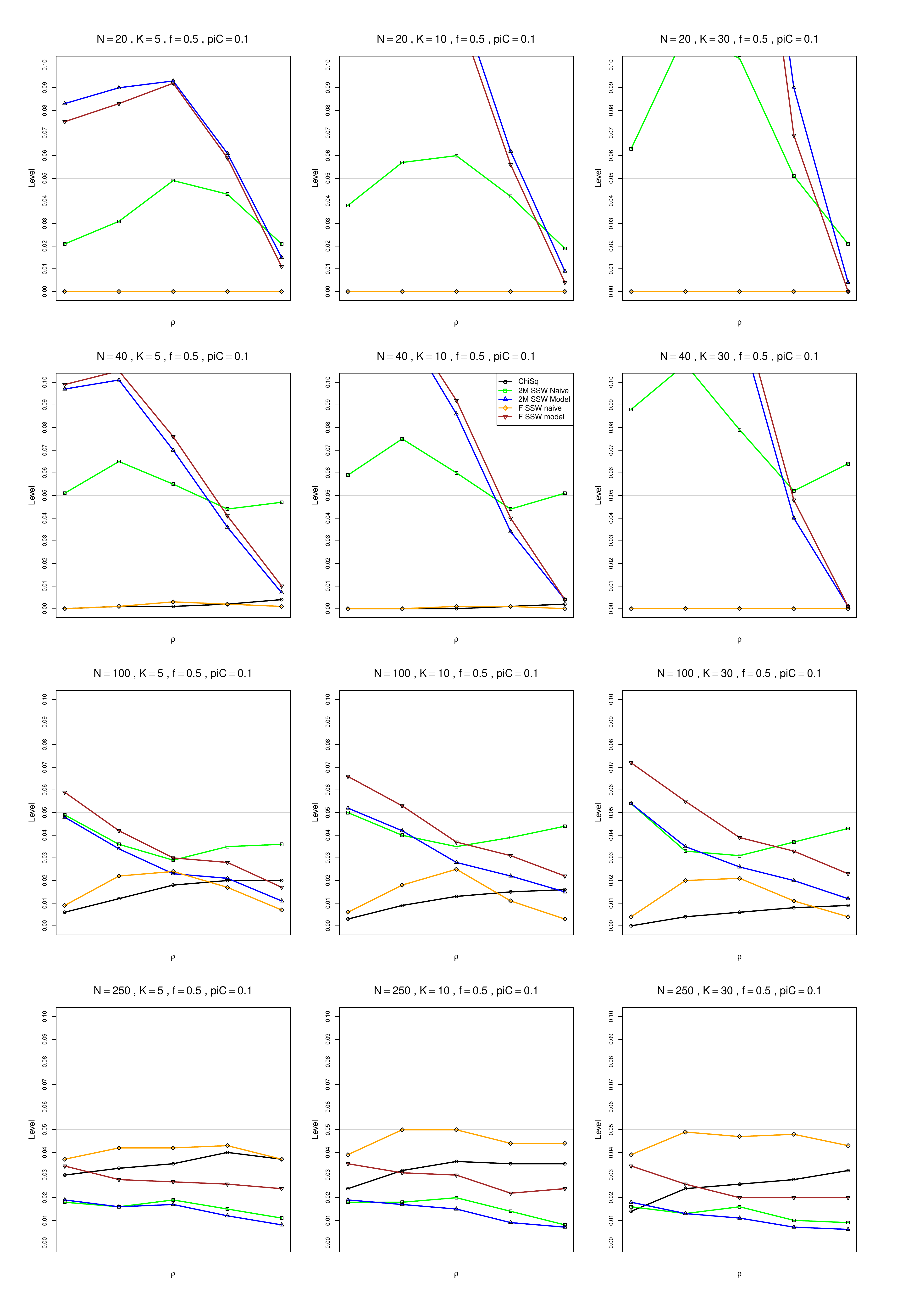}
	\caption{Q for LRR: actual level at $\alpha = .05$ for $p_{iC} = .1$ and $f = .5$, equal sample sizes
		\label{NewQforRR_piC01andq05_equal_sample_sizes}}
\end{figure}

\begin{figure}[t]
	\centering
	\includegraphics[scale=0.33]{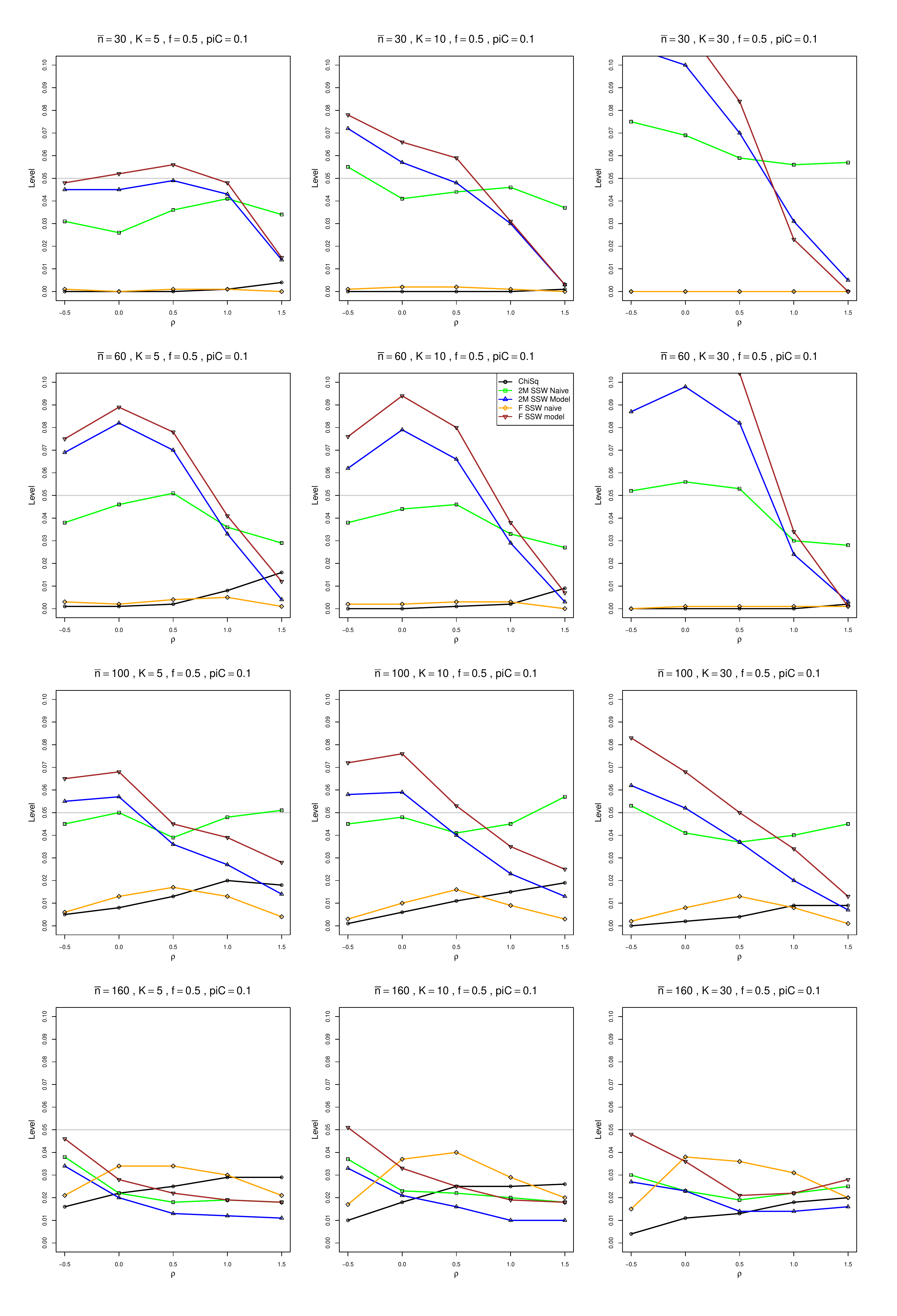}
	\caption{Q for LRR: actual level at $\alpha = .05$ for $p_{iC} = .1$ and $f = .5$, unequal sample sizes
		\label{NewQforRR_piC01andq05_unequal_sample_sizes}}
\end{figure}

\begin{figure}[t]
	\centering
	\includegraphics[scale=0.33]{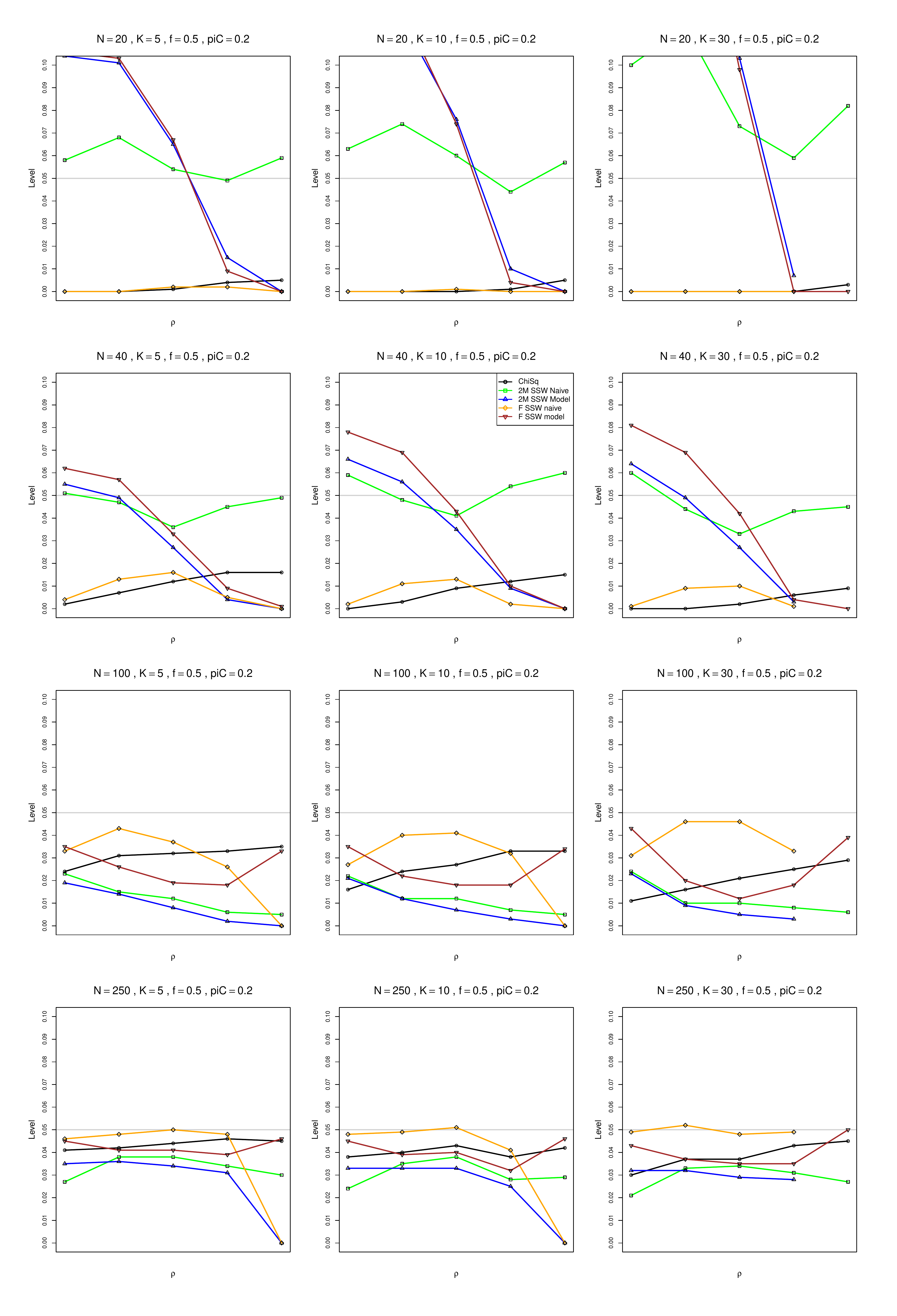}
	\caption{Q for LRR: actual level at $\alpha = .05$ for $p_{iC} = 0.2$ and $f = .5$, equal sample sizes
		\label{NewQforRR_piC02andq05_equal_sample_sizes}}
\end{figure}

\begin{figure}[t]
	\centering
	\includegraphics[scale=0.33]{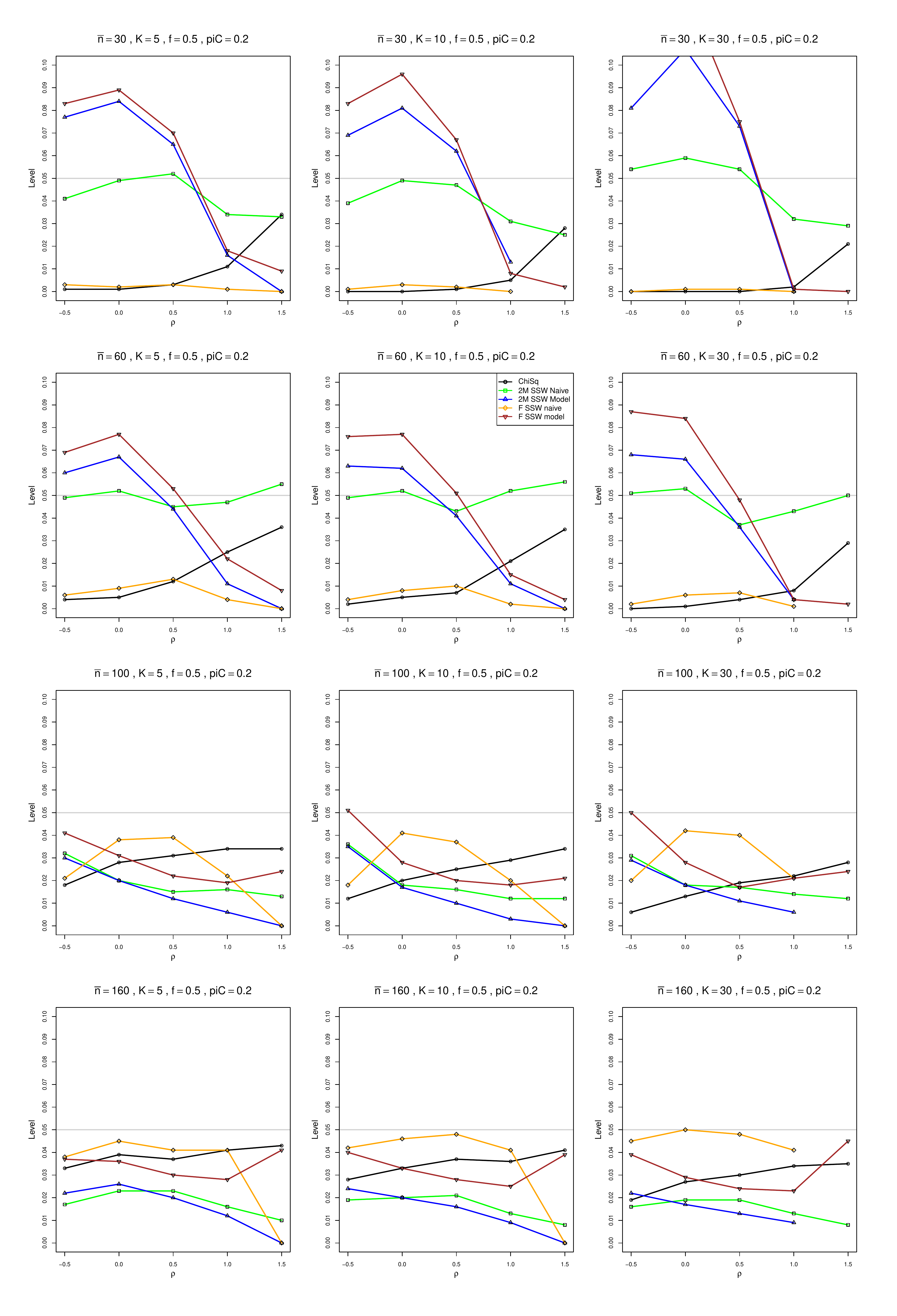}
	\caption{Q for LRR: actual level at $\alpha = .05$ for $p_{iC} = .2$ and $f = .5$, unequal sample sizes
		\label{pNewQforRR_piC02andq05_unequal_sample_sizes}}
\end{figure}

\begin{figure}[t]	\centering
	\includegraphics[scale=0.33]{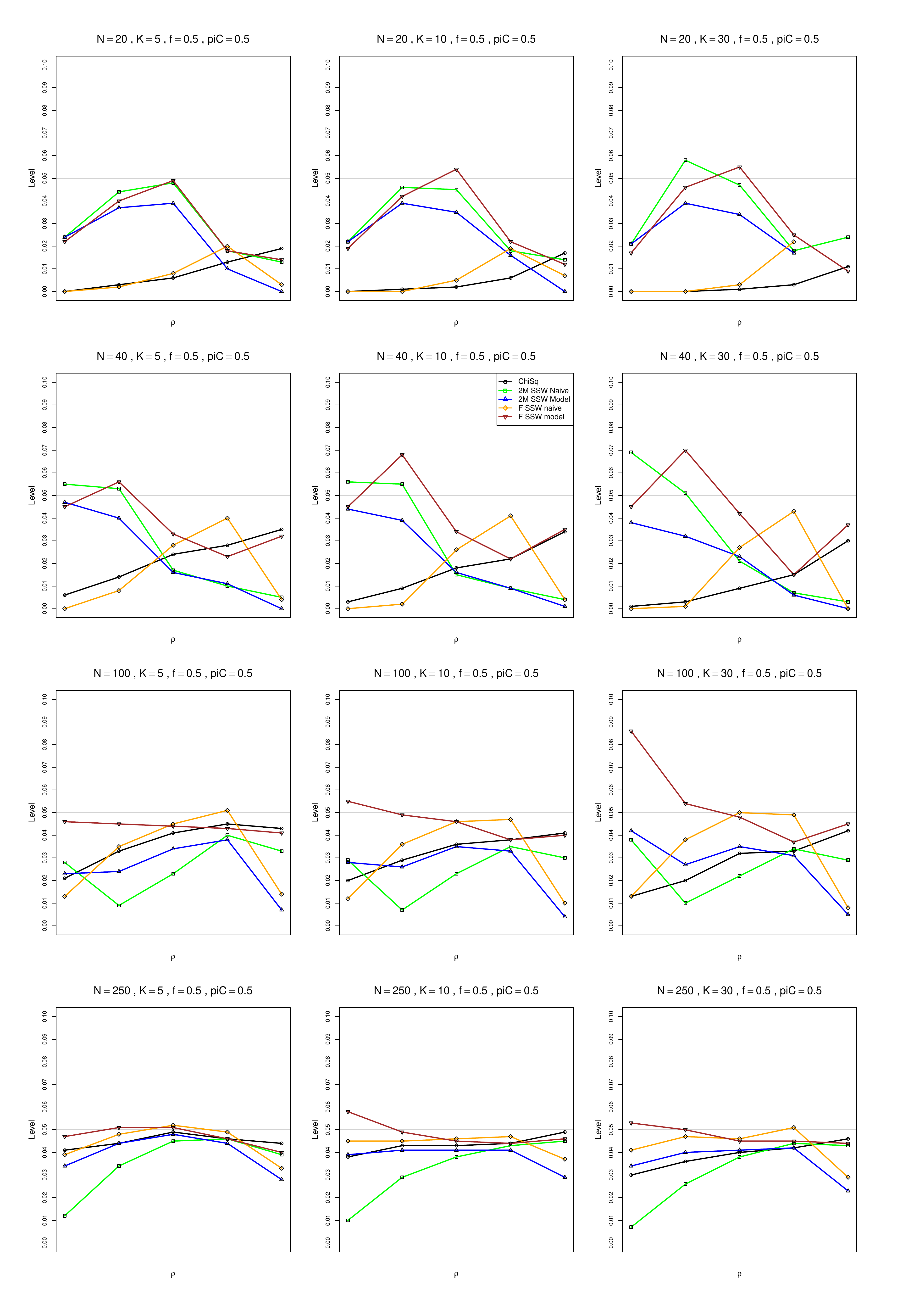}
	\caption{Q for LRR: actual level at $\alpha = .05$ for $p_{iC} = .5$ and $f = .5$, equal sample sizes
		\label{NewQforRR_piC05_equal_sample_sizes}}
\end{figure}

\begin{figure}[t]
	\centering
	\includegraphics[scale=0.33]{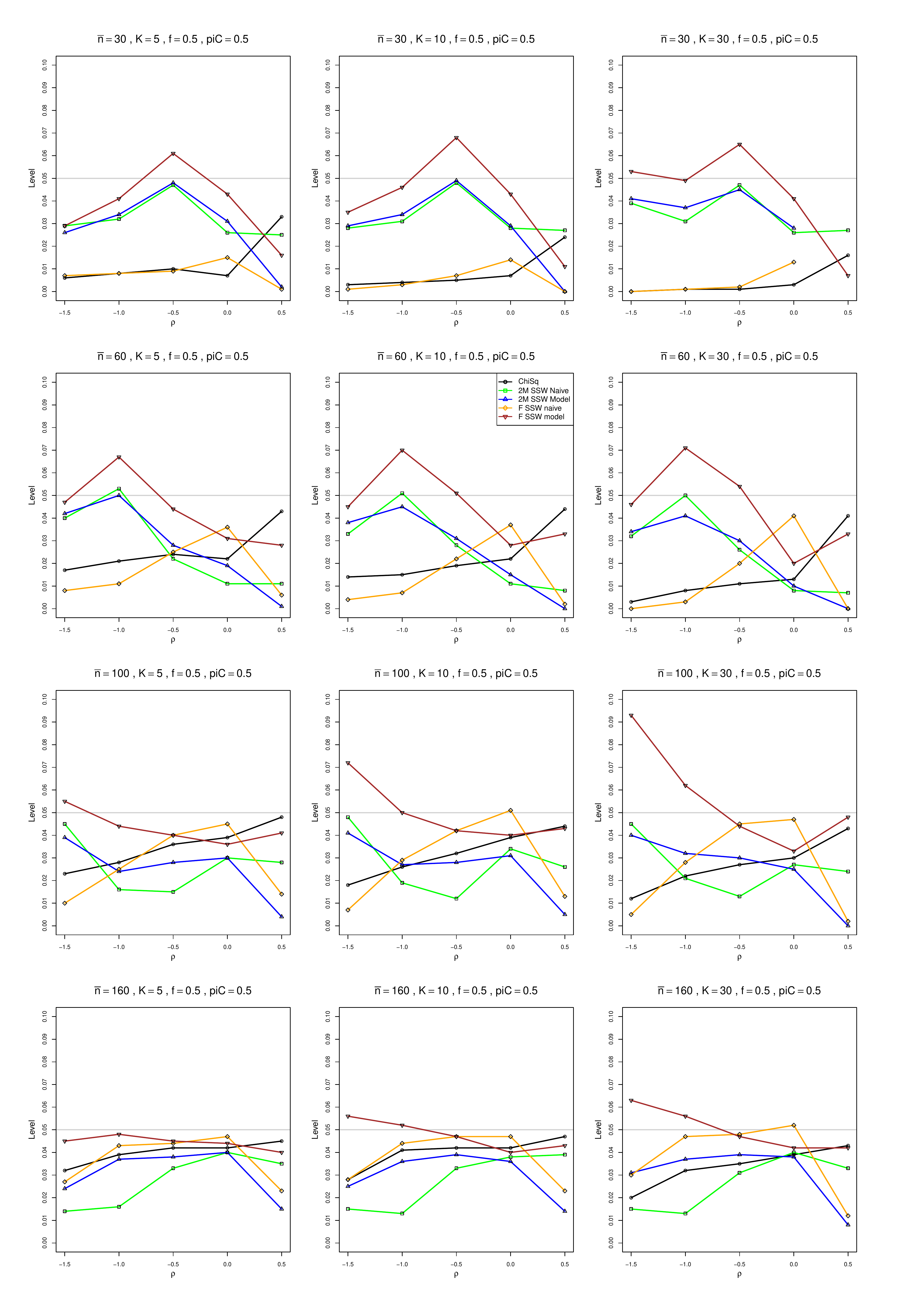}
	\caption{Q for LRR: actual level at $\alpha = .05$ for $p_{iC} = .5$ and $f = .5$, unequal sample sizes
		\label{NewQforRR_piC05andq05_unequal_sample_sizes}}
\end{figure}

\clearpage
\setcounter{figure}{0}
\setcounter{section}{0}
\renewcommand{\thefigure}{F.\arabic{figure}}

\section*{Appendix F: Plots of error in the level of the test for heterogeneity of RD for six approximations for the null distribution of $Q$}

Each figure corresponds to a value of the probability of an event in the Control arm $p_{iC}$  (= .1, .2, .5), a value of $\Delta$, and a choice of equal or unequal sample sizes ($n$ or $bar{n}$). To facilitate comparisons, we used the same pairs ($p_{iC},\; p_{iT}$)
as for LRR. The values of $p_{iT}$ are (.06, .10, .16, .27, .44) when $p_{iC} = .1$,
(.12, .20, .33, .54, .90) when $p_{iC} = .2$, and (.12, .18, .30, .50, .82) when  $p_{iC} = .5$.  The  values of $\Delta$ are ($-.04$, 0, .06, .17, .34), ($-.08$, 0, .13, .34, .70) and ($-.38$, $-.32$, $-.20$, 0, .32), respectively.

The fraction of each study's sample size in the Control arm  $f$ is held constant at 0.5.

For each combination of a value of $n$ (= 20, 40, 100, 250) or  $\bar{n}$ (= 30, 60, 100, 160) and a value of $K$ (= 5, 10, 30), a panel plots, versus the nominal upper-tail areas ( = .001, .0025, .005, .01, .025, .05, .1, .25, .5 and the complementary values .75, \ldots, .999), the difference between the achieved level and the nominal level for six approximations to the null distribution of Q: \\
\begin{itemize}
\item ChiSq (Chi-square approximation with $K-1$ df, inverse-variance weights)
\item KDB (Kulinskaya-Dollinger-Bj{\o}rkest{\o}l (2011) approximation, inverse-variance weights)
\item 2M SSW na\"{i}ve (Two-moment gamma approximation, na\"{i}ve estimation of $p_{iT}$ from $X_{iT}$ and $n_{iT}$, effective-sample-size weights)
\item 2M SSW model (Two-moment gamma approximation, model-based estimation of $p_{iT}$, effective-sample-size weights)
	\item F SSW na\"{i}ve (Farebrother approximation, na\"{i}ve estimation of $p_{iT}$ from $X_{iT}$ and $n_{iT}$, effective-sample-size weights)
	\item F SSW model (Farebrother approximation, model-based estimation of $p_{iT}$, effective-sample-size weights)
\end{itemize}

\clearpage

\begin{figure}[ht]
	\centering
	\includegraphics[scale=0.33]{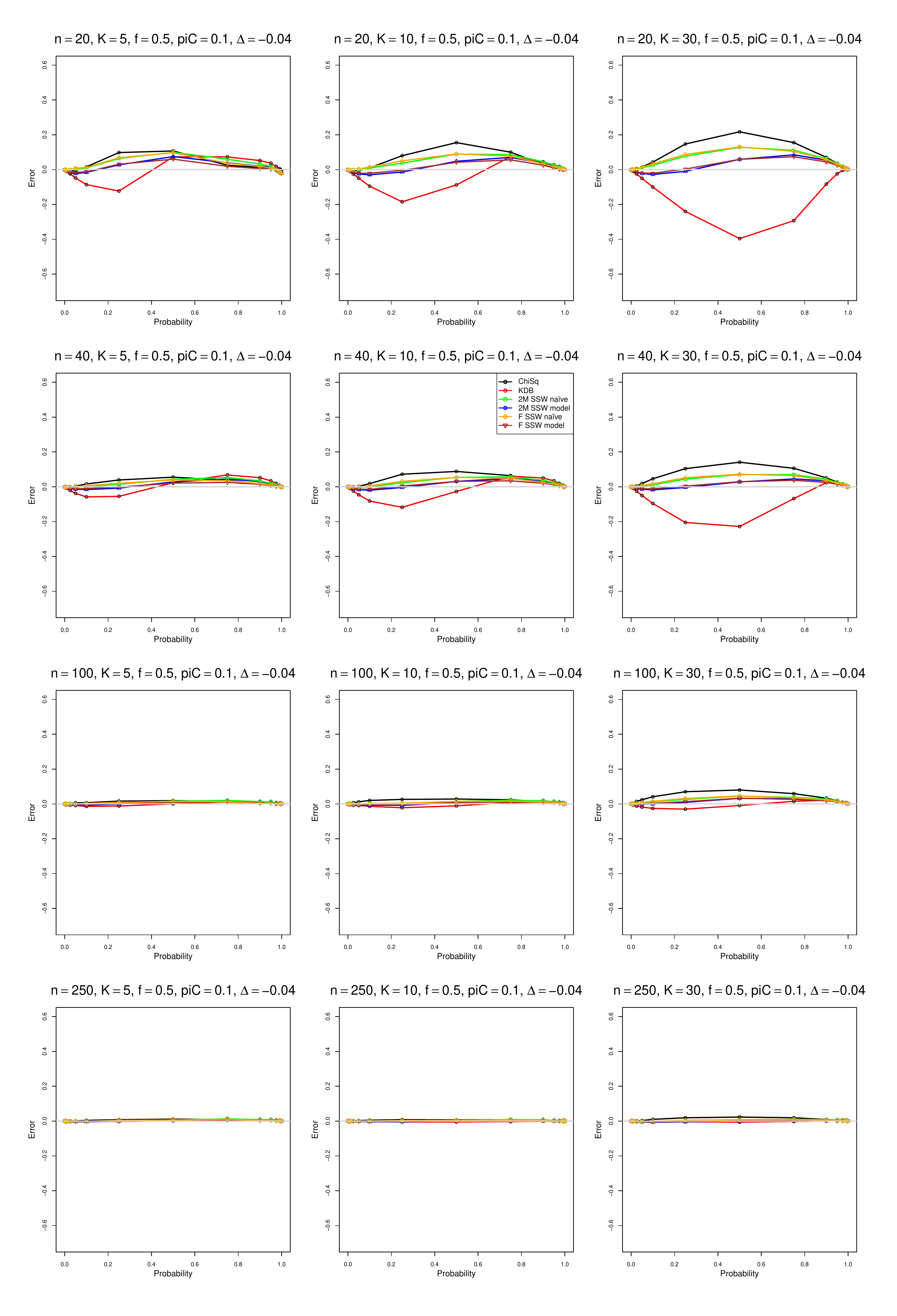}
	\caption{Plots of error in the level of the test for heterogeneity of RD for six approximations for the null distribution of $Q$, $p_{iC} = .1$, $f = .5$, and $\Delta = -0.04$, equal sample sizes}
	\label{PPplot_piC_01theta=-0.04_RD_equal_sample_sizes}
\end{figure}
\begin{figure}[ht]
		\centering
	\includegraphics[scale=0.33]{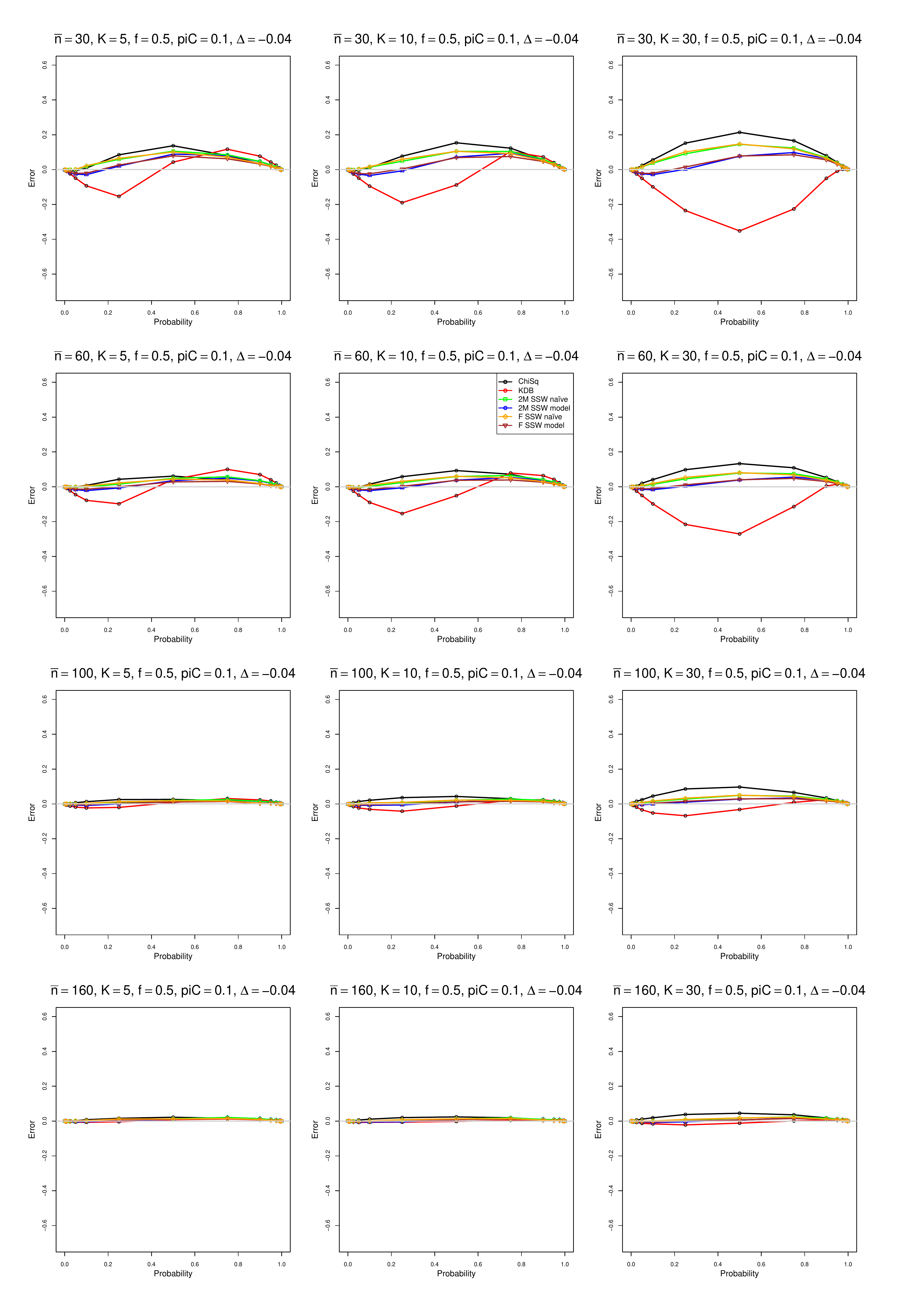}
	\caption{Plots of error in the level of the test for heterogeneity of RD for six approximations for the null distribution of $Q$, $p_{iC} = .1$, $f = .5$, and $\Delta = -0.04$, unequal sample sizes}
	\label{PPplot_piC_01theta=-0.04_RD_unequal_sample_sizes}
\end{figure}
\begin{figure}[ht]
	\centering
	\includegraphics[scale=0.33]{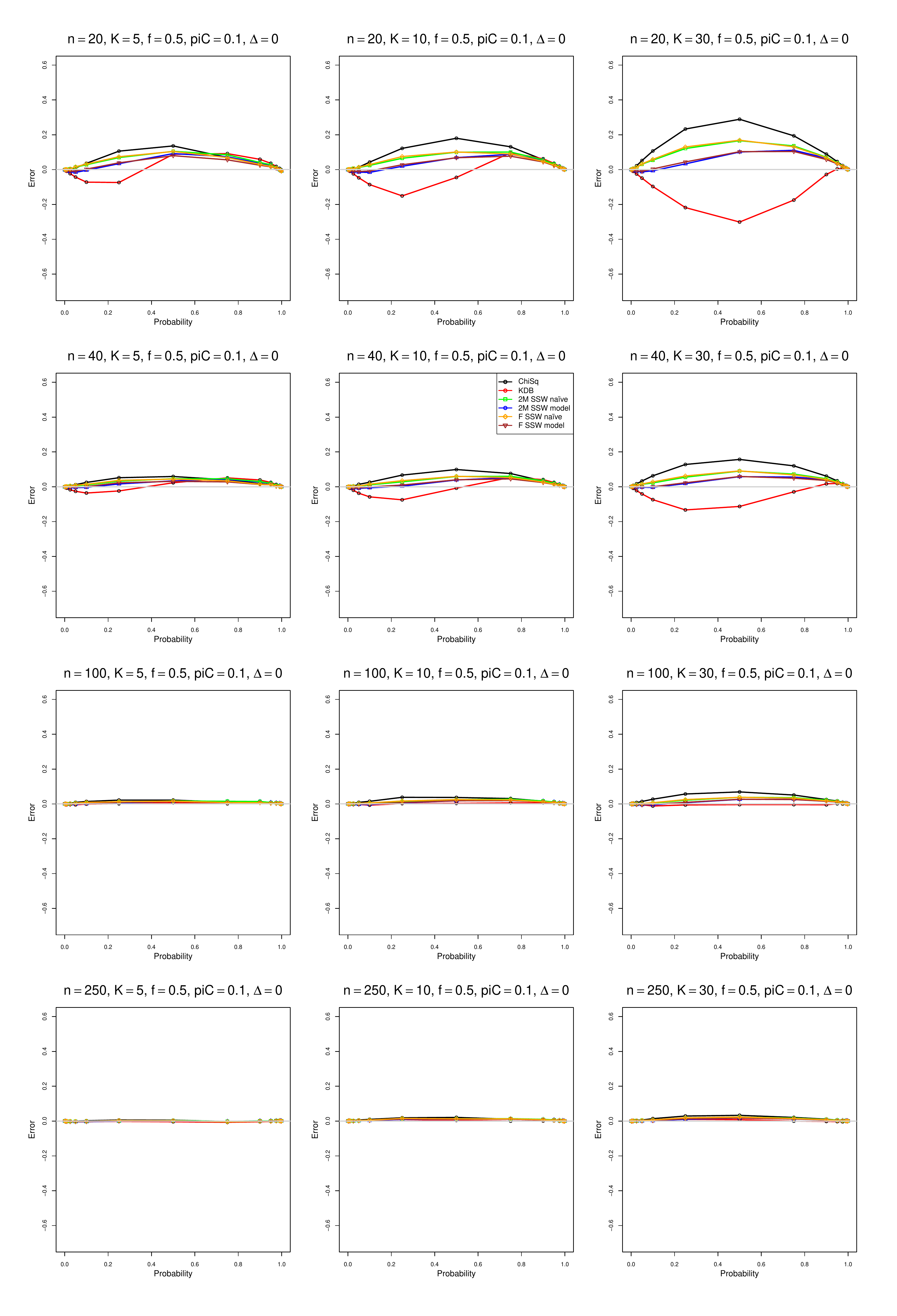}
	\caption{Plots of error in the level of the test for heterogeneity of RD for six approximations for the null distribution of $Q$, $p_{iC} = .1$, $f = .5$, and $\Delta = 0$, equal sample sizes}
	\label{PPplot_piC_01theta=0_RD_equal_sample_sizes}
\end{figure}
\begin{figure}[ht]
		\centering
	\includegraphics[scale=0.33]{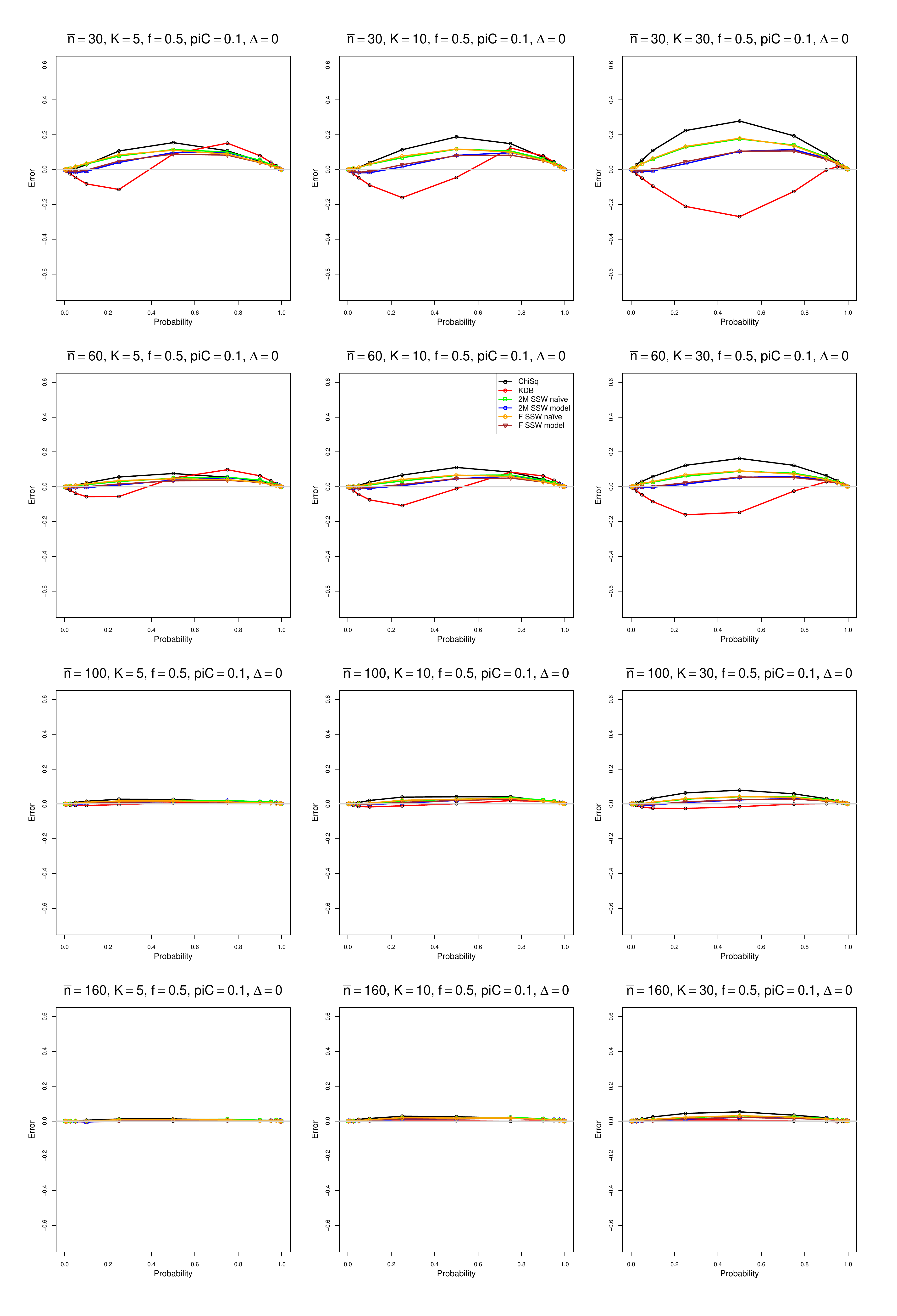}
	\caption{Plots of error in the level of the test for heterogeneity of RD for six approximations for the null distribution of $Q$, $p_{iC} = .1$, $f = .5$, and $\Delta = 0$, unequal sample sizes}
	\label{PPplot_piC_01theta=0_RD_unequal_sample_sizes}
\end{figure}

\begin{figure}[ht]
	\centering
	\includegraphics[scale=0.33]{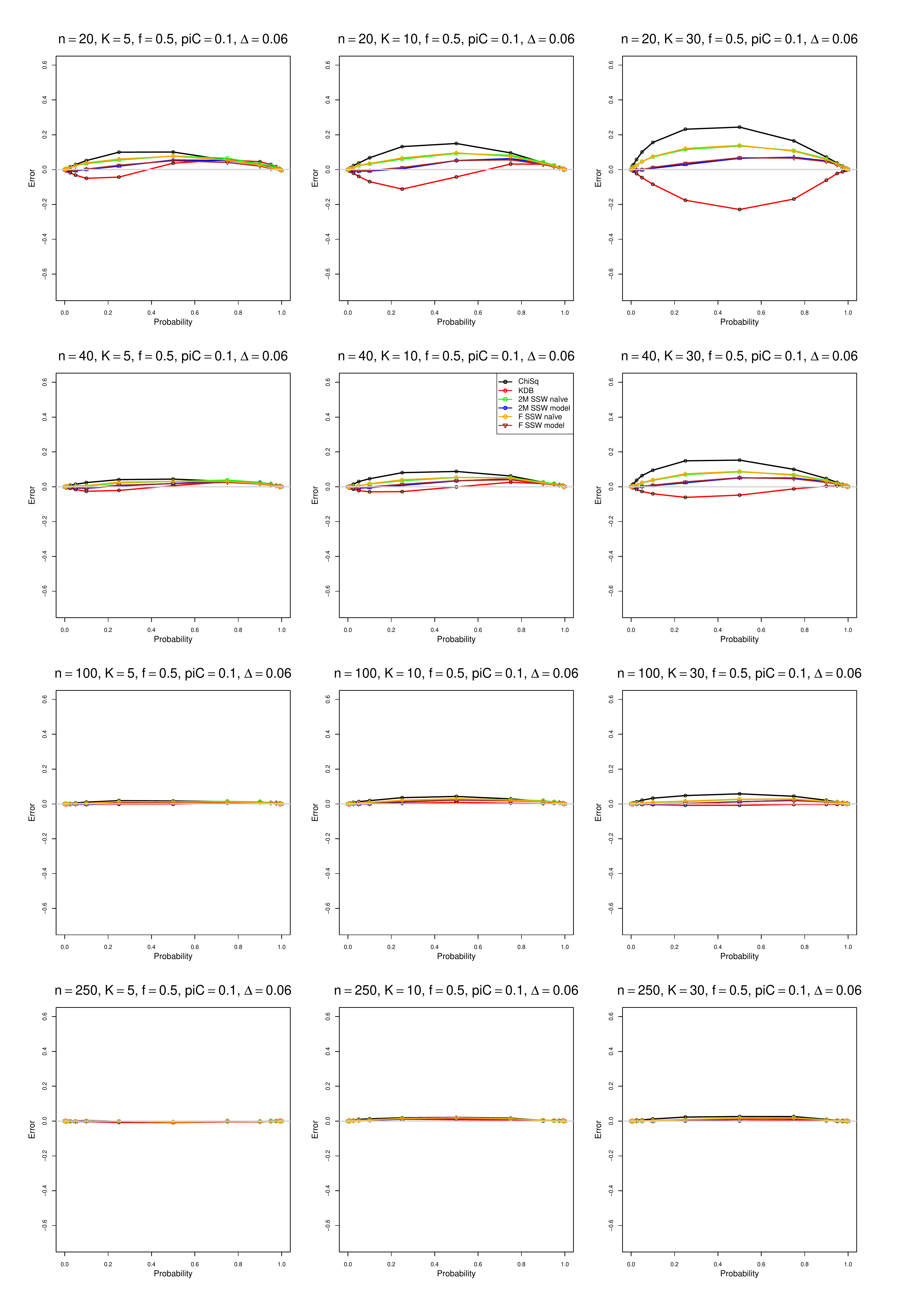}
	\caption{Plots of error in the level of the test for heterogeneity of RD for six approximations for the null distribution of $Q$, $p_{iC} = .1$, $f = .5$, and $\Delta = 0.06$, equal sample sizes}
	\label{PPplot_piC_01theta=0.06_RD_equal_sample_sizes}
\end{figure}
\begin{figure}[ht]
		\centering
	\includegraphics[scale=0.33]{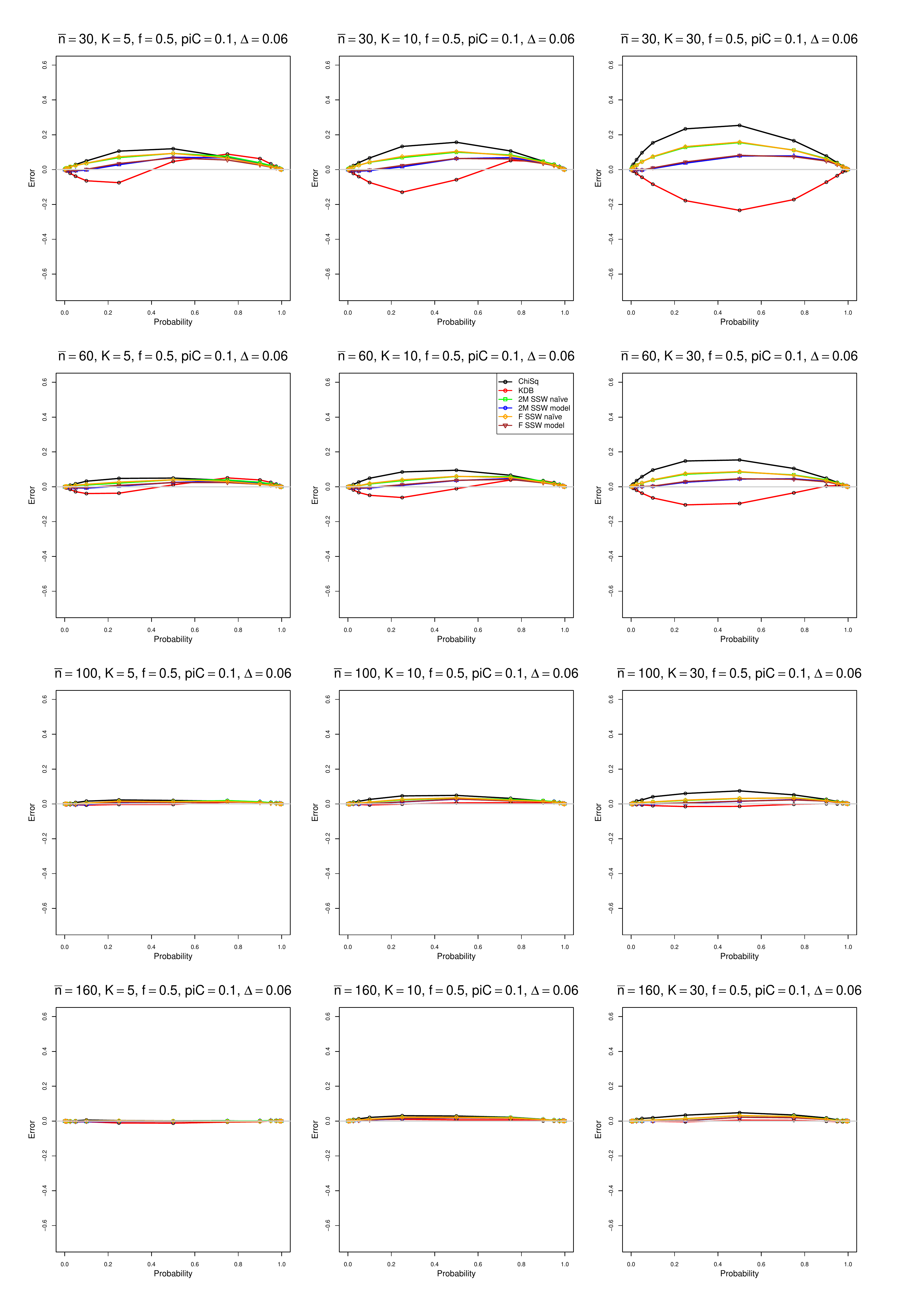}
	\caption{Plots of error in the level of the test for heterogeneity of RD for six approximations for the null distribution of $Q$, $p_{iC} = .1$, $f = .5$, and $\Delta = 0.06$, unequal sample sizes}
	\label{PPplot_piC_01theta=0.06_RD_unequal_sample_sizes}
\end{figure}

\begin{figure}[ht]
	\centering
	\includegraphics[scale=0.33]{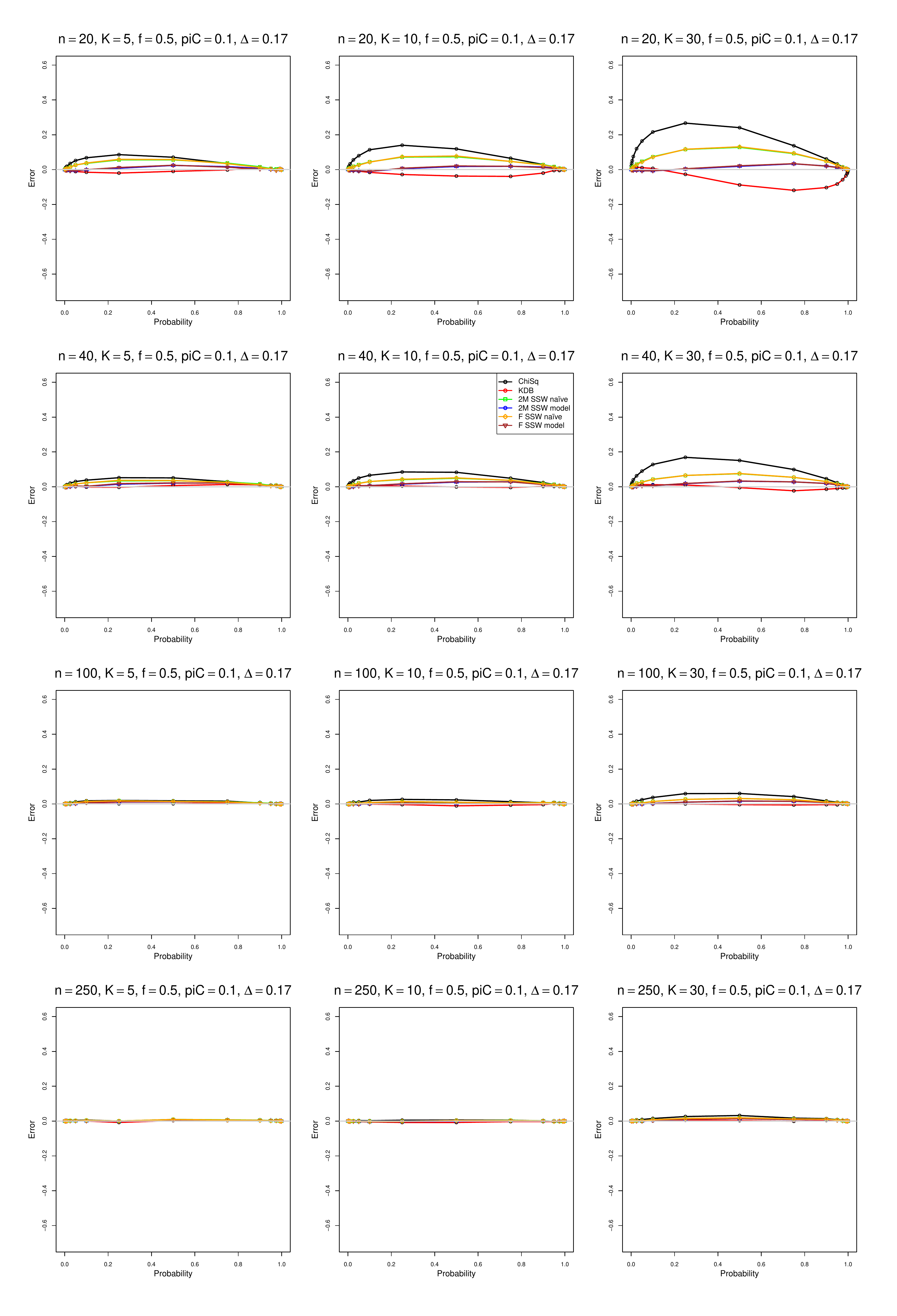}
	\caption{Plots of error in the level of the test for heterogeneity of RD for six approximations for the null distribution of $Q$, $p_{iC} = .1$, $f = .5$, and $\Delta = 0.17$, equal sample sizes}
	\label{PPplot_piC_01theta=0.17_RD_equal_sample_sizes}
\end{figure}
\begin{figure}[ht]
		\centering
	\includegraphics[scale=0.33]{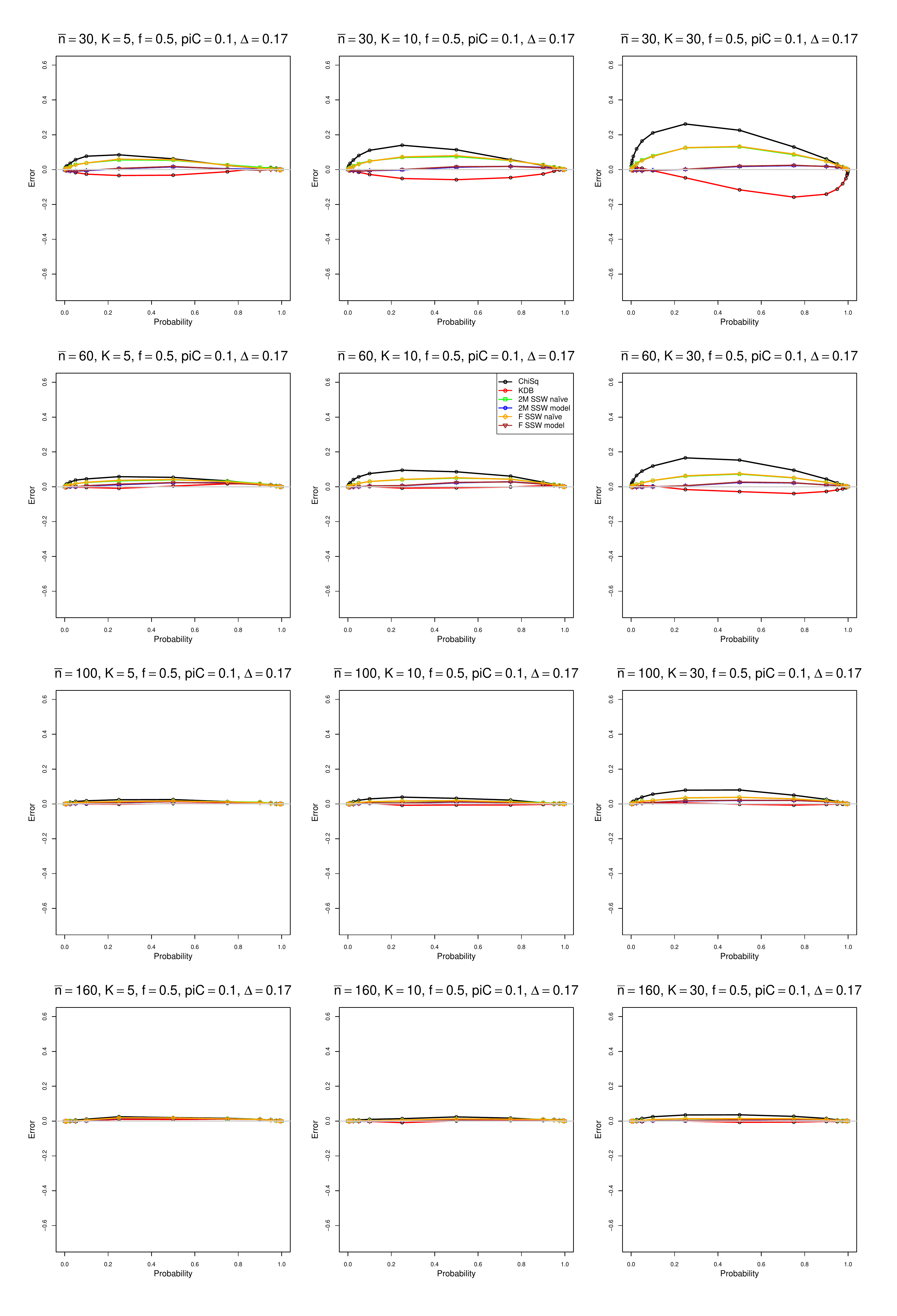}
	\caption{Plots of error in the level of the test for heterogeneity of RD for six approximations for the null distribution of $Q$, $p_{iC} = .1$, $f = .5$, and $\Delta = 0.17$, unequal sample sizes}
	\label{PPplot_piC_01theta=0.17_RD_unequal_sample_sizes}
\end{figure}

\begin{figure}[ht]
	\centering
	\includegraphics[scale=0.33]{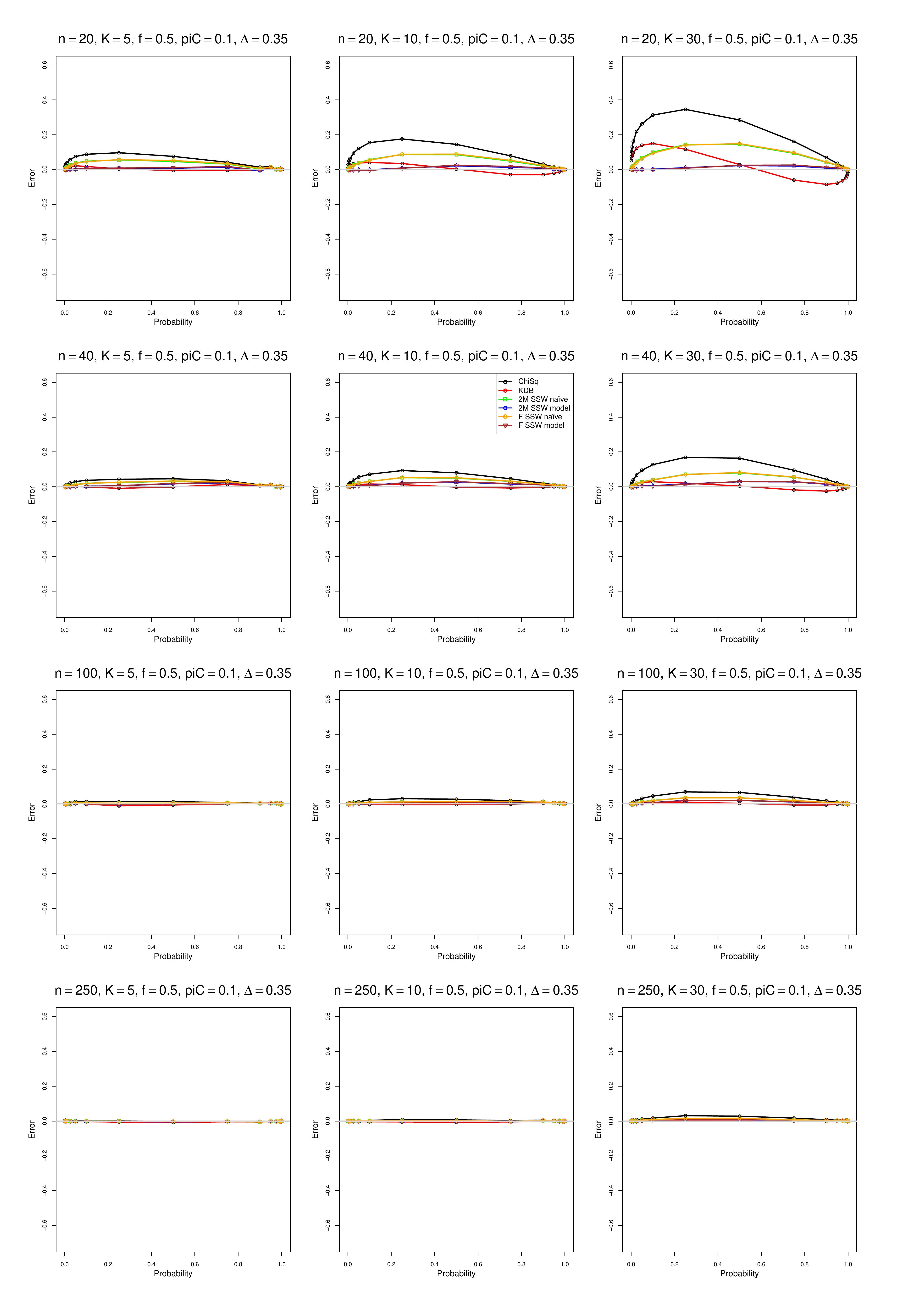}
	\caption{Plots of error in the level of the test for heterogeneity of RD for six approximations for the null distribution of $Q$, $p_{iC} = .1$, $f = .5$, and $\Delta = 0.35$, equal sample sizes}
	\label{PPplot_piC_01theta=0.35_RD_equal_sample_sizes}
\end{figure}
\begin{figure}[ht]
		\centering
	\includegraphics[scale=0.33]{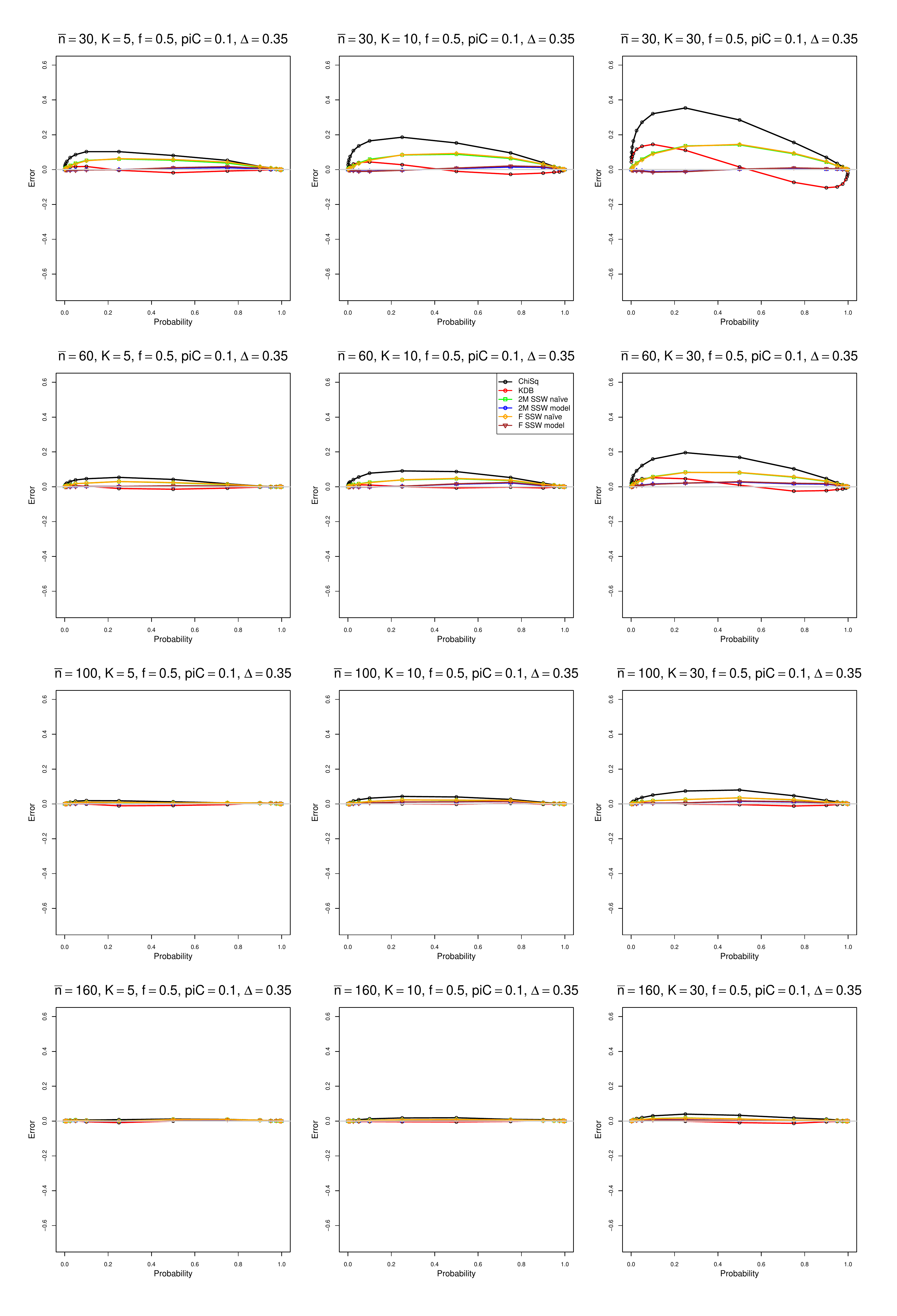}
	\caption{Plots of error in the level of the test for heterogeneity of RD for six approximations for the null distribution of $Q$, $p_{iC} = .1$, $f = .5$, and $\Delta = 0.35$, unequal sample sizes}
	\label{PPplot_piC_01theta=0.35_RD_unequal_sample_sizes}
\end{figure}

\begin{figure}[ht]
	\centering
	\includegraphics[scale=0.33]{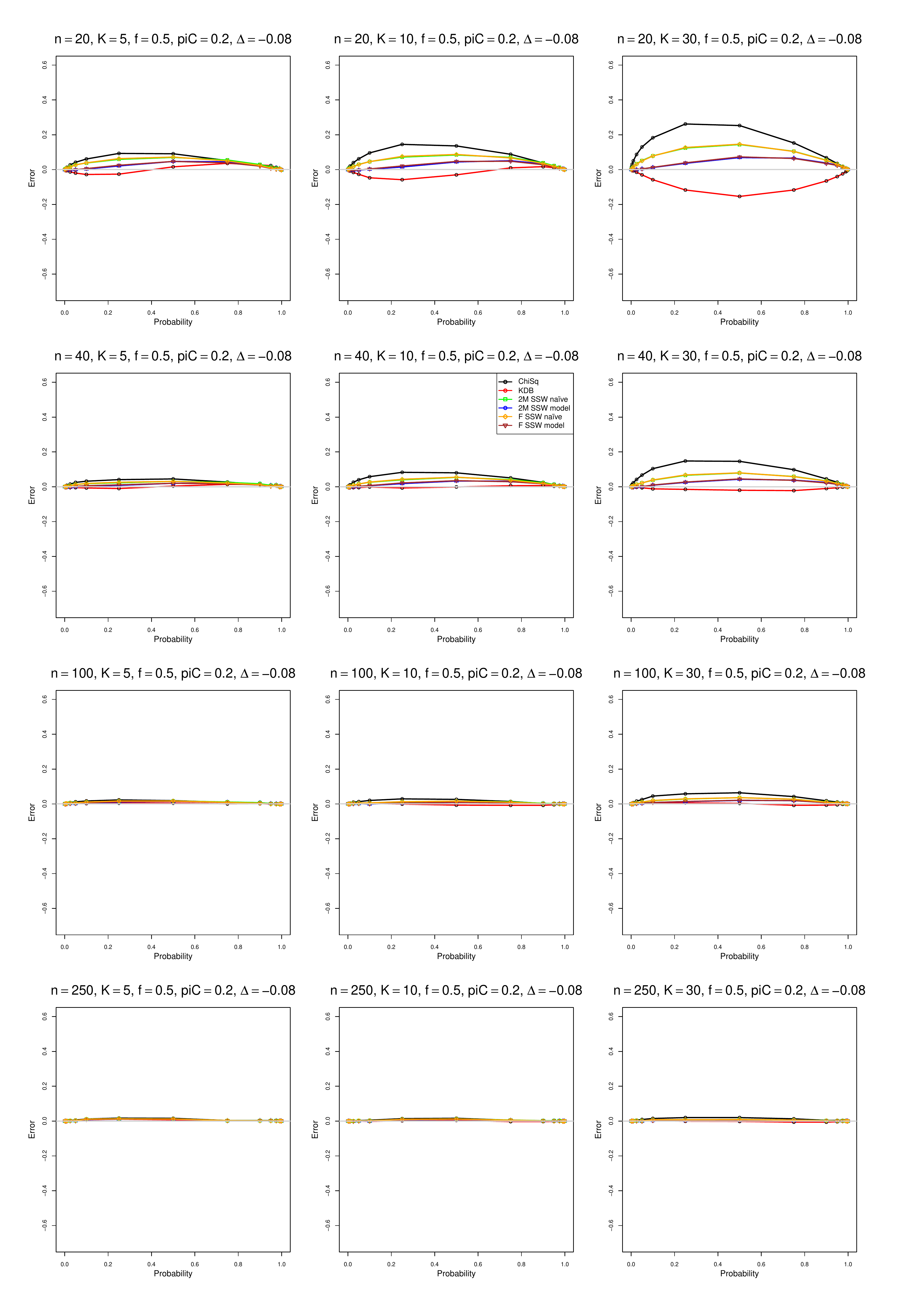}
	\caption{Plots of error in the level of the test for heterogeneity of RD for six approximations for the null distribution of $Q$, $p_{iC} = .2$, $f = .5$, and $\Delta = -0.08$, equal sample sizes}
	\label{PPplot_piC_02theta=-0.08_RD_equal_sample_sizes}
\end{figure}
\begin{figure}[ht]
		\centering
	\includegraphics[scale=0.33]{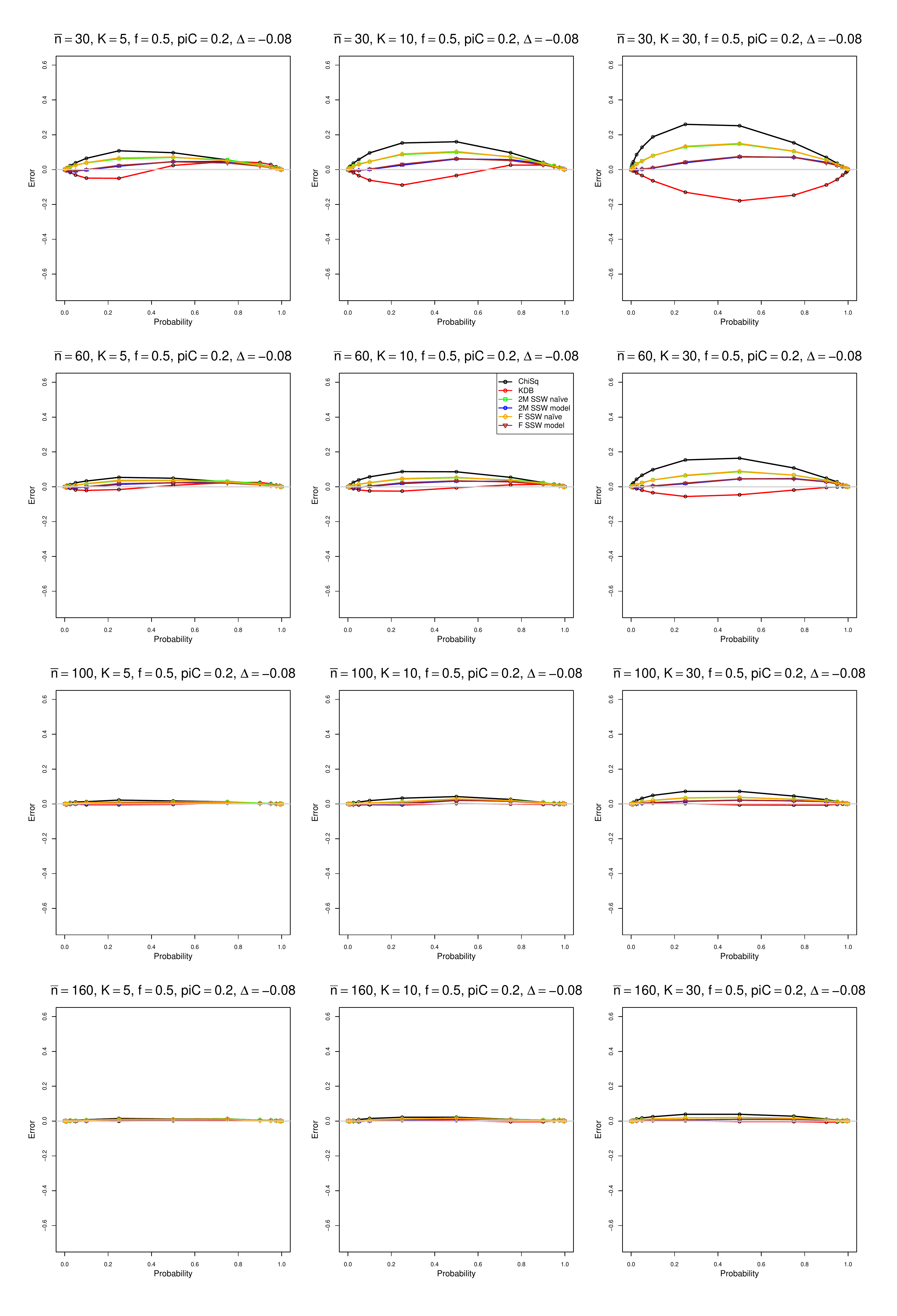}
	\caption{Plots of error in the level of the test for heterogeneity of RD for six approximations for the null distribution of $Q$, $p_{iC} = .2$, $f = .5$, and $\Delta = -0.08$, unequal sample sizes}
	\label{PPplot_piC_02theta=-0.08_RD_unequal_sample_sizes}
\end{figure}
\begin{figure}[ht]
	\centering
	\includegraphics[scale=0.33]{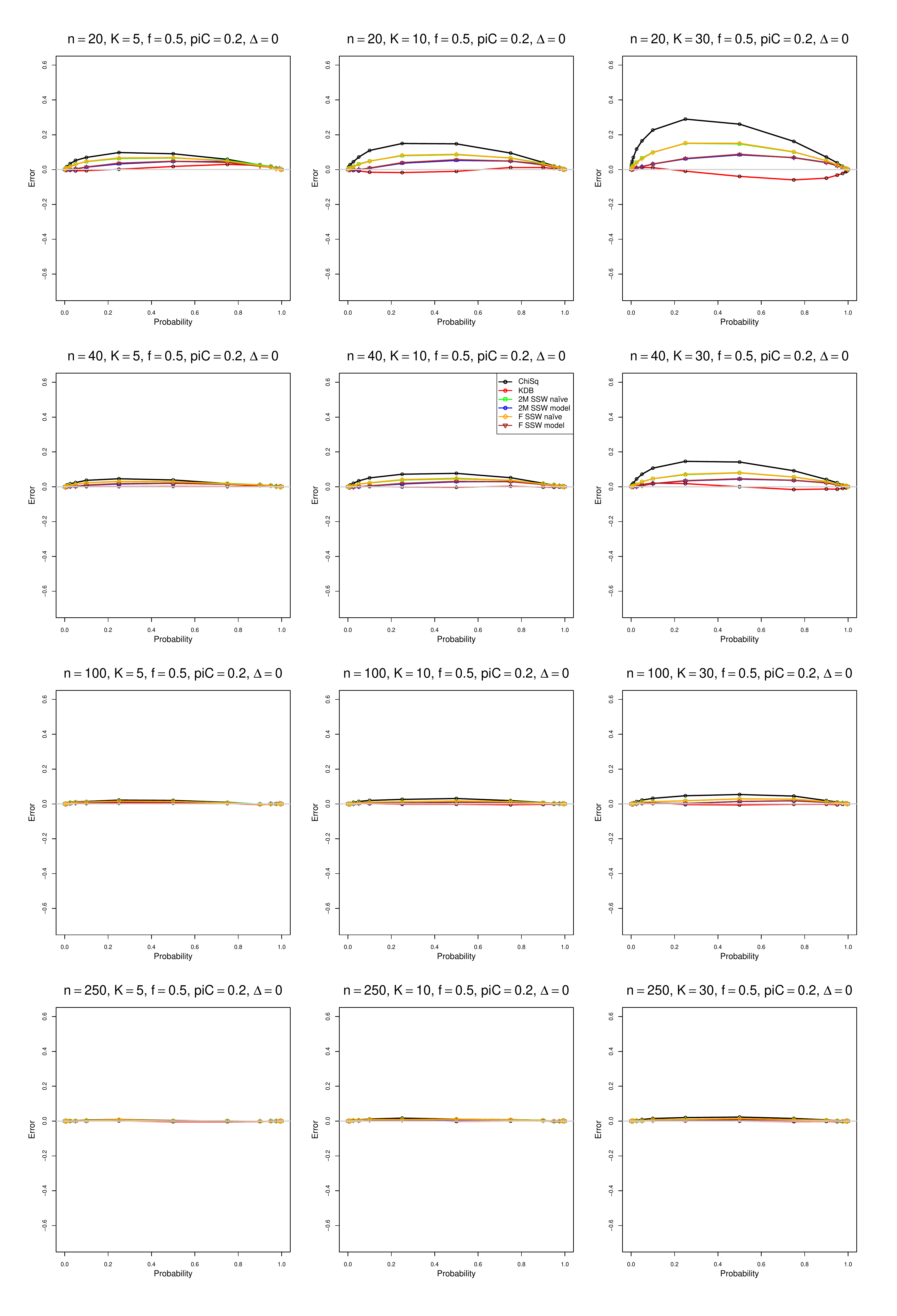}
	\caption{Plots of error in the level of the test for heterogeneity of RD for six approximations for the null distribution of $Q$, $p_{iC} = .2$, $f = .5$, and $\Delta = 0$, equal sample sizes}
	\label{PPplot_piC_02theta=0_RD_equal_sample_sizes}
\end{figure}
\begin{figure}[ht]
		\centering
	\includegraphics[scale=0.33]{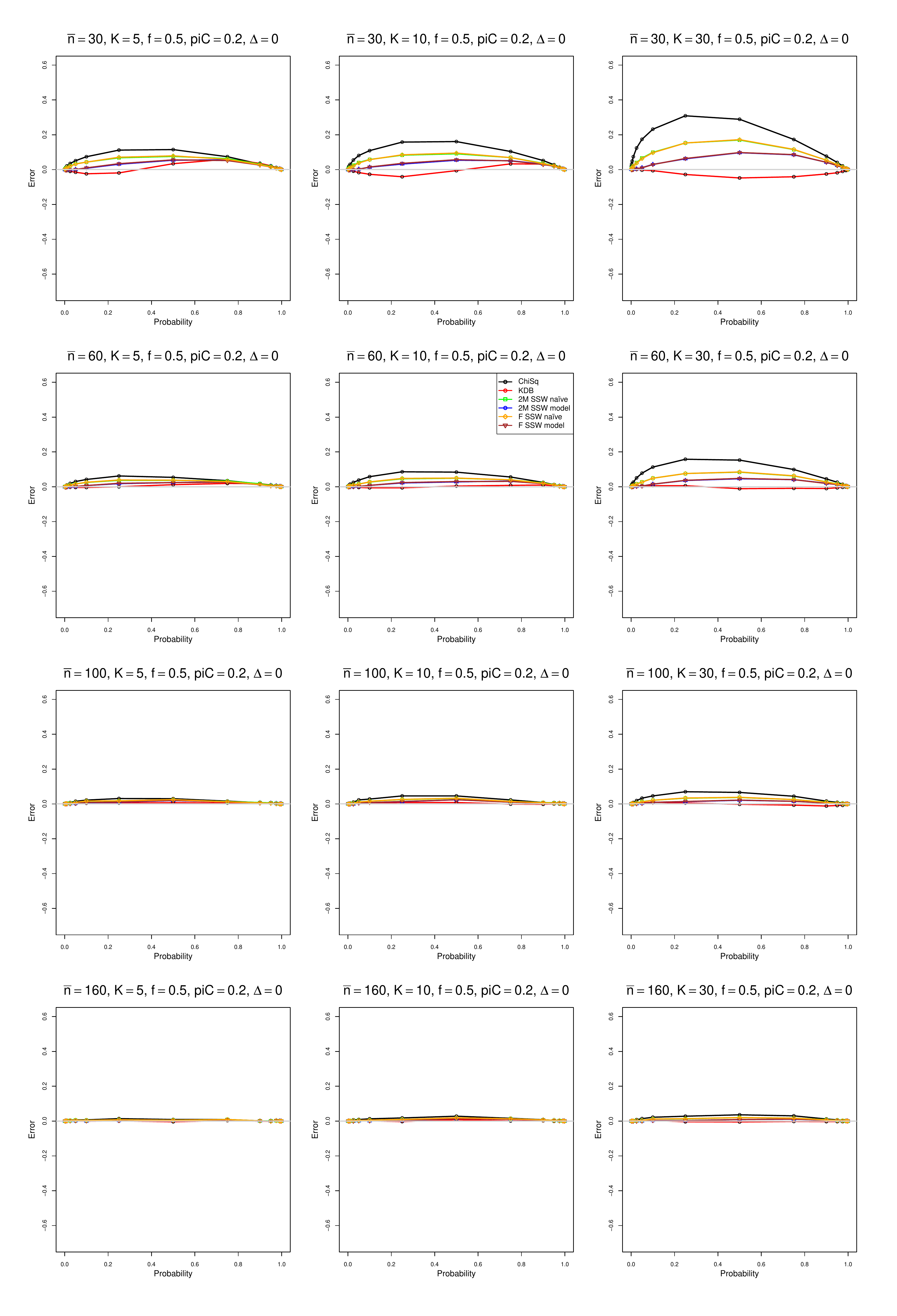}
	\caption{Plots of error in the level of the test for heterogeneity of RD for six approximations for the null distribution of $Q$, $p_{iC} = .2$, $f = .5$, and $\Delta = 0$, unequal sample sizes}
	\label{PPplot_piC_02theta=0_RD_unequal_sample_sizes}
\end{figure}

\begin{figure}[ht]
	\centering
	\includegraphics[scale=0.33]{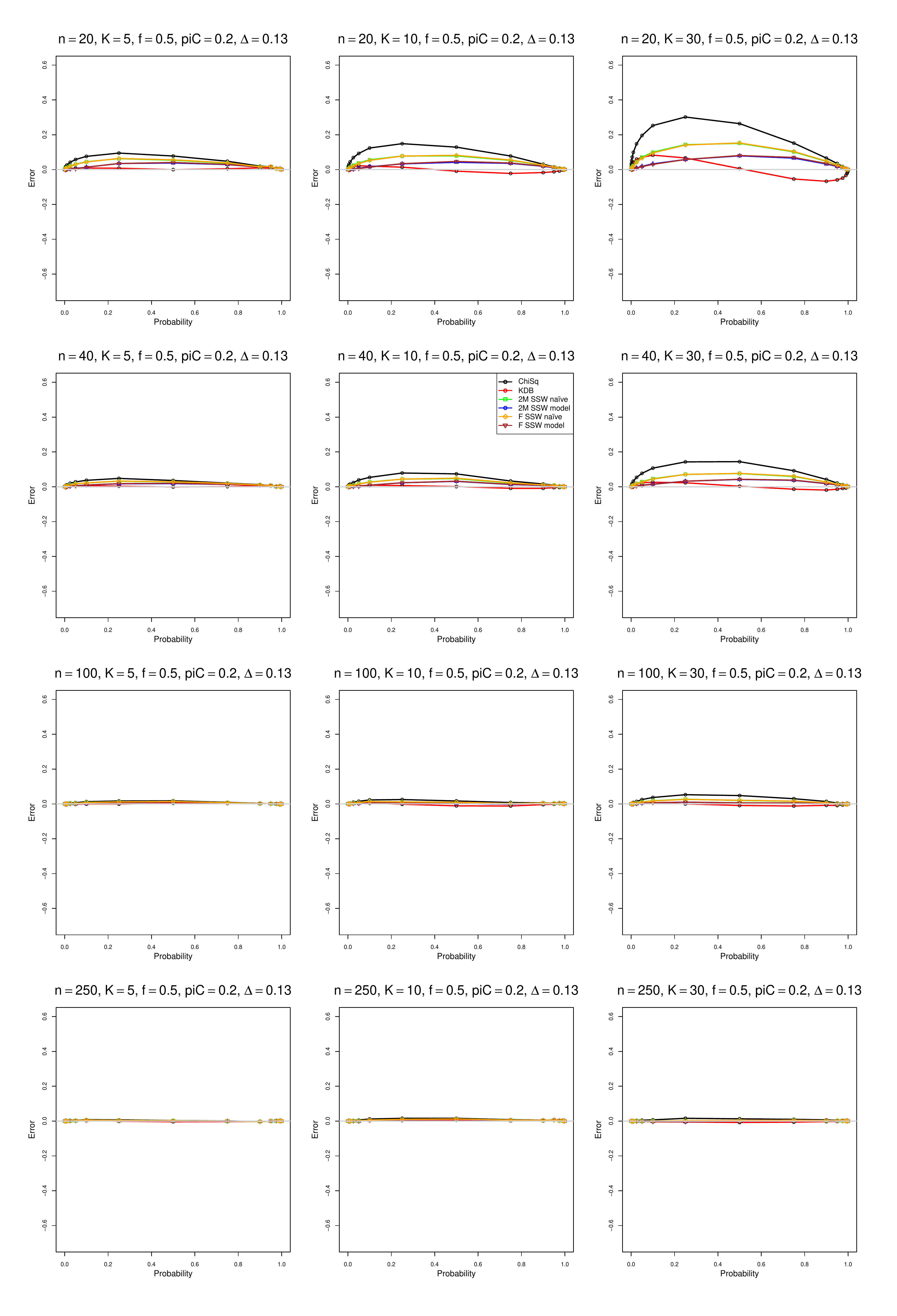}
	\caption{Plots of error in the level of the test for heterogeneity of RD for six approximations for the null distribution of $Q$, $p_{iC} = .2$, $f = .5$, and $\Delta = 0.13$, equal sample sizes}
	\label{PPplot_piC_02theta=0.13_RD_equal_sample_sizes}
\end{figure}
\begin{figure}[ht]
		\centering
	\includegraphics[scale=0.33]{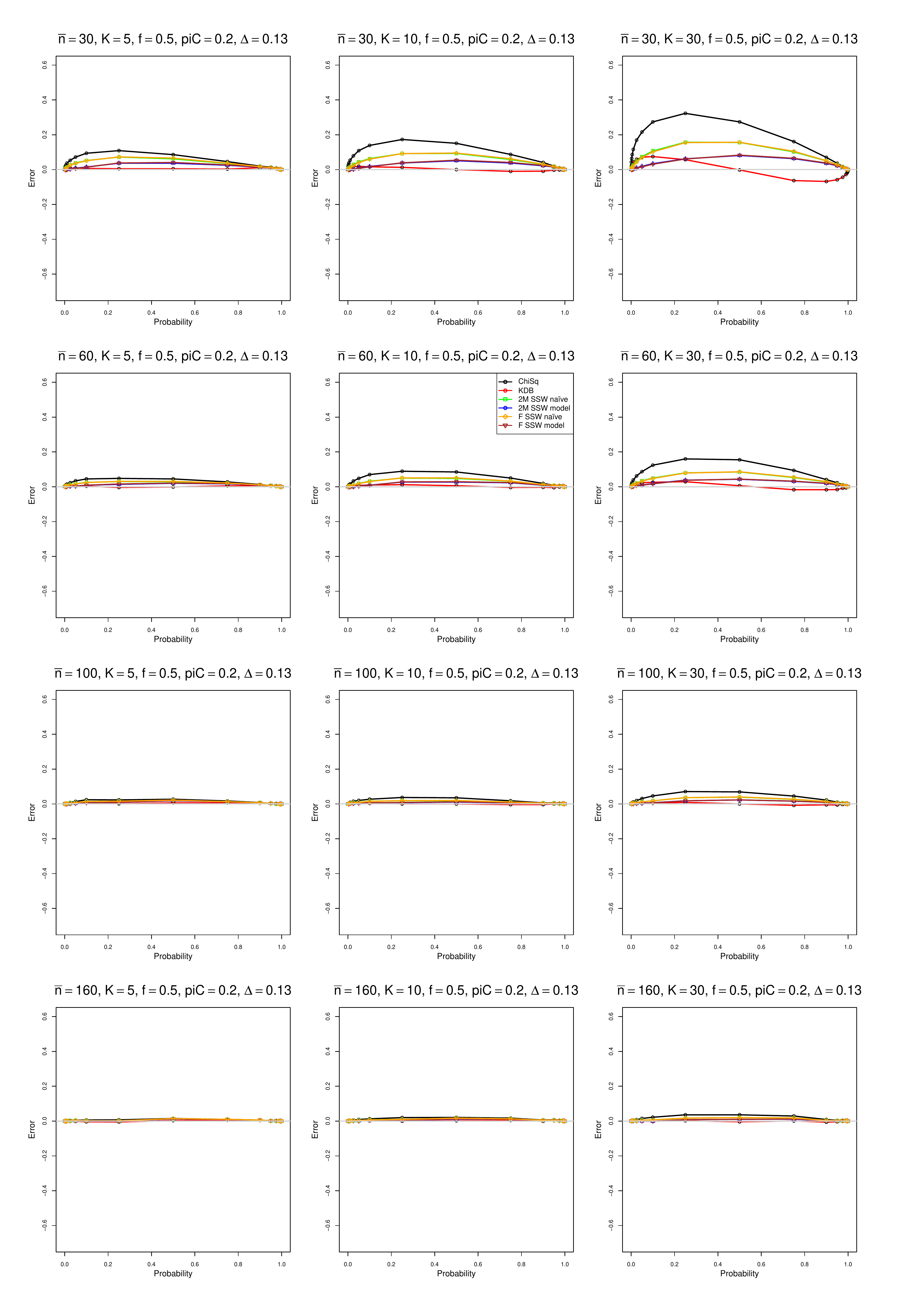}
	\caption{Plots of error in the level of the test for heterogeneity of RD for six approximations for the null distribution of $Q$, $p_{iC} = .2$, $f = .5$, and $\Delta = 0.13$, unequal sample sizes}
	\label{PPplot_piC_02theta=0.13_RD_unequal_sample_sizes}
\end{figure}

\begin{figure}[ht]
	\centering
	\includegraphics[scale=0.33]{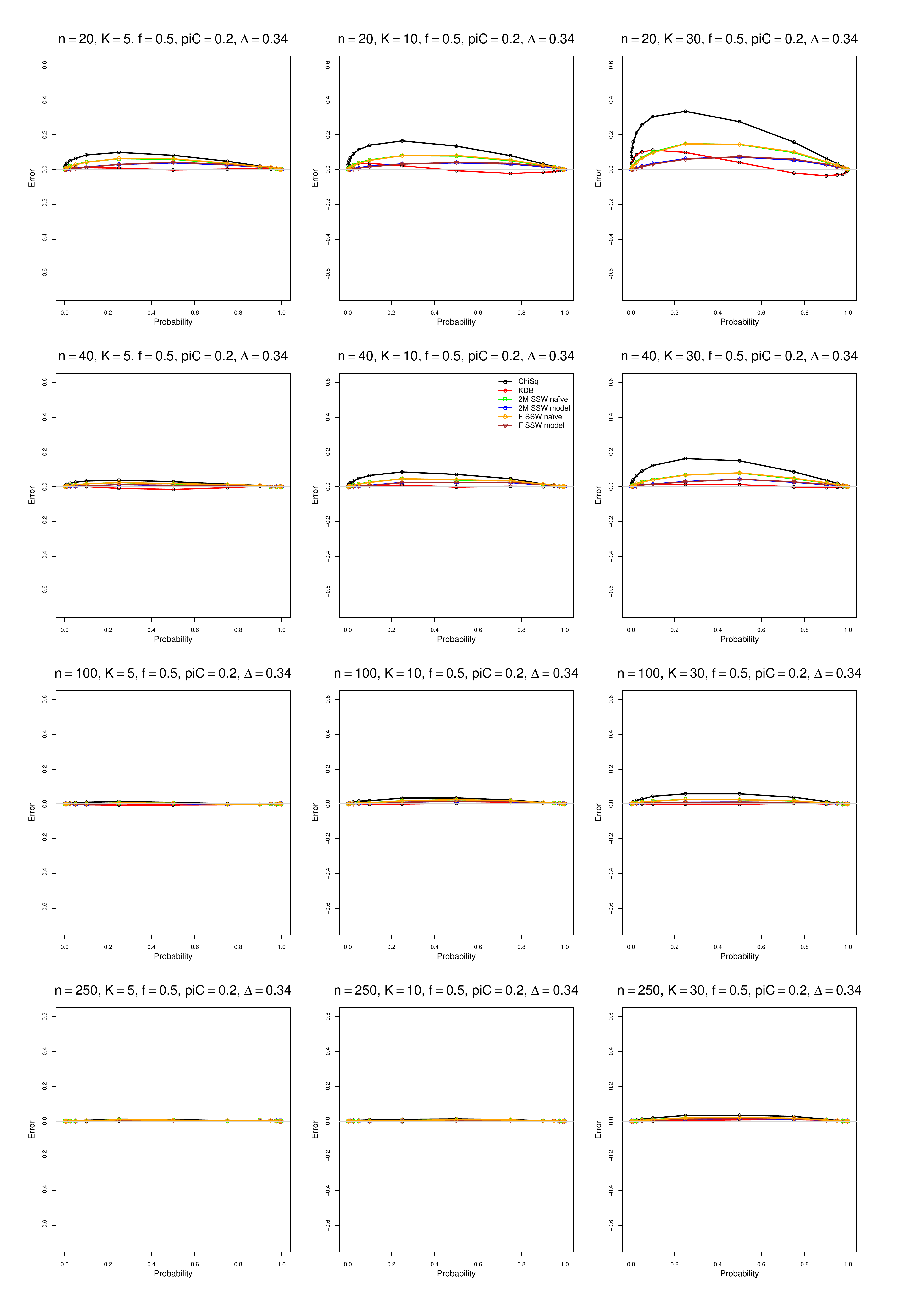}
	\caption{Plots of error in the level of the test for heterogeneity of RD for six approximations for the null distribution of $Q$, $p_{iC} = .2$, $f = .5$, and $\Delta = 0.34$, equal sample sizes}
	\label{PPplot_piC_02theta=0.34_RD_equal_sample_sizes}
\end{figure}
\begin{figure}[ht]
		\centering
	\includegraphics[scale=0.33]{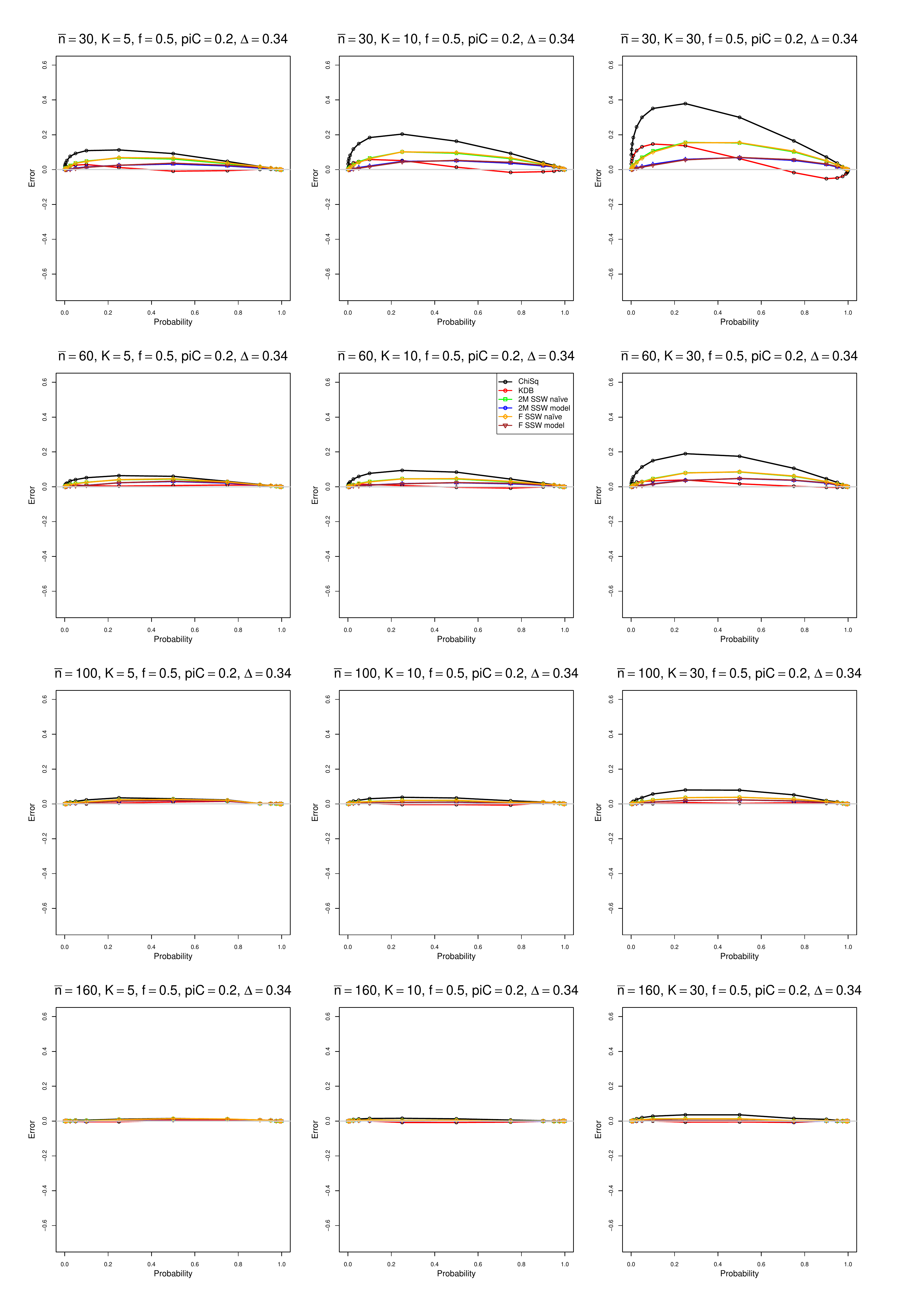}
	\caption{Plots of error in the level of the test for heterogeneity of RD for six approximations for the null distribution of $Q$, $p_{iC} = .2$, $f = .5$, and $\Delta = 0.34$, unequal sample sizes}
	\label{PPplot_piC_02theta=0.34_RD_unequal_sample_sizes}
\end{figure}

\begin{figure}[ht]
	\centering
	\includegraphics[scale=0.33]{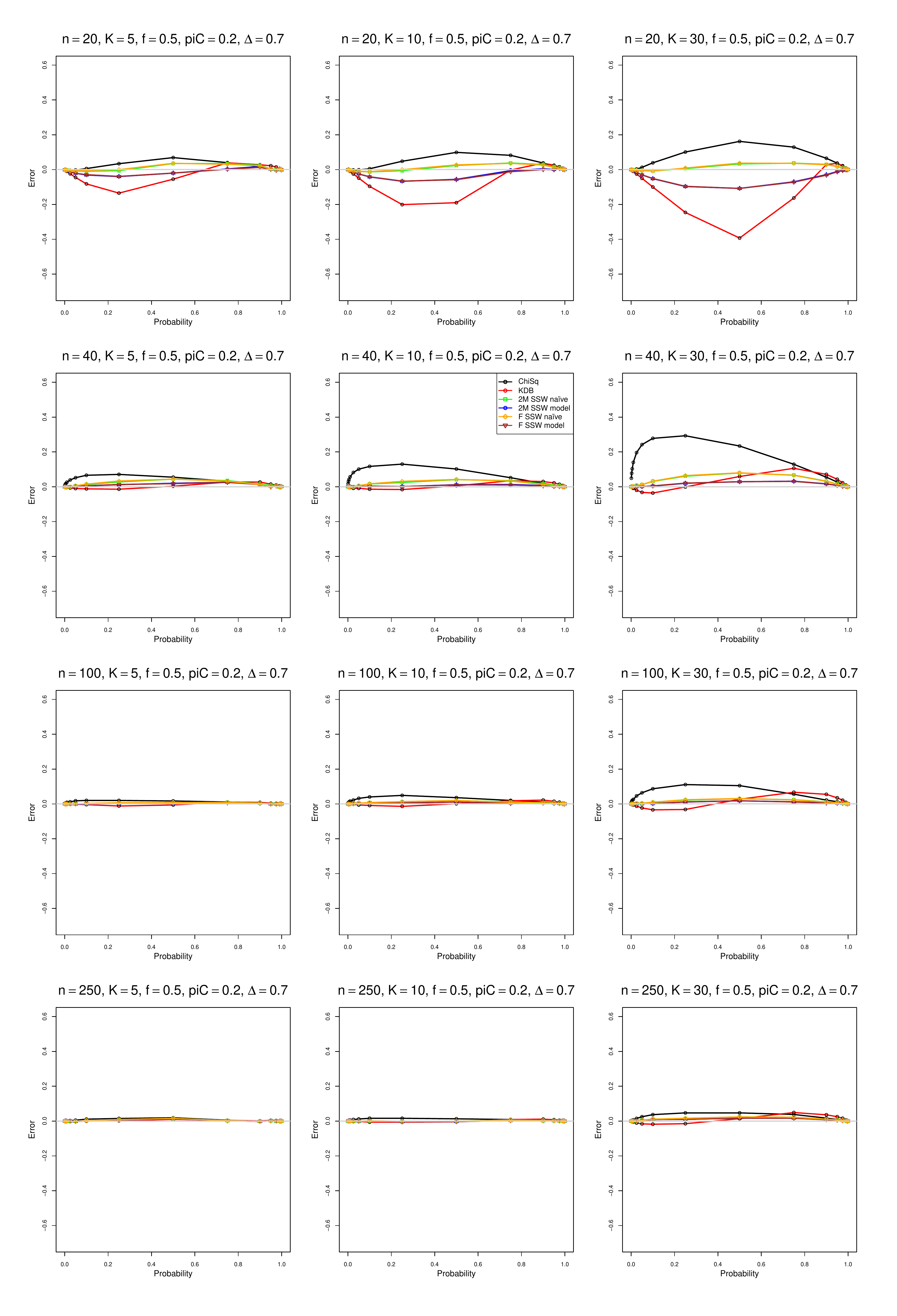}
	\caption{Plots of error in the level of the test for heterogeneity of RD for six approximations for the null distribution of $Q$, $p_{iC} = .2$, $f = .5$, and $\Delta = 0.7$, equal sample sizes}
	\label{PPplot_piC_02theta=0.7_RD_equal_sample_sizes}
\end{figure}
\begin{figure}[ht]
		\centering
	\includegraphics[scale=0.33]{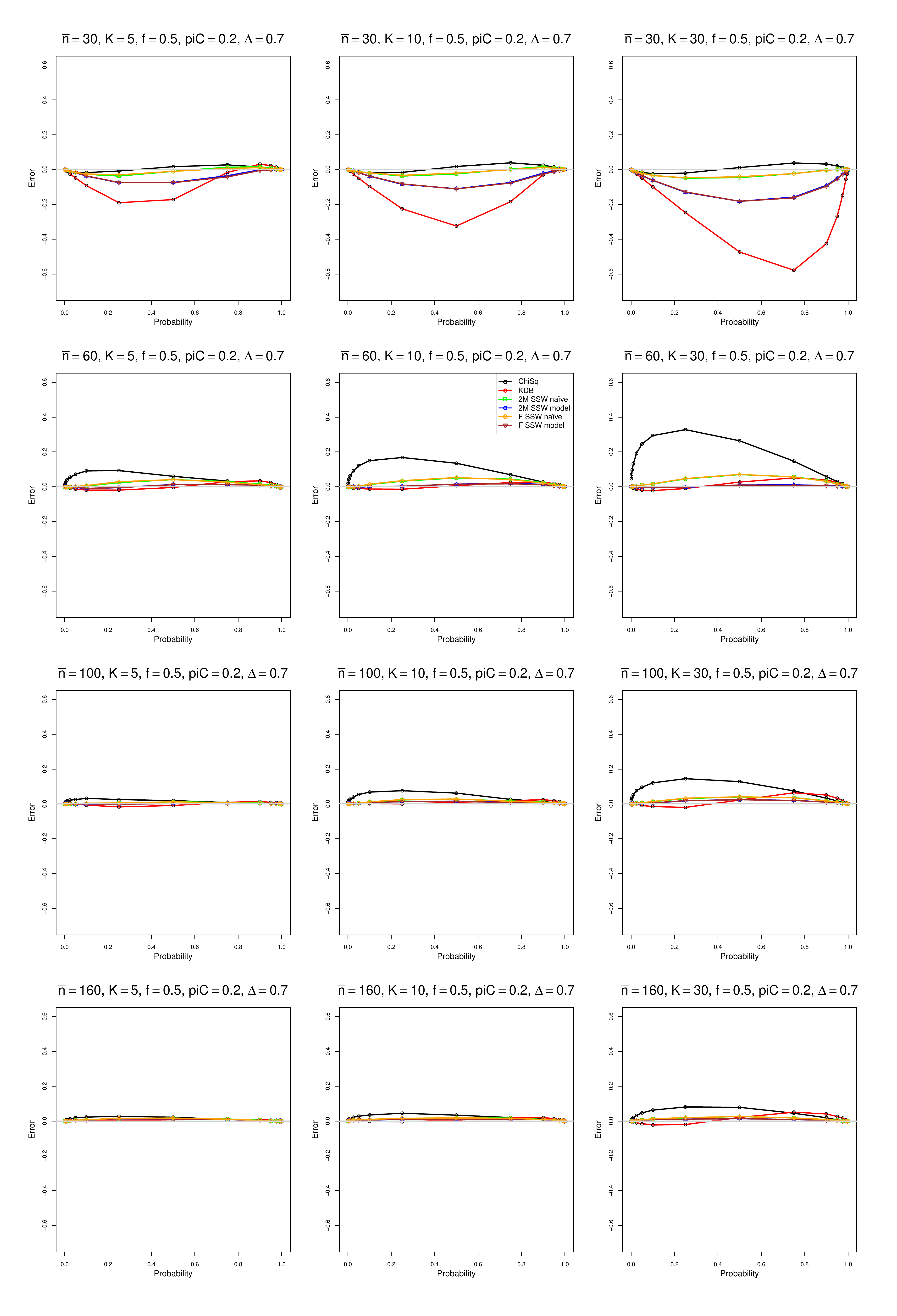}
	\caption{Plots of error in the level of the test for heterogeneity of RD for six approximations for the null distribution of $Q$, $p_{iC} = .2$, $f = .5$, and $\Delta = 0.7$, unequal sample sizes}
	\label{PPplot_piC_02theta=0.7_RD_unequal_sample_sizes}
\end{figure}


\begin{figure}[ht]
	\centering
	\includegraphics[scale=0.33]{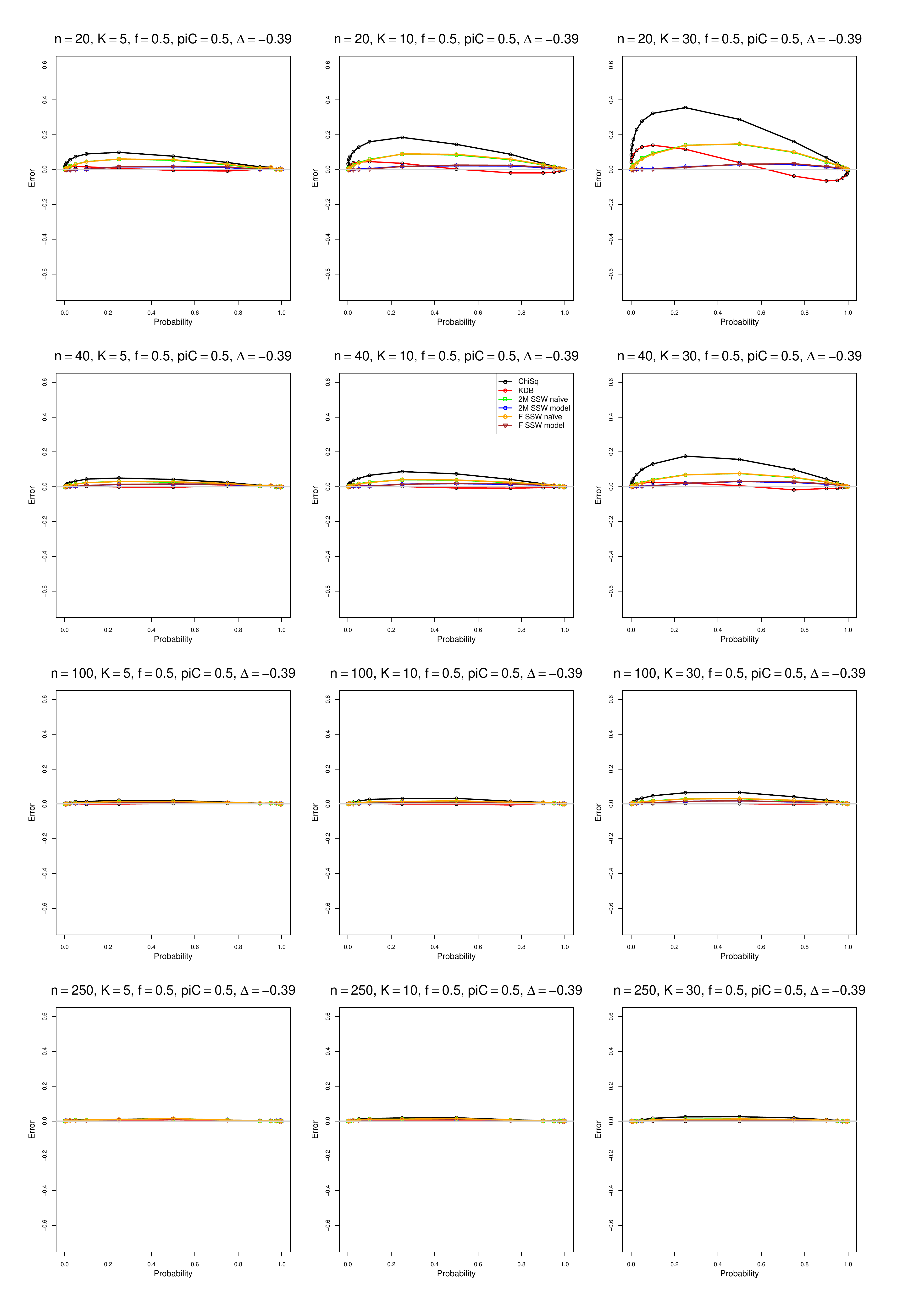}
	\caption{Plots of error in the level of the test for heterogeneity of RD for six approximations for the null distribution of $Q$, $p_{iC} = .5$, $f = .5$, and $\Delta = -0.39$, equal sample sizes}
	\label{PPplot_piC_05theta=-0.39_RD_equal_sample_sizes}
\end{figure}

\begin{figure}[ht]
		\centering
	\includegraphics[scale=0.33]{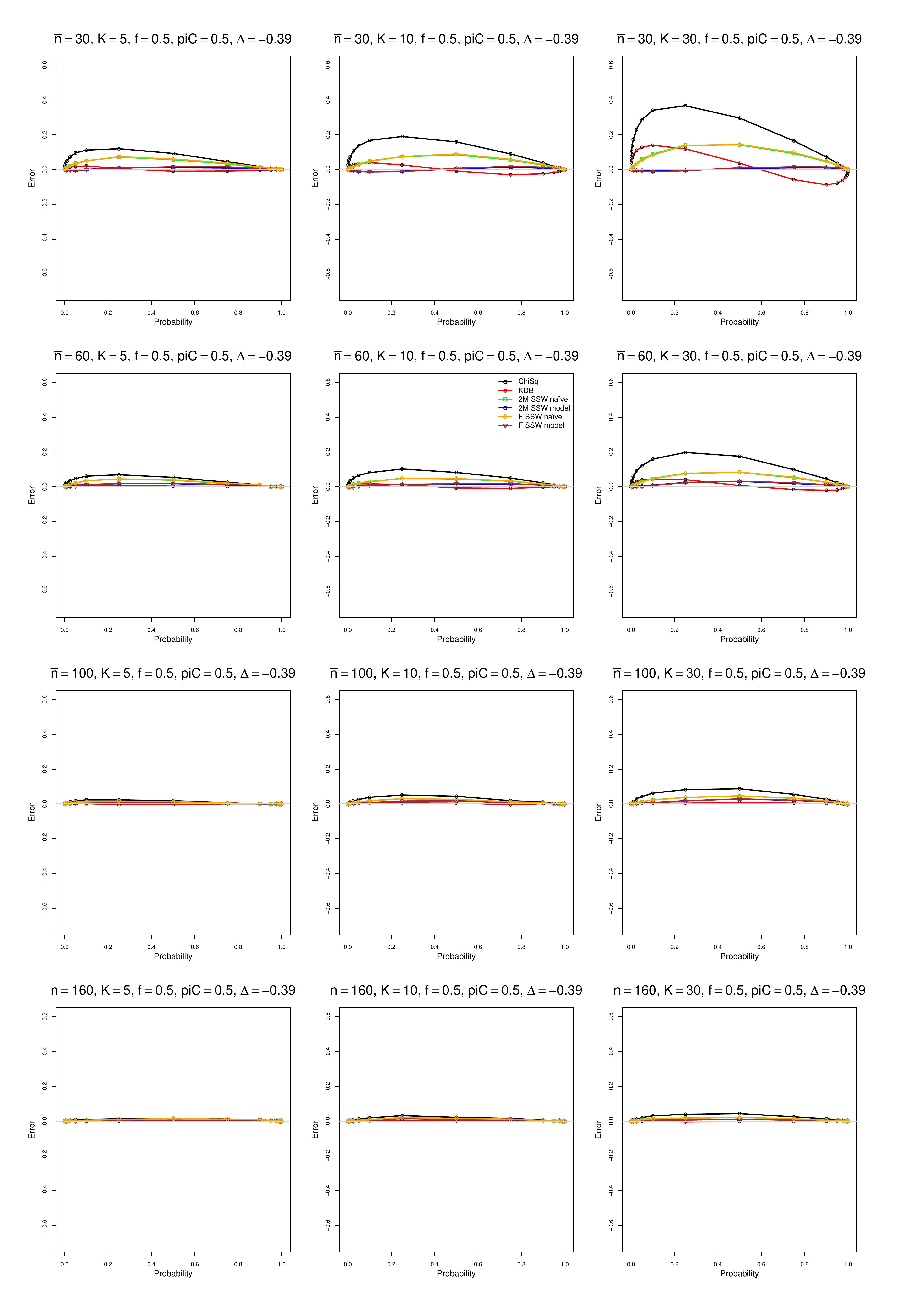}
	\caption{Plots of error in the level of the test for heterogeneity of RD for six approximations for the null distribution of $Q$, $p_{iC} = .5$, $f = .5$, and $\Delta = -0.39$, unequal sample sizes}
	\label{PPplot_piC_05theta=-0.39_RD_unequal_sample_sizes}
\end{figure}

\begin{figure}[ht]
	\centering
	\includegraphics[scale=0.33]{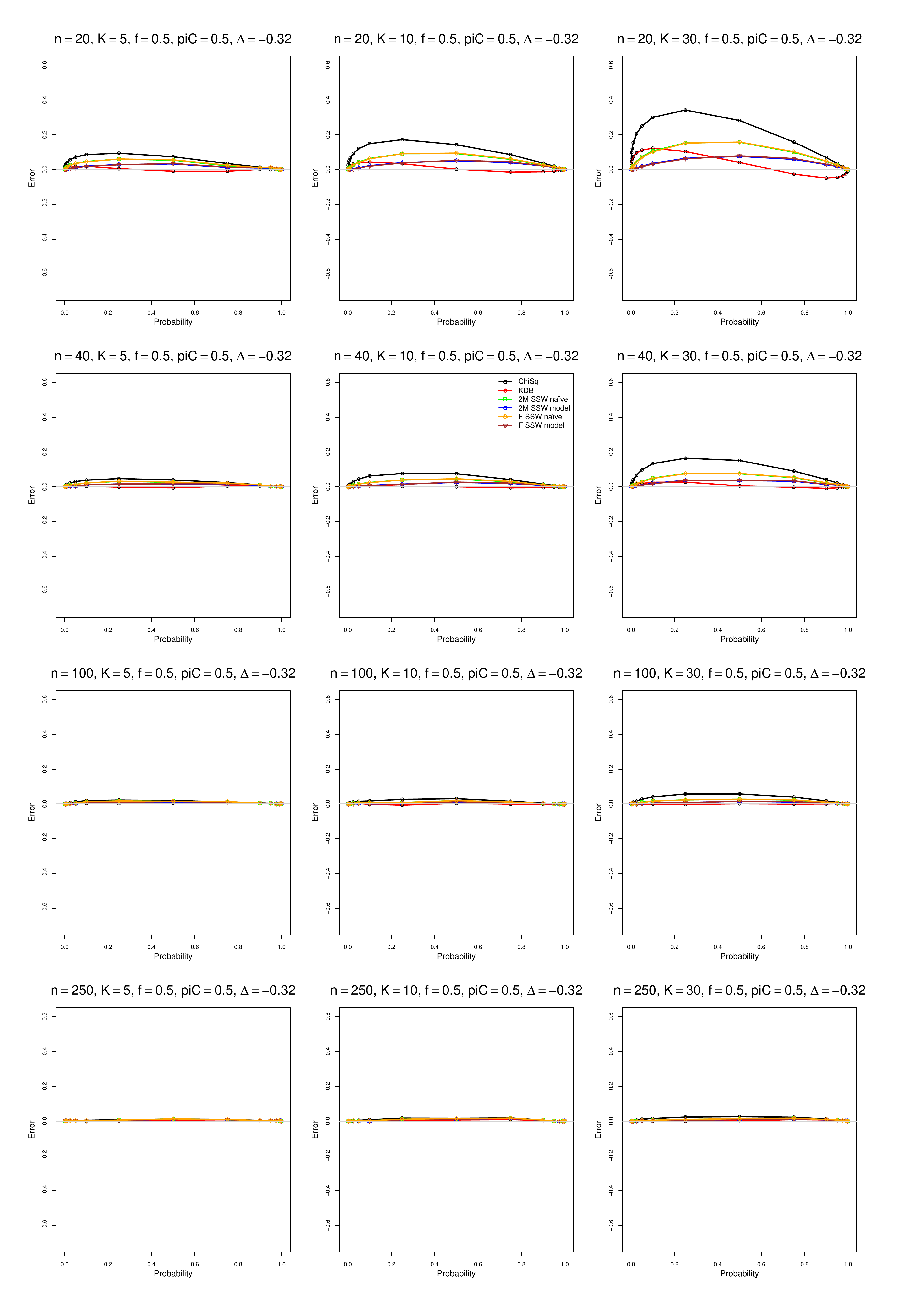}
	\caption{Plots of error in the level of the test for heterogeneity of RD for six approximations for the null distribution of $Q$, $p_{iC} = .5$, $f = .5$, and $\Delta = -0.32$, equal sample sizes}
	\label{PPplot_piC_05theta=-0.32_RD_equal_sample_sizes}
\end{figure}
\begin{figure}[ht]
		\centering
	\includegraphics[scale=0.33]{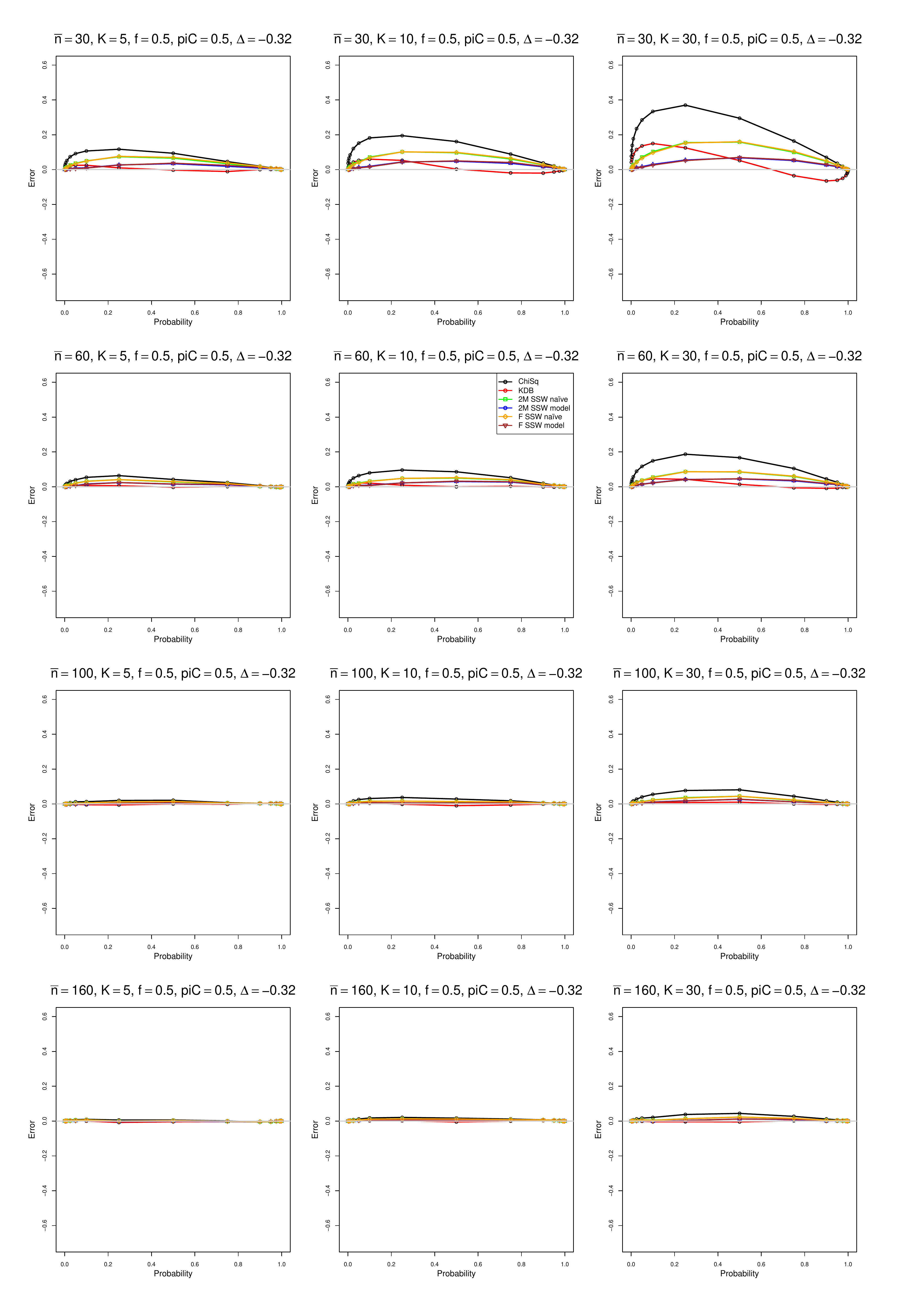}
	\caption{Plots of error in the level of the test for heterogeneity of RD for six approximations for the null distribution of $Q$, $p_{iC} = .5$, $f = .5$, and $\Delta = -0.32$, unequal sample sizes}
	\label{PPplot_piC_05theta=-0.32_RD_unequal_sample_sizes}
\end{figure}

\begin{figure}[ht]
	\centering
	\includegraphics[scale=0.33]{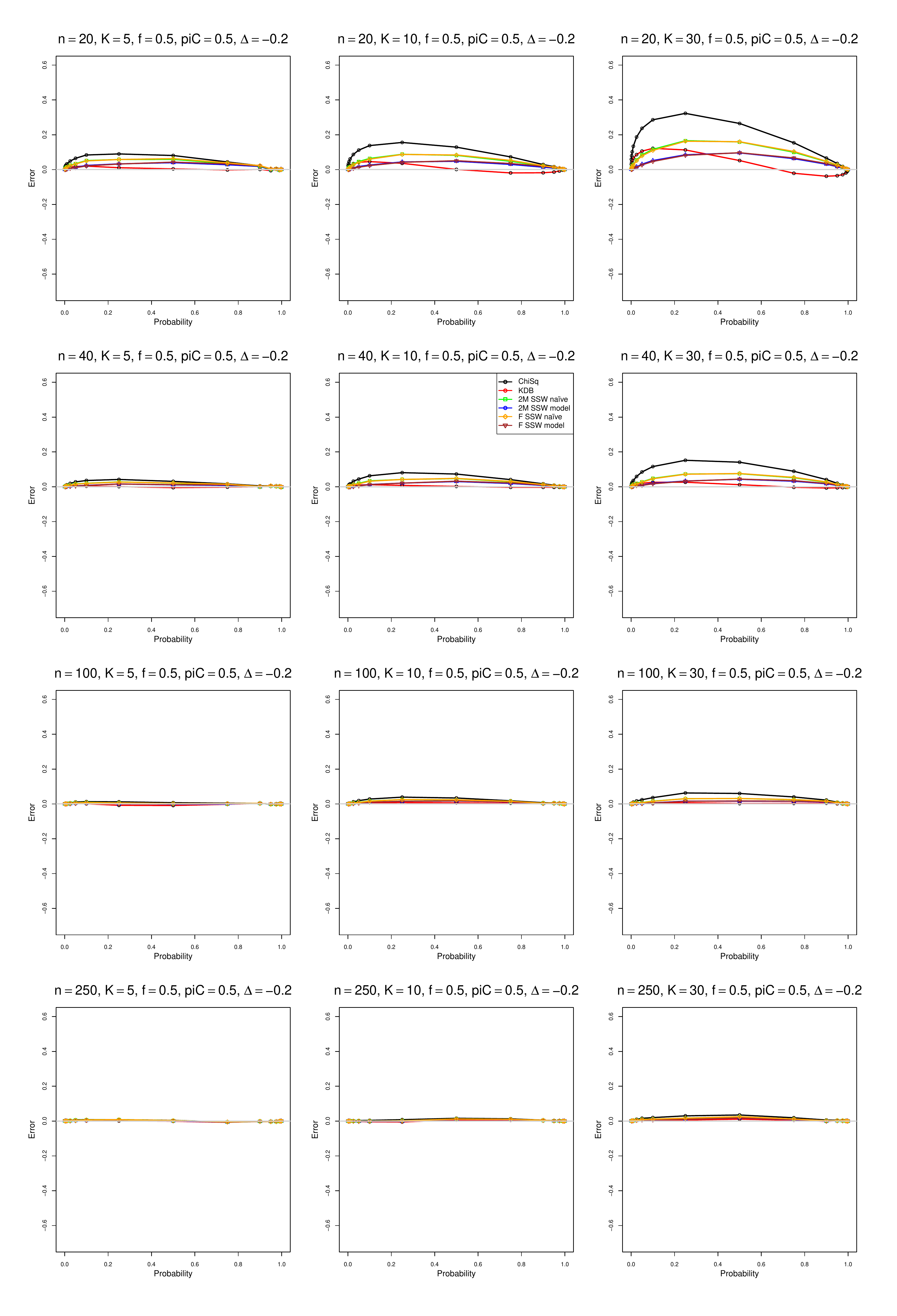}
	\caption{Plots of error in the level of the test for heterogeneity of RD for six approximations for the null distribution of $Q$, $p_{iC} = .5$, $f = .5$, and $\Delta = -0.2$, equal sample sizes}
	\label{PPplot_piC_05theta=-0.2_RD_equal_sample_sizes}
\end{figure}
\begin{figure}[ht]
		\centering
	\includegraphics[scale=0.33]{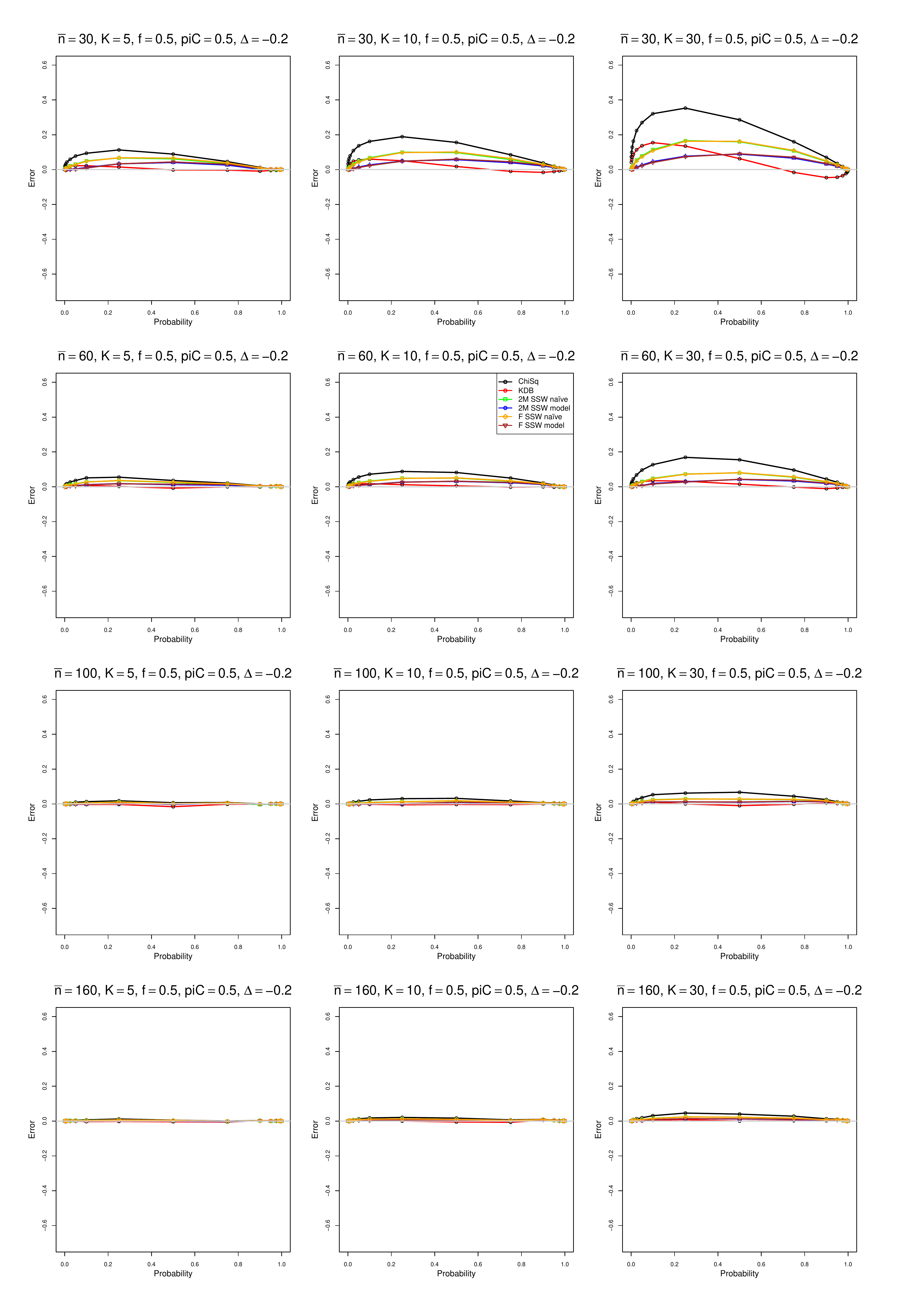}
	\caption{Plots of error in the level of the test for heterogeneity of RD for six approximations for the null distribution of $Q$, $p_{iC} = .5$, $f = .5$, and $\Delta = -0.2$, unequal sample sizes}
	\label{PPplot_piC_05theta=-0.2_RD_unequal_sample_sizes}
\end{figure}

\begin{figure}[ht]
	\centering
	\includegraphics[scale=0.33]{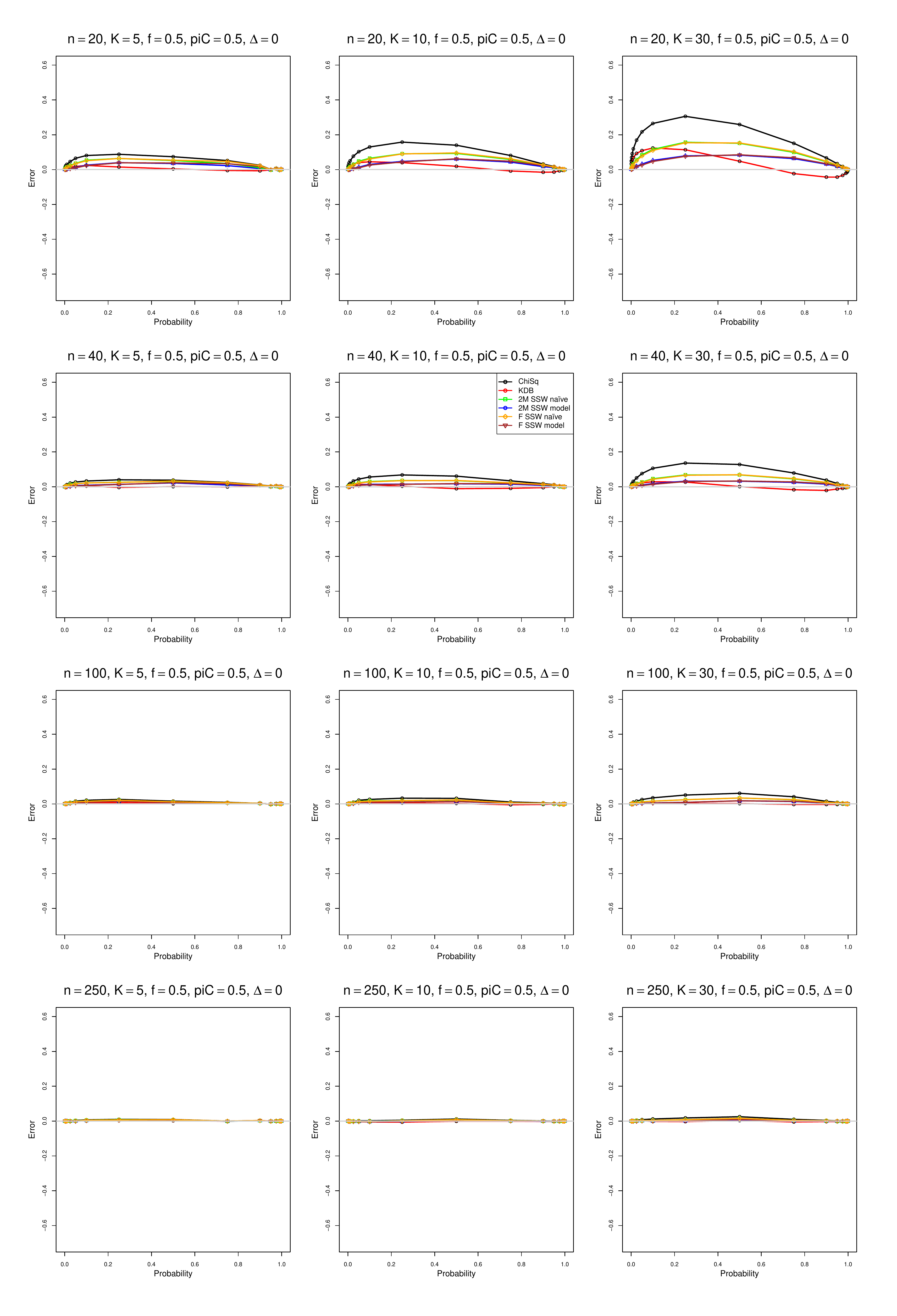}
	\caption{Plots of error in the level of the test for heterogeneity of RD for six approximations for the null distribution of $Q$, $p_{iC} = .5$, $f = .5$, and $\Delta = 0$, equal sample sizes}
	\label{PPplot_piC_05theta=0_RD_equal_sample_sizes}
\end{figure}

\begin{figure}[ht]
		\centering
	\includegraphics[scale=0.33]{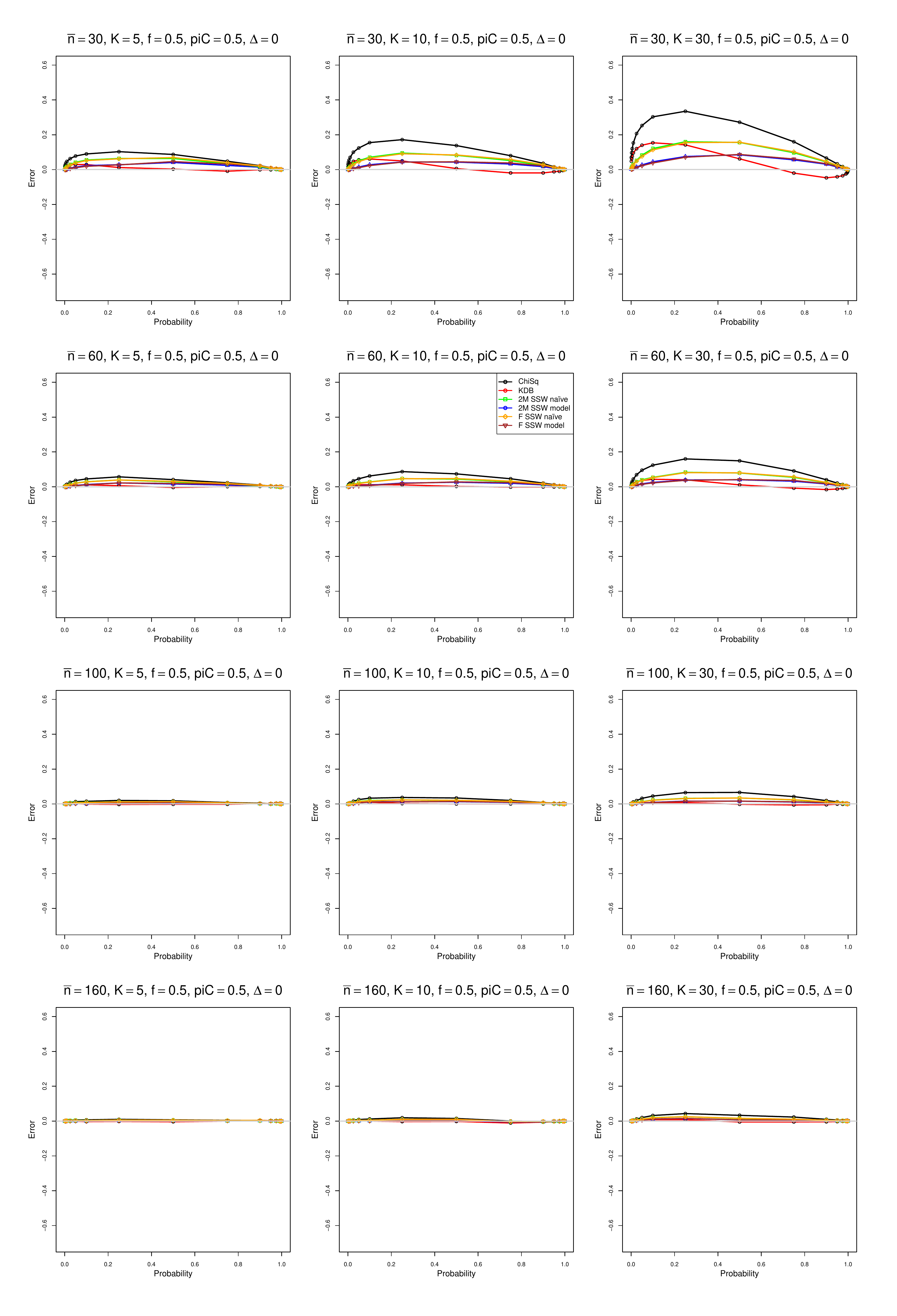}
	\caption{Plots of error in the level of the test for heterogeneity of RD for six approximations for the null distribution of $Q$, $p_{iC} = .5$, $f = .5$, and $\Delta = 0$, unequal sample sizes}
	\label{PPplot_piC_05theta=0_RD_unequal_sample_sizes}
\end{figure}

\begin{figure}[ht]
	\centering
	\includegraphics[scale=0.33]{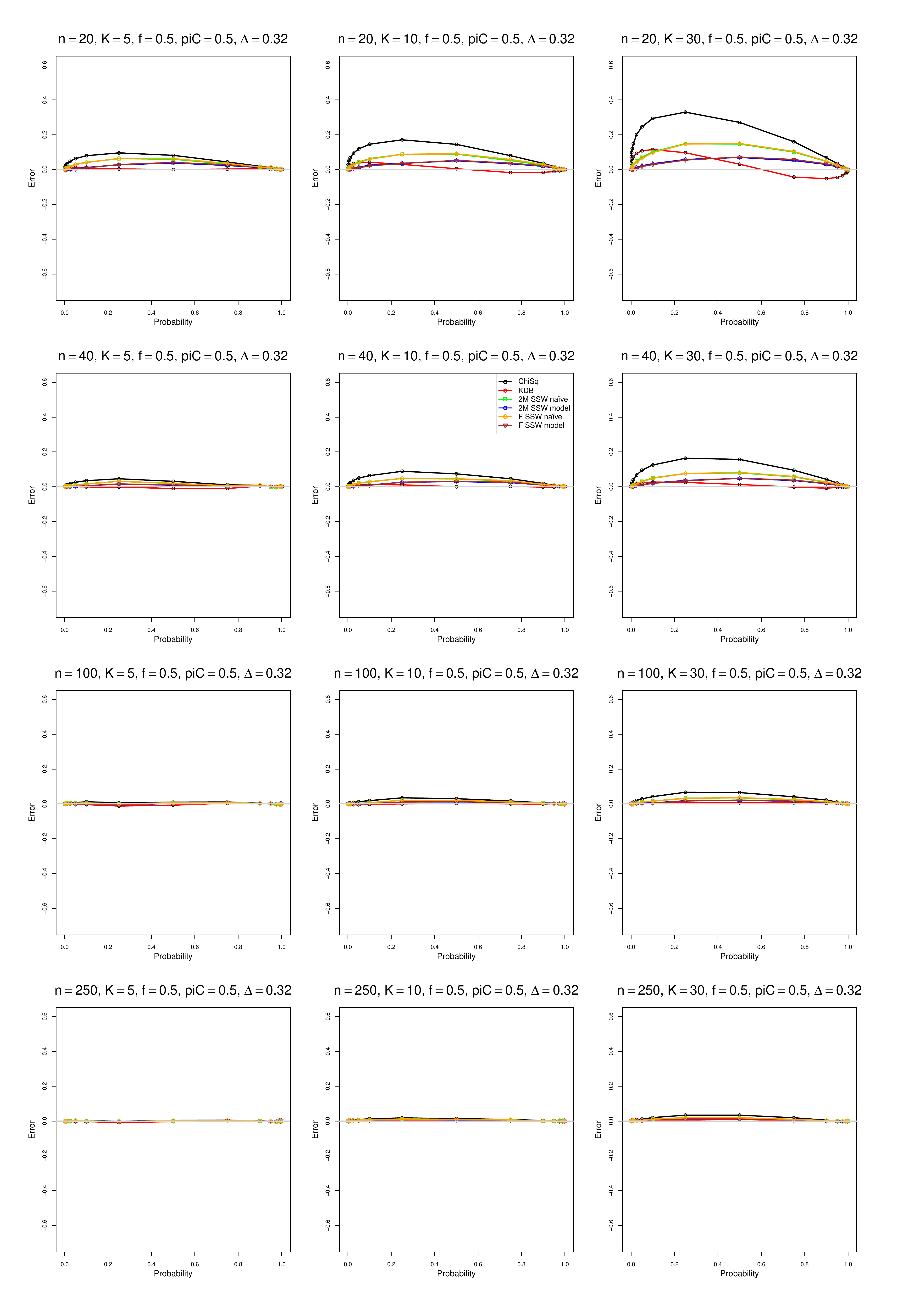}
	\caption{Plots of error in the level of the test for heterogeneity of RD for six approximations for the null distribution of $Q$, $p_{iC} = .5$, $f = .5$, and $\Delta = 0.32$, equal sample sizes}
	\label{PPplot_piC_05theta=0.32_RD_equal_sample_sizes}
\end{figure}
\begin{figure}[ht]
		\centering
	\includegraphics[scale=0.33]{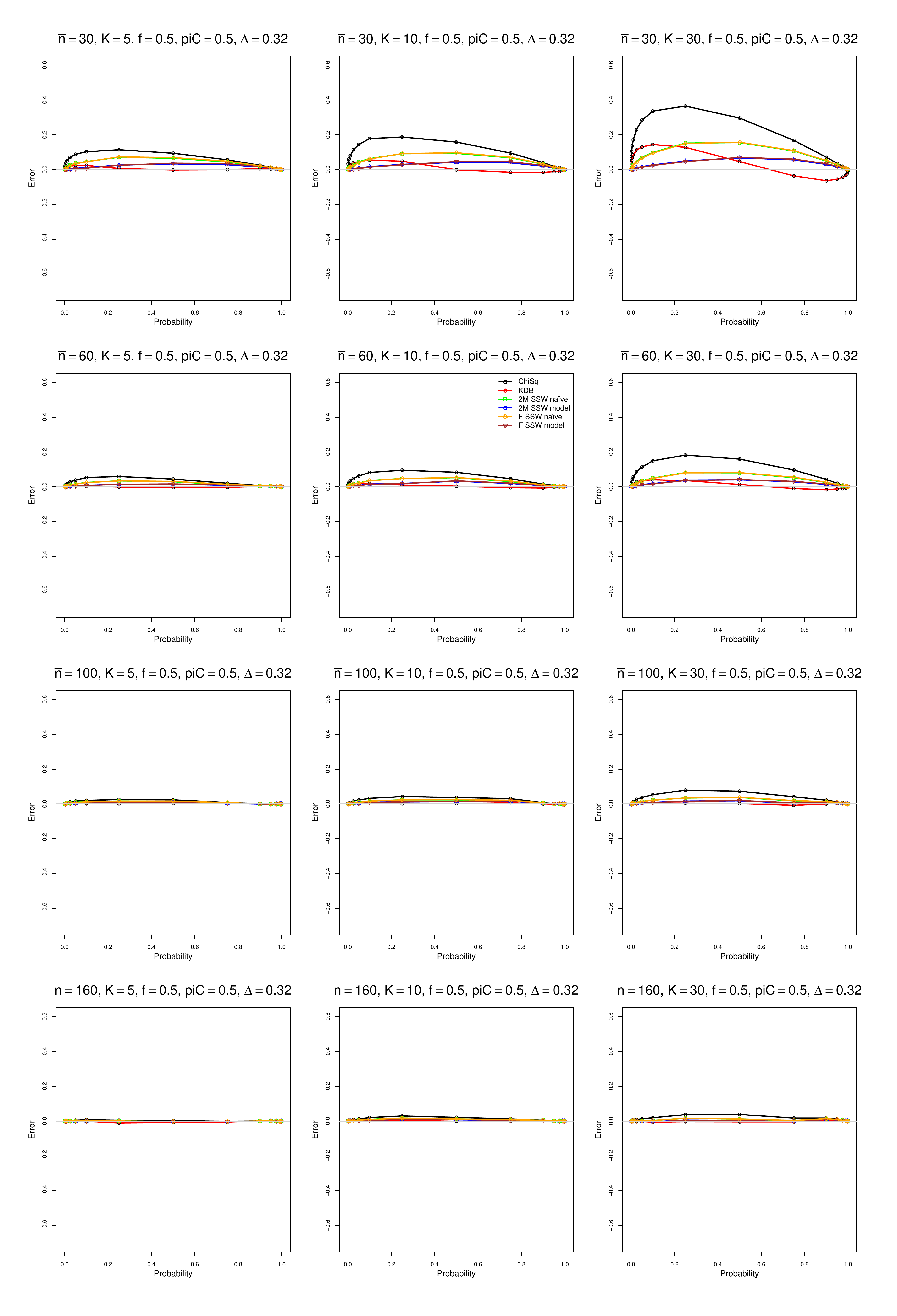}
	\caption{Plots of error in the level of the test for heterogeneity of RD for six approximations for the null distribution of $Q$, $p_{iC} = .5$, $f = .5$, and $\Delta = 0.32$, unequal sample sizes}
	\label{PPplot_piC_05theta=0.32_RD_unequal_sample_sizes}
\end{figure}

\clearpage
\section*{Appendix G: Empirical level at $\alpha = .05$, vs $\Delta$, of the test for heterogeneity of RD ($\tau^2 = 0$ versus $\tau^2 > 0$) based on approximations for the null distribution of $Q$}

Each figure corresponds to a value of the probability of an event in the Control arm $p_{iC}$  (= .1, .2, .5) and a choice of equal or unequal sample sizes ($n$ or $bar{n}$). \\
The fraction of each study's sample size in the Control arm $f$ is held constant at 0.5.

For each combination of a value of $n$ (= 20, 40, 100, 250) or $\bar{n}$ (= 30, 60, 100, 160) and a value of $K$ (= 5, 10, 30), a panel plots the empiical level versus $\Delta$.   To facilitate comparisons, the pairs ($p_{iC},\; p_{iT}$) for RD are the same as for LRR. The values of $p_{iT}$ are (.06, .10, .16, .27, .44) when $p_{iC} = .1$,
(.12, .20, .33, .54, .90) when  $p_{iC} = .2$, and (.12, .18, .30, .50, .82) when  $p_{iC} = .5$.  The  values of $\Delta$ are ($-.04$, 0, .06, .17, .34), ($-.08$, 0, .13, .34, .70), and ($-.38$, $-.32$, $-.20$, 0, .32), respectively.\\
The approximations for the distribution of $Q$ are
\begin{itemize}
\item ChiSq (Chi-square approximation with $K-1$ df, inverse-variance weights)
\item KDB (Kulinskaya-Dollinger-Bj{\o}rkest{\o}l (2011) approximation, inverse-variance weights)
\item 2M SSW na\"{i}ve (Two-moment gamma approximation, na\"{i}ve estimation of $p_{iT}$ from $X_{iT}$ and $n_{iT}$, effective-sample-size weights)
\item 2M SSW model (Two-moment gamma approximation, model-based estimation of $p_{iT}$, effective-sample-size weights)
\item F SSW na\"{i}ve (Farebrother approximation, na\"{i}ve estimation of $p_{iT}$ from $X_{iT}$ and $n_{iT}$, effective-sample-size weights)
\item F SSW model (Farebrother approximation, model-based estimation of $p_{iT}$, effective-sample-size weights)
\end{itemize}

\clearpage
\setcounter{figure}{0}
\renewcommand{\thefigure}{G.\arabic{figure}}

\begin{figure}[t]
	\centering
	\includegraphics[scale=0.33]{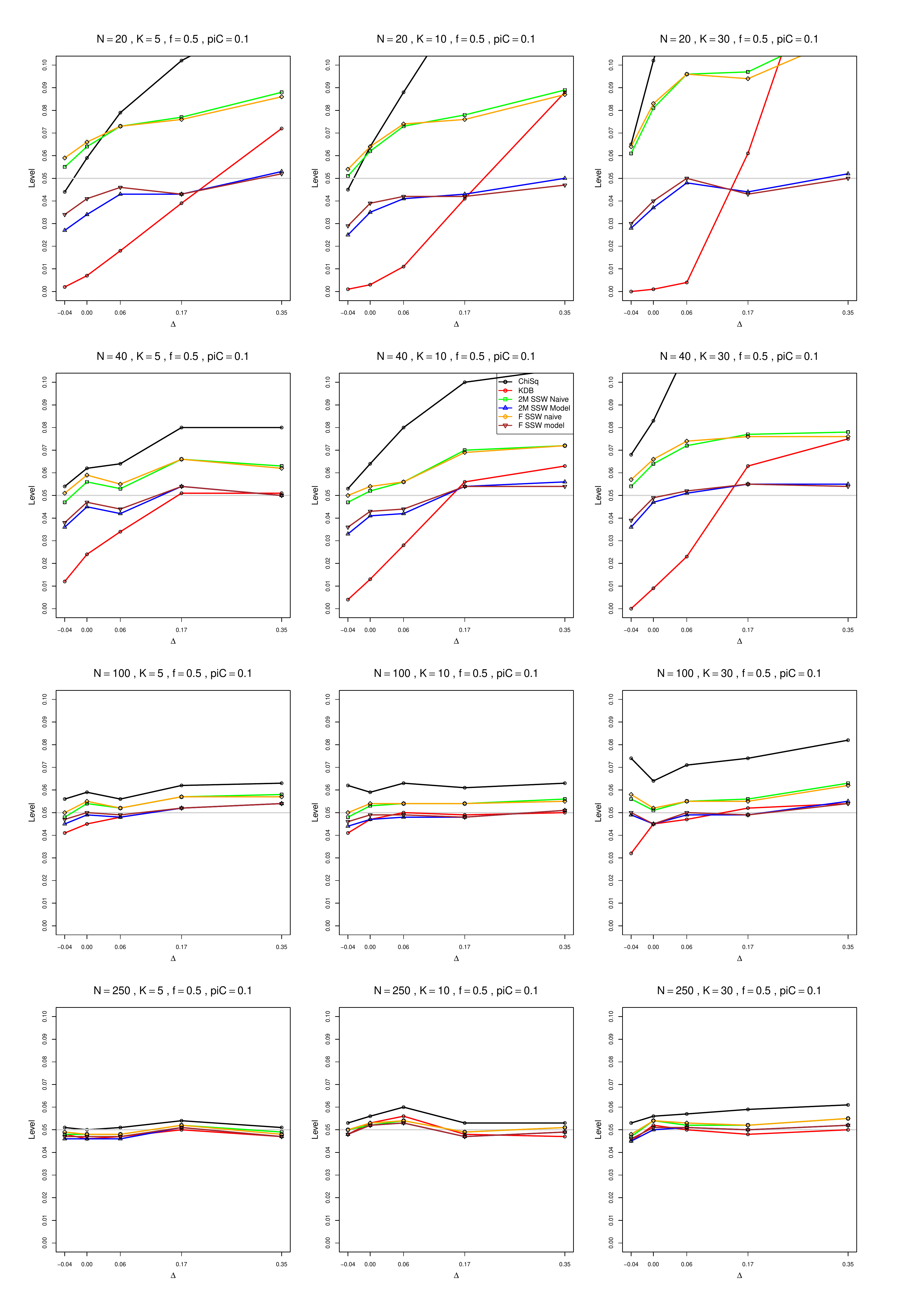}
	\caption{Q for RD: actual level at $\alpha = .05$ for $p_{iC} = .1$ and $f = .5$, equal sample sizes
		\label{NewQforRD_piC01andq05_equal_sample_sizes}}
\end{figure}

\begin{figure}[t]
	\centering
	\includegraphics[scale=0.33]{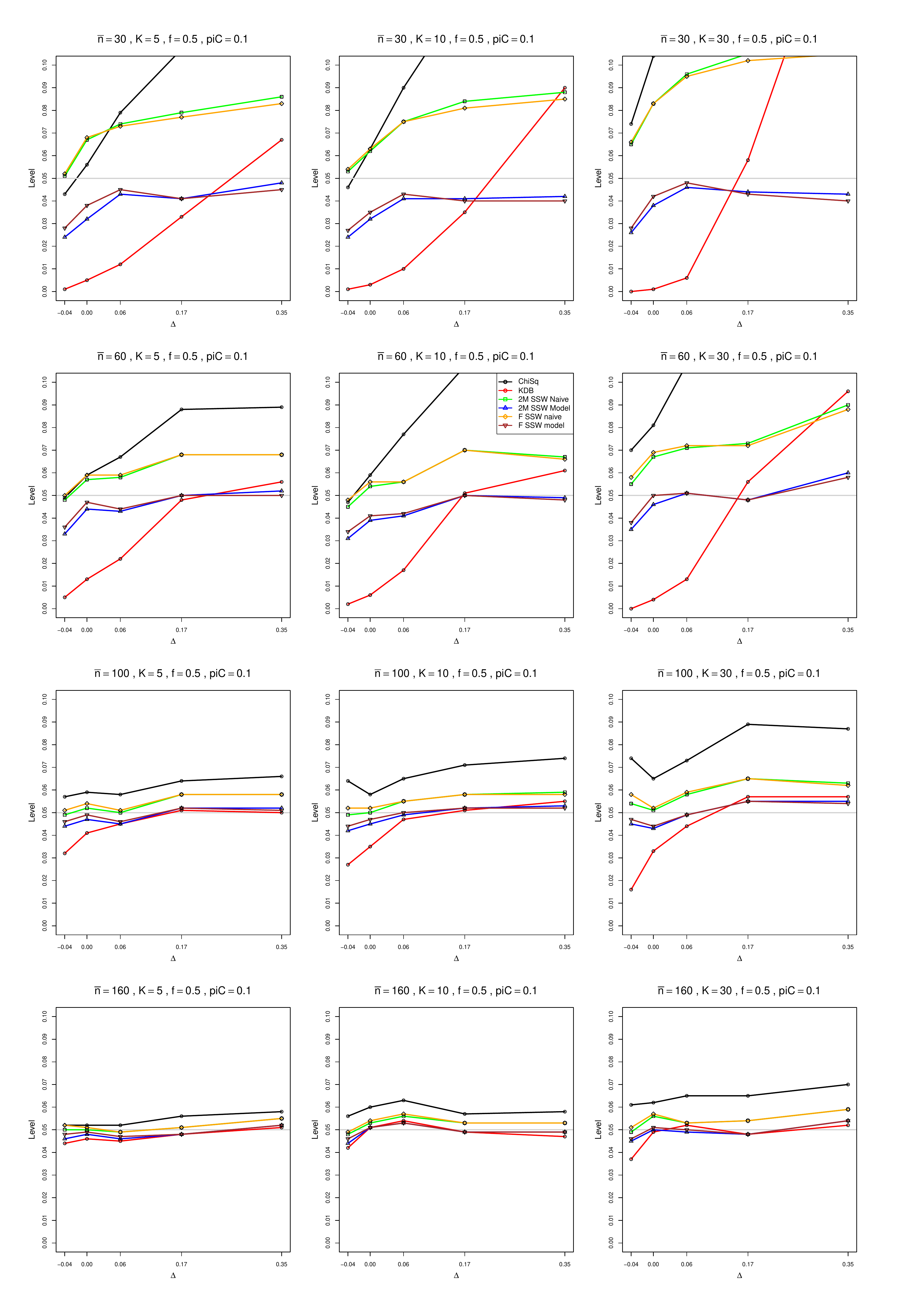}
	\caption{Q for RD: actual level at $\alpha = .05$ for $p_{iC} = .1$ and $f = .5$, unqual sample sizes
		\label{NewQforRD_piC01andq05_unequal_sample_sizes}}
\end{figure}

\begin{figure}[t]
	\centering
	\includegraphics[scale=0.33]{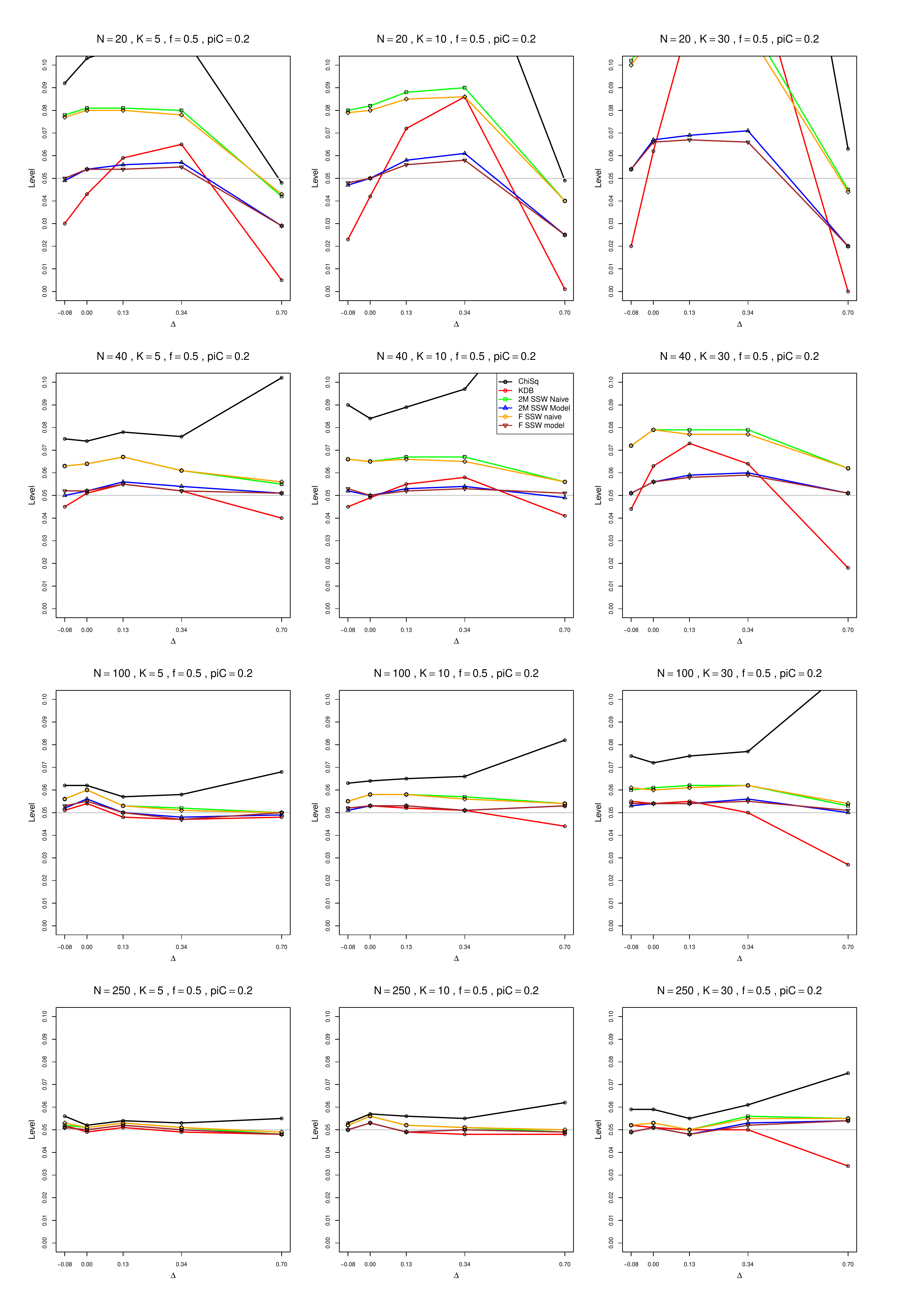}
	\caption{Q for RD: actual level at $\alpha = .05$ for $p_{iC }= .2$ and $f = .5$, equal sample sizes
		\label{NewQforRD_piC02andq05_equal_sample_sizes}}
\end{figure}

\begin{figure}[t]
	\centering
	\includegraphics[scale=0.33]{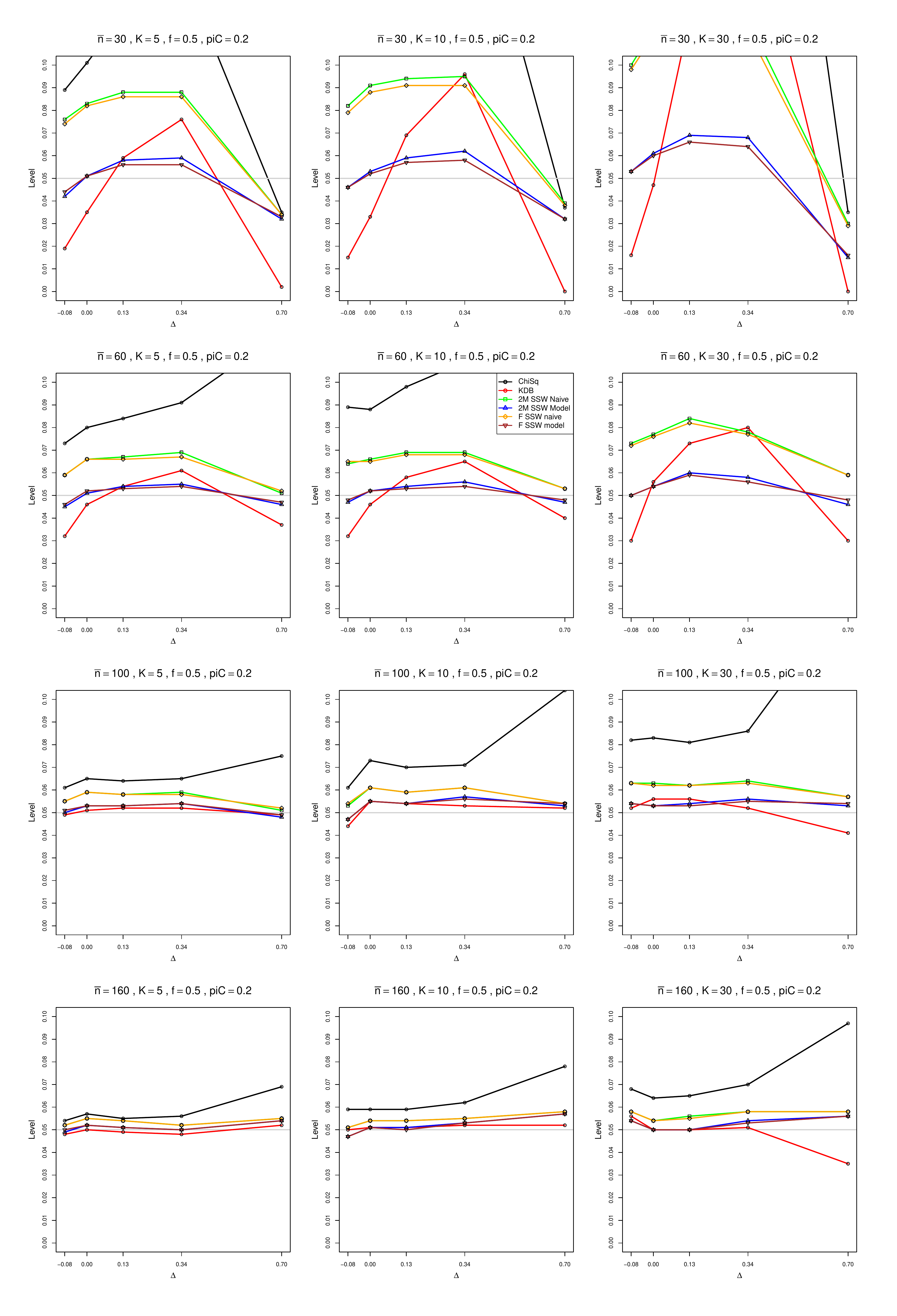}
	\caption{Q for RD: actual level at $\alpha = .05$ for $p_{iC} = .2$ and $f = .5$. unequal sample sizes
		\label{pNewQforRR_piC02andq05_unequal_sample_sizes}}
\end{figure}

\begin{figure}[t]
	\centering
	\includegraphics[scale=0.33]{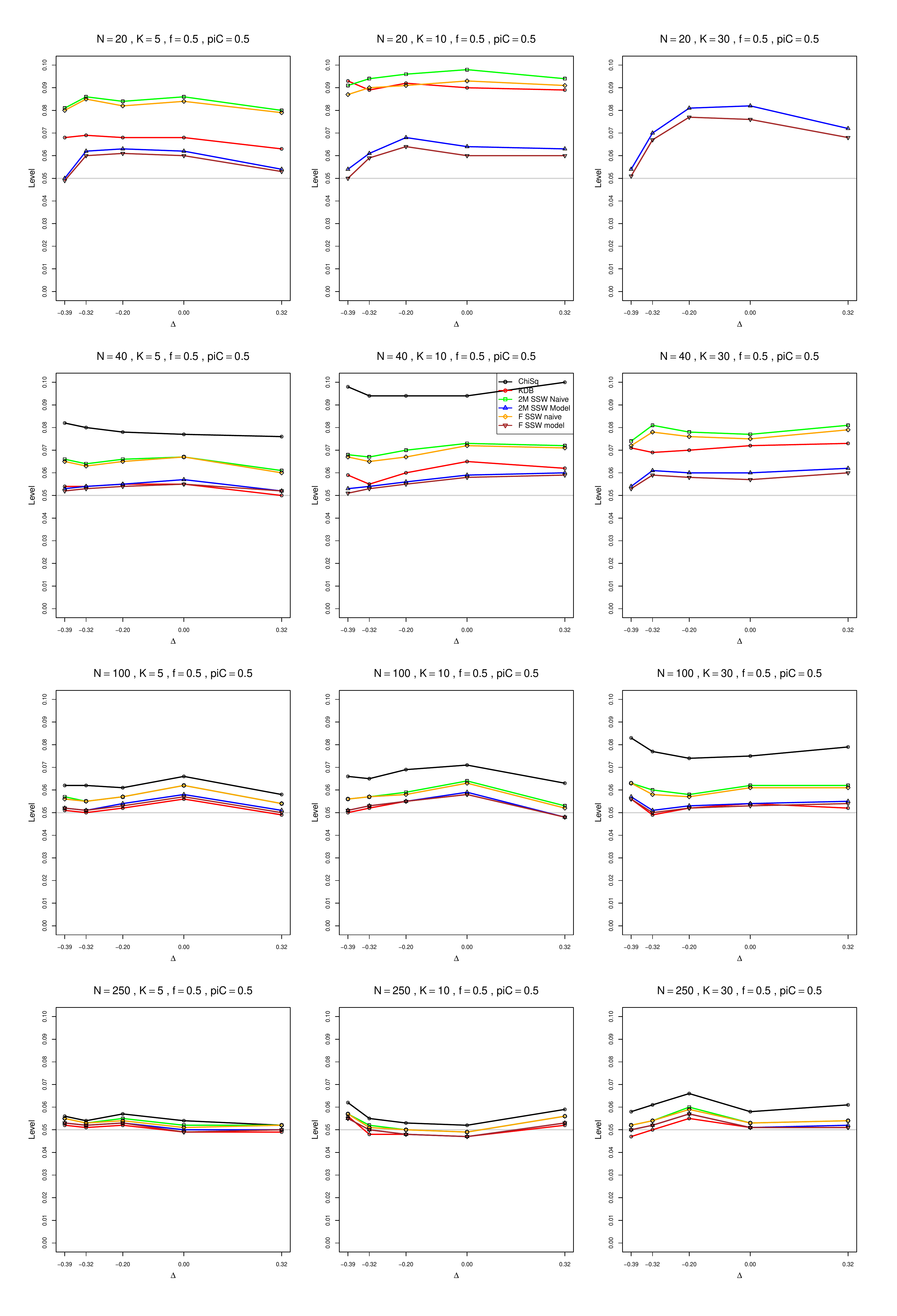}
	\caption{Q for RD: actual level at $\alpha = .05$ for $p_{iC} = .5$ and $f = .5$, equal sample sizes
		\label{NewQforRD_piC05_equal_sample_sizes}}
\end{figure}

\begin{figure}[t]
	\centering
	\includegraphics[scale=0.33]{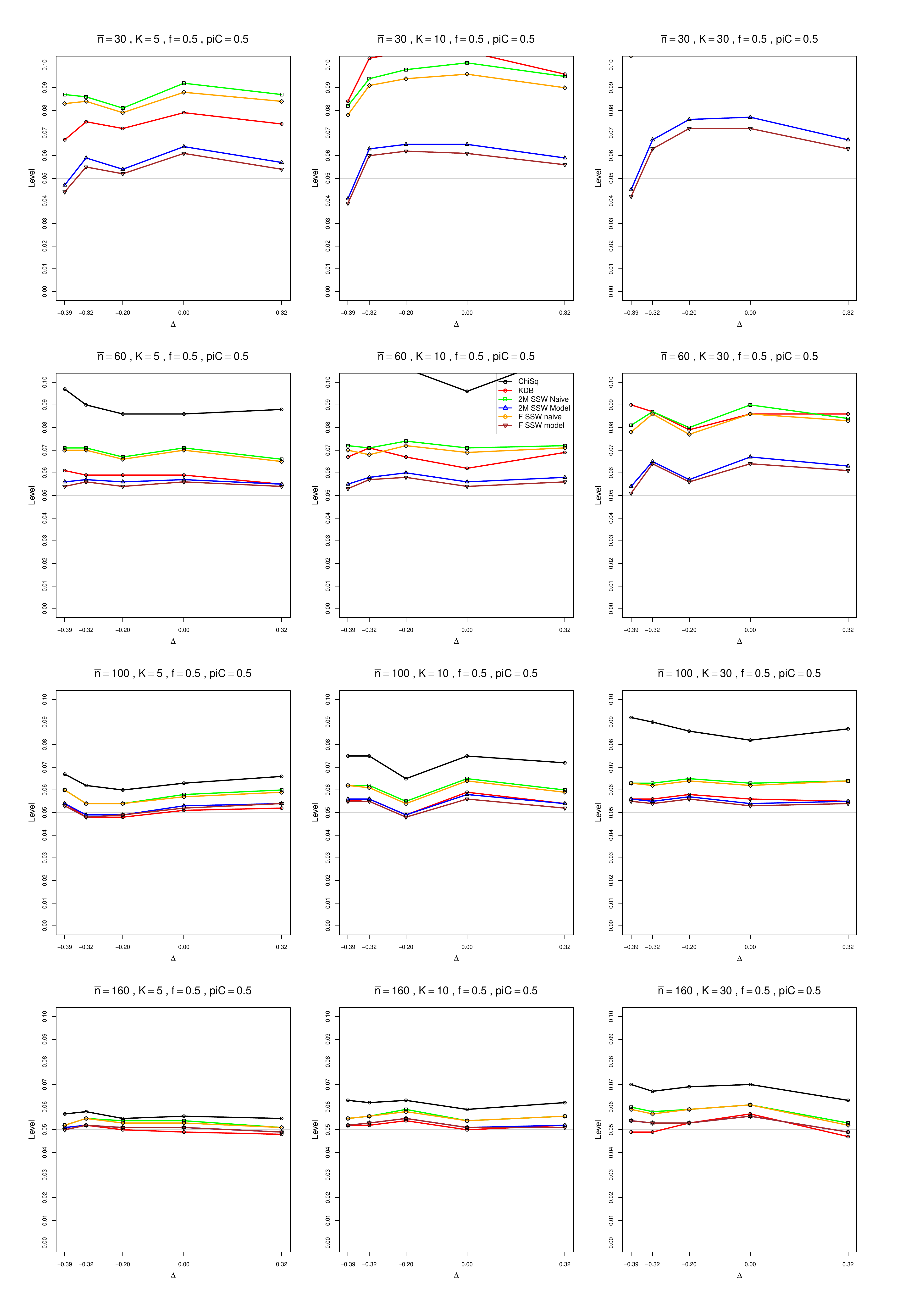}
	\caption{Q for RD: actual level at $\alpha = .05$ for $p_{iC }= .5$ and $f = .5$, unequal sample sizes
		\label{NewQforRD_piC05andq05_unequal_sample_sizes}}
\end{figure}


\end{document}